\documentclass[prd,preprint]{revtex4-1}%
\usepackage[dvips]{graphicx}
\usepackage{color}
\usepackage[colorlinks=true, pdfstartview=FitV, linkcolor=red,
citecolor=blue,urlcolor=blue]{hyperref}
\usepackage{appendix}
\usepackage{float,amsmath,amstext,amsthm,amssymb,amsfonts,slashed,epigraph}
\usepackage[nouppercase]{scrpage2}
\usepackage{makeidx}
\usepackage{caption}
\usepackage{subcaption}
\usepackage{bm}
\usepackage{enumerate}
\usepackage[latin1]{inputenc}
\usepackage{changepage}
\usepackage{float}
\usepackage{amsmath}
\usepackage{amsfonts}
\usepackage{amssymb}%
\newcommand{\ee}{\end{equation}}
\newcommand{\beq}{\begin{equation}}
\newcommand{\eeq}{\end{equation}}
\newcommand{\bea}{\begin{eqnarray}}
\newcommand{\eea}{\end{eqnarray}}

\graphicspath{{}}
\begin{document}

\title{Matrix model for deconfinement in a $SU(N_{c})$ gauge theory in $2+1$ dimensions}

\author{P. Bicudo}
\affiliation{CFTP, Dep. F\'{\i}sica, 
Instituto Superior T\'ecnico, Universidade T\'ecnica de Lisboa, 
Av. Rovisco Pais, 1049-001 Lisboa, Portugal}

\author{Robert D. Pisarski}
\affiliation{Department of Physics, 
Brookhaven National Laboratory, Upton, NY 11973 \\
and RIKEN/BNL, Brookhaven National Laboratory, Upton, NY 11973}

\author{E. Seel}
\affiliation{Institute of Theoretical Physics, 
J. W. Goethe University, Max-von-Laue Str. 1, D-60438, 
Frankfurt am Main, Germany}

\begin{abstract}
We use an effective matrix model to study deconfinement in a pure $SU(N_{c})$
gauge theory, without quarks, in $d=2+1$ dimensions. Expanding about a
constant background $\mathcal{A}_{0}$ field we construct an effective
potential for the eigenvalues of the thermal Wilson line for general $N_{c}$
and in the large-$N_{c}$ limit. The numerical results are presented using one,
two, and four free parameters, which are determined by fitting directly to the
lattice data for the pressure. The matrix model shows a good agreement with
numerical lattice simulations for the pressure and the interaction measure,
starting from the perturbative limit up to the critical temperature. For the
pressure, the details of $\mathcal{A}_{0}$-dependent nonperturbative terms are
relevant only in a narrow transition region below $\sim1.2\,T_{d}$. This is
also the range where the Polyakov loop deviates notably from one. In
accordance with the lattice results we find that, up to a trivial factor
$N_{c}^{2}-1$, there is only a mild dependence on the number of colors.

\end{abstract}
\maketitle


\section{Motivation \label{intro}}

Nowadays increasing attention is being devoted to the study of the
deconfinement phase transition in QCD. There are different, complementary
approaches to the underlying physics, which is strictly nonperturbative:
numerical simulations on the lattice, the construction of various effective
theories, and lastly, results from the collisions of heavy ions at
ultrarelativistic energies. In this work we investigate the key aspects of
deconfinement by using an effective matrix model for three-dimensional pure
$SU(N_{c})$ gauge theories.

The matrix model respects the $Z(N_{c})$ center symmetry of the underlying
pure glue theory. It is valid over a wide temperature range, starting from the
perturbative limit up to the critical temperature for deconfinement $T_{d}$.
The degrees of freedom are given by the eigenvalues of the thermal Wilson
line, and the parameters of the model are obtained by fitting the pressure as
computed within lattice QCD in Ref. \cite{Caselle:2011mn}.

It is known {that in} pure gauge theory the Polyakov loop approaches unity in
the perturbative limit, and vanishes below the deconfining temperature $T_{d}%
$. This behavior can be modeled by constructing an effective theory for the
eigenvalues of the thermal Wilson line. Assuming a constant background
$\mathcal{A}_{0}$ field for the timelike component of the gluonic vector
potential, one first computes the perturbative potential. In order to drive
the transition to confinement, one then constructs additional $\mathcal{A}%
_{0}$-dependent and -independent nonperturbative contributions. This is a
reasonable approach provided that the expectation value of the Polyakov loop
is dominated by the classical configuration for the background $\mathcal{A}%
_{0}$ field, and not by quantum fluctuations around $\mathcal{A}_{0}$. This
assumption is certainly valid for an infinite number of colors, where
$\mathcal{A}_{0}$ represents a master field for deconfinement
\cite{Gopakumar:1994iq}. Namely, in the large-$N_{c}$ limit, the vacuum is
dominated by a single master field at any temperature\textbf{, }since quantum
fluctuations are typically suppressed by powers of $1/N_{c}^{2}$.
Nevertheless, we find that this approach works reasonably well even for two
colors, both in $d=3+1$ \cite{Dumitru:2010mj, Dumitru:2012fw}, and in $d=2+1$,
see Ref. \cite{Bicudo:2013yza}.

A more complete model for deconfinement should ideally consider $SU(N_{c})$
gauge theories coupled to light quarks in $d=3+1$ space-time dimensions.
However, lattice results provide convincing evidence that the QCD transition
is mainly driven by the dynamics of gluons which are the dominant degrees of
freedom in the deconfined phase \cite{Karsch:2001cy}. Moreover, $SU(N_{c})$
gauge theories in three and four space-time dimensions are closely related and
share many important features, like asymptotic freedom, and a
confinement-deconfinement phase transition at a critical temperature $T_{d}$
\cite{Teper:1998te, Johnson:2000qz, Meyer:2003wx}. For these reasons,
three-dimensional pure glue theories are widely used both on the lattice and
in effective theories, in order to obtain a better understanding of the QCD
transition from a broader perspective, see e.g. Refs.\ \cite{Teper:1998te,
Johnson:2000qz, Lucini:2002wg, Meyer:2003wx, Bursa:2005tk, Bialas:2008rk,
Liddle:2008kk, Caselle:2011fy, Bialas:2012qz}.

An important question is the dependence of thermodynamic properties on the
number of colors. In this sense, the large-$N_{c}$ limit is of particular
importance, since it serves as a suitable basis for various effective
approaches to QCD, see e.g. Ref. \cite{Moshe:2003xn} and refs. therein for a
detailed discussion. Matrix models, for instance, are motivated by general
expectations that in the large-$N_{c}$ limit all correlation functions of
gauge-invariant operators factorize, and consequently the functional integral
is dominated by a single master gauge field \cite{Gopakumar:1994iq}.

Recent lattice simulations of pure gauge theories at nonzero temperature
indicate that, except for a trivial proportionality to the number of gluons
$N_{c}^{2}-1$, the thermodynamic observables in $d=3+1$ are essentially
independent of $N_{c}$, and that $SU\left(  3\right)  _{c}$ is already
sufficiently close to the large-$N_{c}$ limit \cite{Boyd:1996bx,
Lucini:2001ej, Teper:2002kh, Umeda:2008bd, Borsanyi:2011zm, Borsanyi:2012vn,
Lucini:2012gg}. This observation is of particular importance, because it
supports the validity of analytical techniques and effective theories based on
large-$N_{c}$ approximations.

In $d=3+1$, the matrix model has been extensively studied for various
$SU(N_{c})$ groups and also in the large-$N_{c}$ limit \cite{Pisarski:2000eq,
Meisinger:2001cq, Meisinger:2001fi, Dumitru:2003hp, Oswald:2005vr,
Pisarski:2006hz, Pisarski:2006yk, Hidaka:2008dr, Hidaka:2009hs, Hidaka:2009xh,
Hidaka:2009ma, Dumitru:2010mj, Dumitru:2012fw, Pisarski:2012bj,
Kashiwa:2012wa, Sasaki:2012bi, Ruggieri:2012ny, Diakonov:2012dx,
Ogilvie:2012is, Kashiwa:2013rm, Lin:2013qu}, showing that one- and two-
parameter models provide an overall good agreement with the lattice data for
the pressure and interaction measure. In Ref. \cite{Bicudo:2013yza} we applied
the matrix model to study deconfinement in three-dimensional pure gauge theory
for the special case $SU(N_{c}=2)$. As in four dimensions, we find that the
model works reasonably well even for two colors. Similar to $d=3+1$, we
demonstrated that in three-dimensional $SU(2)$ theory, for the pressure the
details of the matrix model become relevant only in a narrow transition
region, from $T_{d}$ to $\sim1.2\,T_{d}$. This is also the range where the
Polyakov loop notably deviates from one. The 't Hooft loop, on the other
contrary, is sensitive to the details of the model in a much wider region, up
to $4.0\,T_{d}$.

In this work we extend the study of Ref. \cite{Bicudo:2013yza} to general
$SU(N_{c})$ groups in $d=2+1$. The thermodynamics of three-dimensional pure
glue theories was studied on the lattice by many authors, see e.g. Refs.
\cite{Bialas:2000ev, Bialas:2004gx, Bialas:2008rk, Liddle:2008kk,
Bialas:2009pt, Caselle:2011fy, Caselle:2011mn, Bialas:2012qz, Lucini:2012gg}.
An important observation is that in $d=2+1$ the behavior of thermodynamical
quantities, like pressure $p\left(  T\right)  $, energy density $\epsilon
\left(  T\right)  $, and interaction measure $\Delta\left(  T\right)
=\epsilon\left(  T\right)  -2p\left(  T\right)  $, looks similar to that in
four dimensions. As in $d=3+1$, lattice results show that in three dimensions
there is only a small dependence on the number of colors \cite{Lucini:2002wg}.

Remarkably, in three dimensions, the value for the interaction measure scaled
by $T^{2}T_{d}$ and divided by the number of gluons, $N_{c}^{2}-1$, assumes
approximately the same constant value on the lattice for $1.2\,T_{d}\lesssim
T\lesssim10\,T_{d}$ for all $N_{c}$ \cite{Caselle:2011mn},%
\begin{equation}
\frac{\epsilon\left(  T\right)  -2p\left(  T\right)  }{\left(  N_{c}%
^{2}-1\right)  T^{2}T_{d}}\sim const\text{ .} \label{trc}%
\end{equation}
To a good approximation, this implies that, except for a narrow region near
the phase transition, $T_{d}\leq T\lesssim1.2T_{d}$, the pressure divided by
$N_{c}^{2}-1$, is dominated by a constant term times $T^{2}T_{d}$,%
\begin{equation}
\frac{p\left(  T\right)  }{\left(  N_{c}^{2}-1\right)  T^{2}T_{d}}\sim
const\text{ .} \label{pc}%
\end{equation}
Similar lattice results are obtained in four dimensions, where $p^{d=3+1}%
\left(  T\right)  \sim\left(  N_{c}^{2}-1\right)  T^{2}T_{d}^{2}$ for
$1.2\,T_{d}\lesssim T\lesssim4.0\,T_{d}$.

In $d=2+1$ lattice calculations can be performed with a great precision,
keeping artifacts well under control. Therefore, comparing our results to the
lattice data will provide a crucial test for the validity of the matrix model.

This paper is organized as follows: in Sec. \ref{model} we construct the
confining and deconfining vacua in the presence of a constant background
field. In Sec. \ref{veff2} we derive the effective potential in $d=2+1$ for an
$SU(N_{c})$ gauge group as a sum of a perturbative and a nonperturbative part.
In Sec. \ref{ua} we present the analytical and numerical solution for the
potential for general $N_{c}$ and in the large-$N_{c}$ limit using a so-called
uniform eigenvalue ansatz. In Sec. \ref{results} we show the pressure, the
interaction measure, and the Polyakov loop for $N_{c}=2,3,4,5,6$, and
compare\ to the recent lattice data of Ref. {\cite{Caselle:2011mn}. In Sec.
\ref{meta} we address the question of possible metastable solutions in the
perturbative potential. Conclusions and outlook are given in Sec.
\ref{summary}.}

\section{The confined and deconfined vacua \label{model}}

\bigskip\begin{figure}[t!]
\begin{center}
\includegraphics[width=0.45\textwidth]{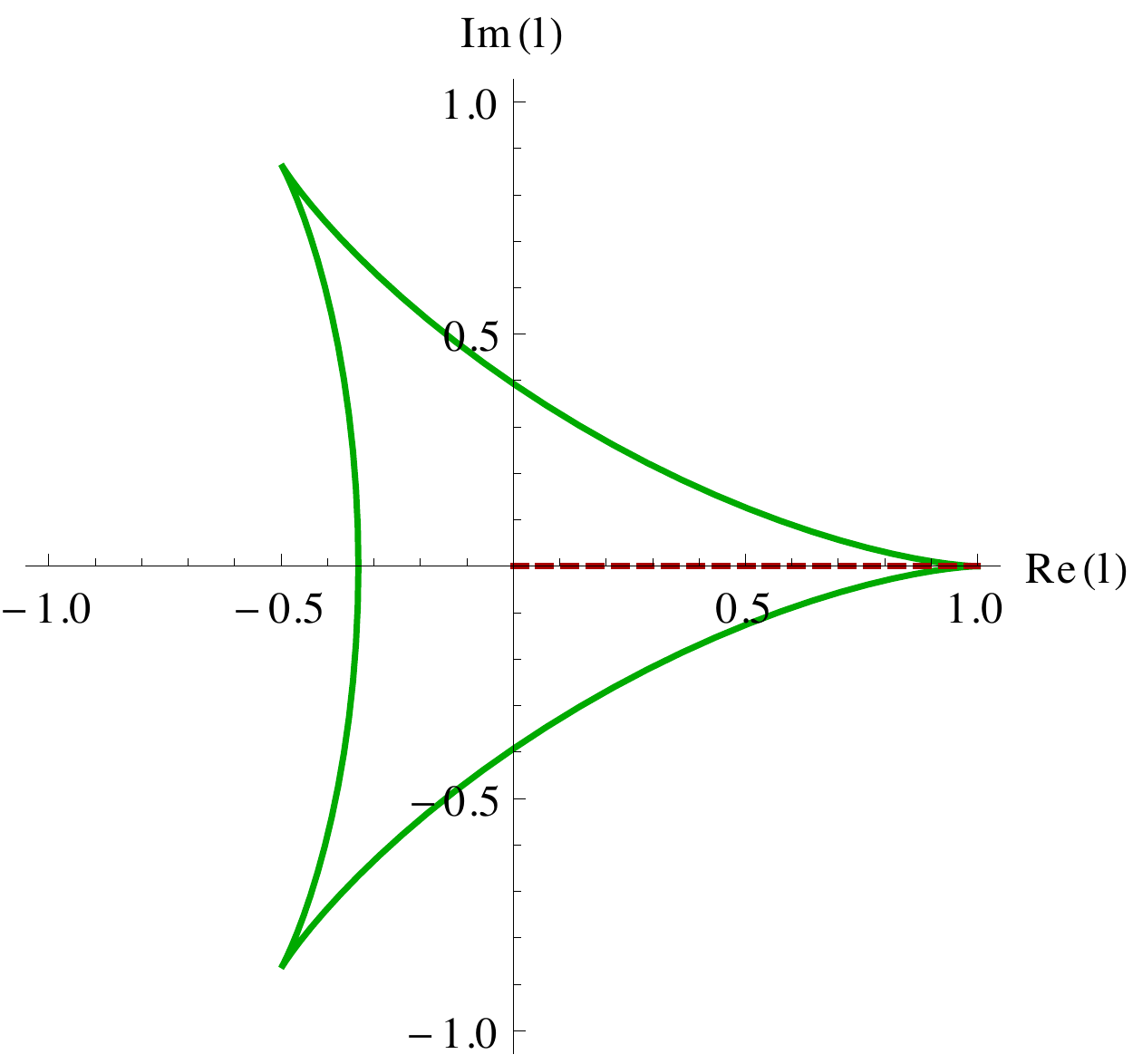} \hspace{10pt}
\includegraphics[width=0.45\textwidth]{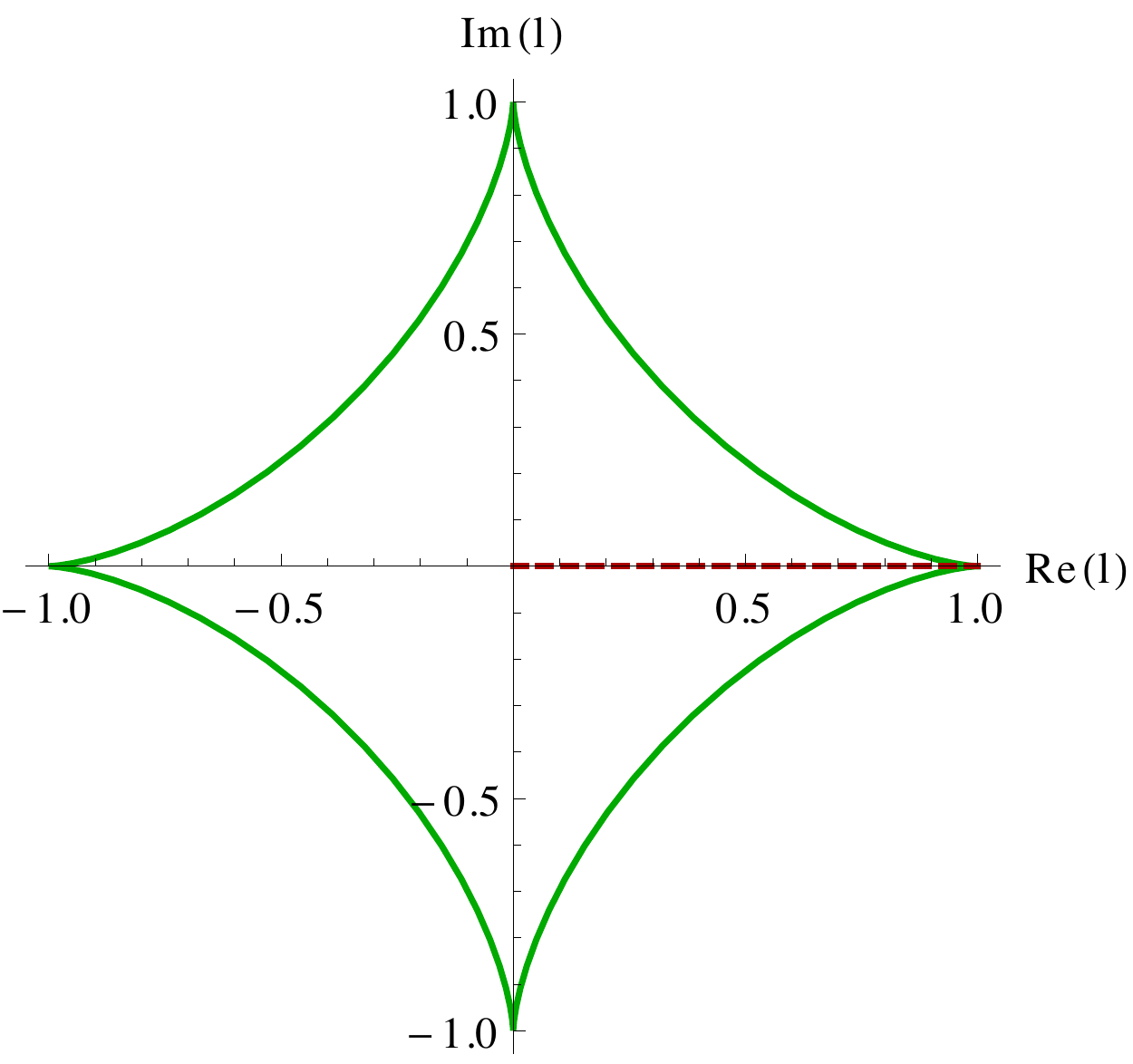}
\includegraphics[width=0.45\textwidth]{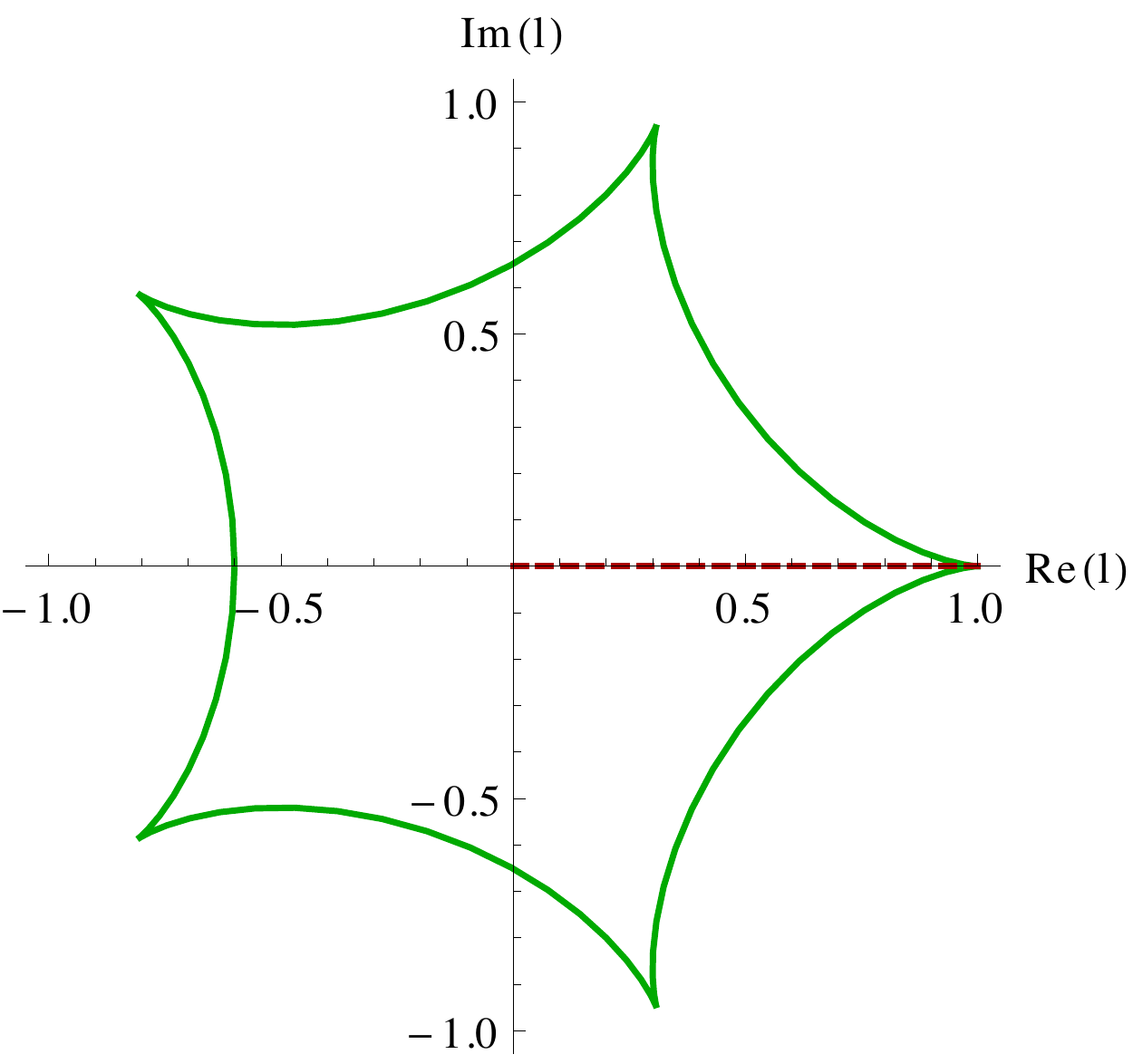} \hspace{10pt}
\includegraphics[width=0.45\textwidth]{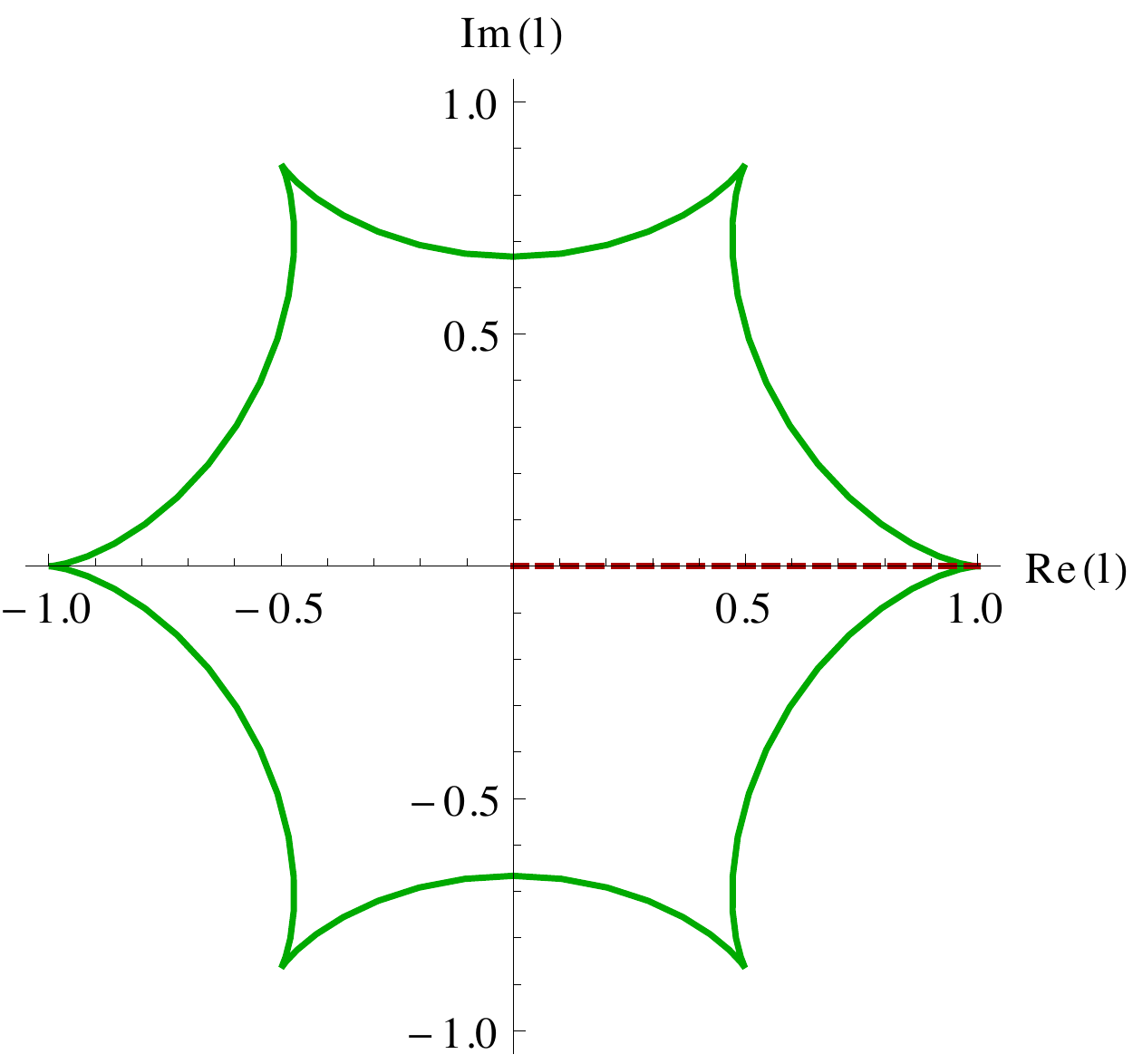}
\end{center}
\caption{Boundaries for values of the Polyakov loop for $N_{c}=3$ (left
top)$,4$ (right top)$,5$ (left bottom)$,6$ (right bottom). The dashed lines
indicate the region between the perturbative vacuum at $l=1$ and the confining
vacuum at $l=0$.}%
\label{flowers}%
\end{figure}

In the absence of dynamical quarks, the pure gauge theory exhibits a
deconfinement phase transition at a critical temperature $T_{d},$ related to
the breaking of the $Z\left(  N_{c}\right)  $ center symmetry of the
$SU(N_{c})$ gauge group. The appropriate gauge-invariant order parameter for
this transition is given by the Polyakov loop $l$, defined as the trace of the
thermal Wilson line $\mathbf{L}$,
\begin{equation}
l=\frac{1}{N_{c}}\operatorname{tr}\mathbf{L}\text{ ,}%
\end{equation}
where%
\begin{equation}
\mathbf{L=}\text{ }\mathcal{P}\exp\left[  ig\int_{0}^{1/T}\mathcal{A}%
_{0}\left(  \tau,\mathbf{x}\right)  d\tau\right]  \text{ .}%
\end{equation}
In the deconfined phase, $T\gg T_{d}$, the allowed vacua of the pure
$SU(N_{c})$ gauge theory exhibit an $N_{c}$-fold degeneracy, where the thermal
expectation value of the Polyakov loop is given by one of the $N_{c}-$th roots
of unity,%

\begin{equation}
\left\langle l\right\rangle =\exp\left(  i\frac{2\pi m}{N_{c}}\right)  \equiv
l_{m}\ ,\ \text{\ \ \ \ }m=0,1,2,...,\text{ }N_{c}-1\text{ .} \label{dv}%
\end{equation}
These vacuum states are degenerate and can be transformed into each other by
global $Z\left(  N_{c}\right)  $ rotations,
\begin{equation}
Z\left(  N_{c}\right)  :\text{ \ \ }l_{m_{1}}\rightarrow l_{\left(
m_{1}+m_{2}\right)  \operatorname{mod}N_{c}}\ \text{.}%
\end{equation}
Consequently, when choosing one particular state, the $Z\left(  N_{c}\right)
$ center symmetry is spontaneously broken. Defining the Polyakov loop to be
real, which is always possible by global $Z\left(  N_{c}\right)  $
transformations, the expectation value of the order parameter approaches unity
in the perturbative vacuum, $\left\langle l\right\rangle \rightarrow1$ as
$T\rightarrow\infty$.

In the confined phase, $T<T_{d}$, the $Z\left(  N_{c}\right)  $ center
symmetry is restored, and the expectation value of the Polyakov loop vanishes,
$\left\langle l_{c}\right\rangle =0$. This behavior of the order parameter is
also confirmed by lattice-QCD calculations, where the expectation value of the
renormalized Polyakov loop is zero in the confined and nonzero in the
deconfined phase, approaching unity in the perturbative limit.

In general, the Polyakov loop is a complex-valued quantity. Only for the
special case $N_{c}=2,$\ the order parameter is real, assuming values between
$-1$\ and $+1$. In Fig. \ref{flowers} we plot the boundaries of $l$\ for
different numbers of colors. Depending on the parametrization of
$\mathcal{A}_{0}$, which is an element of the Lie algebra of the $SU(N_{c}%
)$\ gauge group, the Polyakov loop takes certain values within the solid
lines. The degenerate ground states in the deconfined phase reside at the
corners,\textbf{ }while the confining vacuum $l_{c}$ corresponds to the origin
of the diagrams. It is obtained by computing the average of the deconfined
vacua,\textbf{ }%
\begin{equation}
l_{c}\equiv{\frac{1}{N_{c}}}\sum_{m=0}^{N_{c}-1}l_{m}\ \text{,}%
\end{equation}
and is therefore automatically invariant under $Z\left(  N_{c}\right)  $ transformations.

In this paper the numerical results are obtained using a uniform eigenvalue
ansatz \cite{Dumitru:2012fw}. Within this ansatz, $\mathcal{A}_{0}$\ is
parameterized in such a way that the Polyakov loop always assumes values along
the real axis in Fig. \ref{flowers}.

\subsection{Background field}

Given the known behavior of the Polyakov loop, we can model the phase
transition and the deconfined phase, $T\geq T_{d}$, by constructing an
effective theory for the eigenvalues of the thermal Wilson line. The simplest
ansatz is to assume a constant, but nonzero value for the time component
$\mathcal{A}_{0}$ of the vector potential. Due to $SU(N_{c})$ gauge
transformations, it is always possible to write the background $\mathcal{A}%
_{0}$ field as
\begin{equation}
\mathcal{A}_{0}=\;\frac{2\pi\,T}{g}\mathbf{\,q}\text{ ,} \label{a0n}%
\end{equation}
where $\mathbf{q}$ is a traceless diagonal matrix in the $SU(N_{c})$ Lie
algebra. The eigenvalues $q_{j}$ are constrained only by the unimodularity,
\begin{equation}
\sum_{j=1}^{N_{c}}q_{j}=0\text{ , \ \ }j=1,...,N_{c}\text{ ,} \label{q0}%
\end{equation}
and the number of independent $q_{j}$'s corresponds to the rank of the
$SU(N_{c})$ group,$\ r=N_{c}-1$.

In the presence of the background $\mathcal{A}_{0}$ field (\ref{a0n}), the
Wilson line is given by a diagonal matrix,%
\begin{align}
\mathbf{L}  &  =\mathcal{P}\exp\left[  ig\int_{0}^{1/T}\mathcal{A}_{0}\left(
\tau,\mathbf{x}\right)  d\tau\right]  =\exp\left(  i2\pi\mathbf{q}\right)
\label{w}\\
&  =\operatorname{diag}\left(  e^{i2\pi q_{1}},e^{i2\pi q_{2}},\cdots,e^{i2\pi
q_{_{N_{c}}}}\right)  \text{ .}\nonumber
\end{align}
The Polyakov loop is\textbf{\ }the trace over the Wilson line,
\begin{equation}
l=\frac{1}{N_{c}}\operatorname{tr}\mathbf{L=}\text{ }\frac{1}{N_{c}}\sum
_{j=1}^{N_{c}}e^{i2\pi q_{j}}\text{ .}%
\end{equation}

\subsubsection{Perturbative vacuum}

The degenerate ground states (\ref{dv}) in the deconfined phase can be
described by an appropriate choice for the matrix $\mathbf{q}$,%
\begin{align}
l_{m}  &  =\exp\left(  i\frac{2\pi m}{N_{c}}\right) \\
&  =\frac{1}{N_{c}}\operatorname{tr}\mathbf{L}_{m}=\frac{1}{N_{c}%
}\operatorname{tr}\exp\left(  i2\pi\mathbf{q}_{m}\right)  \ \text{,\ \ \ \ }%
m=0,1,..,\text{ }N_{c}-1\text{ ,}\nonumber
\end{align}
where $\mathbf{q}_{m}$ are $N_{c}$ traceless diagonal matrices of the form%
\begin{equation}
\mathbf{q}_{m}={\frac{1}{N_{c}}}\text{diag}(m,\cdots,m,m-N_{c},\cdots
,m-N_{c})\text{ ,} \label{hc}%
\end{equation}
with $N_{c}-m$ entries $m$, and $m$ entries $m-N_{c}$. The $\mathbf{q}_{m}$
are also referred to as hypercharges \cite{Dumitru:2012fw}, and serve as
generators for the elements of the $Z\left(  N_{c}\right)  $ center of the
$SU(N_{c})$ gauge group,
\begin{equation}
\exp(i2\pi\,\mathbf{q}_{m})=\exp\left(  i\frac{2\pi\,m}{N_{c}}\right)
\mathbf{1}\equiv z_{m}\text{ ,}%
\end{equation}
where $\mathbf{1}$ is an $N_{c}\times N_{c}$ unit matrix. Thus, under global
$Z\left(  N_{c}\right)  $ rotations the hypercharges are transformed into each
other,%
\begin{equation}
Z\left(  N_{c}\right)  :\text{ \ \ }\mathbf{q}_{m_{1}}\rightarrow
\mathbf{q}_{\left(  m_{1}+m_{2}\right)  \operatorname{mod}N_{c}}\ \text{.}%
\end{equation}
The perturbative vacuum where the Polyakov loop is equal to one, $l=1$, is
obtained for
\begin{equation}
\mathbf{q}_{0}=0\text{ .}%
\end{equation}

\subsubsection{Confining vacuum}

In the confining vacuum $\mathbf{q}_{c}$, the Polyakov loop must vanish,
$l_{c}=0$. The confining vacuum is constructed as the average of all
degenerate ground states (\ref{hc}), \
\begin{align}
\mathbf{q}_{c}  &  \equiv{\frac{1}{N_{c}}}\sum_{m=0}^{N_{c}-1}\mathbf{q}%
_{m}\label{yc}\\
&  =\frac{1}{2N_{c}}\left(
\begin{array}
[c]{cccc}%
N_{c}-1 & 0 & \cdots & 0\\
0 & N_{c}-3 & \cdots & 0\\
\vdots & \vdots & \ddots & \vdots\\
0 & \cdots & 0 & -(N_{c}-1)
\end{array}
\right)  \text{ ,}\nonumber
\end{align}
and is therefore is automatically $Z\left(  N_{c}\right)  $ invariant. The
eigenvalues of $\mathbf{q}_{c}$ are separated by a constant spacing,
\begin{equation}
q_{j}-q_{j+1}=\frac{1}{N_{c}}\text{ .}%
\end{equation}
This implies that the eigenvalues of the Wilson line,
\begin{equation}
\mathbf{L}_{c}=\exp(i2\pi\,\mathbf{q}_{c})=\left(
\begin{array}
[c]{cccc}%
e^{i\pi(N_{c}-1)/N_{c}} & 0 & \cdots & 0\\
0 & e^{i\pi(N_{c}-3)/N_{c}} & \cdots & 0\\
\vdots & \vdots & \ddots & \vdots\\
0 & \cdots & 0 & e^{-i\pi(N_{c}-1)/N_{c}}%
\end{array}
\right)  \text{ ,} \label{wlc}%
\end{equation}
are equally distributed about the unit circle, with a spacing $2\pi/N_{c}$.
Thus, the confining vacuum is characterized by a uniform, i.e., complete
repulsion of eigenvalues.

\section{The Potential \label{veff2}}

\begin{figure}[t!]
\begin{center}
\includegraphics[width=0.6\textwidth]{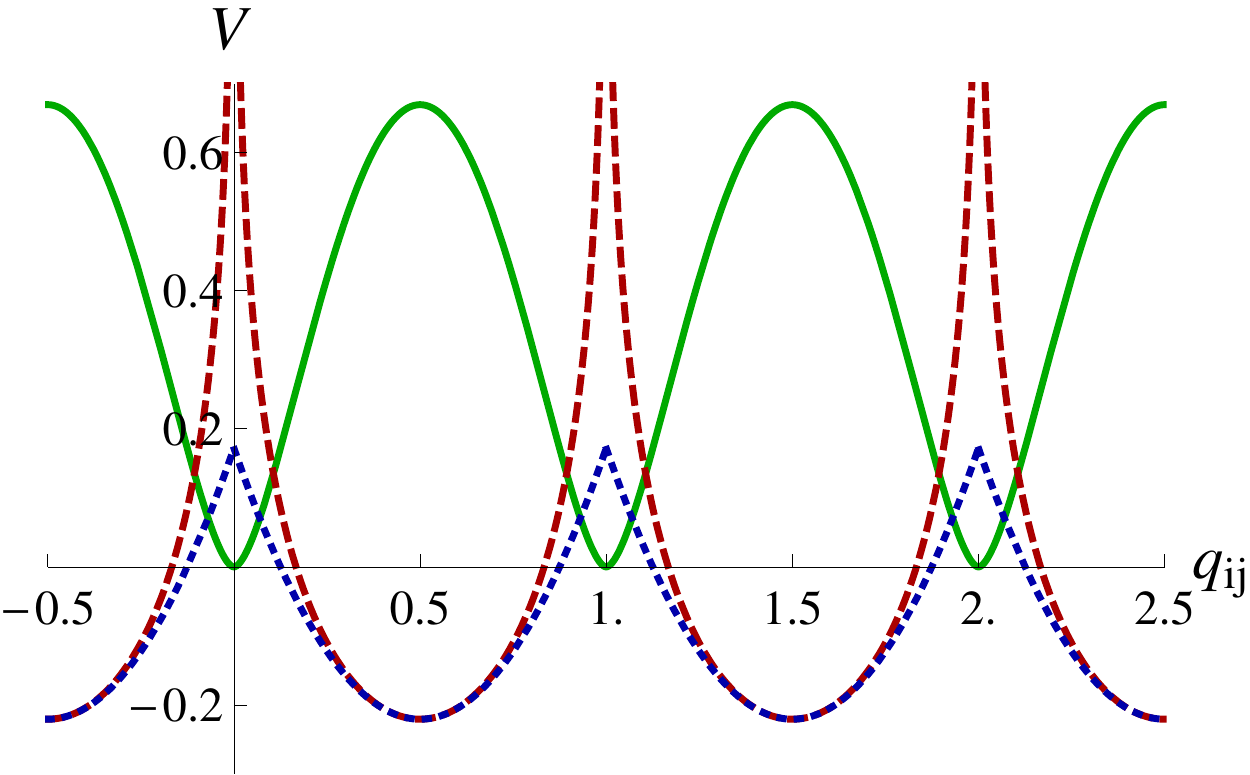}
\end{center}
\caption{The perturbative potential $\tilde{V}_{pt}$ (solid line), the
Vandermonde nonperturbative term $\check{V}_{Vdm}$ (dashed line), and the
linear nonperturbative term $\tilde{V}_{lin}$ (dotted line) as a function of
$q_{ij}$.}%
\label{ptall}%
\end{figure}

Provided that the expectation value of the Polyakov loop near the critical
temperature $T_{d}$ is dominated by the classical configuration of the
background $\mathcal{A}_{0}$ field (\ref{a0n}), we can model the deconfinement
phase transition by introducing a potential for $\mathbf{q}$. In this section
we construct the effective potential as a sum of two parts: the perturbative
potential, $V_{pt}(\mathbf{q})$, and nonperturbative contributions
$V_{npt}(\mathbf{q}).$

\subsection{Parametrization of the $SU(N_{c})$ generators}

Before proceeding, we need to introduce a suitable parametrization for the
generators of the $SU(N_{c})$ Lie algebra. It is convenient to choose the same
basis as in Ref. \cite{Bhattacharya:1992qb} comprised of $N_{c}-1$ diagonal
matrices, and $N_{c}^{2}-N_{c}$ off-diagonal matrices.

The diagonal generators can be chosen\textbf{\ }as%
\begin{equation}
t_{j}=\frac{1}{\sqrt{2j\left(  j-1\right)  }}\left(
\begin{array}
[c]{ccc}%
\mathbf{1}_{j-1} & \cdots & 0\\
\vdots & -\left(  j-1\right)  & \vdots\\
0 & \cdots & 0
\end{array}
\right)  \text{ ,} \label{diag}%
\end{equation}
where $j=2,...,N_{c}$. For the off-diagonal generators it is useful to
introduce a ladder basis,
\begin{equation}
\left(  t_{j,i}^{+}\right)  _{mn}=\frac{1}{\sqrt{2}}\delta_{jn}\delta
_{im}\text{ ,\ \ }\left(  t_{j,i}^{-}\right)  _{mn}=\frac{1}{\sqrt{2}}\text{
}\delta_{jm}\delta_{in}\text{ , \ \ }1\leq i<j\leq N_{c}\text{ .}
\label{off-diag}%
\end{equation}
The generators in Eqs. (\ref{diag}) and (\ref{off-diag}) form an orthogonal
set, with the normalization for the diagonal generators,
\begin{equation}
\operatorname{tr}\left(  t_{i}t_{j}\right)  =\frac{1}{2}\delta_{ij}\text{ ,}
\label{norm1}%
\end{equation}
and for the off-diagonal generators,%
\begin{equation}
\operatorname{tr}\left(  t_{i,j}^{+}t_{i^{\prime},j^{\prime}}^{-}\right)
=\frac{1}{2}\delta_{ii^{\prime}}\delta_{jj^{\prime}},\ \ \ \operatorname{tr}%
\left(  t_{i,j}^{+}t_{i^{\prime},j^{\prime}}^{+}\right)  =\operatorname{tr}%
\left(  t_{i,j}^{-}t_{i^{\prime},j^{\prime}}^{-}\right)  =0\text{ .}
\label{norm2}%
\end{equation}

\subsection{Perturbative potential \label{vpert}}

The perturbative potential is computed by expanding about the classical
background $\mathcal{A}_{0}$ field (\ref{a0n}), see e.g. Refs.
\cite{Bhattacharya:1990hk, Bhattacharya:1992qb, KorthalsAltes:1996xp,
Dumitru:2012fw}. To one-loop order in $d=2+1$\textbf{,} the result is given
by
\begin{equation}
V_{pt}(\mathbf{q})=\frac{T}{2\mathcal{V}}\operatorname{tr}\ln\left[
-D^{2}(\mathbf{q})\right]  \text{ ,} \label{v1}%
\end{equation}
see Ref. \cite{KorthalsAltes:1996xp}, where $D_{\mu}$ is the covariant
derivative in the adjoint representation in the presence of the background
$\mathcal{A}_{0}$ field,
\begin{align}
D_{\mu}(\mathbf{q})  &  =\partial_{\mu}-ig\left[  \mathcal{A}_{\mu}%
\mathbf{,}\text{ }\right] \\
&  =\partial_{\mu}-i2\pi T\delta_{\mu,0}\left[  \mathbf{q,}\text{ }\right]
\text{ ,}\nonumber
\end{align}
and $D^{2}$ is the gauge-covariant d'Alembertian,%
\begin{equation}
D^{2}(\mathbf{q})=\left(  \partial_{0}-i2\pi T\left[  \mathbf{q,}\text{
}\right]  \right)  ^{2}+{\boldsymbol{\partial}}^{2}\text{ .}%
\end{equation}
The trace in Eq. (\ref{v1}) is over all momenta and color degrees of freedom.
In order to evaluate the color trace, one has to sum over all commutators
between $\mathbf{q}$ and the generators of the $SU(N_{c})$ Lie algebra defined
in Eqs. (\ref{diag}) and (\ref{off-diag}). This can be done using the
normalizations in Eqs. (\ref{norm1}) and (\ref{norm2}).

Since $\mathbf{q}$ is a diagonal matrix, it commutes with all diagonal
generators,%
\begin{equation}
\left[  \mathbf{q,}\text{ }t_{j}\right]  =0\text{ , \ }j=2,...,N_{c}\text{ .}%
\end{equation}
Thus, for each degree of freedom along diagonal generators the potential is as
in zero background field, i.e., it has no $\mathbf{q}$-dependence,%
\begin{equation}
V_{pt}^{diag}=\frac{T}{2\mathcal{V}}\operatorname{tr}\ln\left(  k_{0}%
^{2}+\mathbf{k}^{2}\right)  \text{ , }%
\end{equation}
where\textbf{\ }$\mathbf{k}$ is the two-dimensional vector in momentum space,
and $k_{0}=2\pi Tn,$ $n=0,\pm1,\pm2,...$ is the Matsubara frequency for
bosonic fields.

For all off-diagonal generators the commutator\ with $\mathbf{q}$ is nonzero,%
\begin{equation}
\left[  \mathbf{q,}\text{ }t_{j,i}^{\pm}\right]  =\pm q_{ij}t_{j,i}^{\pm
}\text{ ,}%
\end{equation}
where we introduced the notation
\begin{equation}
q_{ij}\equiv q_{i}-q_{j}\text{ .} \label{q}%
\end{equation}
Therefore, the propagators for the degrees of freedom along the ladder
operators $t_{j,i}^{\pm}$ are as in zero background field, except that $k_{0}$
is shifted by a constant amount to $k_{0}^{\pm}=2\pi T\left(  n\pm
q_{ij}\right)  ,$%
\begin{equation}
V_{pt}^{off-diag}\left(  q_{ij}\right)  =\frac{T}{2\mathcal{V}}%
\operatorname{tr}\ln\left[  \left(  k_{0}^{\pm}\right)  ^{2}+\mathbf{k}%
^{2}\right]  \text{ .}%
\end{equation}
Summing over all $SU(N_{c})$\ generators, the full perturbative potential to
one-loop order is
\begin{align}
V_{pt}(q_{ij})  &  =\frac{T}{2\mathcal{V}}\left\{  (N_{c}-1)\operatorname{tr}%
\ln\left(  -\partial_{0}^{2}-{\boldsymbol{\partial}}^{2}\right)  +\sum_{1\leq
i<j\leq N_{c}}\operatorname{tr}\ln\left[  -\left(  \partial_{0}\pm i2\pi
Tq_{ij}\right)  ^{2}-{\boldsymbol{\partial}}^{2}\right]  \right\} \nonumber\\
&  =\frac{T}{2\mathcal{V}}\left\{  (N_{c}-1)\operatorname{tr}\ln\left(
k_{0}^{2}+\mathbf{k}^{2}\right)  +\sum_{1\leq i<j\leq N_{c}}\operatorname{tr}%
\ln\left[  \left(  k_{0}^{\pm}\right)  ^{2}+\mathbf{k}^{2}\right]  \right\}
\;\text{,} \label{vpt}%
\end{align}
The first term in Eq. (\ref{vpt}) is from diagonal modes $t_{j}$,\ while the
second term comes from the off-diagonal modes $t_{i,j}^{\pm}$. As demonstrated
in Ref. \cite{Bicudo:2013yza}, the sum integrals in Eq. (\ref{vpt}) can be
evaluated via contour integration \cite{Gross:1980br},
\begin{align}
\operatorname{tr}\ln\left(  k_{0}^{2}+\mathbf{k}^{2}\right)   &
=-\frac{\mathcal{V}\,T^{2}}{\pi}\zeta\left(  3\right)  \text{ ,}\\
\operatorname{tr}\ln\left[  \left(  k_{0}^{\pm}\right)  ^{2}+\mathbf{k}%
^{2}\right]   &  =2\mathcal{V}\int\frac{d^{2}\mathbf{k}}{\left(  2\pi\right)
^{2}}\ln\left(  1-e^{-\left\vert \mathbf{k}\right\vert /T\pm i2\pi q_{ij}%
}\right) \nonumber\\
&  =-\frac{\mathcal{V}\,T^{2}}{\pi}\text{$\mathrm{Li}$}_{3}\left(  e^{\pm
i2\pi q_{ij}}\right)  \text{ ,}%
\end{align}
where we used the polylogarithm function,%
\begin{equation}
\mathrm{Li}_{j}(z)=\sum_{n=1}^{\infty}\;\frac{z^{n}}{n^{j}}\text{ ,}%
\end{equation}
and the Riemann zeta function,
\begin{equation}
\zeta(j)=\sum_{n=1}^{\infty}\;\frac{1}{n^{j}}\text{ .}%
\end{equation}
The final result for the one-loop perturbative potential is then given by%
\begin{equation}
V_{pt}(q_{ij})=-\frac{T^{3}}{2\pi}\left\{  (N_{c}-1)\zeta\left(  3\right)
+\sum_{1\leq i<j\leq N_{c}}\left[  \text{$\mathrm{Li}$}_{3}\left(  e^{i2\pi
q_{ij}}\right)  +\text{$\mathrm{Li}$}_{3}\left(  e^{-i2\pi q_{ij}}\right)
\right]  \right\}  \;\text{,} \label{vpt_1}%
\end{equation}
where $\zeta\left(  3\right)  =$ $\mathrm{Li}_{3}\left(  1\right)  $.

The result for $V_{pt}(q_{ij})$, Eq. (\ref{vpt_1}), can also be written as a
sum of the zero-field contribution, independent of $q_{ij}$,
\begin{equation}
V_{pt}(0)=-\left(  N_{c}^{2}-1\right)  \frac{T^{3}}{2\pi}\;\zeta(3)\ \text{,}
\label{vc2}%
\end{equation}
and the quantum correction,%
\begin{align}
V_{pt}^{qu}(q_{ij})  &  =-\frac{T^{3}}{2\pi}\left\{  \sum_{1\leq i<j\leq
N_{c}}\left[  \text{$\mathrm{Li}$}_{3}\left(  e^{i2\pi q_{ij}}\right)
+\text{$\mathrm{Li}$}_{3}\left(  e^{-i2\pi q_{ij}}\right)  -2\zeta\left(
3\right)  \right]  \right\} \label{vpt2}\\
&  \equiv T^{3}\sum_{1\leq i<j\leq N_{c}}\tilde{V}_{pt}^{qu}(q_{ij}%
)\ \text{.}\nonumber
\end{align}
Note that the quantum correction is automatically zero in the perturbative
vacuum $\mathbf{q}_{0}=0$. The zero-field contribution (\ref{vc2}) corresponds
to the free energy of an ideal gas of $N_{c}^{2}-1$ massless gluons.

\subsection{Nonperturbative potential}

In order to model the deconfinement phase transition at $T=T_{d},$ it is
necessary to add nonperturbative terms to the perturbative potential. In
general, the nonperturbative contributions may include similar functions of
$\mathbf{q}$\ to the ones found at one-loop order.

\subsubsection{$\mathbf{q}$-independent term}

Since the lattice data for $(\epsilon-2p)/(T_{d}T^{2})$ is constant in the
region $1.2\,T_{d}\lesssim T\lesssim10T_{d}$ \cite{Bialas:2008rk,
Caselle:2011mn} for all $N_{c}$, see Eq. (\ref{trc}), we must certainly
include a nonperturbative term which is independent of $\mathbf{q}$,%
\begin{equation}
\sim T_{d}T^{2}\left(  N_{c}^{2}-1\right)  \frac{\zeta\left(  3\right)  }%
{2\pi}\text{ ,} \label{ind}%
\end{equation}
similar to the free energy of an ideal gas of $N_{c}^{2}-1$ massless gluons in
Eq. (\ref{vc2}).

Notably, in $d=2+1$ dimensions\ the one-loop perturbative corrections to the
pressure are $\sim g^{2}T^{2}$ \cite{D'Hoker:1980az, D'Hoker:1981qp,
D'Hoker:1981us}, and so automatically proportional to $T_{d}T^{2}$, because in
three dimensions $g^{2}$ has the dimension of mass. The results of numerical
simulations on the lattice are nevertheless surprising, since it is not
natural to expect that perturbation theory at one-loop order is dominant down
to temperatures as low as $\sim1.2\,T_{d}$. Moreover, on the lattice there is
no evidence of higher-order perturbative contributions. The two-loop order
terms would be $\sim g^{4}T$, while the three-loop order contributions are
independent of temperature, $\sim g^{6}$. In detail, perturbation theory is
more involved, including logarithms of $g^{2}/T$ \cite{D'Hoker:1980az,
D'Hoker:1981qp, D'Hoker:1981us}.

\subsubsection{$\mathbf{q}$-dependent term}

In addition, we may also add a $\mathbf{q}$-dependent term similar to the
one-loop quantum correction (\ref{vpt2}),%
\begin{equation}
\sim T_{d}T^{2}\sum_{1\leq i<j\leq N_{c}}{\frac{1}{2\pi}}\left[
\text{$\mathrm{Li}$}_{3}\left(  e^{i2\pi q_{ij}}\right)  +\text{$\mathrm{Li}$%
}_{3}\left(  e^{-i2\pi q_{ij}}\right)  -2\zeta\left(  3\right)  \right]
\text{ .} \label{dep}%
\end{equation}
The temperature dependence $\sim T_{d}T^{2}$ is a necessary condition for the
$\mathbf{q}$-independent term (\ref{ind}), but for the $\mathbf{q}$-dependent
term this is manifestly an assumption.

\subsubsection{Vandermonde determinant}

Finally, in order to avoid another phase transition above $T_{d}$, we need to
add a Vandermonde-like term to the potential, see Ref. \cite{Bicudo:2013yza}.
It is interesting to note that in our matrix model a Vandermonde term can be
introduced using different approaches.

\vspace{0.5cm} \noindent\textit{Leading order in a mass expansion} \newline

One possibility is to consider a mass parameter in the gluon propagator, and
then to expand the one-loop determinant (\ref{v1}) to leading order in the
mass \cite{Bicudo:2013yza},
\begin{equation}
\frac{T}{\mathcal{V}}\mathrm{tr}\ln\left(  -D^{2}+m^{2}\right)  =\frac
{T}{\mathcal{V}}\mathrm{tr}\ln\left(  -D^{2}\right)  +m^{2}\frac
{T}{\mathcal{V}}\mathrm{tr}\left(  \frac{1}{-D^{2}}\right)  +\text{ ... .}
\label{det}%
\end{equation}
The first term in the expansion reproduces the one-loop order perturbative
potential, while the second term corresponds to the Vandermonde determinant.

The trace over momentum in Eq. (\ref{det}) can be evaluated via contour
integration,%
\begin{align}
V_{Vdm}(q_{ij})  &  =m^{2}\frac{T}{\mathcal{V}}\mathrm{tr}\left(  \frac
{1}{-D^{2}}\right)  =m^{2}T\sum_{1\leq i<j\leq N_{c}}\frac{\,1}{2\pi}\left[
\mathrm{Li}_{1}(e^{i2\pi q_{ij}})+\mathrm{Li}_{1}(e^{-i2\pi q_{ij}})\right]
\label{vdm22}\\
&  =-m^{2}T\sum_{1\leq i<j\leq N_{c}}{\frac{1}{2\pi}}\left\{  \ln\left[
1-\exp\left(  i2\pi q_{ij}\right)  \right]  +\ln\left[  1-\exp\left(  -i2\pi
q_{ij}\right)  \right]  \right\} \nonumber\\
&  =-m^{2}T\sum_{1\leq i<j\leq N_{c}}{\frac{1}{\pi}}\ln\left[  2\sin(\pi
q_{ij})\right] \nonumber\\
&  \equiv m^{2}T\sum_{1\leq i<j\leq N_{c}}\tilde{V}_{Vdm}(q_{ij})\text{ .}%
\end{align}
The function $\tilde{V}_{Vdm}(q_{ij})$ is divergent at $q_{ij}=0,1$ and has a
minimum at $q_{ij}=1/2$, see Fig. \ref{ptall}. Moreover, due to the $Z\left(
N_{c}\right)  $ symmetry, $\tilde{V}_{Vdm}(q_{ij})$ is periodic in
$q_{ij}\rightarrow q_{ij}+1$, see Sec. \ref{zN}. Identifying the mass
parameter with the Debye screening mass $m_{D}$, in three dimensions
$m_{D}^{2}\sim g^{2}T,$ see e.g. Refs. \cite{Kajantie:1997tt, Rischke:2003mt}.
Thus, the overall temperature dependence of the Vandermonde term is
$g^{2}T^{2}\sim T_{d}T^{2}$,%
\begin{equation}
V_{Vdm}(q_{ij})=T_{d}T^{2}\sum_{1\leq i<j\leq N_{c}}\tilde{V}_{Vdm}%
(q_{ij})\text{ ,}%
\end{equation}
where $T_{d}$ acts just as a mass parameter.

\vspace{0.5cm} \noindent\textit{Two-loop quantum correction} \newline

Performing a mass expansion is just one possibility to derive the Vandermonde
determinant. The same term can also be found at two-loop order which may
already include nonperturbative contributions. From perturbative calculations
in four dimensions \cite{Bhattacharya:1992qb, Belyaev:1989bj, Enqvist:1990ae},
we know that the two-loop corrections to the potential in a constant
background $\mathcal{A}_{0}$ field (\ref{a0n}) include terms similar to the
ones obtained at one-loop order, as well as terms similar to the second
derivative of the one-loop result with respect to $2\pi q_{ij}$.

Similar arguments should apply in $d=2+1$. Thus we derive
\begin{equation}
\sum_{1\leq i<j\leq N_{c}}\frac{d^{2}}{4\pi^{2}(d\,{q_{ij}})^{2}}{\frac
{-1}{2\pi}}\left[  \text{$\mathrm{Li}$}_{3}(e^{i2\pi q_{ij}}%
)+\text{$\mathrm{Li}$}_{3}(e^{-i2\pi q_{ij}})\right]  =\sum_{1\leq i<j\leq
N_{c}}\frac{\,1}{2\pi}\left[  \mathrm{Li}_{1}(e^{i2\pi q_{ij}})+\mathrm{Li}%
_{1}(e^{-i2\pi q_{ij}})\right]  \ .
\end{equation}
This result is identical to the Vandermonde term obtained via the mass
expansion, Eq. (\ref{vdm22}).

\subsubsection{Linear term}

In order to derive a regularized version of the Vandermonde term, we expand
the function $\tilde{V}_{Vdm}(q_{ij})$ around its minimum at $q_{ij}=1/2$,
keeping only terms up to order $\left(  q_{ij}-1{/2}\right)  ^{2}$,%
\begin{equation}
\tilde{V}_{Vdm}(q_{ij})=-{\frac{1}{\pi}}\ln\left[  2\sin(\pi q_{ij})\right]
\rightarrow-{\frac{\ln2}{\pi}}+{\frac{\pi}{2}}\left[  q_{ij}-\text{floor}%
(q_{ij})-{\frac{1}{2}}\right]  ^{2}\text{ ,}%
\end{equation}
where $q_{ij}-\text{floor}(q_{ij})$ returns the decimal part of $q_{ij}$. This
ensures that the periodicity in $q_{ij}\rightarrow q_{ij}+1$ is maintained,
see Fig. \ref{ptall}. We refer to this term as linear, since it gives a
contribution linear in $q_{ij}$ for small $q_{ij}$ when expanding around the
minimum. Including the temperature dependence and the sum, the linear
nonperturbative term is%
\begin{align}
V_{lin}(q_{ij})  &  ={T_{d}T^{2}}\sum_{1\leq i<j\leq N_{c}}\left\{
-{\frac{\ln2}{\pi}}+{\frac{\pi}{2}}\left[  q_{ij}-\text{floor}(q_{ij}%
)-{\frac{1}{2}}\right]  ^{2}\right\}  \text{ }\label{Vlin}\\
&  \equiv{T_{d}T^{2}}\sum_{1\leq i<j\leq N_{c}}\tilde{V}_{lin}(q_{ij})\text{
.}\nonumber
\end{align}

\vspace{0.5cm} \noindent\textit{Adjoint Higgs phase} \noindent

The need for a Vandermonde or a linear contribution to the potential is the
following: for static fields, $\mathcal{A}_{0}$ couples to the spatial degrees
of freedom $\mathcal{A}_{i}$ as an adjoint scalar. When $\mathbf{q}$, and
consequently $\mathcal{A}_{0}$, develops a nonvanishing expectation value, the
system is in an adjoint Higgs phase \cite{Pisarski:2006hz, Unsal:2008ch}.
While there is no gauge-invariant order parameter for an adjoint Higgs phase,
there can still be a first-order transition from a truly perturbative phase,
where $\langle\mathbf{q}\rangle=0$, to one where $\langle\mathbf{q}\rangle
\neq0$. This scenario would correspond to a second phase transition, at a
temperature higher than $T_{d}$. Though\textbf{\ }possible, lattice
calculations find no evidence of such a second phase transition.

This can be avoided by adding a Vandermonde or a linear term to the effective
potential, which will generate a nonzero expectation value for $\mathcal{A}%
_{0}$ at any temperature, obviating such a second phase transition. As a
consequence, the theory is in an adjoint Higgs phase for all $T$. In practice,
however, for the parameters of the model determined by fitting the lattice
data, the condensate is very small except near the critical temperature
$T_{d}$, see Sec. \ref{results}.

\subsubsection{Full nonperturbative potential}

Summing all contributions, the nonperturbative potential with the Vandermonde
determinant (\ref{vdm22}) takes the form
\begin{align}
V_{npt}(q_{ij})  &  ={T_{d}T^{2}}C_{1}\sum_{1\leq i<j\leq N_{c}}{\frac{1}%
{2\pi}}\left[  \mathrm{Li}_{1}(e^{i2\pi q_{ij}})+\mathrm{Li}_{1}(e^{-i2\pi
q_{ij}})\right] \nonumber\\
&  +T_{d}T^{2}C_{2}\sum_{1\leq i<j\leq N_{c}}{\frac{1}{2\pi}}\left[
\text{$\mathrm{Li}$}_{3}\left(  e^{i2\pi q_{ij}}\right)  +\text{$\mathrm{Li}$%
}_{3}\left(  e^{-i2\pi q_{ij}}\right)  -2\zeta\left(  3\right)  \right]
\nonumber\\
&  +T_{d}T^{2}C_{3}\,\left(  N_{c}^{2}-1\right)  \frac{\zeta\left(  3\right)
}{2\pi}\ \text{,} \label{vnpt_A}%
\end{align}
where we introduced three parameters $C_{1},\,C_{2},\,C_{3}$ which will be
determined in Sec. \ref{fixparam}. For the linear term (\ref{Vlin}) the full
nonperturbative potential is obtained by replacing in Eq. (\ref{vnpt_A}) the
term $\sim C_{1}$\ as
\begin{equation}
{\frac{1}{2\pi}}\left[  \mathrm{Li}_{1}(e^{i2\pi q_{ij}})+\mathrm{Li}%
_{1}(e^{-i2\pi q_{ij}})\right]  \rightarrow-{\frac{\ln2}{\pi}}+{\frac{\pi}{2}%
}\left[  q_{ij}-\text{floor}(q_{ij})-{\frac{1}{2}}\right]  ^{2}\text{ .}
\label{vnpt_B}%
\end{equation}

\subsection{Effective Potential}

The effective potential is given by the sum of the perturbative term
(\ref{vpt_1}) plus the nonperturbative contributions (\ref{vnpt_A}) and
(\ref{vnpt_B}),
\begin{equation}
V_{eff}\left(  \mathbf{q}\right)  =V_{pt}(\mathbf{q})+V_{npt}(\mathbf{q}%
)\text{ .}%
\end{equation}
In the presence of the Vandermonde term (\ref{vdm22}) we derive
\begin{align}
V_{eff}(q_{ij})  &  ={T_{d}T^{2}}C_{1}\sum_{1\leq i<j\leq N_{c}}{\frac{1}%
{2\pi}}\left[  \mathrm{Li}_{1}(e^{i2\pi q_{ij}})+\mathrm{Li}_{1}(e^{-i2\pi
q_{ij}})\right] \nonumber\\
&  -T^{3}\left(  1-\frac{T_{d}}{T}C_{2}\right)  \sum_{1\leq i<j\leq N_{c}%
}{\frac{1}{2\pi}}\left[  \text{$\mathrm{Li}$}_{3}\left(  e^{i2\pi q_{ij}%
}\right)  +\text{$\mathrm{Li}$}_{3}\left(  e^{-i2\pi q_{ij}}\right)
-2\zeta\left(  3\right)  \right] \nonumber\\
&  -T^{3}\left(  1-\frac{T_{d}}{T}C_{3}\right)  \left(  N_{c}^{2}-1\right)
\frac{\zeta\left(  3\right)  }{2\pi}\ \text{.} \label{veffa}%
\end{align}
For the linear term (\ref{Vlin}) the effective potential is given by replacing
the term $\sim C_{1}$ in Eq. (\ref{veffa}) as%
\begin{equation}
{\frac{1}{2\pi}}\left[  \mathrm{Li}_{1}(e^{i2\pi q_{ij}})+\mathrm{Li}%
_{1}(e^{-i2\pi q_{ij}})\right]  \rightarrow-{\frac{\ln2}{\pi}}+{\frac{\pi}{2}%
}\left[  q_{ij}-\text{floor}(q_{ij})-{\frac{1}{2}}\right]  ^{2}\text{ .}%
\end{equation}

\subsection{$Z\left(  N_{c}\right)  $ symmetry \label{zN}}

The effective potential (\ref{veffa}) is a function of $e^{i2\pi q_{ij}}$,
where $e^{i2\pi q_{j}}$ are the eigenvalues of the thermal Wilson line
$\mathbf{L}$ (\ref{w}). They are the fundamental variables of the matrix
model. While the Wilson line is gauge variant, its eigenvalues are gauge invariant.

Applying a global $Z\left(  N_{c}\right)  $ transformation to the Wilson line,
$\mathbf{L=}$ $\exp\left(  i2\pi\mathbf{q}\right)  $\textbf{,}
\begin{equation}
Z\left(  N_{c}\right)  :\mathbf{L\rightarrow}\text{ }z_{m}\mathbf{L=}\text{
}\exp\left(  i\frac{2\pi m}{N_{c}}\right)  \mathbf{L}\text{ ,}\ \ 0\leq m\leq
N_{c}-1\text{ ,}%
\end{equation}
the eigenvalues undergo a uniform rotation along the unit circle by a constant
angle $2\pi m/N_{c}$\textbf{,}%
\begin{equation}
Z\left(  N_{c}\right)  :\text{\ }e^{i2\pi q_{j}}\rightarrow e^{i2\pi\left(
q_{j}+\frac{m}{N_{c}}\right)  }\text{\ .} \label{cp}%
\end{equation}
Due to the periodicity in $e^{i2\pi q_{j}}\rightarrow$ $e^{i2\pi\left(
q_{j}+1\right)  }$, it is sufficient to restrict the eigenvalues of the matrix
$\mathbf{q}$ to lie in the interval $-1\leq q_{j}\leq1.$ Thus, the $Z\left(
N_{c}\right)  $ transformation (\ref{cp}) is associated with a constant shift
of the matrix elements $q_{j}$,%
\begin{equation}
Z\left(  N_{c}\right)  :q_{j}\rightarrow\left\{
\begin{array}
[c]{c}%
\text{ }q_{j}+\frac{m}{N_{c}}\text{ \ \ \ \ \ \ \ if \ \ }q_{j}+\frac{m}%
{N_{c}}\leq1\\
q_{j}+\frac{m}{N_{c}}-1\text{ \ \ \ if \ \ }q_{j}+\frac{m}{N_{c}}>1
\end{array}
\right.  \text{ .} \label{cp1}%
\end{equation}
This implies that the differences $q_{ij}$ $=q_{i}-q_{j}$ either vanish or
equal $\pm1,$ leaving $e^{i2\pi q_{ij}}$ invariant,
\begin{equation}
Z\left(  N_{c}\right)  :q_{ij}\rightarrow\left\{
\begin{array}
[c]{c}%
q_{ij}\\
q_{ij}\pm1
\end{array}
\right.  \text{ , \ \ }e^{i2\pi q_{ij}}\rightarrow e^{i2\pi q_{ij}}\text{
\ .\ } \label{zn}%
\end{equation}
Therefore, our effective potential is manifestly $Z\left(  N_{c}\right)  $
symmetric and periodic in $q_{ij}\rightarrow q_{ij}+1$. Note that in the
confining vacuum the $Z\left(  N_{c}\right)  $ transformation corresponds to
cyclic permutations of the eigenvalues of the Wilson line $\mathbf{L}%
_{c}\mathbf{=}$ $\exp\left(  i2\pi\mathbf{q}_{c}\right)  $ (\ref{wlc}) and of
the matrix $\mathbf{q}_{c}$ (\ref{s1}).

\section{Uniform eigenvalue ansatz \label{ua}}

\begin{figure}[t!]
\begin{center}
\includegraphics[width=0.45\textwidth]{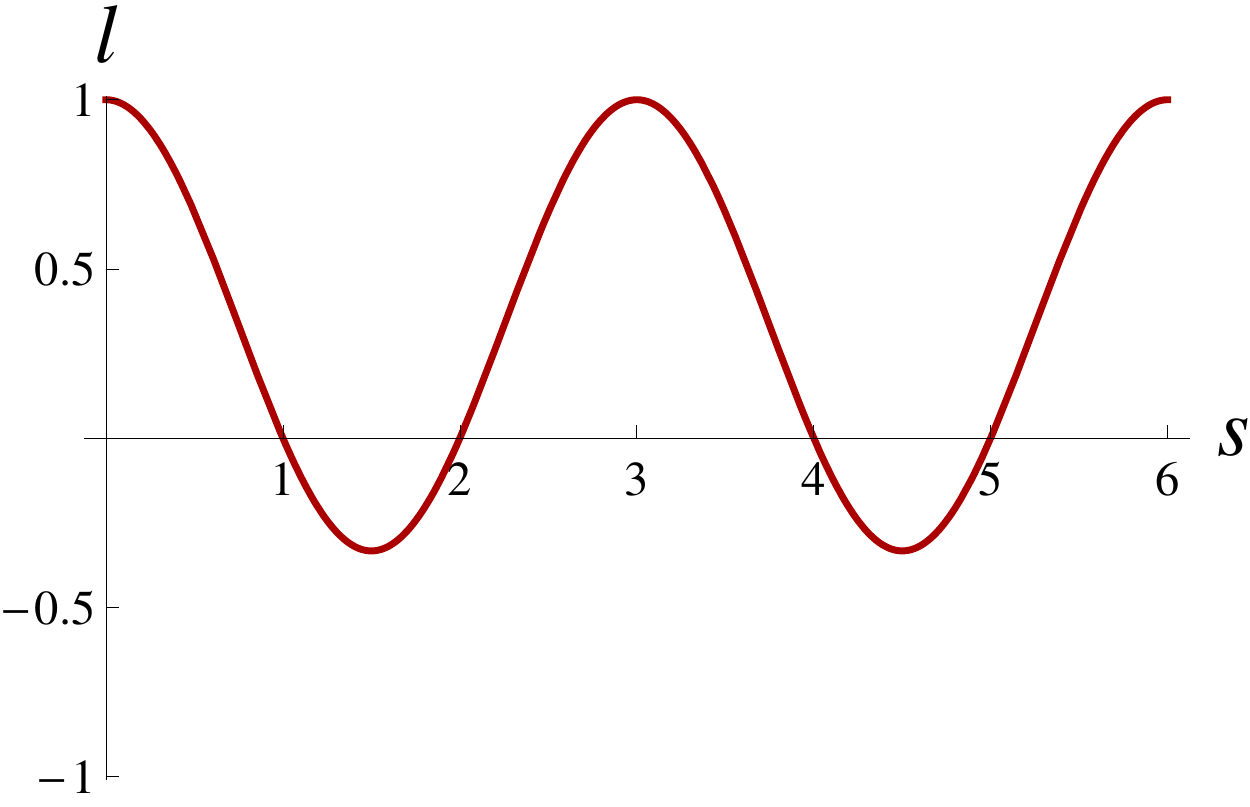}
\hspace{10pt}
\includegraphics[width=0.45\textwidth]{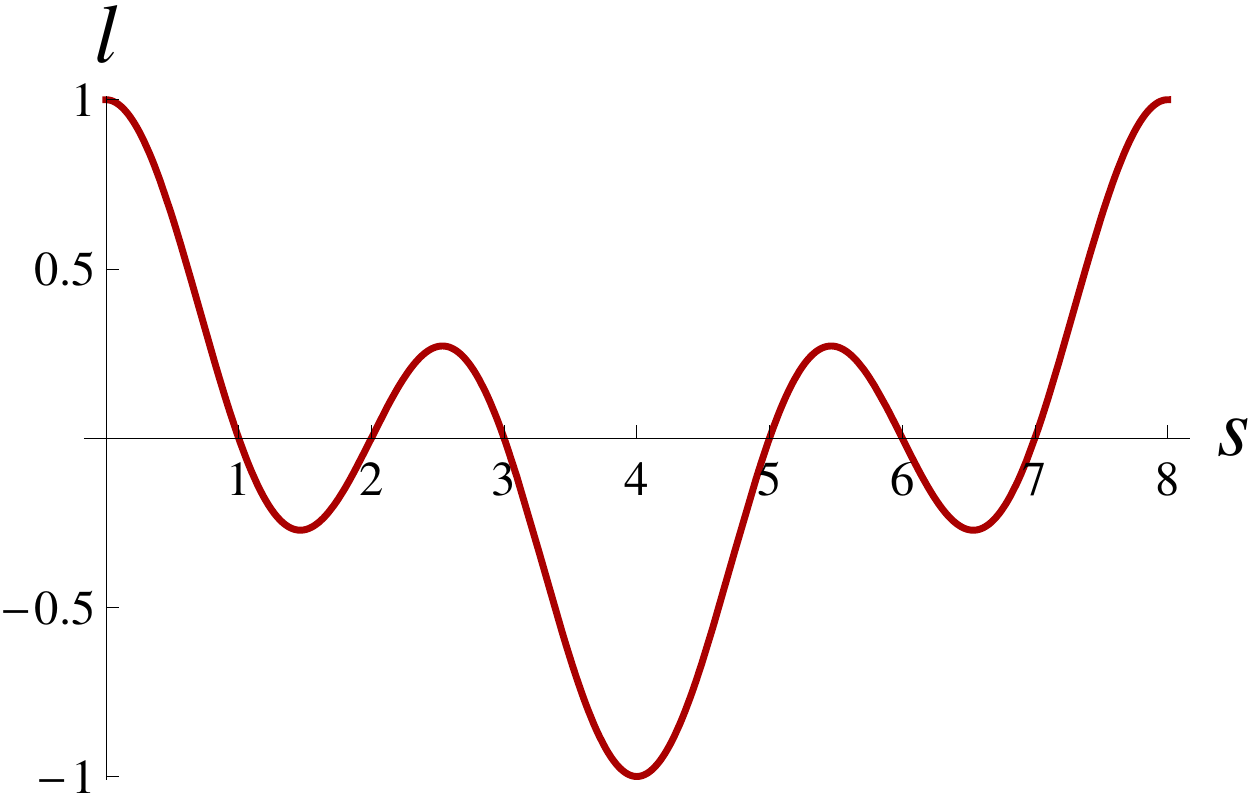}
\includegraphics[width=0.45\textwidth]{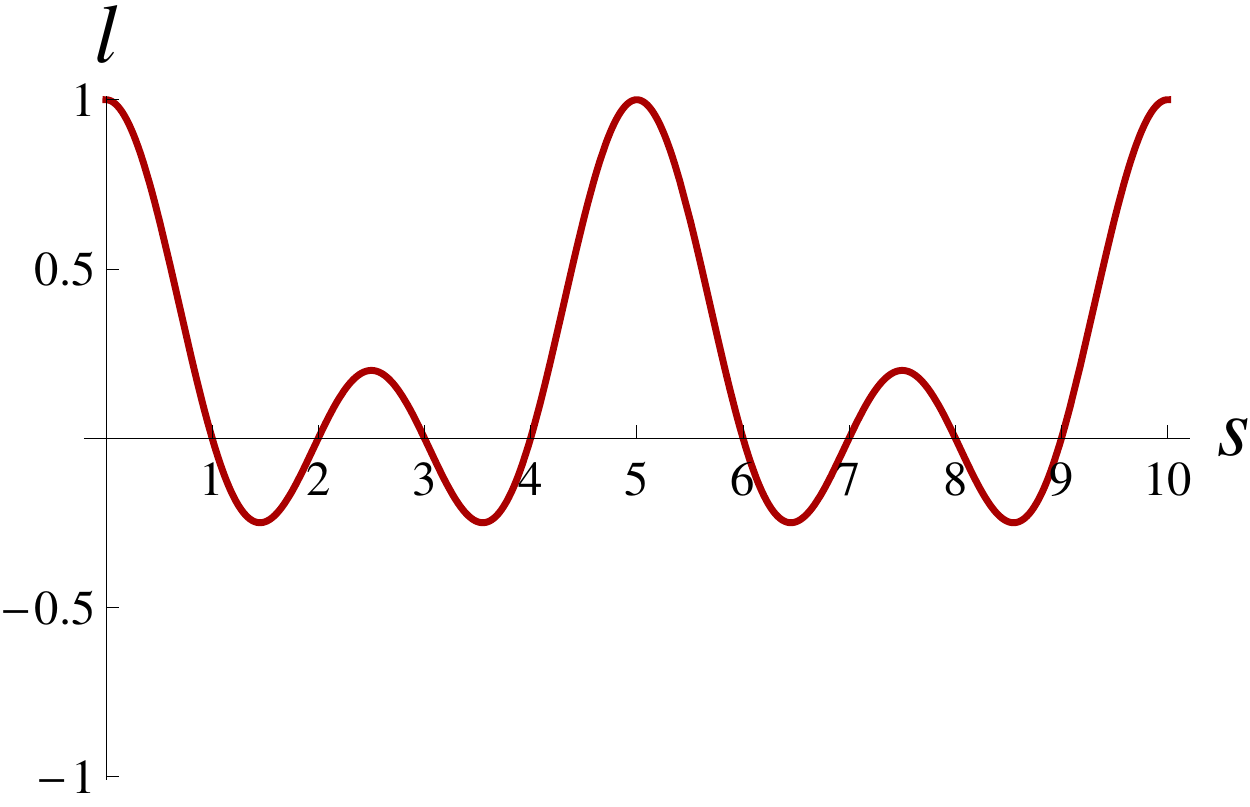}
\hspace{10pt}
\includegraphics[width=0.45\textwidth]{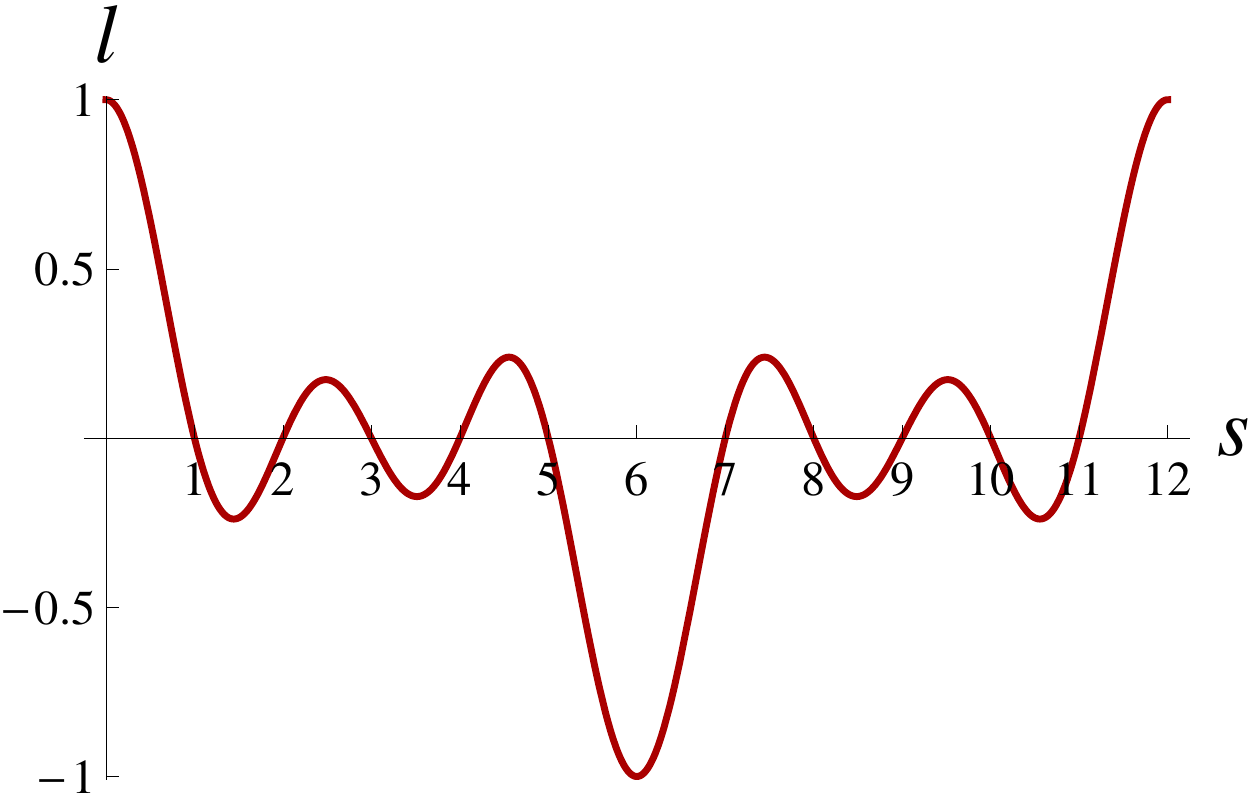}
\end{center}
\caption{ The Polyakov loop as a function of $s$\ in the uniform eigenvalue
Ansatz for $N_{c}=3$\ (left top), $N_{c}=4$\ (right top), $N_{c}=5$\ (left
bottom), $N_{c}=6$\ (right bottom). }%
\label{pl}%
\end{figure}

In this section we present the numerical and analytical solution for the
potential using a uniform eigenvalue Ansatz \cite{Dumitru:2012fw}, which
guarantees that the Polyakov loop is always real.

We are interested in studying the region between the perturbative and
confining vacuum, $0\leq l\leq1$, indicated by the dashed lines in Fig.
\ref{flowers}. In this case, a suitable parametrization of the matrix
$\mathbf{q}$ is given by a straight line from $\mathbf{q}_{0}=0$ to
$\mathbf{q}_{c}$ (\ref{yc}),%
\begin{equation}
\mathbf{q}(s)=s\mathbf{q}_{c}\ \text{, \ \ \ }\ 0\leq s\leq1\text{ ,}
\label{s1}%
\end{equation}
where $\mathbf{q}(s)$ is a diagonal matrix with the eigenvalues
\begin{equation}
q_{j}\left(  s\right)  =\frac{N_{c}-2j+1}{2N_{c}}s\text{ .} \label{ansatz}%
\end{equation}
For an even number of colors, there are $M$ pairs of eigenvalues, $\pm q_{j}$,
\ $j=1,...,M$, where $N_{c}=2M$. For odd $N_{c}$, there are again $M$ pairs of
eigenvalues, $\pm q_{j}$, \ $j=1,...,M,$ where $N_{c}=2M+1$, and the remaining
eigenvalue is zero. All $q_{j}$'s have the constant spacing
\begin{equation}
q_{j}-q_{j+1}=\frac{s}{N_{c}}\text{ .}%
\end{equation}
The eigenvalues of the associated Wilson line are distributed about the unit
circle, with a spacing $2\pi s/N_{c}$,
\begin{equation}
\mathbf{L}=\exp\left(  i2\pi s\,\mathbf{q}_{c}\right)  \text{ .} \label{wlu}%
\end{equation}
In the large-$N_{c}$ limit, this ansatz generates a uniform eigenvalue density
$q_{j}\rightarrow q(x)$, where $x=j/N_{c}$, see Sec. \ref{LN}.

\begin{figure}[t!]
\begin{center}
\includegraphics[width=0.6\textwidth]{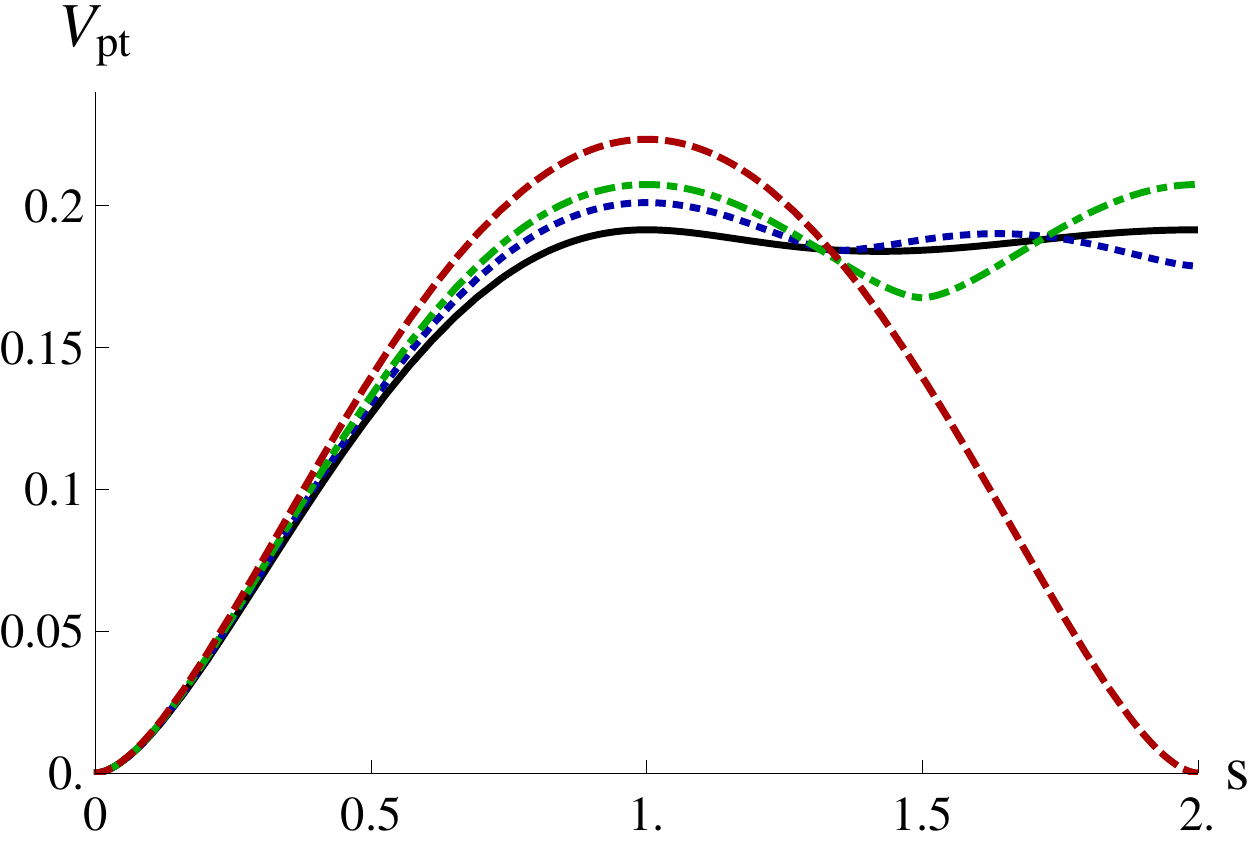}
\end{center}
\caption{The perturbative potential $V_{pt}(s)$ as a function of $s$ in the
uniform eigenvalue ansatz. The plots are shown for $N_{c}=2$ (dashed line),
$N_{c}=3$ (dashed-dotted line), $N_{c}=4$ (dotted line), and in the
large-$N_{c}$ limit (solid line).}%
\label{ptall2}%
\end{figure}We stress that in $d=3+1$ dimensions the uniform eigenvalue Ansatz
applies for two or three colors, but is not an exact solution for four or more
colors, see Ref. \cite{Dumitru:2012fw}. It is, however, rather close to the
exact numerical solution. For $N_{c}=4$ to $N_{c}=7$ in $d=3+1$ dimensions,
for instance, the difference between the uniform eigenvalue Ansatz and the
exact solution is less than $\sim1\%$ for all thermodynamic quantities and for
the expectation value of the Polyakov loop, even at $T_{d}$ where the
differences are naturally greatest. This is our main motivation to employ the
uniform Ansatz. A general parametrization of the background $\mathcal{A}_{0}$
field is discussed in Sec. \ref{meta}.

\subsection{Polyakov loop}

Taking the trace of the Wilson line (\ref{wlu}) gives the Polyakov loop for
even $N_{c}$,
\begin{align}
l(s)  &  =\frac{1}{N_{c}}\sum_{j=1}^{N_{c}/2}2\cos\left[  2\pi q_{j}\left(
s\right)  \right] \label{pl1}\\
&  =\frac{1}{N_{c}}\sum_{j=1}^{N_{c}/2}2\cos\left(  \pi\frac{N_{c}-2j+1}%
{N_{c}}s\right)  \text{\ ,}\nonumber
\end{align}
and for odd $N_{c}$,%
\begin{align}
l(s)  &  =\frac{1}{N_{c}}\left[  \sum_{j=1}^{\left(  N_{c}-1\right)  /2}%
2\cos\left[  2\pi q_{j}\left(  s\right)  \right]  +1\right] \label{pl2}\\
&  =\frac{1}{N_{c}}\left[  \sum_{j=1}^{\left(  N_{c}-1\right)  /2}2\cos\left(
\pi\frac{N_{c}-2j+1}{N_{c}}s\right)  +1\right]  \text{ ,}\nonumber
\end{align}
where we used Eq. (\ref{ansatz}) for $q_{j}$. The perturbative vacuum, $l=1,$
is realized for $s=0$, while the confining vacuum, $l_{c}=0,$ is at $s_{c}=1$.

In Fig. \ref{pl} we plot the Polyakov loop in the uniform eigenvalue ansatz as
a function of $s$ for different numbers of colors. The function $l(s)$\ is
periodic in $s\rightarrow s+2N_{c}$ and symmetric around $s=0$, and $s=N_{c}$.
For any $s,$\ the Polyakov loop takes values along the real axis within the
solid lines in Fig. \ref{flowers}, and thus represents a possible physical solution.

\subsection{The Potential \label{vnonpert}}

\begin{figure}[t!]
\begin{center}
\includegraphics[width=0.6\textwidth]{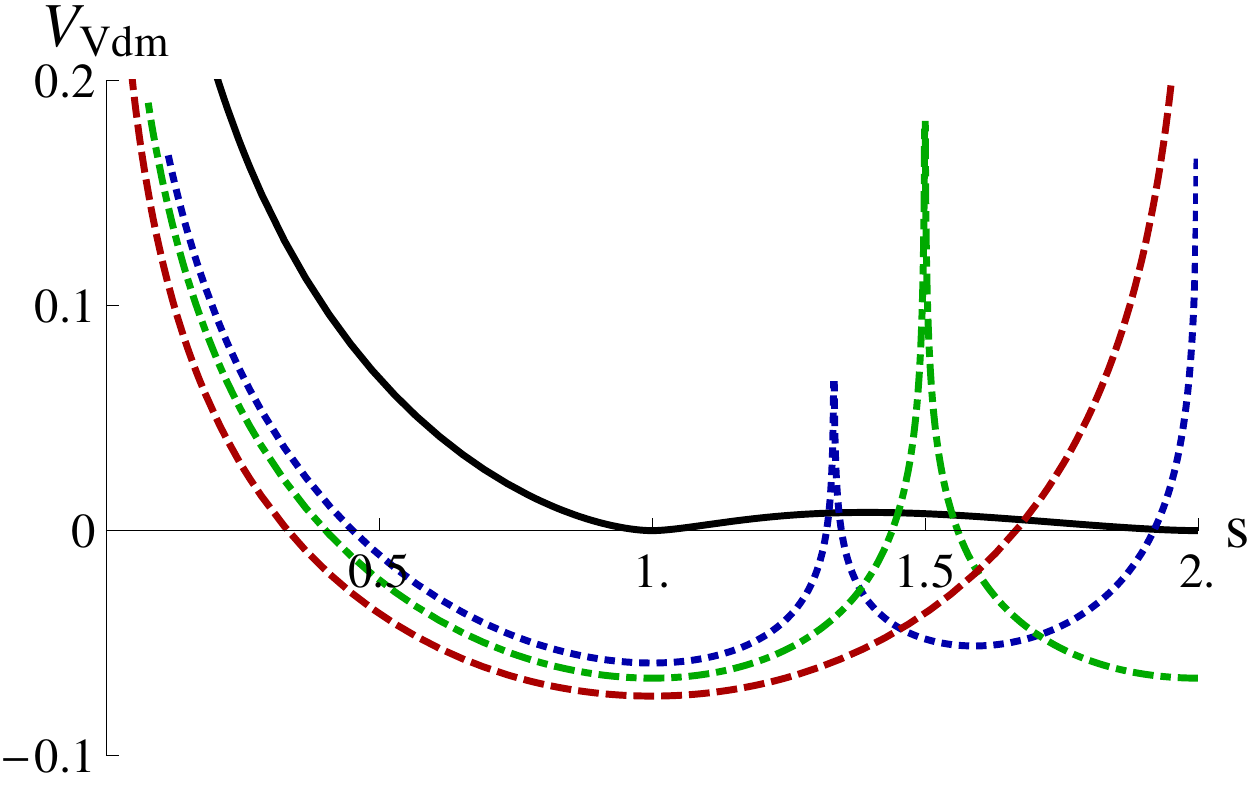}
\end{center}
\caption{The Vandermonde term $V_{Vdm}(s)$ as a function of $s$ for $N_{c}=2$
(dashed line), $N_{c}=3$ (dashed-dotted line), $N_{c}=4$ (dotted line), and in
the large-$N_{c}$ limit (solid line).}%
\label{ptall3}%
\end{figure}

\subsubsection{Perturbative and nonperturbative potential}

\begin{figure}[t!]
\begin{center}
\includegraphics[width=0.6\textwidth]{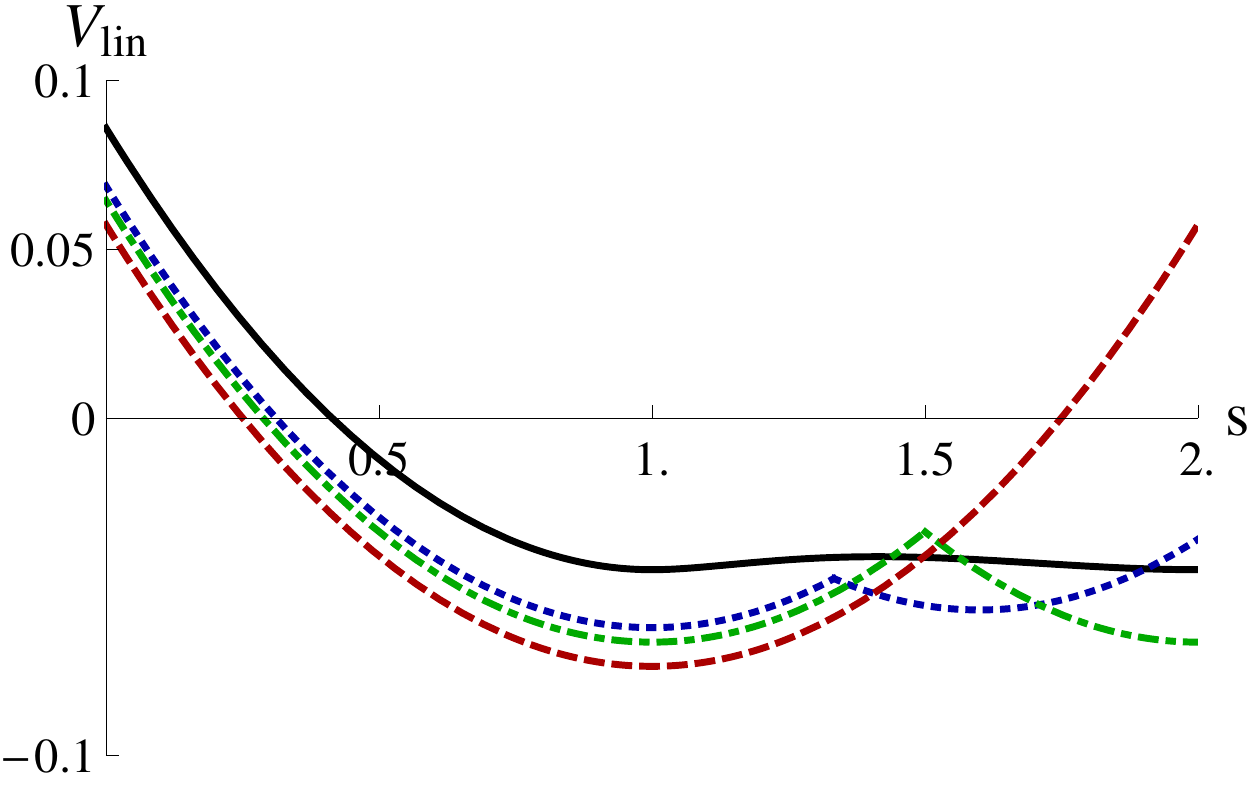}
\end{center}
\caption{The linear term $V_{lin}(s)$ as a function of $s$ for $N_{c}=2$
(dashed line), $N_{c}=3$ (dashed-dotted line), $N_{c}=4$ (dotted line), and in
the large-$N_{c}$ limit (solid line).}%
\label{ptall4}%
\end{figure}

In the uniform eigenvalue ansatz $\mathbf{q}\left(  s\right)  $ (\ref{ansatz}%
), the potential becomes a function of $s$ through the dependence of $q_{ij}$
on $s$,
\begin{equation}
q_{ij}\equiv q_{ij}\left(  s\right)  =\frac{i-j}{N_{c}}s\text{ .} \label{qa}%
\end{equation}
In Figs. \ref{ptall2}, \ref{ptall3}, and \ref{ptall4} we plot the perturbative
potential $V_{pt}(s)$\ (\ref{vpt_1}), the Vandermonde term $V_{Vdm}%
(s)$\ (\ref{vdm22}), and the linear term $V_{lin}(s)$ (\ref{Vlin}) for
different numbers of colors in the region between the perturbative vacuum,
$s=0,$ and the confining vacuum, $s_{c}=1$. The perturbative potential
exhibits a minimum at $s=0$ and a maximum at $s_{c}=1$, while the
nonperturbative terms both have a minimum at the confining vacuum $s_{c}$.

We note that the Vandermonde term is divergent for $s=0$. However, in the
presence of a Vandermonde or a linear term the condensate for $s$ never
identically vanishes. A nonzero condensate for $s$ will also ensure that the
thermodynamical quantities computed at the minimum of the effective potential
remain finite.

\subsubsection{Effective potential \label{veff}}

Before we discuss the effective potential, it is useful to make some remarks
on the summation over the eigenvalues $q_{ij}$. The perturbative and
nonperturbative terms in Eqs. (\ref{vpt_1}), (\ref{vnpt_A}), and
(\ref{vnpt_B}) are even functions in $q_{ij}$ of the form
\begin{equation}
\sum_{1\leq i<j\leq N_{c}}f\left(  \left\vert q_{ij}\right\vert \right)
=\frac{1}{2}\sum_{1\leq i,j\leq N_{c};\text{ }i\neq j}f\left(  \left\vert
q_{ij}\right\vert \right)  \ , \label{sum}%
\end{equation}
In the uniform eigenvalue Ansatz, Eq. (\ref{qa}), this sum can be written as a
function of\ the parameter $s,$%
\begin{align}
\frac{1}{2}\sum_{1\leq i,j\leq N_{c};\text{ }i\neq j}f(\left\vert
q_{ij}\right\vert )  &  =\frac{1}{2}\sum_{1\leq i,j\leq N_{c};\text{ }i\neq
j}f\left(  \frac{\left\vert i-j\right\vert }{N_{c}}s\right) \nonumber\\
&  =\sum_{j=1}^{N_{c}-1}\left(  N_{c}-j\right)  f\left(  \frac{j}{N_{c}%
}s\right)  \ \text{.} \label{sum1}%
\end{align}
It is convenient to decompose the effective potential into two parts,
\begin{equation}
V_{eff}\left(  s\right)  =-\text{ }(N_{c}^{2}-1)T^{3}\left[  \frac
{\zeta\left(  3\right)  }{2\pi}\left(  1-\frac{T_{d}}{T}C_{3}\right)  -\left(
1-\frac{T_{d}}{T}C_{2}\right)  V(s,a)\right]  \ , \label{veff1a}%
\end{equation}
where $V(s,a)$ denotes the $s$-dependent terms.

Using Eq. (\ref{sum1}),\ in the presence of the Vandermonde determinant we
derive
\begin{align}
V(s,a)  &  =\sum_{j=1}^{N_{c}-1}-{\frac{N_{c}-j}{N_{c}^{2}-1}}\ {\frac{1}%
{2\pi}}\,\left[  \text{$\mathrm{Li}$}_{3}\left(  e^{i2\pi\frac{j}{N_{c}}%
s}\right)  +\text{$\mathrm{Li}$}_{3}\left(  e^{-i2\pi\frac{j}{N_{c}}s}\right)
-2\zeta\left(  3\right)  \right] \nonumber\\
&  +a(T)\sum_{j=1}^{N_{c}-1}{\frac{N_{c}-j}{N_{c}^{2}-1}}\ {\frac{1}{2\pi}%
}\,\left[  \text{$\mathrm{Li}$}_{1}\left(  e^{i2\pi\frac{j}{N_{c}}s}\right)
+\text{$\mathrm{Li}$}_{1}\left(  e^{-i2\pi\frac{j}{N_{c}}s}\right)  \right]
\ , \label{veff3}%
\end{align}
where
\begin{equation}
a(T)=\frac{C_{1}}{\frac{T}{T_{d}}-C_{2}}\ \label{vA}%
\end{equation}
is a temperature-dependent parameter.

For the linear nonperturbative term (\ref{Vlin}), in the relevant region
between $s=0$ and $s_{c}=1$, the effective potential is obtained by replacing
the term $\sim a(T)$ in $V(s,a)$ as%
\begin{equation}
{\frac{1}{2\pi}}\left[  \text{$\mathrm{Li}$}_{1}\left(  e^{i2\pi\frac{j}%
{N_{c}}s}\right)  +\text{$\mathrm{Li}$}_{1}\left(  e^{-i2\pi\frac{j}{N_{c}}%
s}\right)  \right]  \rightarrow-{\frac{\ln2}{\pi}}+{\frac{\pi}{2}}\left(
\frac{j}{N_{c}}s-{\frac{1}{2}}\right)  ^{2}\text{ .}%
\end{equation}
\begin{figure}[t!]
\begin{center}
\includegraphics[width=0.45\textwidth]{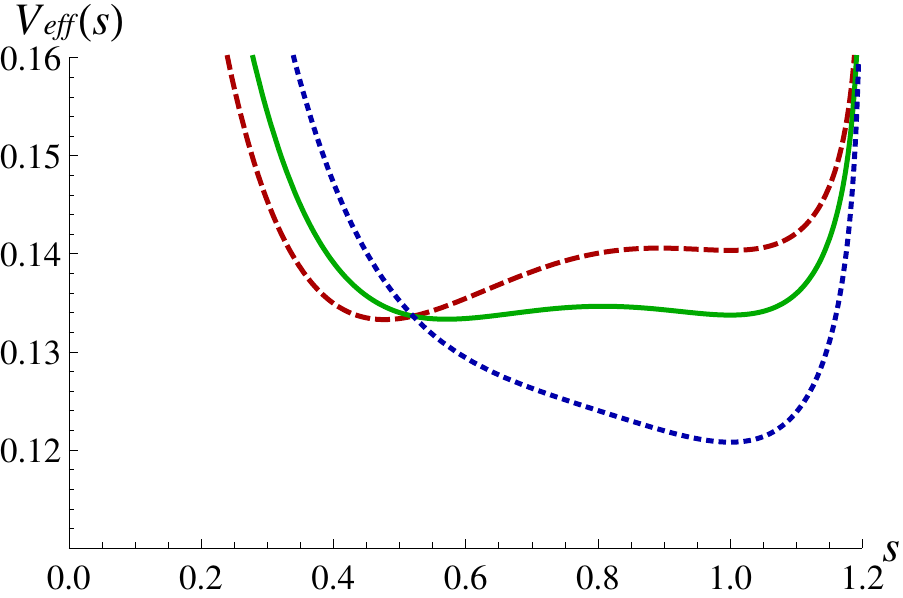}
\hspace{10pt}
\includegraphics[width=0.45\textwidth]{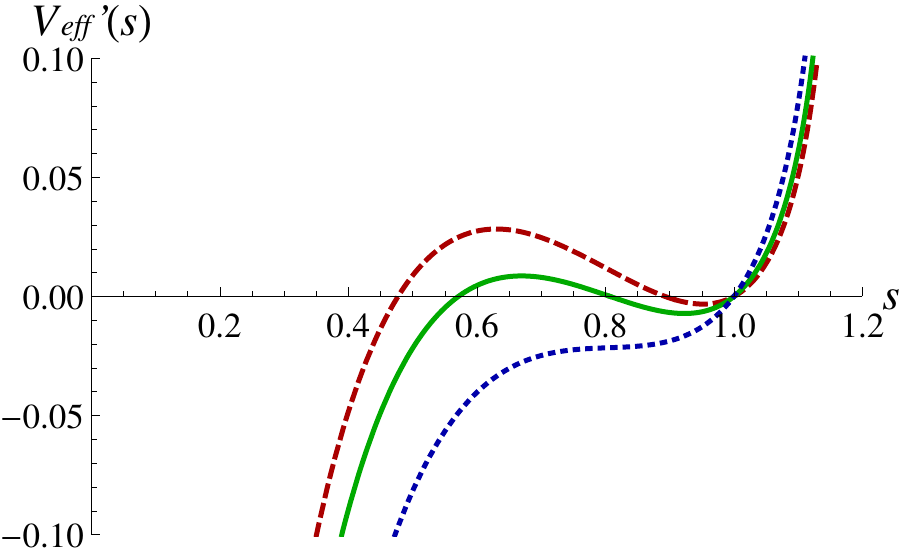}
\end{center}
\caption{The effective potential $V_{eff}(s)$ for $N_{c}=6$, and its
derivative $V_{eff}^{\prime}(s)$ as a function of the variable $s.$ The plots
are obtained by using the Vandermonde term for three different values of $a:$
$a<a_{d}$ (dashed line) represents the semi-QGP, at $a=a_{d}$ (solid line) the
phase transition to confinement takes place, and for $a>a_{d}$ (dotted line)
the system is in the confined phase.}%
\label{veffmin6}%
\end{figure}

\subsection{Large-$N_{c}$ Limit \label{LN}}

In order to study the large-$N_{c}$ limit, we introduce the variable
\begin{equation}
x=j/N_{c}\ .
\end{equation}
Then, the eigenvalues become a function of $x,$
\begin{equation}
q_{j}\rightarrow q(x)\ ,
\end{equation}
and the sum over $j$ can be written as an integral over $x$,
\begin{equation}
\sum_{j=1}^{N_{c}-1}{\frac{N_{c}-j}{N_{c}^{2}-1}}f\left(  \frac{j}{N_{c}%
}s\right)  \underset{N_{c}\rightarrow\infty}{=}\int_{0}^{1}dx\,\left(
1-x\right)  f\left(  xs\right)  \ . \label{int}%
\end{equation}
Using this integral, the $s$-dependent part of the effective potential with
the Vandermonde term in Eq. (\ref{veff3}) takes the form
\begin{align}
V(s,a)  &  =-\int_{0}^{1}dx{\frac{\left(  1-x\right)  }{2\pi}}\left[
\text{$\mathrm{Li}$}_{3}\left(  e^{i2\pi xs}\right)  +\text{$\mathrm{Li}$}%
_{3}\left(  e^{-i2\pi xs}\right)  -2\zeta\left(  3\right)  \right]
\label{veffA}\\
&  +a(T)\int_{0}^{1}dx{\frac{\left(  1-x\right)  }{2\pi}}\left[
\text{$\mathrm{Li}$}_{1}\left(  e^{i2\pi xs}\right)  +\text{$\mathrm{Li}$}%
_{1}\left(  e^{-i2\pi xs}\right)  \right] \nonumber\\
&  =-{\frac{1}{2\pi}}\left[  {\frac{\text{$\mathrm{Li}$}_{5}\left(  e^{i2\pi
xs}\right)  +\text{$\mathrm{Li}$}_{5}\left(  e^{-i2\pi xs}\right)  -2\zeta
(5)}{4\pi^{2}s^{2}}}-\zeta(3)\right] \nonumber\\
&  +{\frac{a(T)}{2\pi}}\left[  {\frac{\text{$\mathrm{Li}$}_{3}\left(  e^{i2\pi
xs}\right)  +\text{$\mathrm{Li}$}_{3}\left(  e^{-i2\pi xs}\right)  -2\zeta
(3)}{4\pi^{2}s^{2}}}\right]  \ .\nonumber
\end{align}
For the linear nonperturbative term (\ref{Vlin}), the large-$N_{c}$ limit of
effective potential in the relevant region between $s=0$ and $s_{c}=1$ is
given by replacing the term $\sim a(T)$ in Eq. (\ref{veffA}) as
\begin{equation}
{\frac{1}{2\pi}}{\frac{\text{$\mathrm{Li}$}_{3}\left(  e^{i2\pi xs}\right)
+\text{$\mathrm{Li}$}_{3}\left(  e^{-i2\pi xs}\right)  -2\zeta(3)}{4\pi
^{2}s^{2}}}\rightarrow-{\frac{\ln2}{2\pi}}+{\frac{\pi}{48}}[3-2s(2-s)]\ .
\end{equation}
\begin{table}[t!]
\begin{center}%
\begin{tabular}
[c]{c|cccc}\hline
$N_{c}$ & $s_{d}^{-}$ Vdm & $s_{d}^{-}$ lin & $a_{d}$ Vdm & $a_{d}$
lin\\\hline
$2$ & 1. & 1. & 2.77259 & 2.77259\\
$3$ & 0.795883 & 0.768018 & 1.84148 & 2.33328\\
$4$ & 0.735259 & 0.704746 & 1.55426 & 2.20086\\
$5$ & 0.707102 & 0.677182 & 1.41634 & 2.14137\\
$6$ & 0.691139 & 0.662458 & 1.33576 & 2.10918\\
$\infty$ & 0.638659 & 0.628419 & 1.0354 & 2.03424\\\hline
\end{tabular}
\end{center}
\caption{ The values for $s_{d}^{-}=s(T_{d})$ and $a_{d}=a(T_{d})$ for
different $SU(N_{c})$ groups. We use the notation \textquotedblleft
Vdm\textquotedblright\ for the Vandermonde determinant, and \textquotedblleft
lin\textquotedblleft\ for the linear nonperturbative term. At a first-order
phase transition the minimum of the effective potential jumps at the
deconfinement temperature $T_{d}$ from $s_{d}^{-}<1$ to the confining vacuum
$s_{c}=1.$}%
\label{table1}%
\end{table}

\subsection{The order of the phase transition}

Lattice results of Ref. \cite{Liddle:2008kk} indicate that in $d=2+1$
dimensions the phase transition is of second order for $N_{c}\leq3$, of very
weak first order for $N_{c}=4$, and of stronger first order for $N_{c}\geq
5$\textbf{.} In our matrix model the phase transition is of second order for
$N_{c}=2$, while for $N_{c}\geq3$ the transition is of first order. The reason
for the discrepancy between our model and numerical simulations on the lattice
is due to infrared fluctuations \cite{D'Hoker:1981us} not included in our
effective theory, which render the transition second order for three colors in
$d=2+1$ dimensions.

In order to illustrate the behavior of the effective potential (\ref{veff1a})
near $T_{d}$, in Fig. \ref{veffmin6} we show $V_{eff}\left(  s\right)  $, and
$dV_{eff}(s)/ds$ as a function of $s$ for $N_{c}=6$. Depending on the value of
the temperature-dependent parameter $a\left(  T\right)  $ defined in Eq.
(\ref{vA}), we can describe the transition from deconfinement to confinement.

\begin{enumerate}
\item[(i)] At $a(T)=0$ the system is in the complete QGP phase, with a global
minimum at the perturbative vacuum $s=0$, and a maximum at the confining
vacuum $s_{c}=1$.

\item[(ii)] In the semi-QGP region, $0<a(T)<a_{d}$, there is a global minimum
at $0<s<1,$ and a second extremum at $s>1,$ which corresponds to a second
degenerate minimum for $N_{c}=2$, or to a local minimum for $N_{c}\geq3$.

\item[(iii)] At $a(T)=a_{d}$ the transition to confinement takes place. For a
first-order phase transition, the global minimum jumps at $T_{d}$ from a
critical value $s(T_{d})=s_{d}^{-}$ $<1$ to the confining vacuum at $s_{c}=1.$
\end{enumerate}

The first-order phase transition becomes stronger with increasing $N_{c}$.
This is supported by universality class arguments \cite{vonSmekal:2010du,
Strodthoff:2010dz} and numerical simulations on the lattice, where the first
order of the transition becomes more pronounced for larger $N_{c}$
\cite{Liddle:2008kk}. Accordingly, the value for $s_{d}^{-}$ decreases with
increasing number of colors, and approaches a constant value in the
large-$N_{c}$ limit,%
\begin{equation}
\lim_{N_{c}\rightarrow\infty}s_{d}^{-}\rightarrow0.63\text{ .}%
\end{equation}
In Table \ref{table1} we list the values for $s_{d}^{-}=s(T_{d})$, and
$a_{d}=a(T_{d})$ for different $N_{c}$.

\begin{figure}[t!]
\begin{center}
\includegraphics[width=0.45\textwidth]{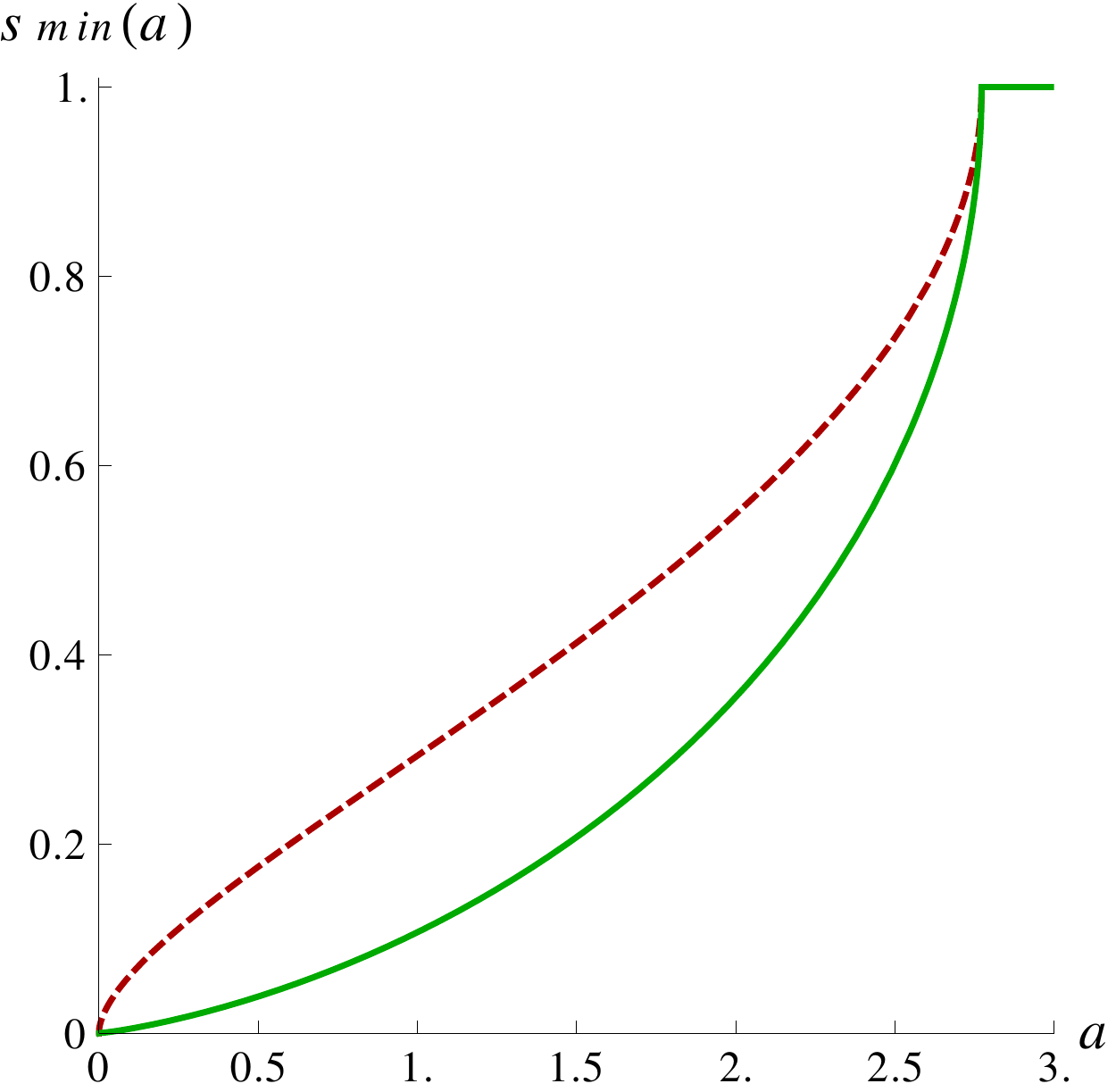}
\hspace{10pt}
\includegraphics[width=0.45\textwidth]{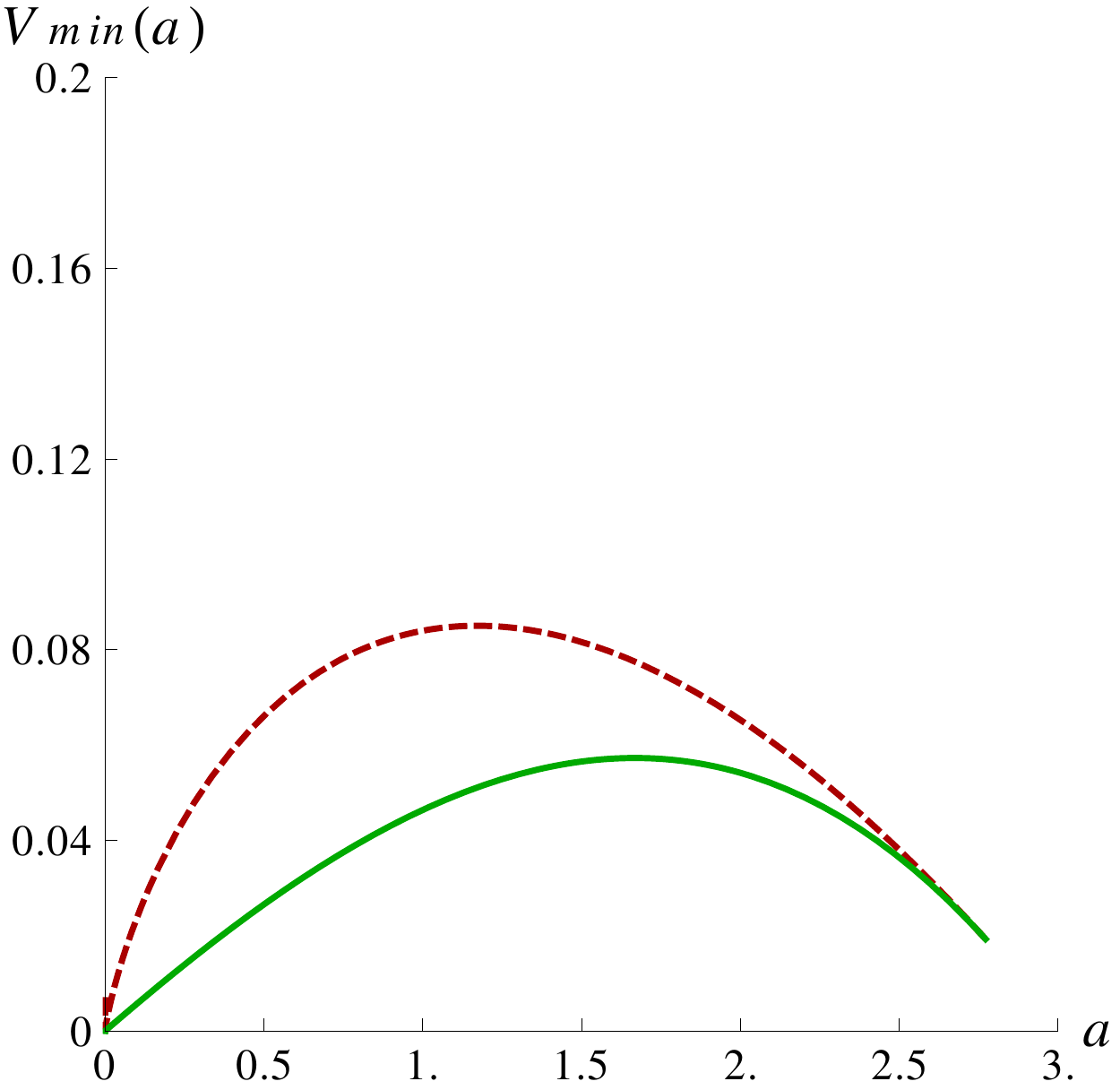}
\includegraphics[width=0.45\textwidth]{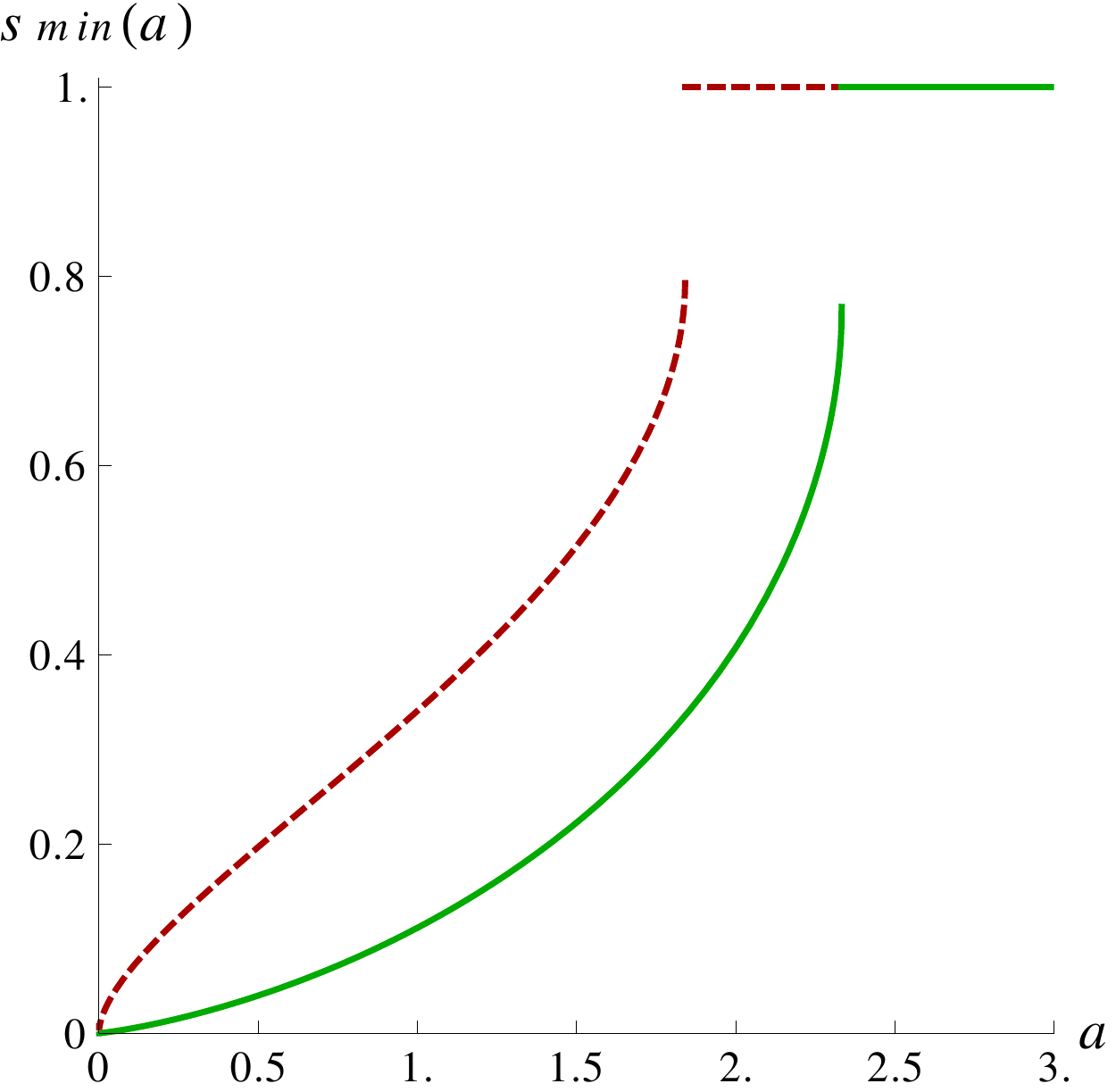}
\hspace{10pt}
\includegraphics[width=0.45\textwidth]{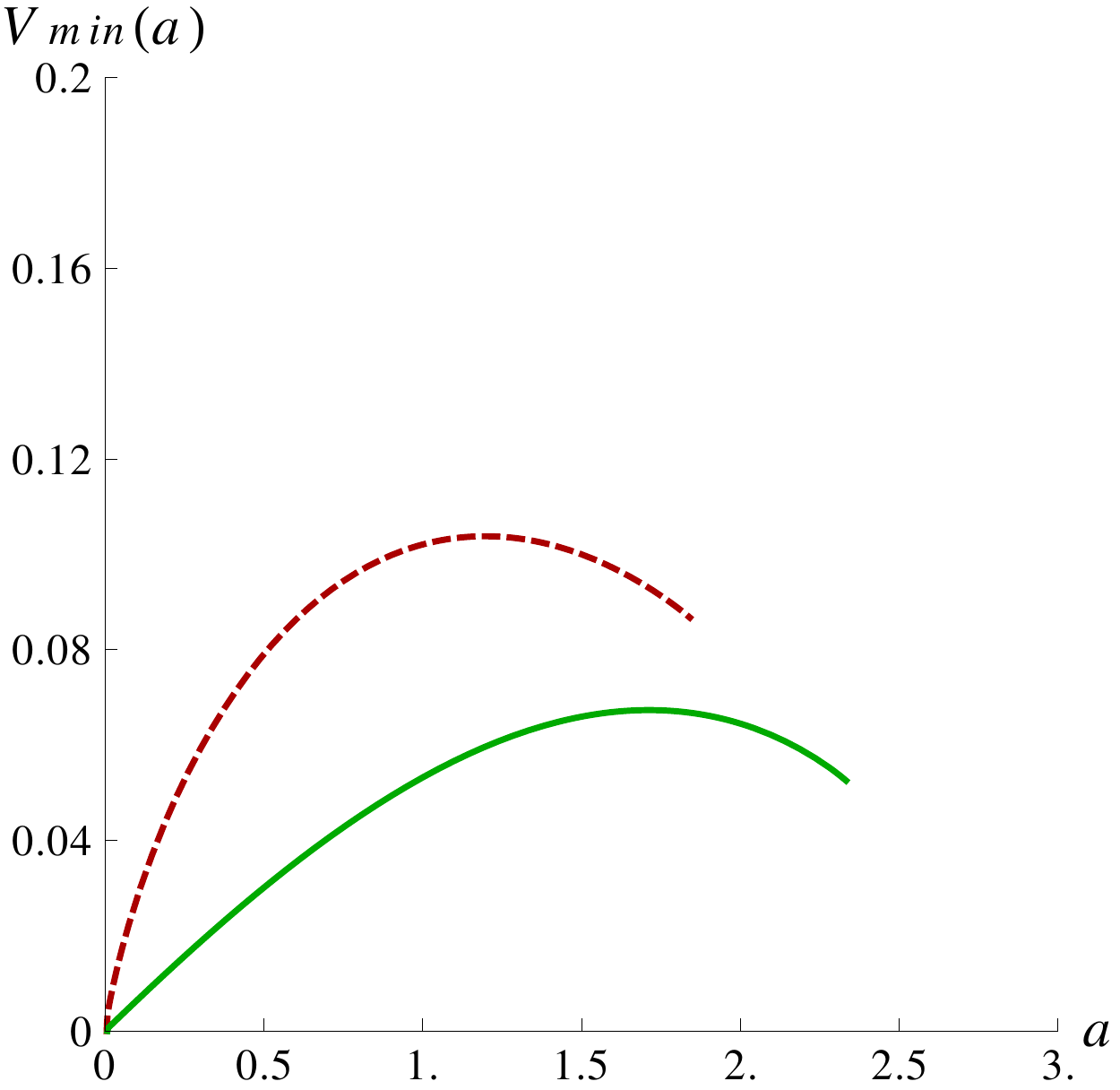}
\end{center}
\caption{Left panel: the minimum of the effective potential as a function of
$a,$ $s_{min}(a),$ for the linear term (solid line), and the Vandermonde
determinant (dashed line). Right panel: the potential at the minimum as a
function of $a,$ $V_{min}(a)$. In the upper panels we show the results for
$N_{c}=2$, and in the lower panels for $N_{c}=3$.}%
\label{qmin23}%
\end{figure}

\subsection{The minimum of the effective potential \label{pT}}

In order to compute the thermodynamic quantities, like pressure and
interaction measure, we first have to determine the minimum of the effective
potential (\ref{veff1a}) as a function of the temperature. The minimum is
obtained by solving numerically the equation
\begin{equation}
\frac{\partial V(s,a)}{\partial s}{\Big|}_{s=s_{min}}=0\ \label{vmin}%
\end{equation}
in the region between the perturbative and confining vacuum, $0\leq a\leq
a_{d}$. This defines the minimum $s_{min}(a)$ as a function of the parameter
$a$.

\begin{figure}[t!]
\begin{center}
\includegraphics[width=0.45\textwidth]{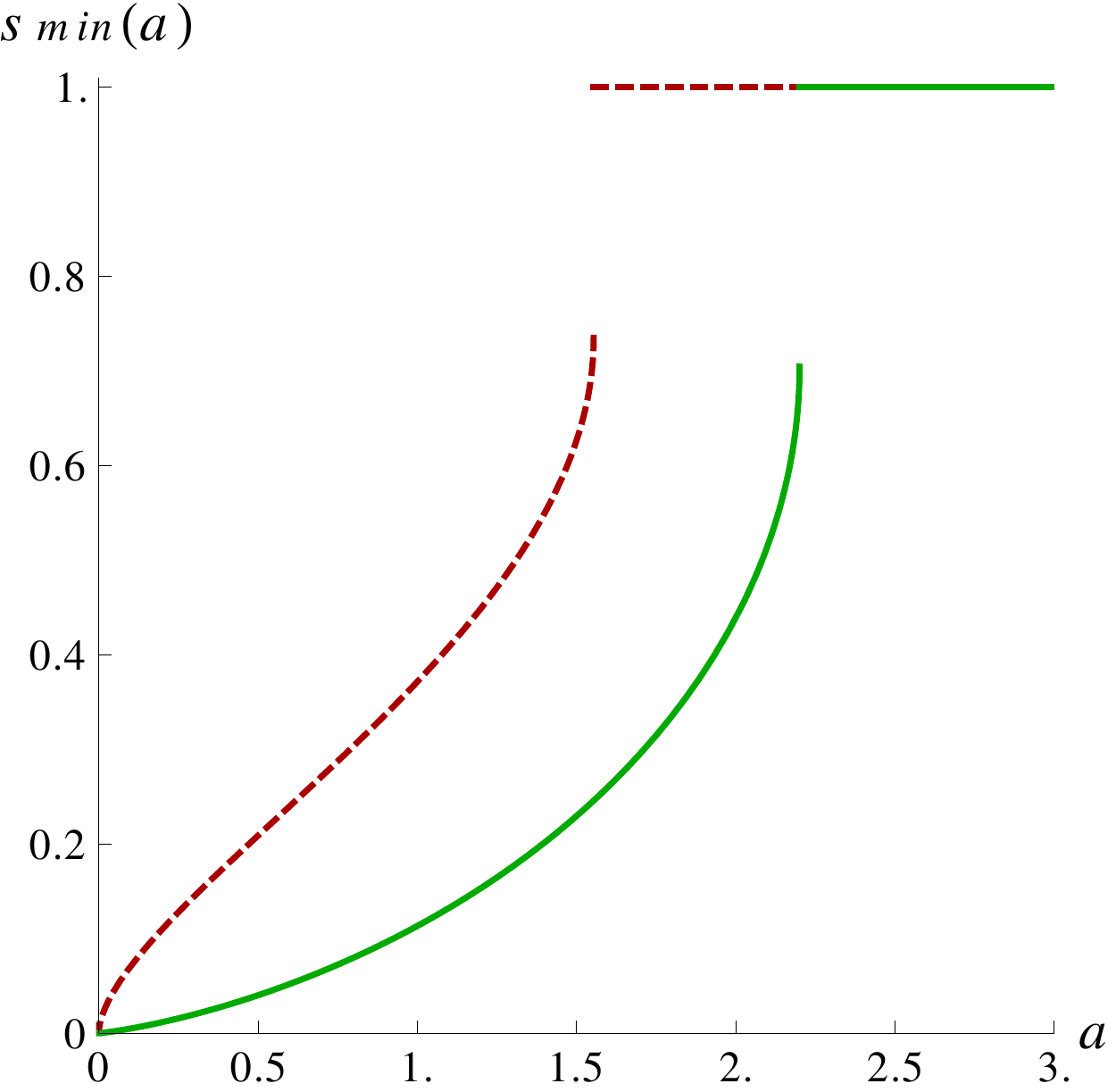}
\hspace{0pt}
\includegraphics[width=0.45\textwidth]{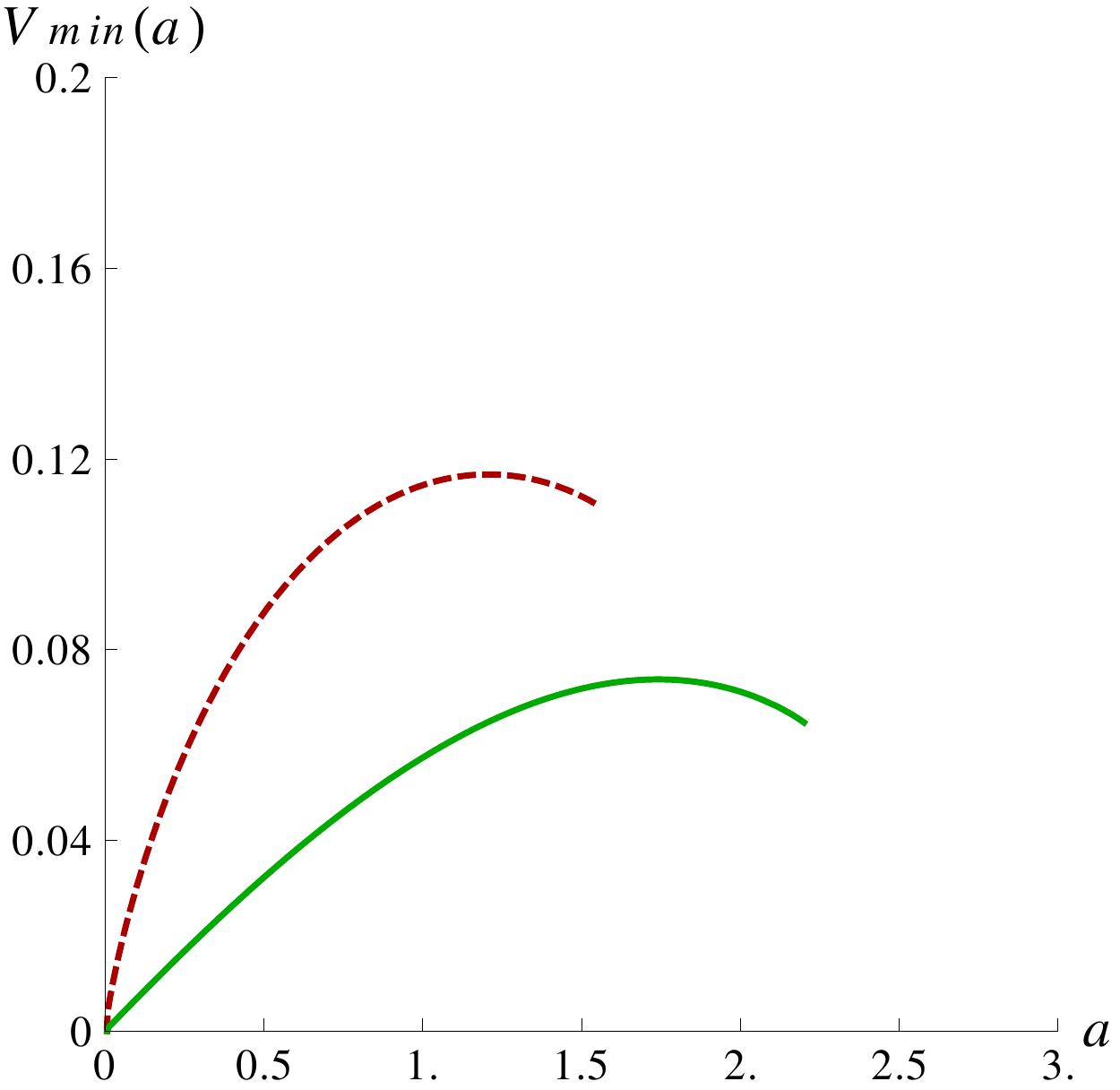}
\includegraphics[width=0.45\textwidth]{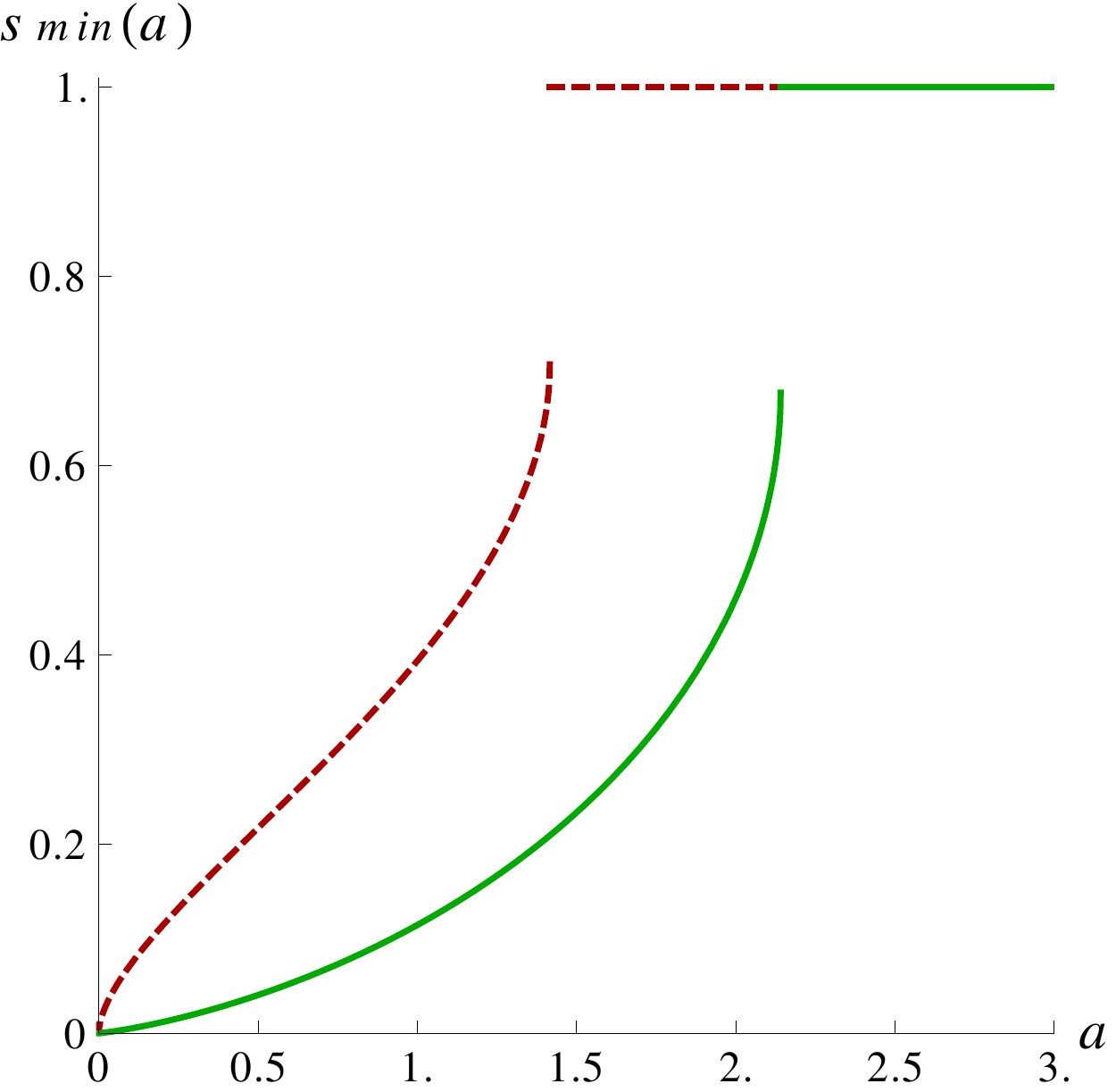}
\hspace{0pt}
\includegraphics[width=0.45\textwidth]{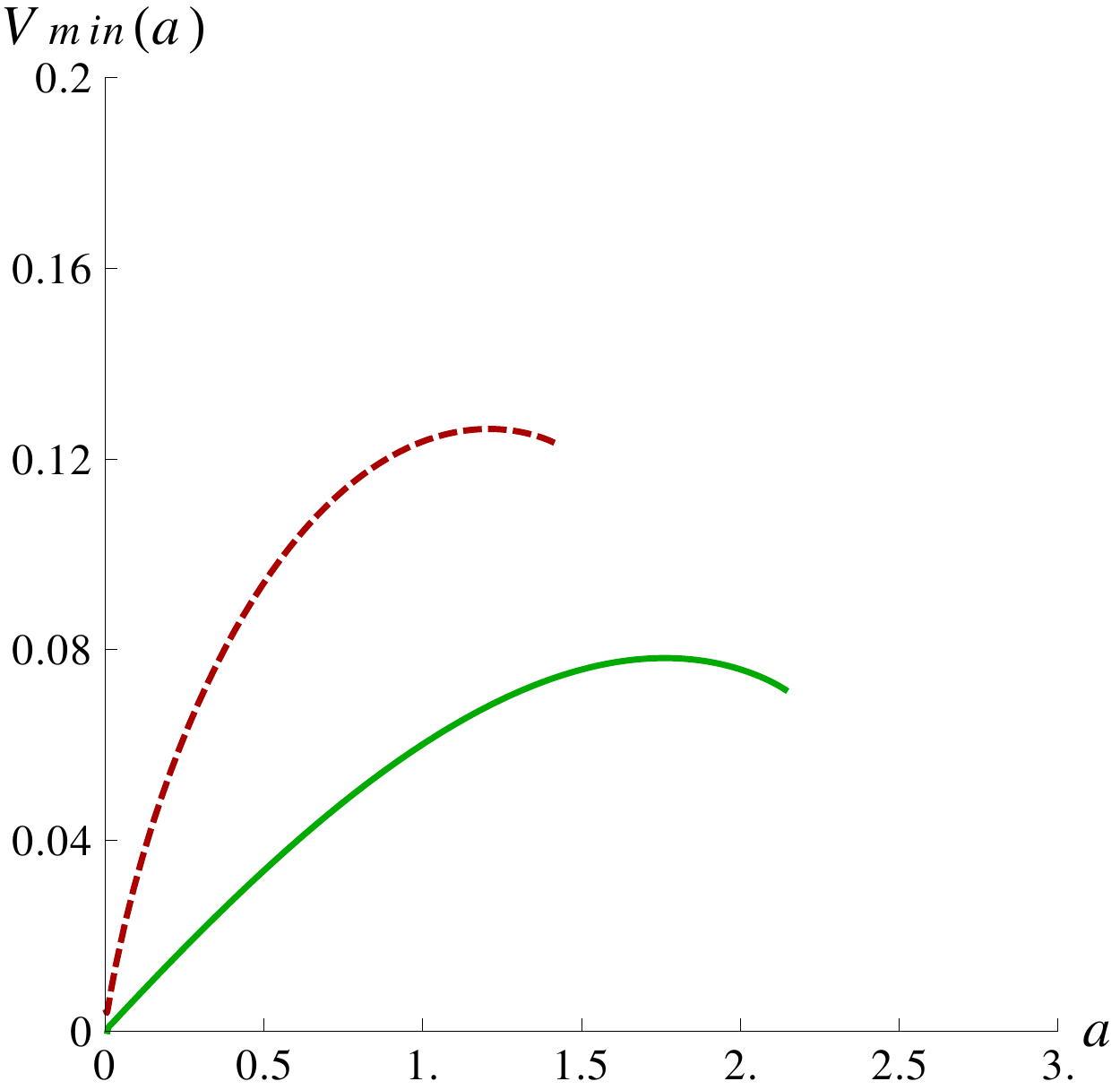}
\end{center}
\caption{Left panel: the minimum of the effective potential as a function of
$a,$ $s_{min}(a),$ for the linear term (solid line), and the Vandermonde
determinant (dashed line). Right panel: the potential at the minimum as a
function of $a,$ $V_{min}(a)$. In the upper panels we show the results for
$N_{c}=4$, and in the lower panels for $N_{c}=5$.}%
\label{qmin45}%
\end{figure}

\begin{figure}[t!]
\begin{center}
\includegraphics[width=0.45\textwidth]{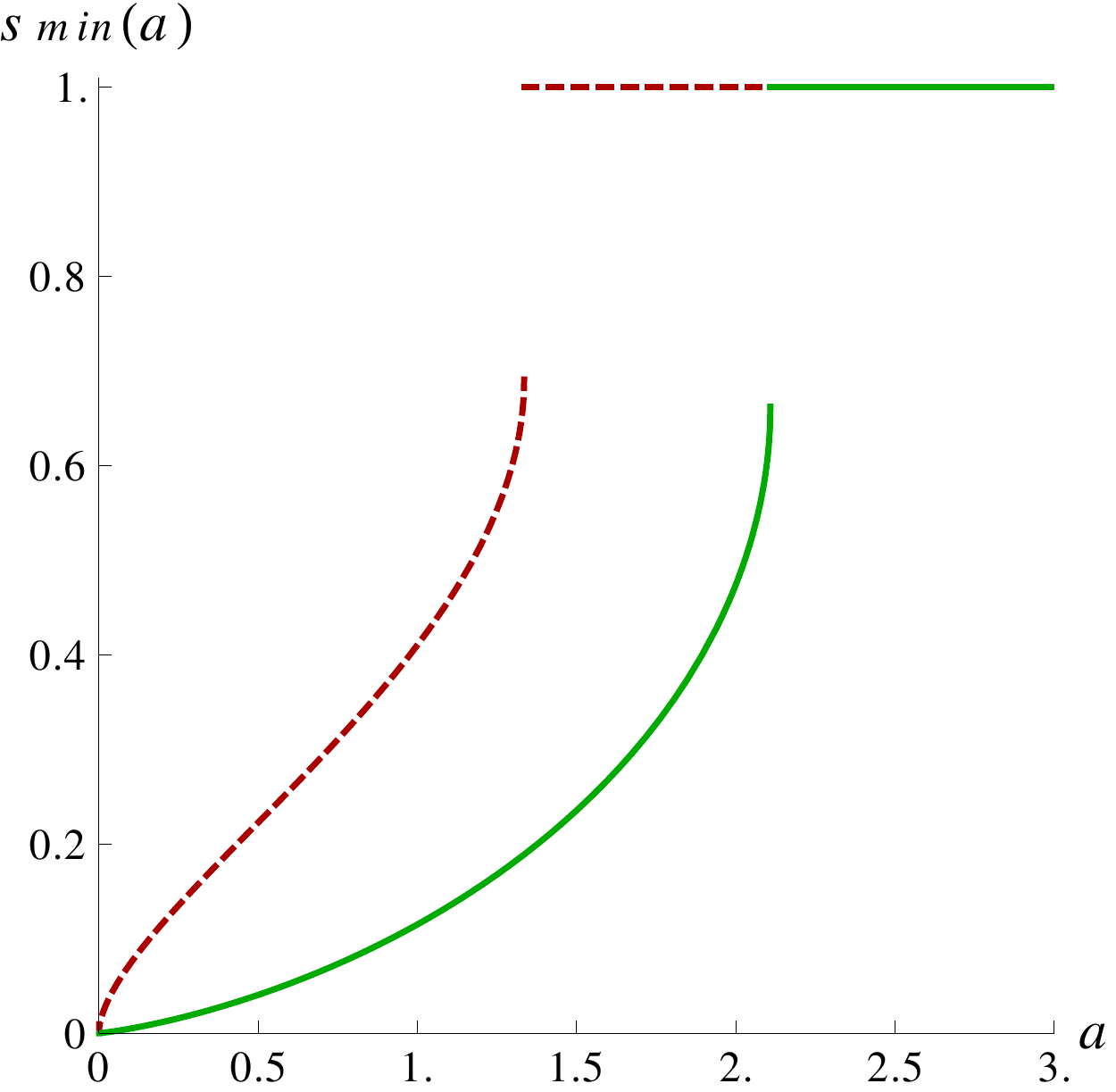}
\includegraphics[width=0.45\textwidth]{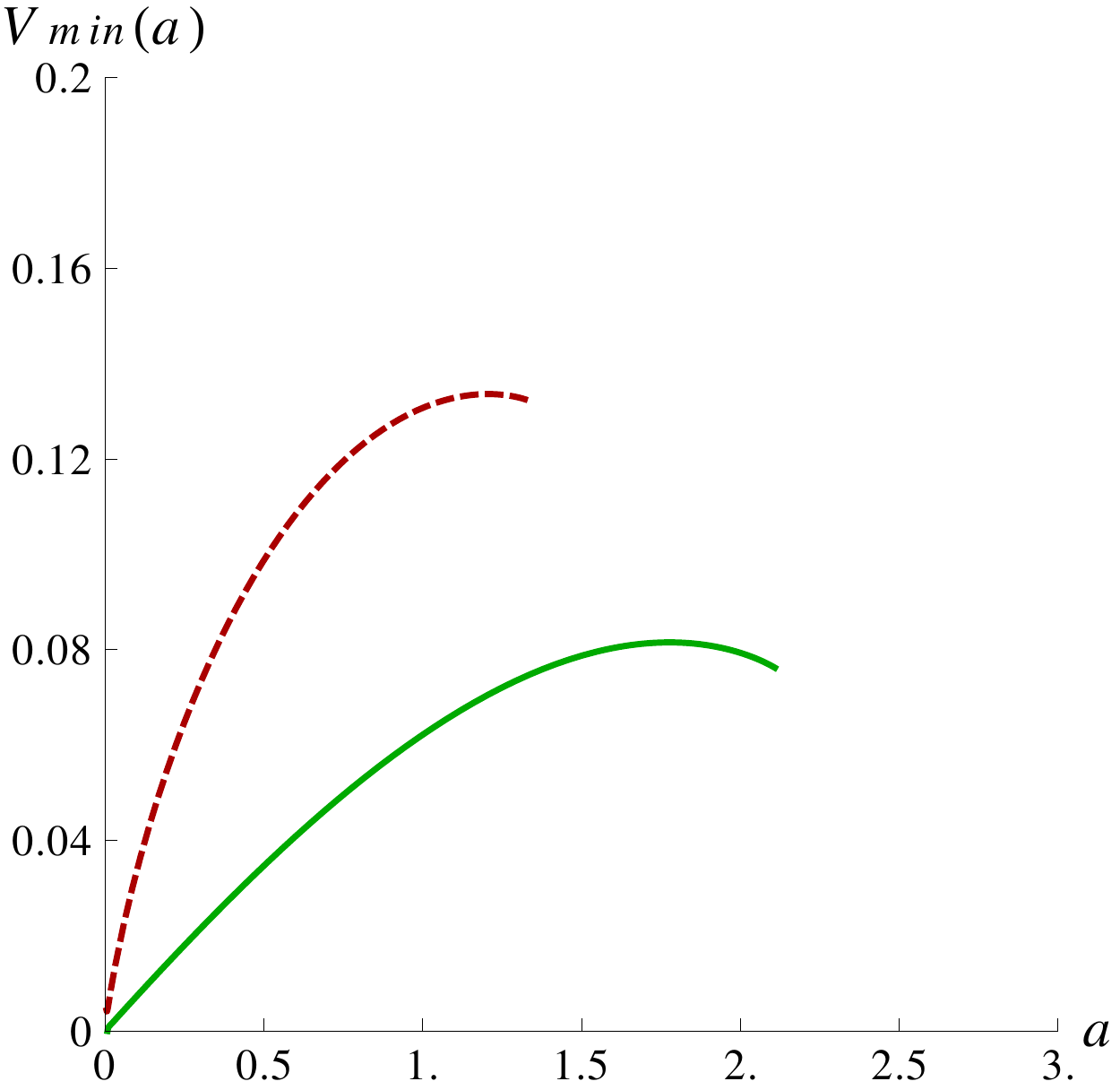}
\end{center}
\caption{Left panel: the minimum of the effective potential for $N_{c}=6$ as a
function of $a,$ $s_{min}(a)$, for the linear term (solid line), and the
Vandermonde determinant (dashed line). Right panel: the potential at the
minimum as a function of $a,$ $V_{min}(a)$.}%
\label{qmin6}%
\end{figure}

Using the solution for $s_{min}(a)$, we can obtain an expression for the
potential at the minimum which depends only on $a,$%
\begin{equation}
V_{min}(a)\equiv V\left[  s_{min}(a),a\right]  \ \text{,} \label{vmina}%
\end{equation}
where $V\left(  s,a\right)  $ is given by Eq. (\ref{veffA}). This will become
useful when we compute the pressure. In Figs. \ref{qmin23}, \ref{qmin45}, and
\ref{qmin6} we plot the solutions for $s_{min}(a)$ and $V_{min}(a)$ as a
function of $a$ for $N_{c}=2,3,4,5,6,$ and in the large-$N_{c}$ limit.

In order to determine the temperature dependence of the minimum, we apply the
definition for the parameter $a(T)$ in Eq. (\ref{vA}),
\begin{align}
s_{min}(T)  &  =s_{min}\left[  a\left(  T\right)  \right]  =s_{min}\left(
\frac{C_{1}}{\frac{T}{T_{d}}-C_{2}}\right)  \ ,\label{sminT}\\
V_{min}(T)  &  =V_{min}\left[  a\left(  T\right)  \right]  =V_{min}\left(
\frac{C_{1}}{\frac{T}{T_{d}}-C_{2}}\right)  \ . \label{Vmin2T}%
\end{align}

\subsection{Pressure and interaction measure}

The pressure is defined as minus the effective potential at the minimum,%
\begin{equation}
p\left(  T\right)  =-V_{eff}\left[  s_{min}\left(  T\right)  \right]  \text{
.}%
\end{equation}
A more straightforward way to compute the pressure is to use directly the
solution for $V_{min}(T)$ (\ref{Vmin2T}),%
\begin{equation}
{\frac{p(T)}{(N_{c}^{2}-1)T^{3}}}=\left(  1-\frac{T_{d}}{T}C_{3}\right)
\text{ }\frac{\zeta\left(  3\right)  }{2\pi}-\left(  1-\frac{T_{d}}{T}%
C_{2}\right)  V_{min}(T)\ , \label{Pressure}%
\end{equation}
From the pressure we can derive the interaction measure,
\begin{equation}
{\frac{\Delta(T)}{(N_{c}^{2}-1)T^{3}}}=T{\frac{d}{dT}}\left[  \frac
{p(T)}{(N_{c}^{2}-1)T^{3}}\right]  \ \text{.}%
\end{equation}
At very high temperatures, the pressure approaches a constant perturbative
limit,%
\begin{equation}
\lim_{T\rightarrow\infty}{\frac{p\left(  T\right)  }{(N_{c}^{2}-1)T^{3}}%
}\rightarrow\frac{\zeta\left(  3\right)  }{2\pi}\ \equiv c\text{ },
\label{Limit}%
\end{equation}
while the interaction measure vanishes,
\begin{equation}
\lim_{T\rightarrow\infty}\frac{\Delta\left(  T\right)  }{(N_{c}^{2}-1)T^{3}%
}\rightarrow{0}\text{ .}%
\end{equation}
Both the pressure and interaction measure in three-dimensional pure
$SU(N_{c})$ gauge theory were computed on the lattice for $N_{c}=2,3,4,5,6$ in
Ref. \cite{Caselle:2011mn}.

\subsubsection{Susceptibility}

\begin{figure}[t!]
\begin{center}
\includegraphics[width=0.45\textwidth]{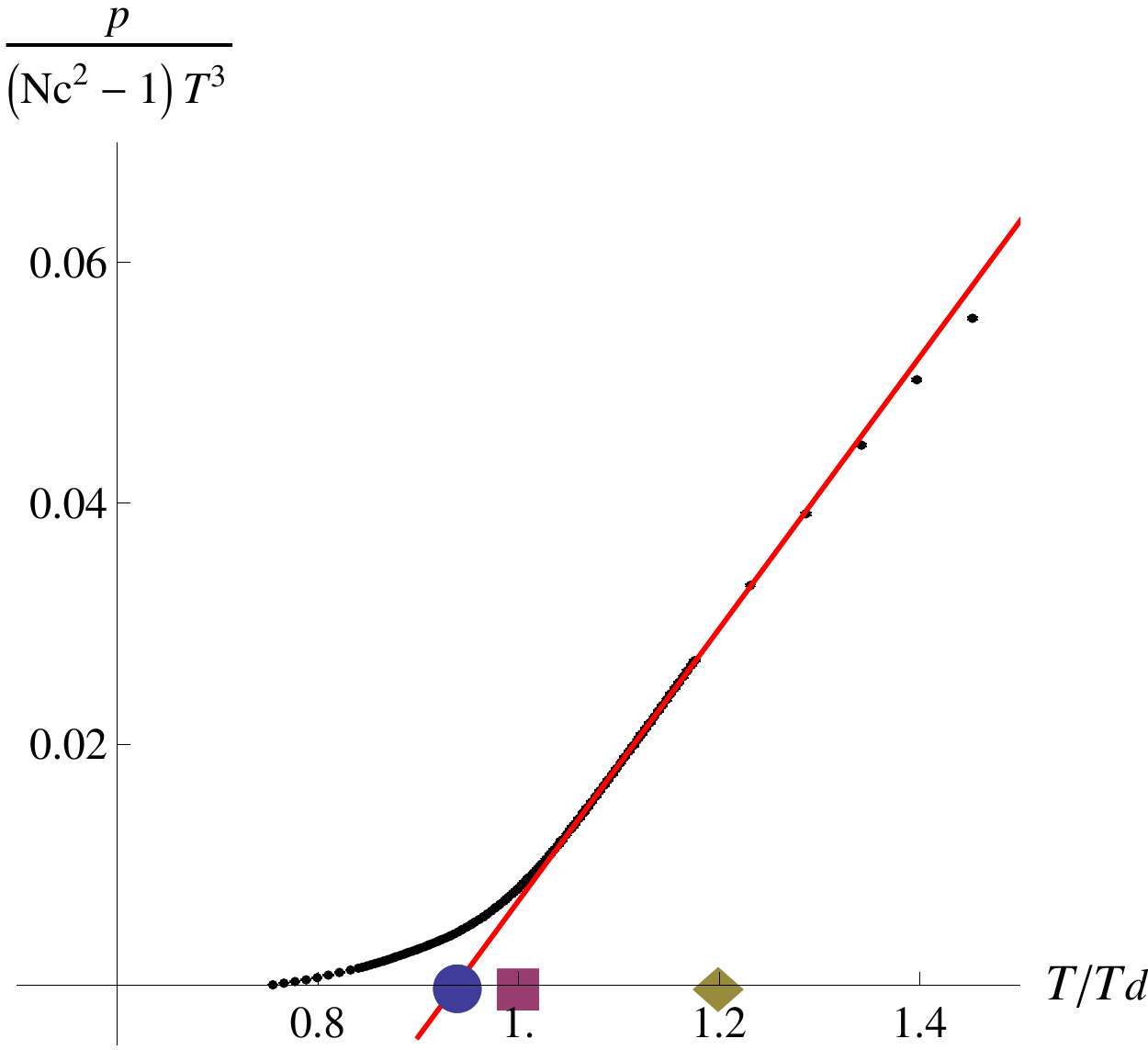}
\includegraphics[width=0.45\textwidth]{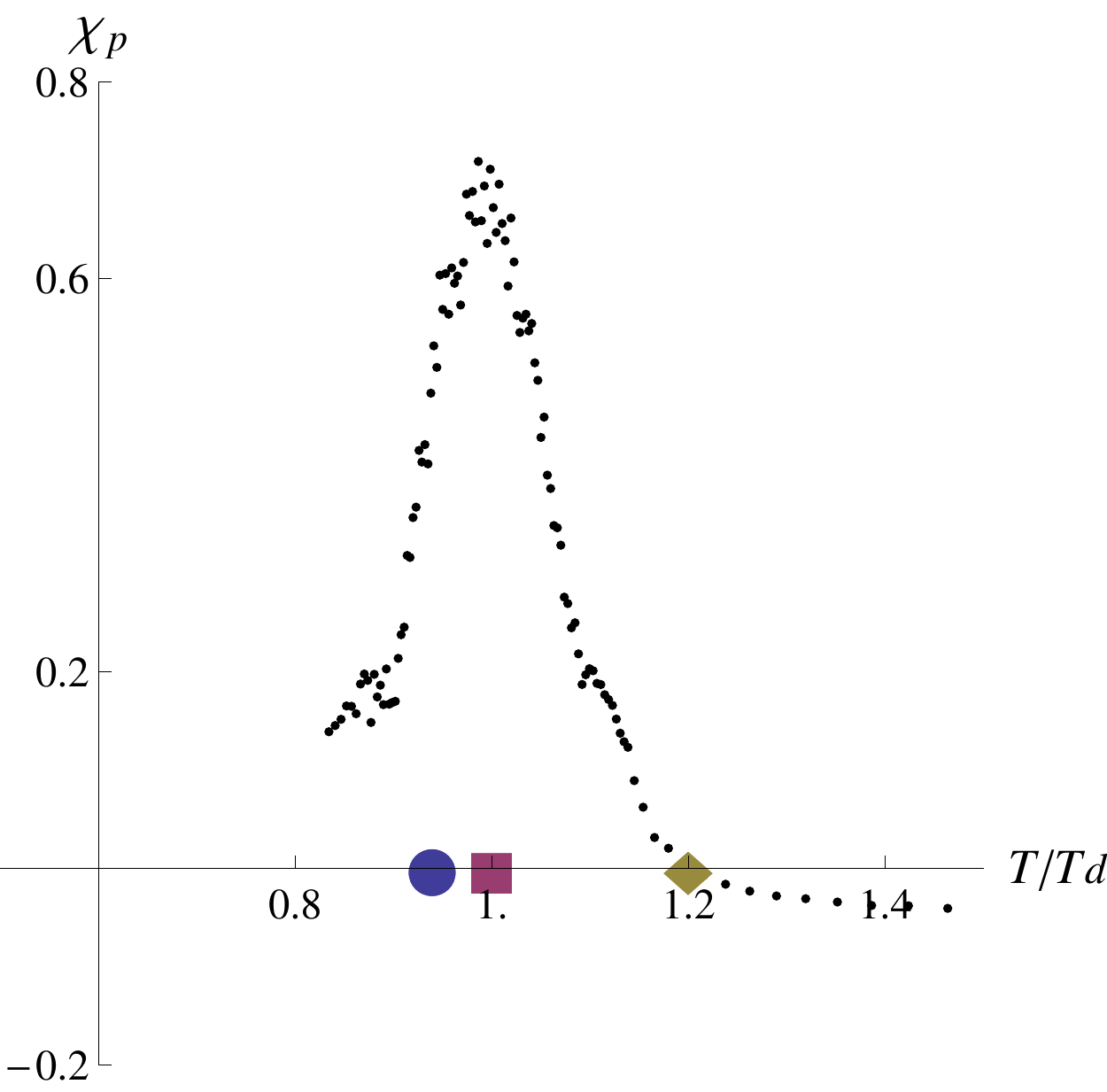}
\end{center}
\caption{ Lattice data of Ref. \cite{Caselle:2011mn} for $SU(2)_{c}$ close to
the critical temperature. Left panel: the pressure with the tangent to the
inflection point. Right panel: the susceptibility $\chi_{p}$ obtained from the
lattice pressure. The critical temperature $T_{d}$ is denoted by a square, the
intercept temperature $T_{i}$ by a circle, and the inflection point $T_{p}$ by
a diamond.}%
\label{td}%
\end{figure}

In order to study the deconfinement temperature $T_{d}$, it is also relevant
to compute the susceptibility. The common way to determine the critical
temperature on the lattice in pure gauge theory is via the susceptibility for
the Polyakov loop, which exhibits a peak at $T_{d}$. An equivalent method is
to look at the susceptibility for the pressure, which may be defined as
\begin{equation}
{\frac{\chi_{p}(T)}{(N_{c}^{2}-1)T^{3}}}\equiv{\frac{d^{2}}{dT^{2}}}\left[
\frac{p(T)}{(N_{c}^{2}-1)T^{3}}\right]  \ \text{.} \label{sc}%
\end{equation}
This quantity also has a peak at the deconfinement temperature $T_{d},$ see
Fig. \ref{td}.

\subsection{Fixing the parameters at $T_{d}$ \label{fixparam}}

\subsubsection{One-parameter model}

The effective potential (\ref{veff1a}) involves three parameters
$C_{1},\,C_{2},\,C_{3}$. Introducing two constraints at the transition
temperature $T_{d}$ leaves only one free parameter, say $C_{1}$, which is
determined by fitting the lattice data for the pressure of Ref.
\cite{Caselle:2011mn}.

\begin{enumerate}
\item[(i)] The first condition is that the transition occurs at $T_{d}$. From
Eq. (\ref{vA}) for the temperature-dependent parameter $a(T)$, we then derive
$C_{2}$ as a function of $C_{1},$
\begin{equation}
a_{d}\equiv a(T_{d})=\frac{C_{1}}{1-C_{2}}\ . \label{c_1}%
\end{equation}

\end{enumerate}

In the deconfined phase, the pressure is proportional to the number of gluons,
$p\left(  T>T_{d}\right)  \sim N_{c}^{2}-1.$ Below the critical temperature,
the gluons are confined in colorless glueballs which are exponentially
suppressed, $p\left(  T<T_{d}\right)  \sim\exp\left(  -m_{G}/T\right)  $,
where $m_{G}$ represents some glueball mass. Thus, in the confined phase the
pressure is by a factor $\sim1/\left(  N_{c}^{2}-1\right)  $ smaller than
above $T_{d}$.

\begin{enumerate}
\item[(ii)] Therefore, as a second condition we impose that the pressure
identically vanishes at the critical temperature,
\begin{equation}
p\left(  T_{d}\right)  =0\text{ .} \label{p00}%
\end{equation}

\end{enumerate}

Using the equation for the pressure (\ref{Pressure}), the second condition
(\ref{p00}) allows to determine $C_{3}$ as a function of $C_{1}$,
\begin{equation}
\ C_{3}=1-\frac{C_{1}V(s_{d}^{-},a_{d})}{(N_{c}^{2}-1)\frac{\zeta\left(
3\right)  }{2\pi}a_{d}}\text{ ,} \label{c_3}%
\end{equation}
where the values for $s_{d}^{-}$ and $a_{d}$ are listed in Table \ref{table1}.
Note that the constraint we impose in Eq. (\ref{p00}) is just an
approximation. Ideally, we should fit the pressure to some sort of hadronic,
i.e., glueball\textbf{\ }resonance gas in the confined phase.

\subsubsection{Two-parameter model}

The effective potential in Eq. (\ref{veff1a}) can be generalized to include
terms $\sim T^{0}$. In the two-parameter model the constants $C_{1}$ and
$C_{2}$ remain unchanged, while $C_{3}$ is replaced by a temperature-dependent
parameter,%
\begin{equation}
C_{3}(T)=C_{3}\left(  \infty\right)  +\frac{C_{3}\left(  T_{d}\right)
-C_{3}\left(  \infty\right)  }{T^{2}/T_{d}^{2}}\text{ ,} \label{Deltac3}%
\end{equation}
which is equivalent to adding a bag constant{\ $B$} to the effective
potential. The two-parameter model is the most general case studied nowadays.

\subsubsection{Four-parameter fit}

We also study a four-parameter fit by extending the two-parameter model by two
additional free parameters: one for the critical temperature $T_{d}$, and one
for the perturbative limit of the pressure $c$ defined in Eq. (\ref{Limit}).
The parameter for $c$ is introduced in order to take into account higher-order
loop contributions to the perturbative potential. In addition, the parameter
for $T_{d}$ accounts for possible glueball effects near the transition, which
are not included in the matrix model. In our model\textbf{,} we fix the
parameters by imposing the constraint that the pressure vanishes at $T_{d}$.
Ideally, however, we should match the pressure to some hadronic resonance gas
below the critical temperature. This gas is made of massive glueballs, and
thus can be described by a series of Boltzmann factors. If there are many
massive glueballs, as in a Hagedorn spectrum, the temperature dependence of
the glueball contribution may become more complicated, including powers of
$T_{H}-T$, where $T_{H}$ is the Hagedorn temperature \cite{Hagedorn:1965st}.
The emergence of a Hagedorn-like spectrum (i.e., an exponential growth in the
number of hadronic states, as a function of their mass) has been studied in
$d=3+1$ and $d=2+1$ in a recent work \cite{Cohen:2011yx}.

In $d=3+1$ the condition $p(T_{d})=0$ provides a good approximation even for
two colors \cite{Dumitru:2012fw}. In $d=2+1$ we find that this simplified
constraint works well for larger $N_{c}$. For small $N_{c},$ however, we need
to relax the condition $p(T_{d})=0$, because the glueball contribution to the
pressure in the confined phase is not negligible.

Figure \ref{td} shows the lattice data for the pressure and the corresponding
susceptibility $\chi_{p}$, defined in Eq. (\ref{sc}) for $N_{c}=2$. Near the
transition at $T_{d}$, we find three different temperatures:

\begin{enumerate}
\item[(i)] First, the critical temperature $T_{d}$ can be determined from the
susceptibility $\chi_{p}$ which has a maximum at $T_{d}$.

\item[(ii)] The inflection point $T_{p}$ corresponds to the temperature where
the derivative of the pressure is maximal. This coincides with the peak of the
interaction measure, and with the point at which the susceptibility vanishes.

\item[(iii)] The best estimate for the temperature $T_{i},$ at which the
pressure vanishes, is found by the intercept of the tangent to the inflection
point $T_{p}$ of the pressure and the $T/T_{d}$\textbf{-}axis.
\end{enumerate}

For $N_{c}=2$ we get $T_{p}\equiv1.2T_{d}>T_{d}$ for the inflection point, and
$T_{i}\equiv0.94T_{d}<T_{d}$ for the intercept temperature. We find that for
small $N_{c}$, the critical temperature $T_{d}$ determined from the lattice
data via the susceptibility does not coincide with the estimated value for the
intercept temperature $T_{i}$, at which the pressure vanishes. This is the
main motivation to allow for a shift in $T_{d}$ introduced in the
four-parameter fit. We stress, however, that our four-parameter fit should be
regarded just as an approximation to a more complete model including an
effective theory for the confined phase.

Note that the beauty and simplicity of these three models including one, two,
and four free parameters is that, as a common denominator, the constraints on
$C_{1},\,C_{2},\,C_{3}$ are maintained. In the two-parameter model, the
constraint on the pressure (\ref{p00}) is now used to determine $C_{3}%
(T_{d}),$ and $C_{3}\left(  \infty\right)  $ is the new free parameter. In the
four-parameter fit\textbf{,} the constraint is applied to $C_{3}(T_{i})$, and
$T_{i}$ is introduced as a new free parameter, together with the perturbative
constant $c.$

\section{Results \label{results}}

\begin{figure}[t!]
\begin{center}
\includegraphics[width=0.49\textwidth]{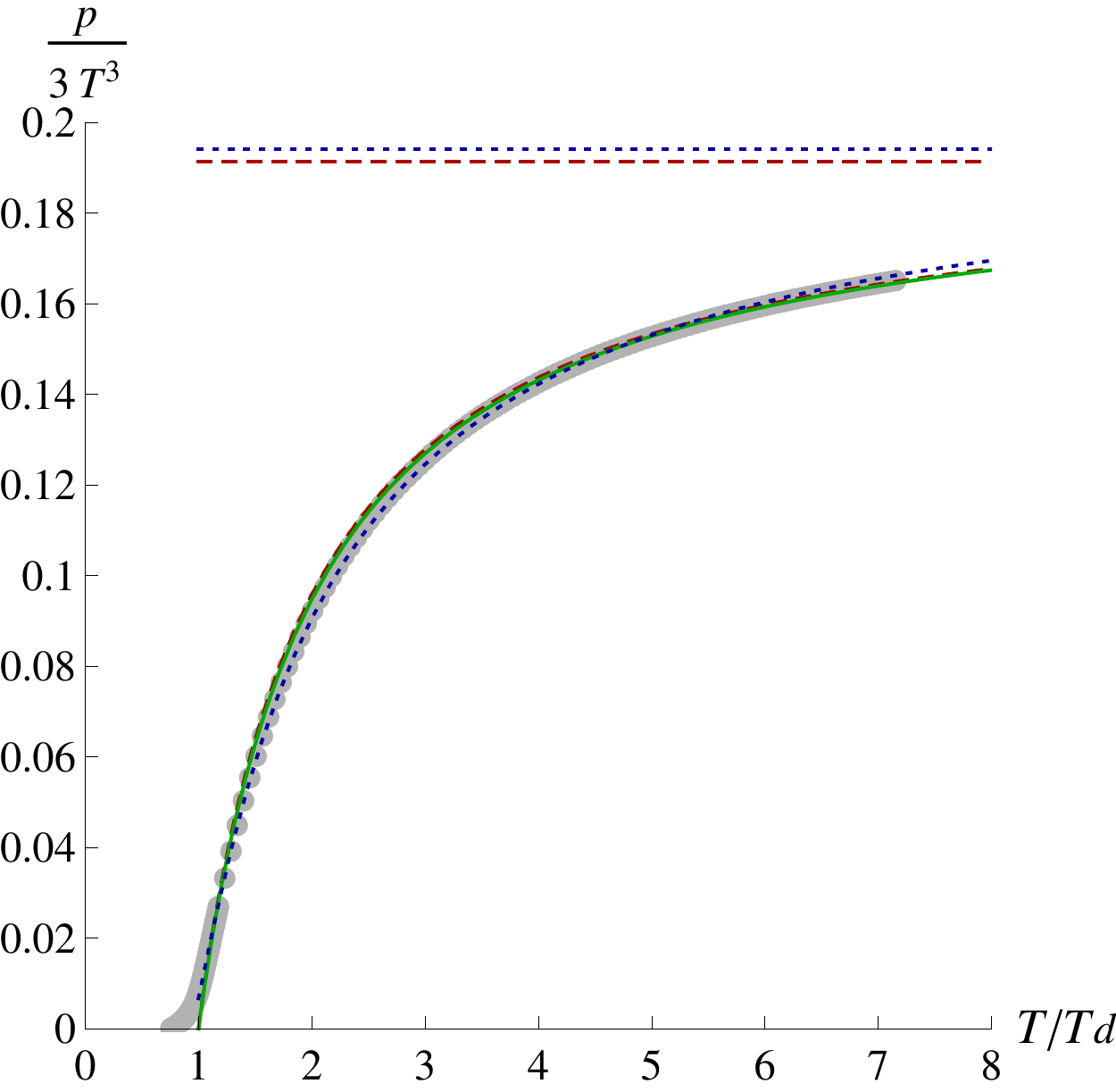}
\includegraphics[width=0.49\textwidth]{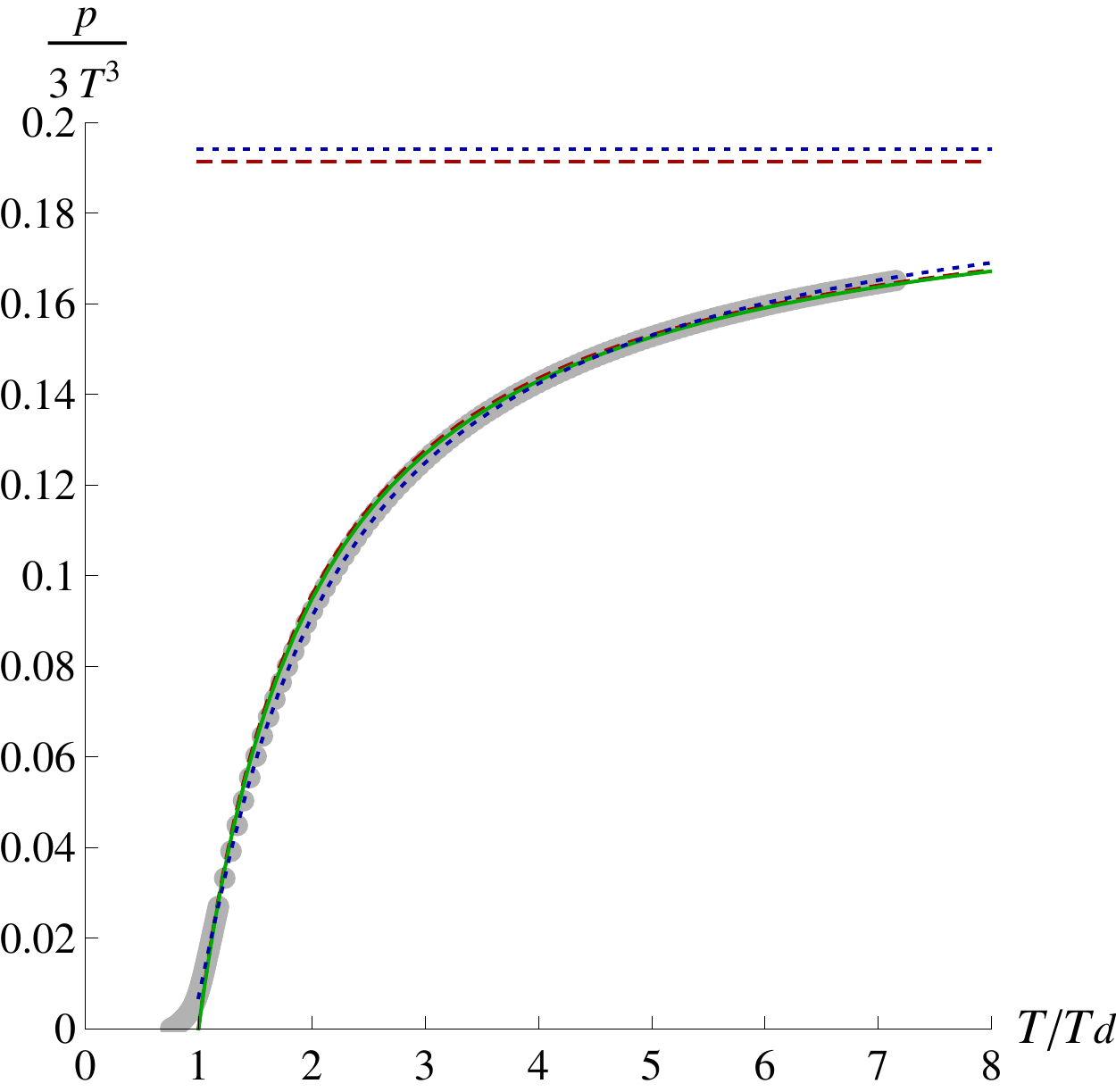}
\includegraphics[width=0.49\textwidth]{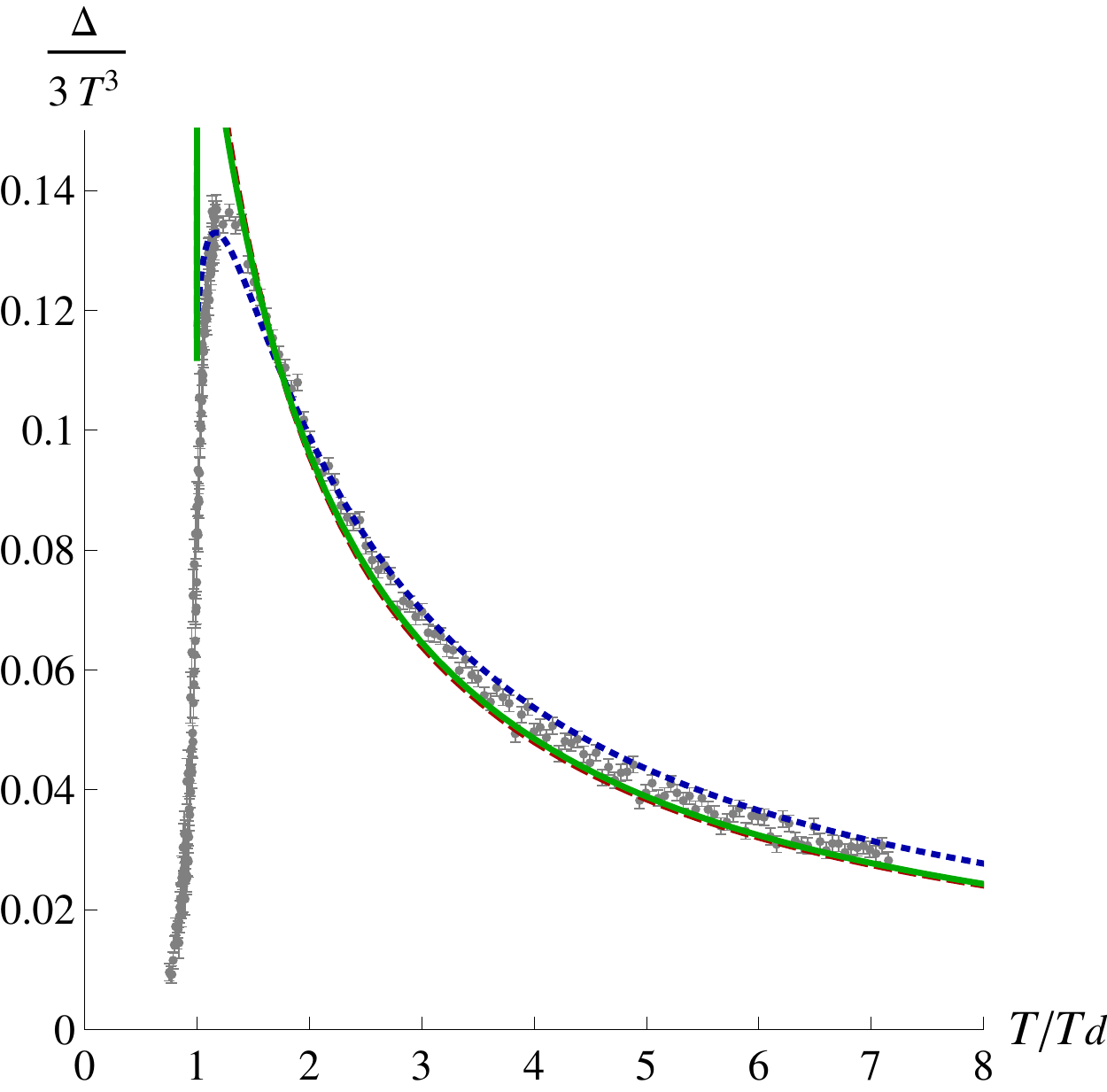}
\includegraphics[width=0.49\textwidth]{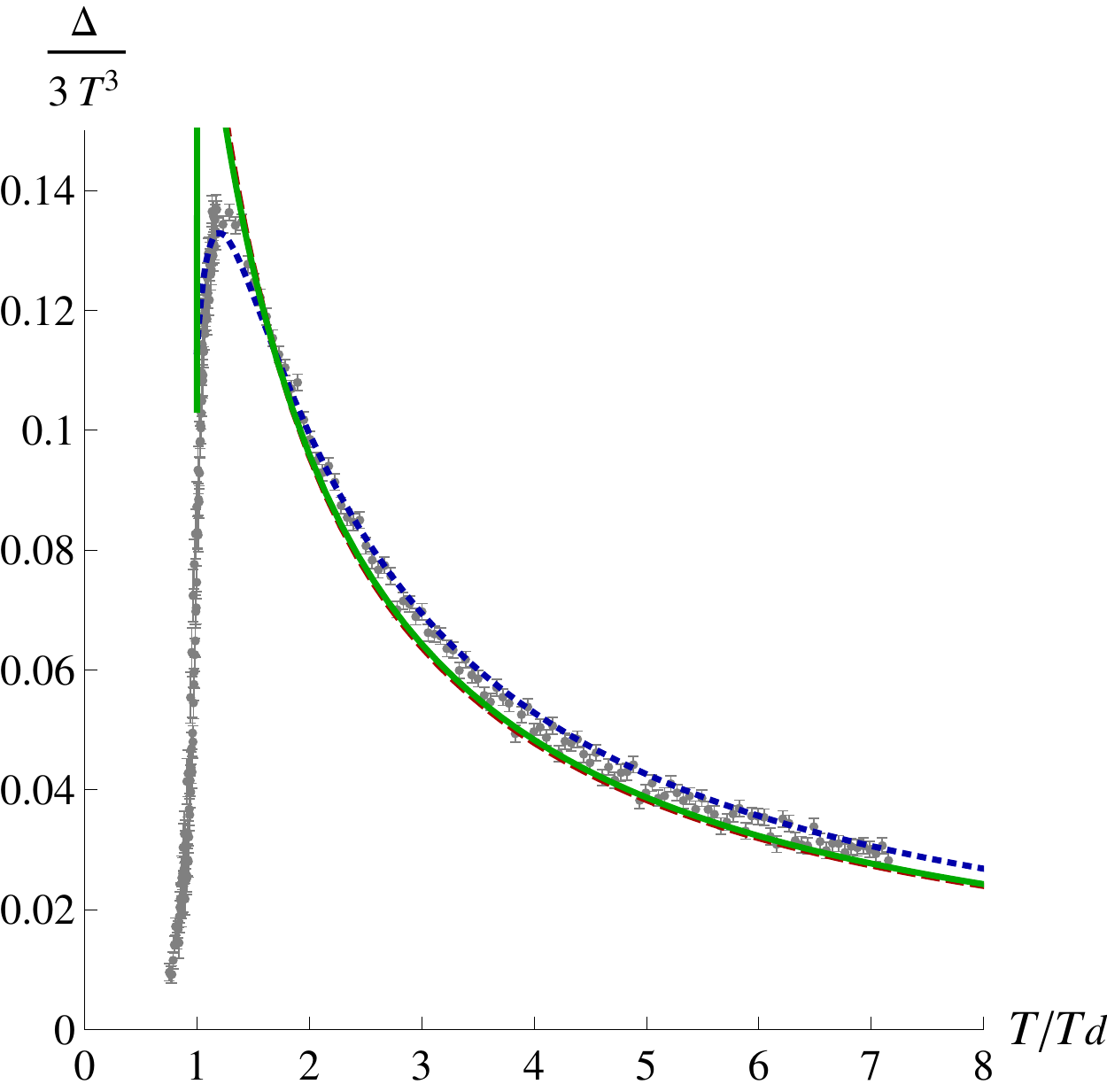}
\end{center}
\caption{Pressure (upper panels) and interaction measure (lower panels) scaled
by $N_{c}^{2}-1=3$ for $SU(N_{c}=2)$. Lattice data, taken from Ref.
\cite{Caselle:2011mn}, are denoted by dots. For the interaction measure we
also show the lattice error bars, while for the pressure the error bars are
smaller than the symbols. The results of the matrix model are obtained by
using the Vandermonde determinant (left panel), and the linear term (right
panel) within the one-parameter model (dashed line), two-parameter model
(solid line), and four-parameter fit (dotted line). The horizontal lines
represent the perturbative limit of the pressure $c=\zeta\left(  3\right)
/2\pi.$}%
\label{ptr2}%
\end{figure}

\begin{figure}[t!]
\begin{center}
\includegraphics[width=0.49\textwidth]{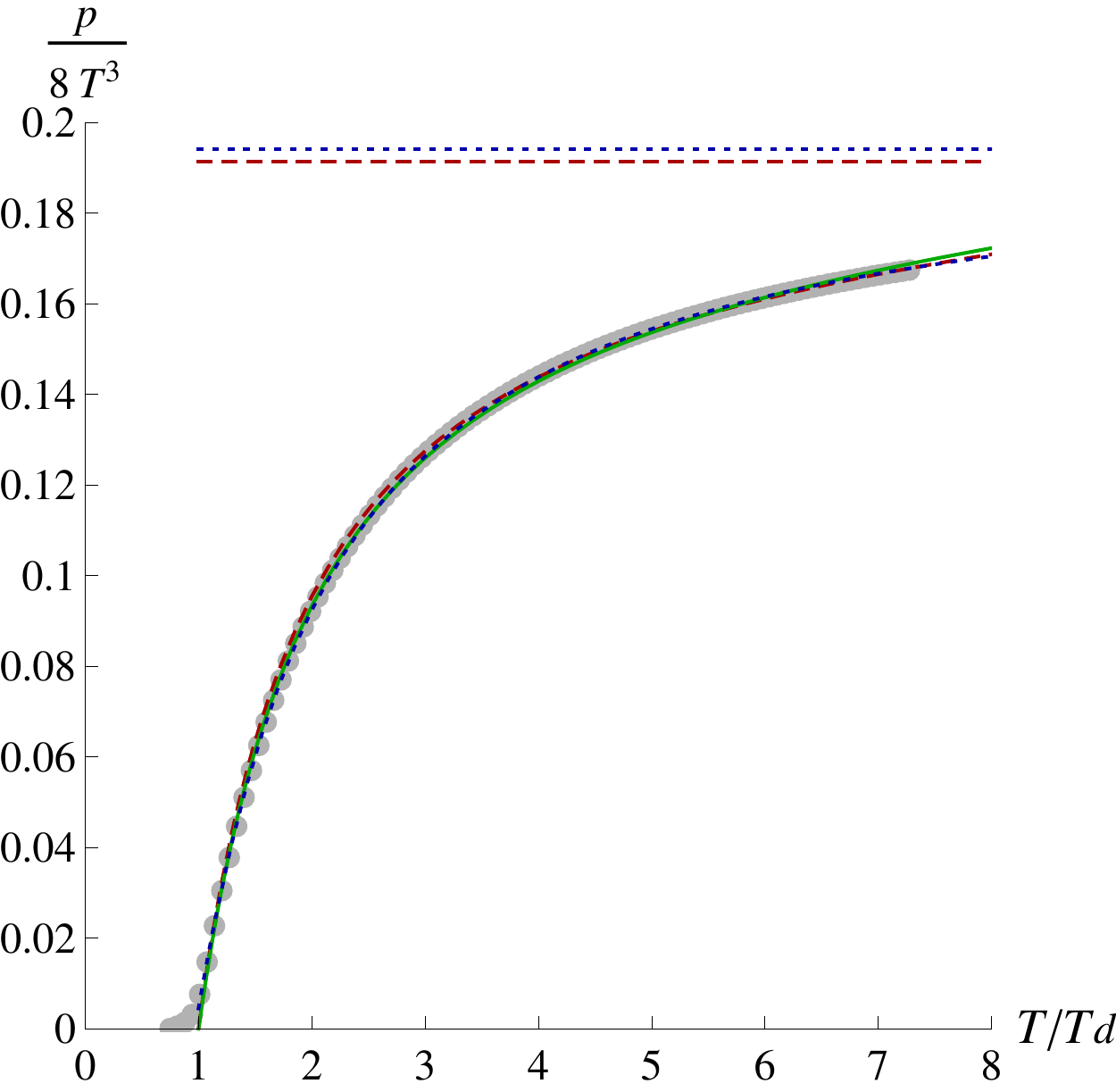}
\includegraphics[width=0.49\textwidth]{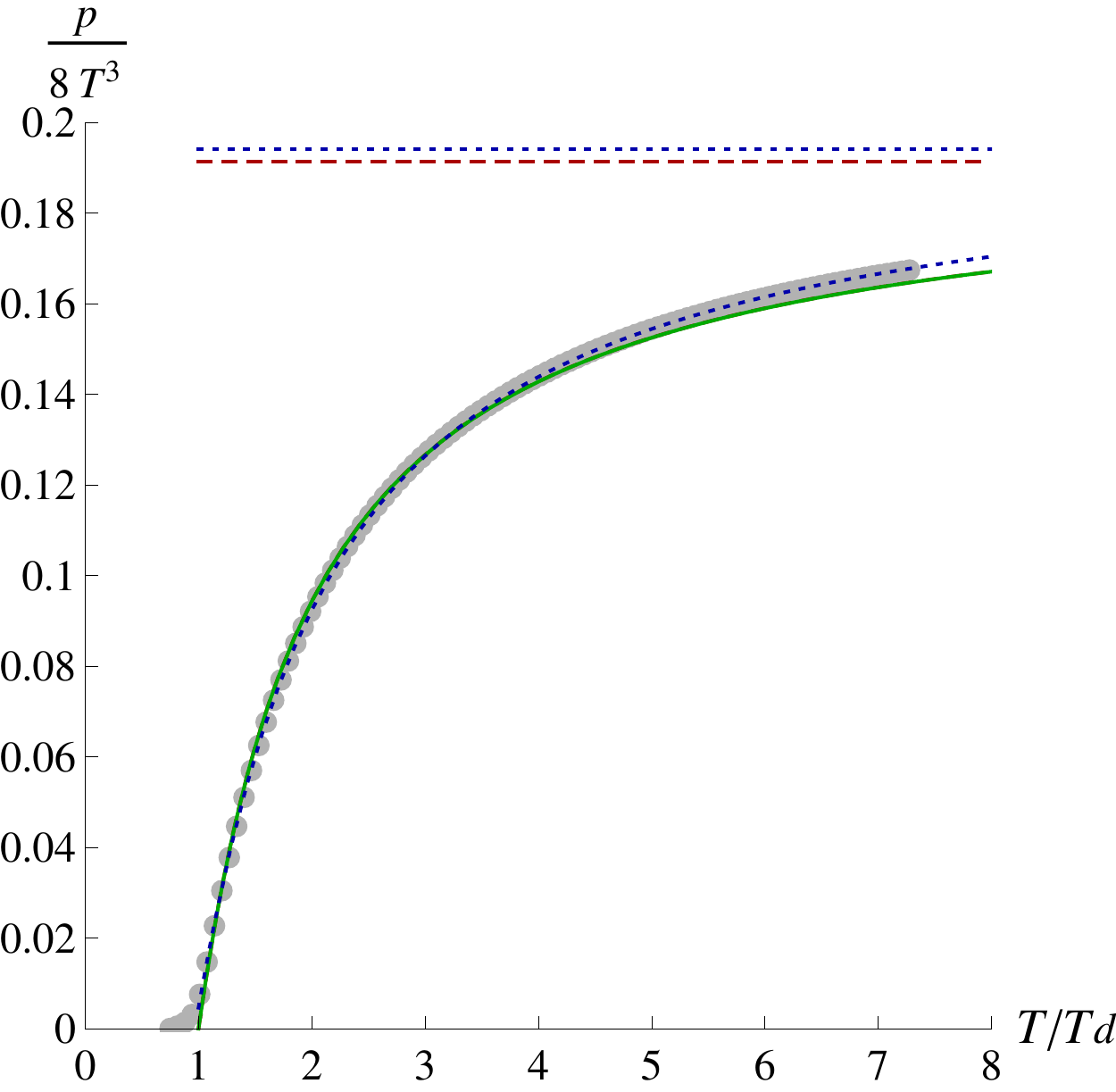}
\includegraphics[width=0.49\textwidth]{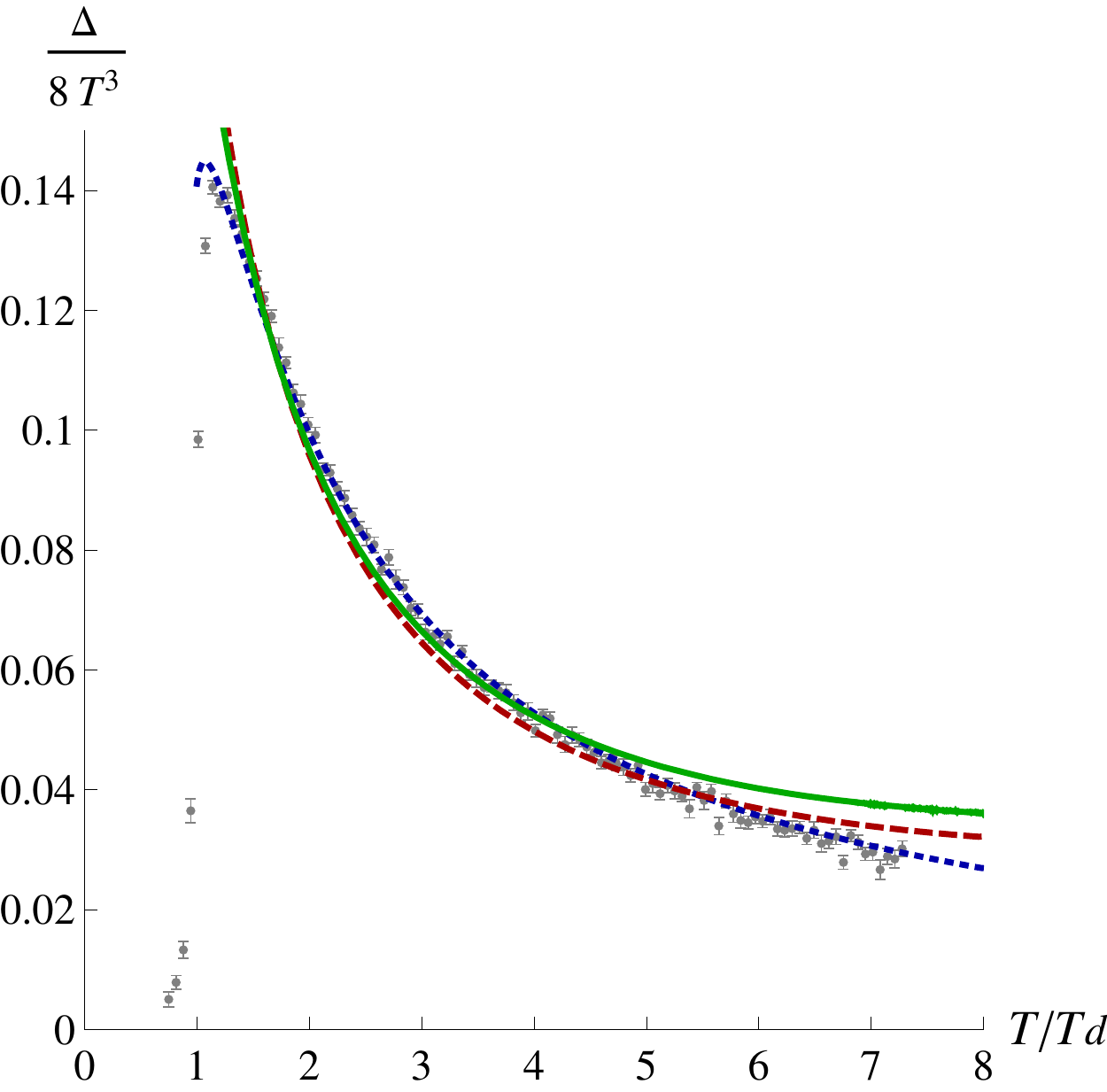}
\includegraphics[width=0.49\textwidth]{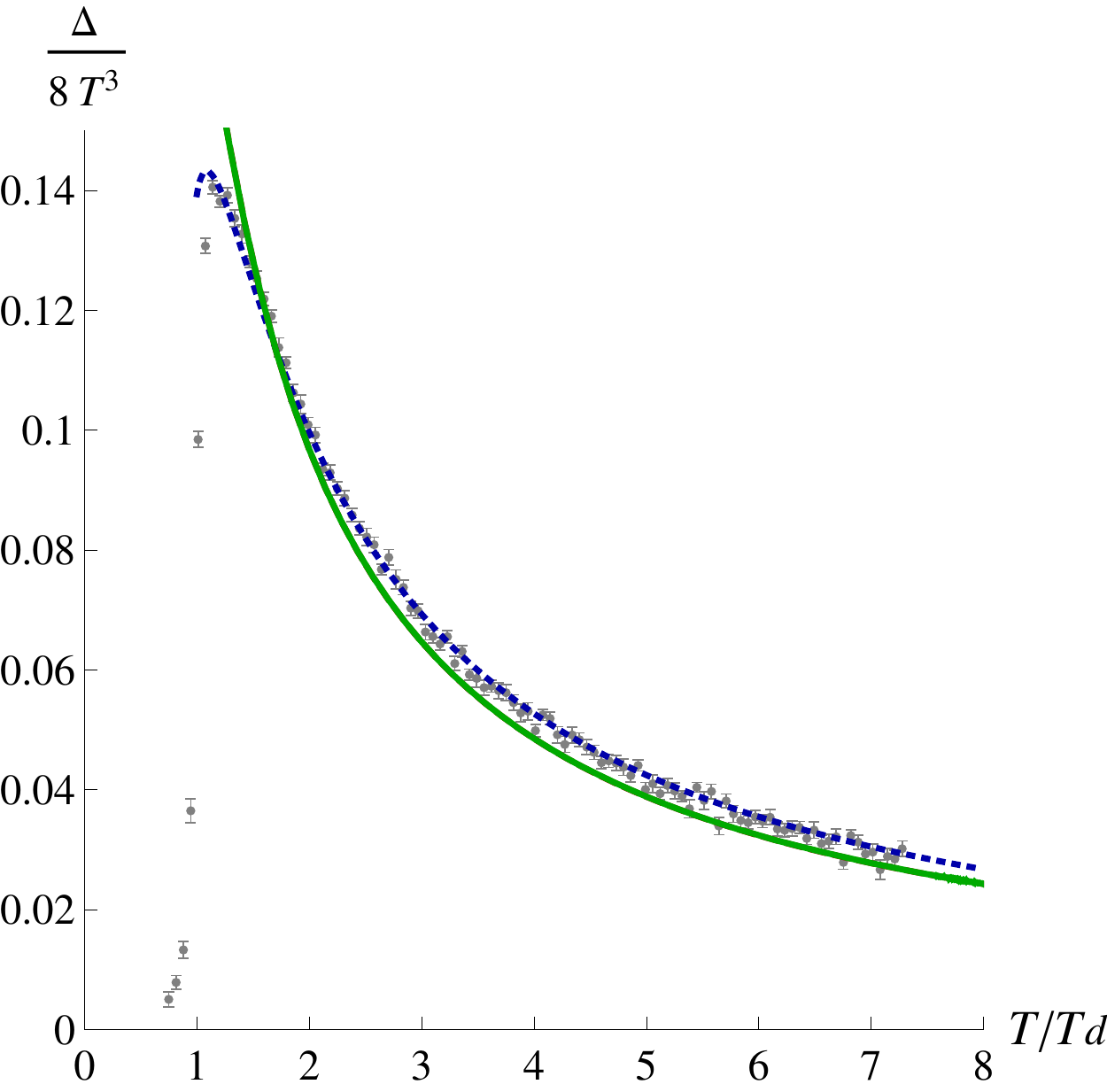}
\end{center}
\caption{Pressure (upper panels) and interaction measure (lower panels) scaled
by $N_{c}^{2}-1=8$ for $SU(N_{c}=3)$. Lattice data are denoted by dots. The
results of the matrix model are obtained by using the Vandermonde determinant
(left panel), and the linear term (right panel) within the one-parameter model
(dashed line), two-parameter model (solid line), and four-parameter fit
(dotted line). The horizontal lines represent the Stefan-Boltzmann limit of
the pressure $c=\zeta\left(  3\right)  /2\pi$.}%
\label{ptr3}%
\end{figure}

\begin{figure}[t!]
\begin{center}
\includegraphics[width=0.49\textwidth]{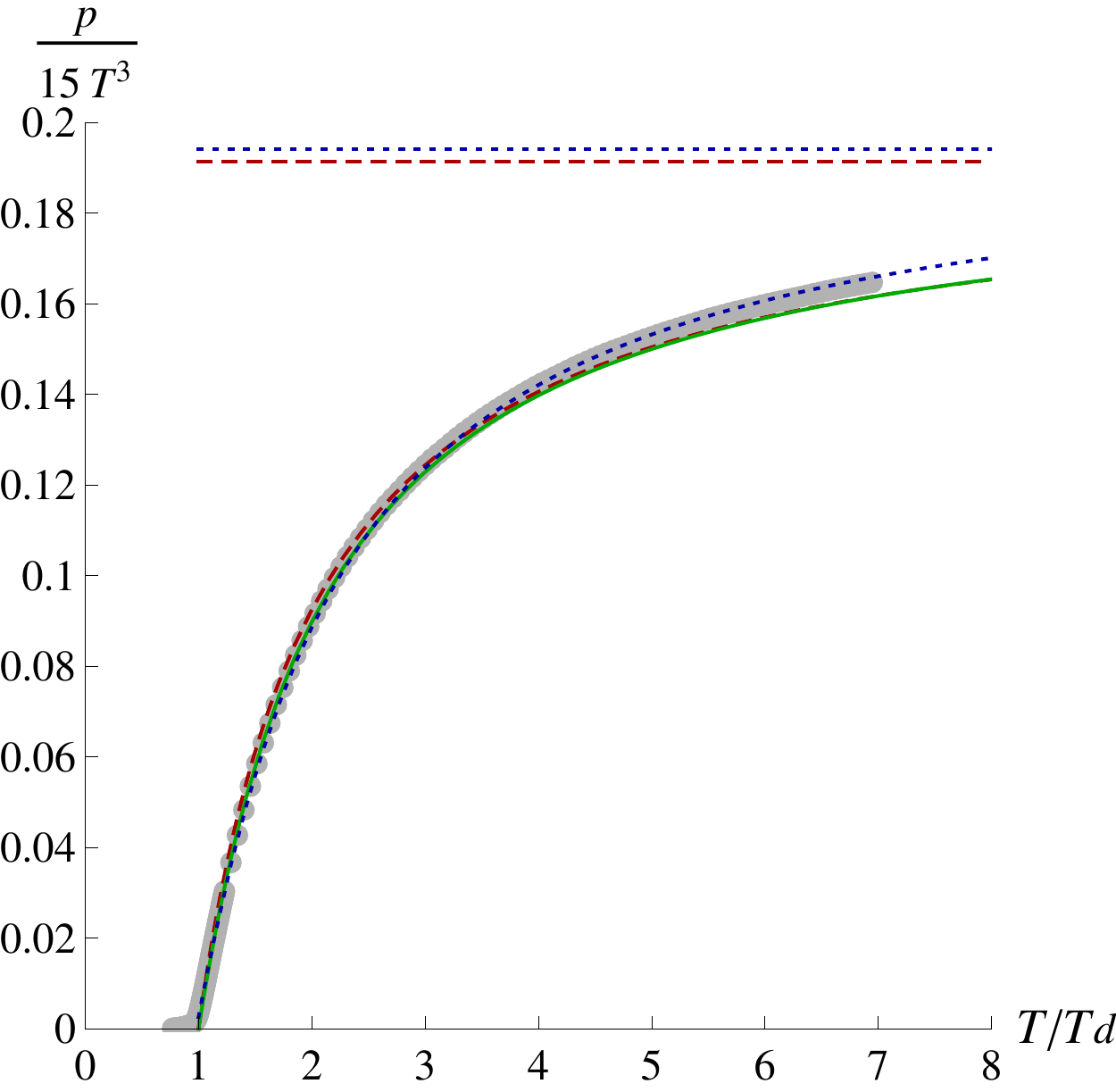}
\includegraphics[width=0.49\textwidth]{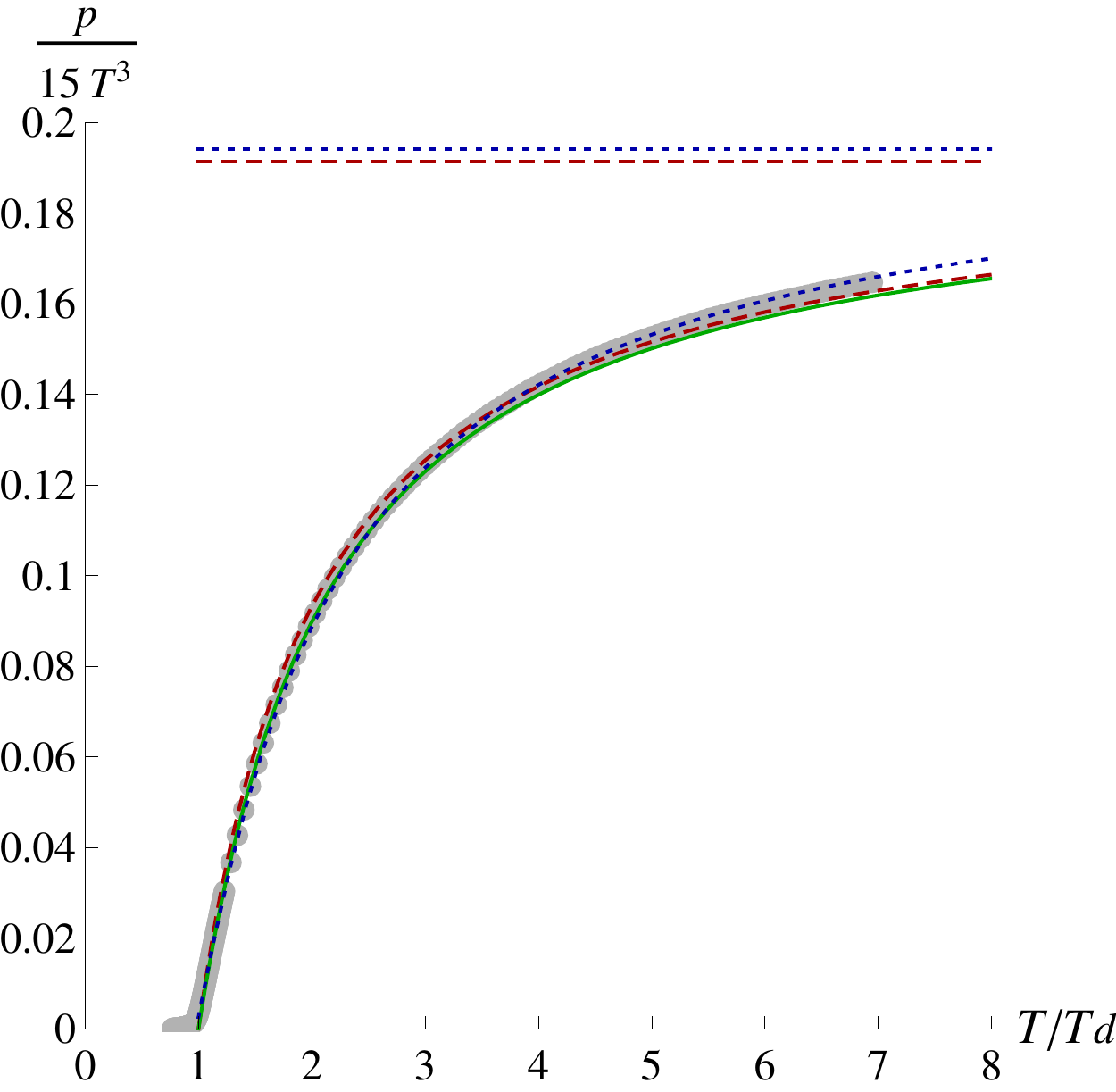}
\includegraphics[width=0.49\textwidth]{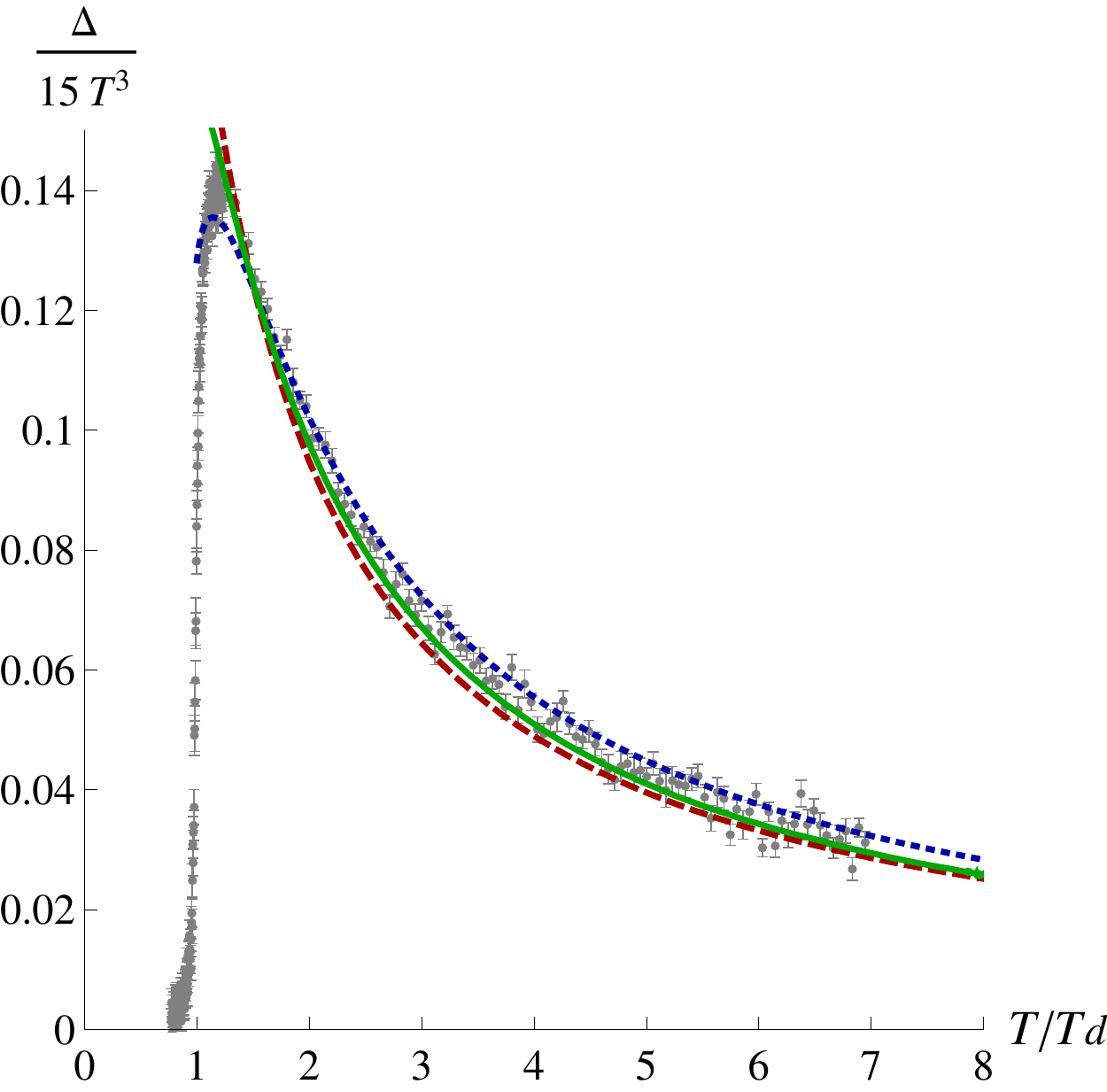}
\includegraphics[width=0.49\textwidth]{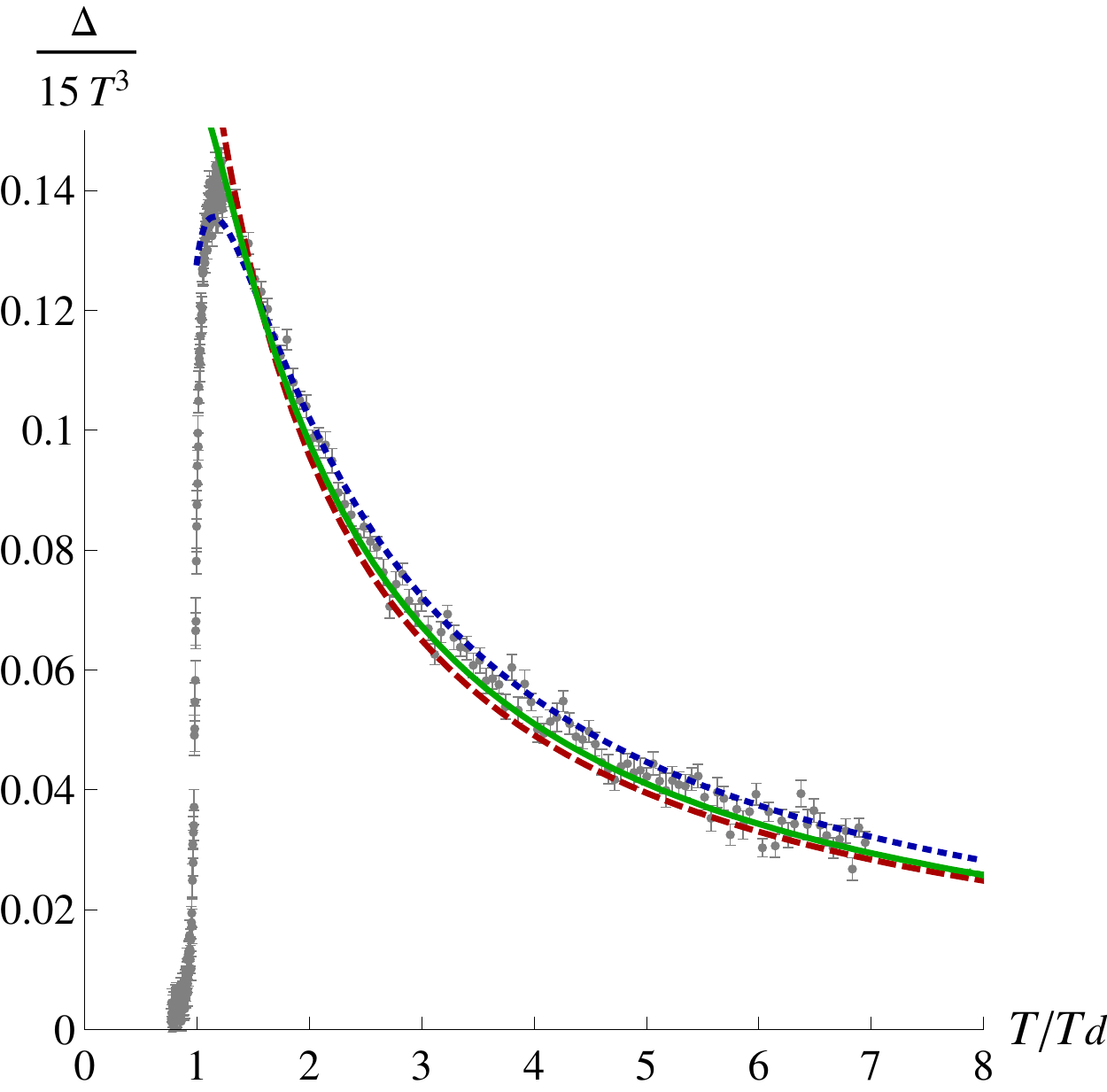}
\end{center}
\caption{Pressure (upper panels) and interaction measure (lower panels) scaled
by $N_{c}^{2}-1=15$ for $SU(N_{c}=4)$. Lattice data \cite{Caselle:2011mn} are
denoted by dots. The results of the matrix model are obtained by using the
Vandermonde determinant (left panel), and the linear term (right panel) within
the one-parameter model (dashed line), two-parameter model (solid line), and
four-parameter fit (dotted line). The horizontal lines represent the
perturbative limit of the pressure $c=\zeta\left(  3\right)  /2\pi$.}%
\label{ptr4}%
\end{figure}

\begin{figure}[t!]
\begin{center}
\includegraphics[width=0.49\textwidth]{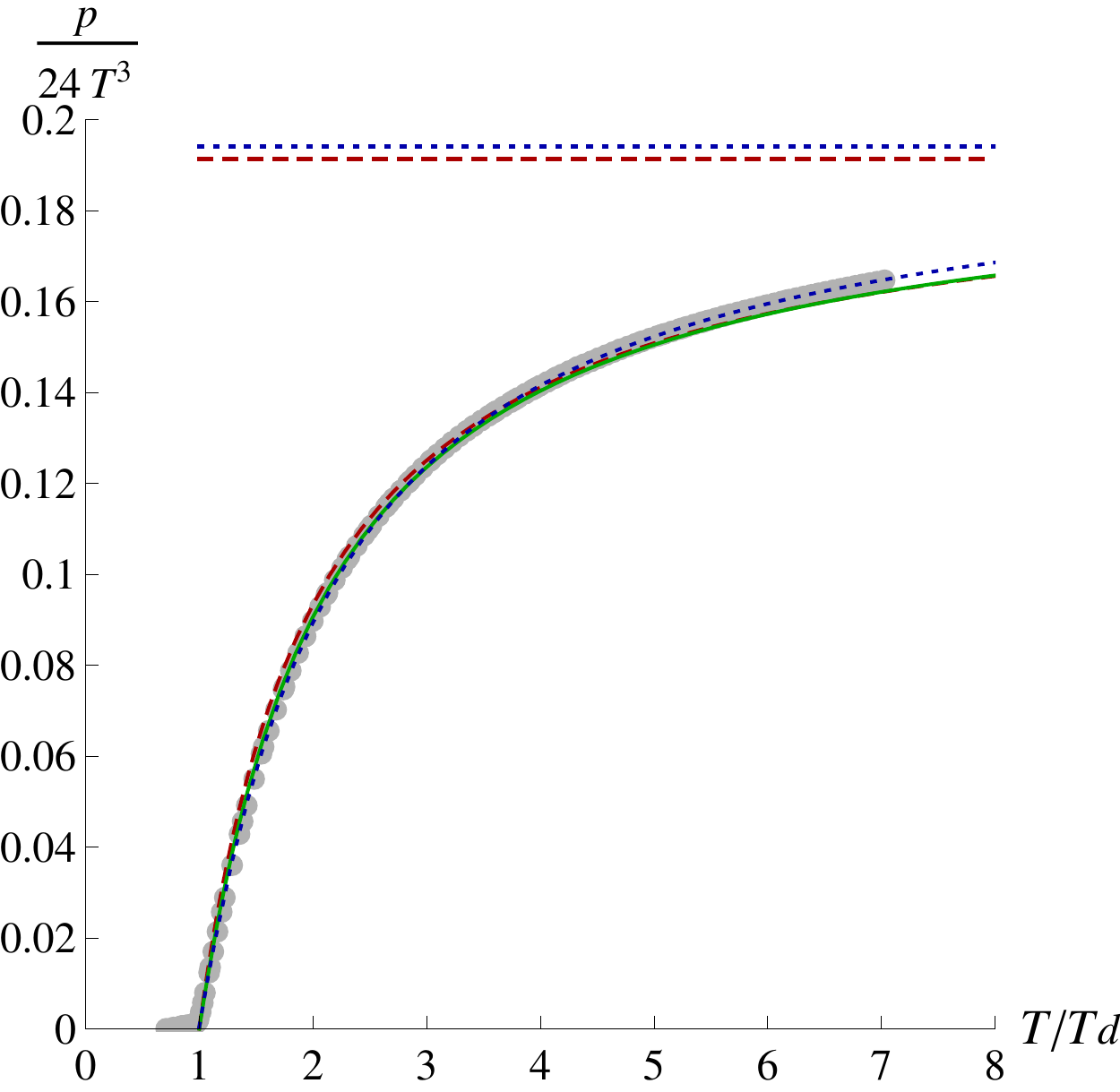}
\includegraphics[width=0.49\textwidth]{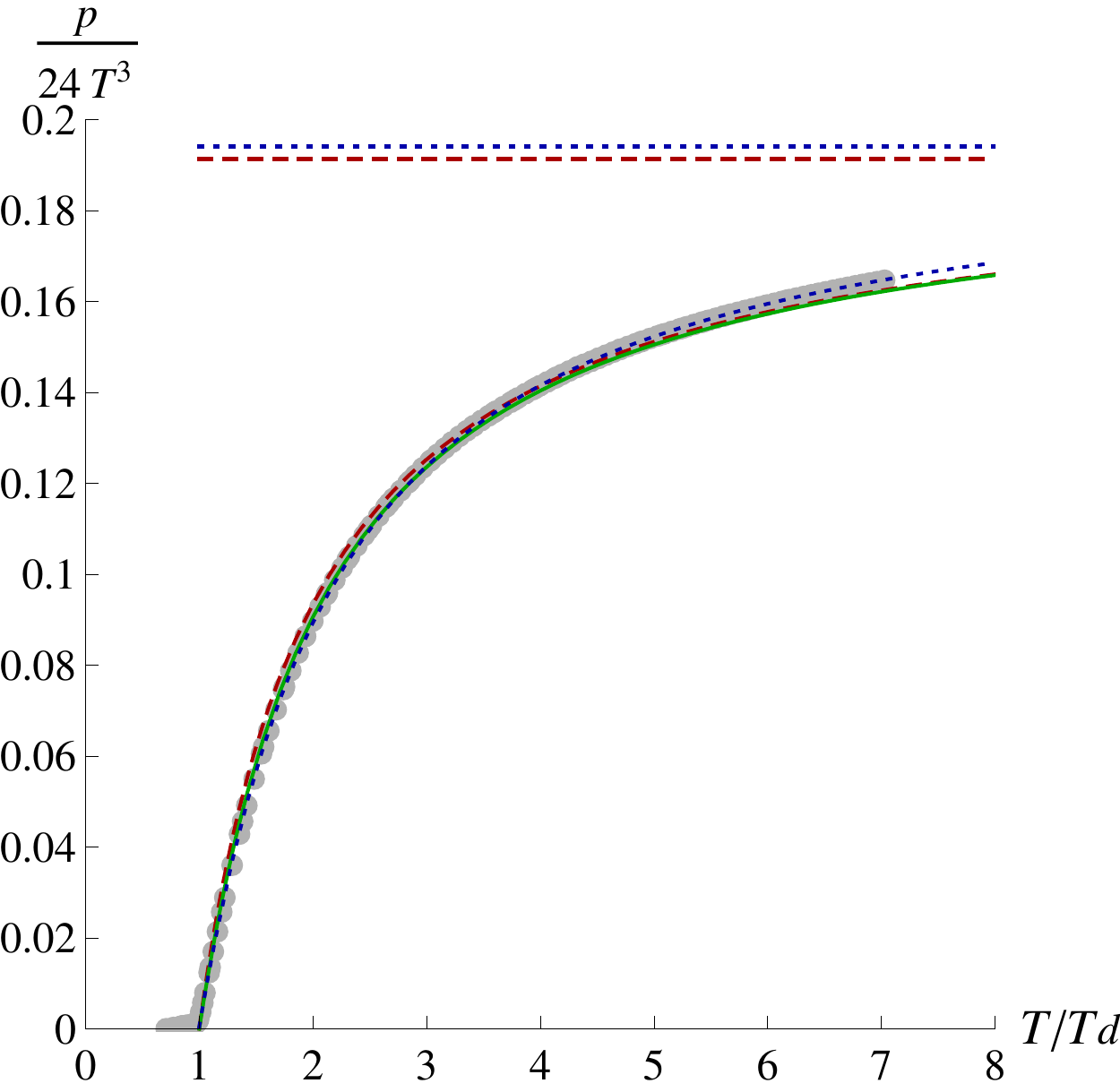}
\includegraphics[width=0.49\textwidth]{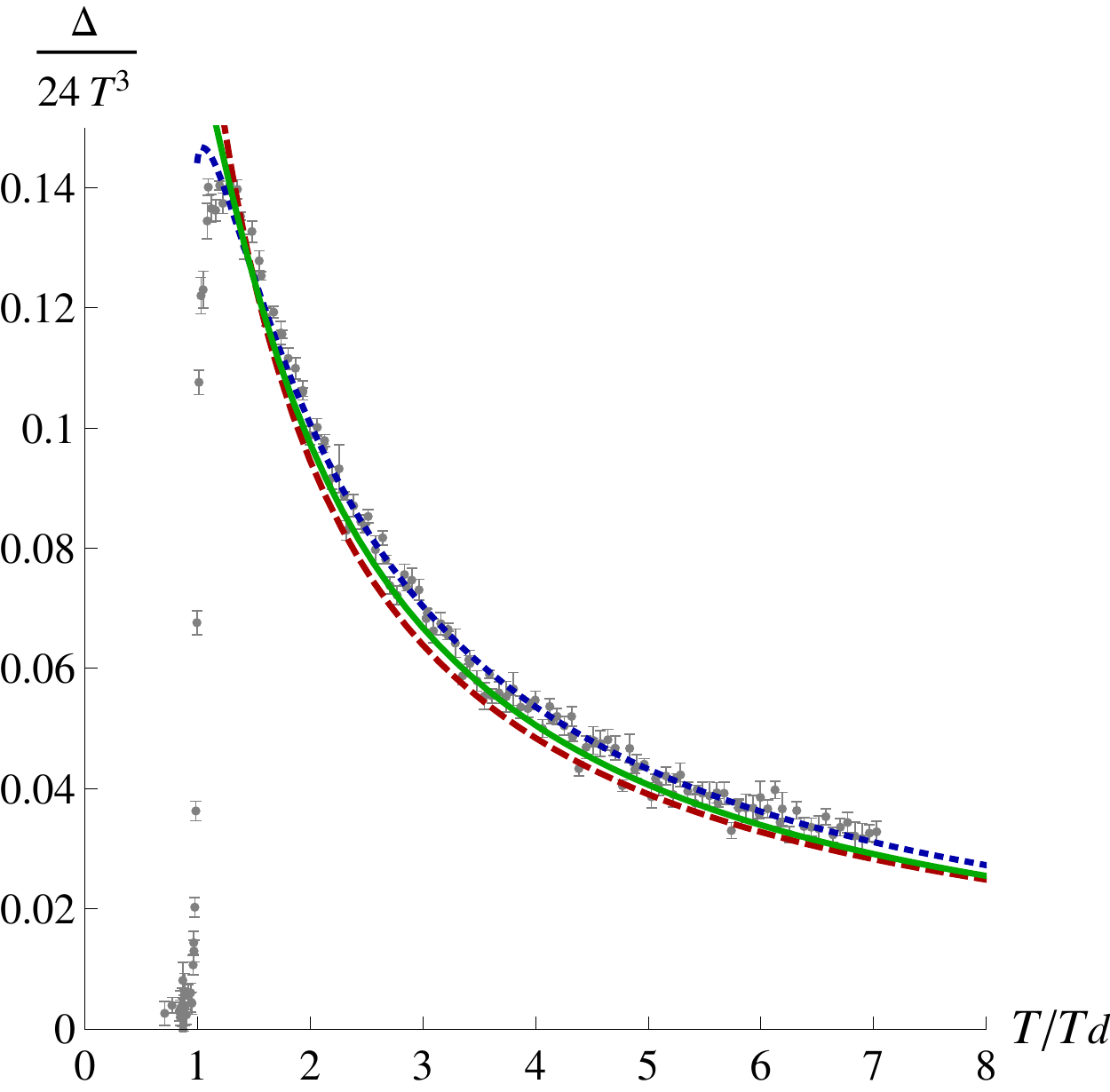}
\includegraphics[width=0.49\textwidth]{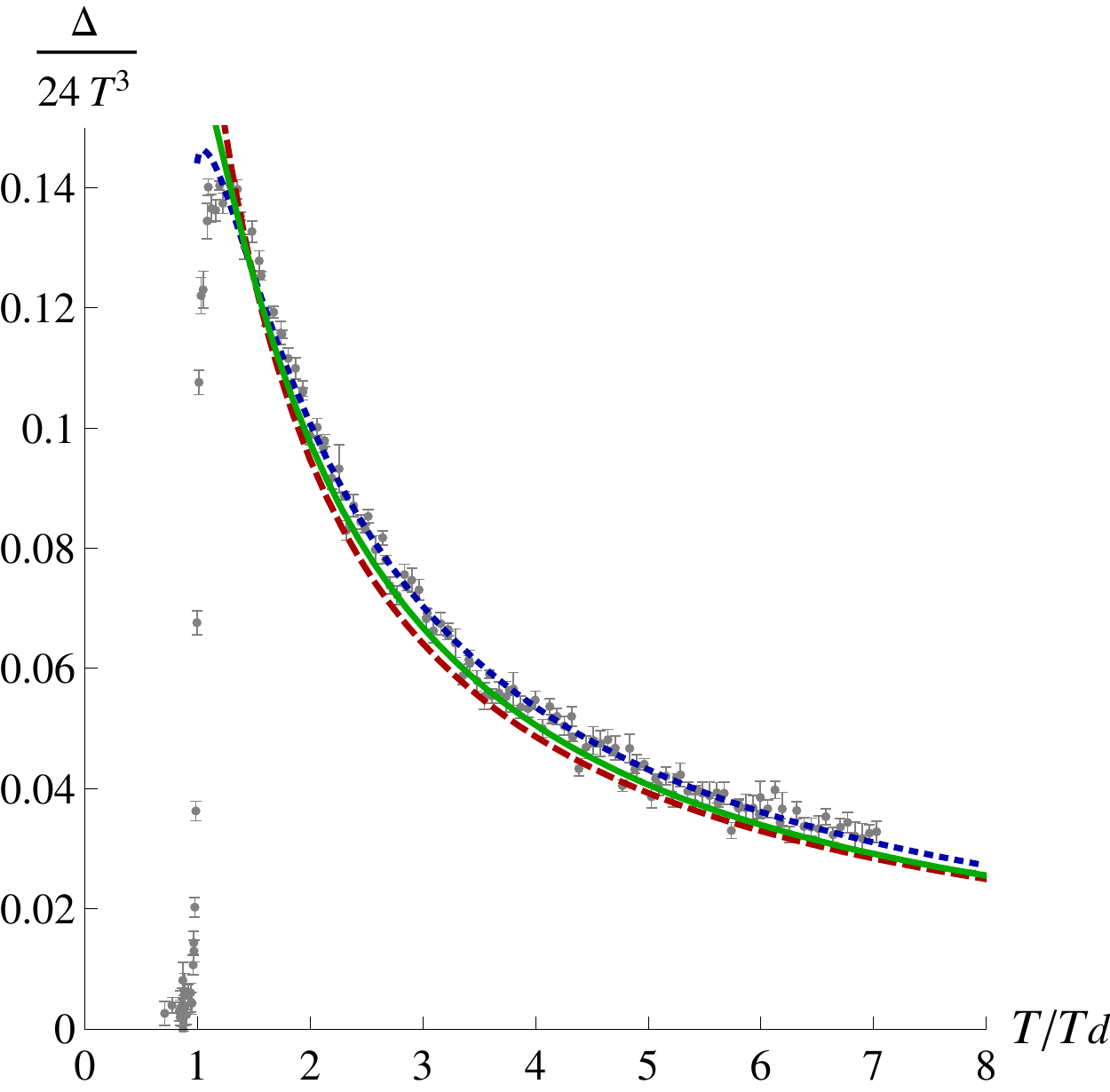}
\end{center}
\caption{Pressure (upper panels) and interaction measure (lower panels) scaled
by $N_{c}^{2}-1=24$ for $SU(N_{c}=5)$. Lattice data \cite{Caselle:2011mn} are
denoted by dots. The results of the matrix model are obtained by using the
Vandermonde determinant (left panel), and the linear term (right panel) within
the one-parameter model (dashed line), two-parameter model (solid line), and
four-parameter fit (dotted line). The horizontal lines represent the
perturbative limit of the pressure $c=\zeta\left(  3\right)  /2\pi$.}%
\label{ptr5}%
\end{figure}

\begin{figure}[t!]
\begin{center}
\includegraphics[width=0.49\textwidth]{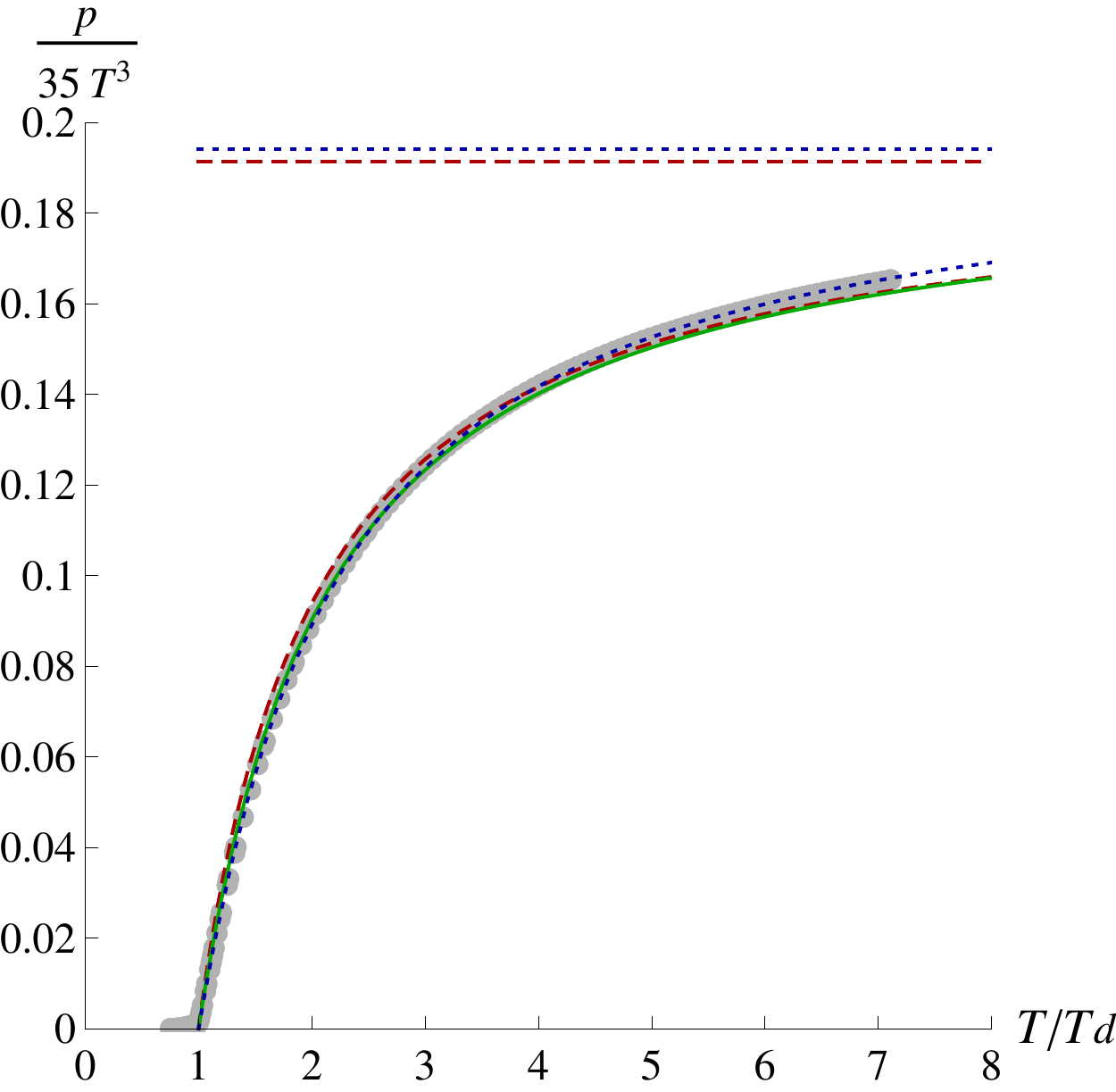}
\includegraphics[width=0.49\textwidth]{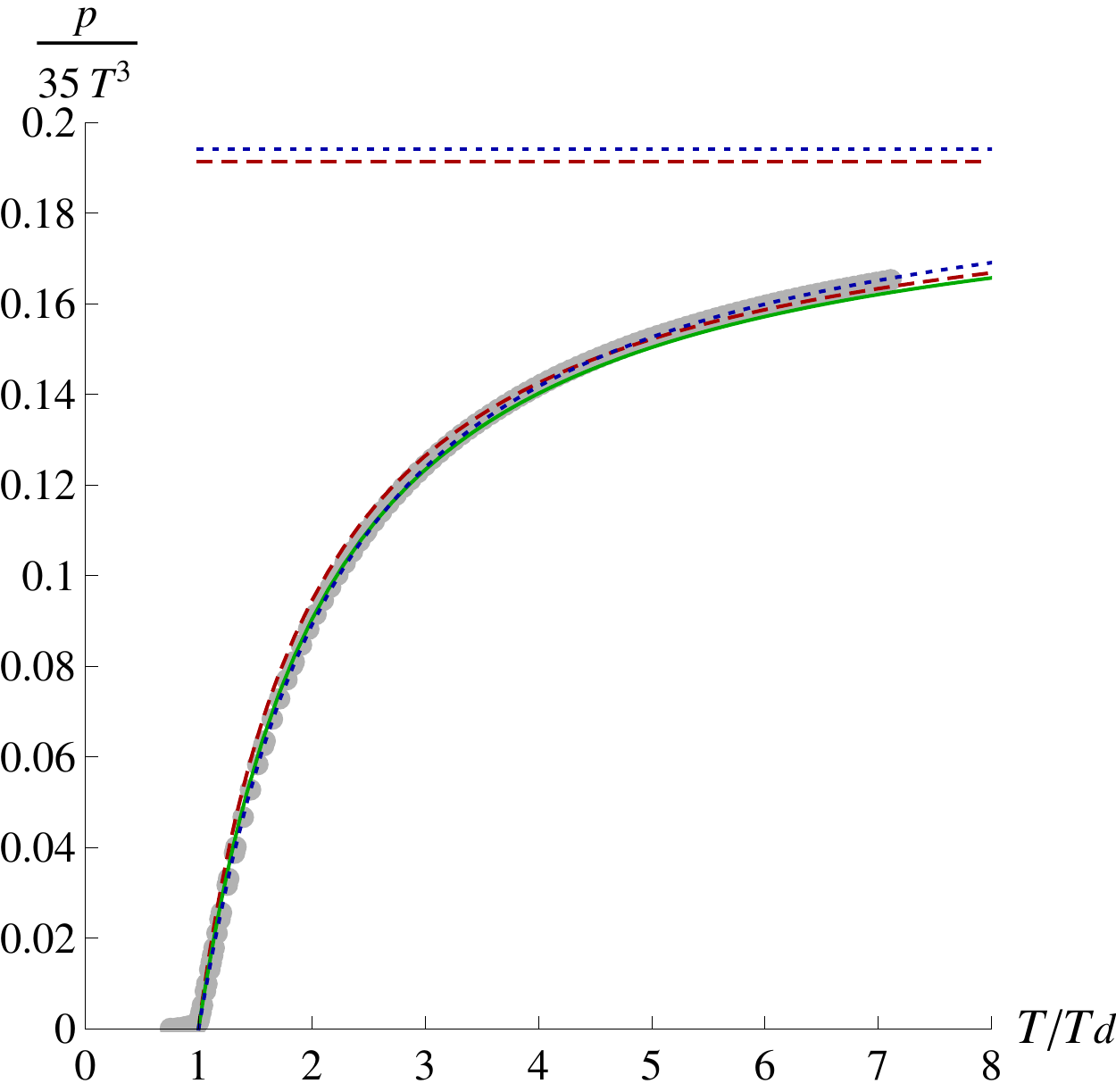}
\includegraphics[width=0.49\textwidth]{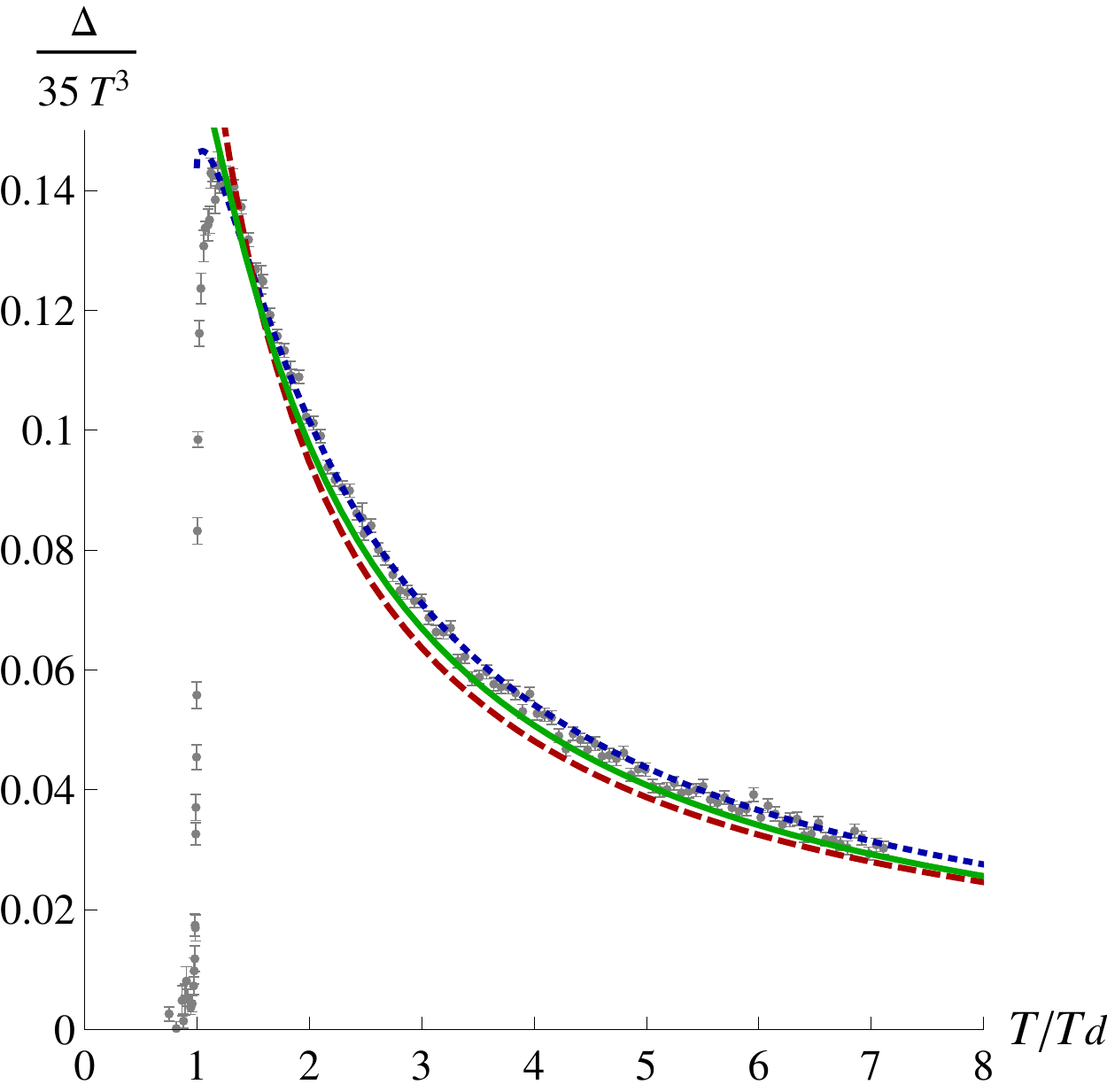}
\includegraphics[width=0.49\textwidth]{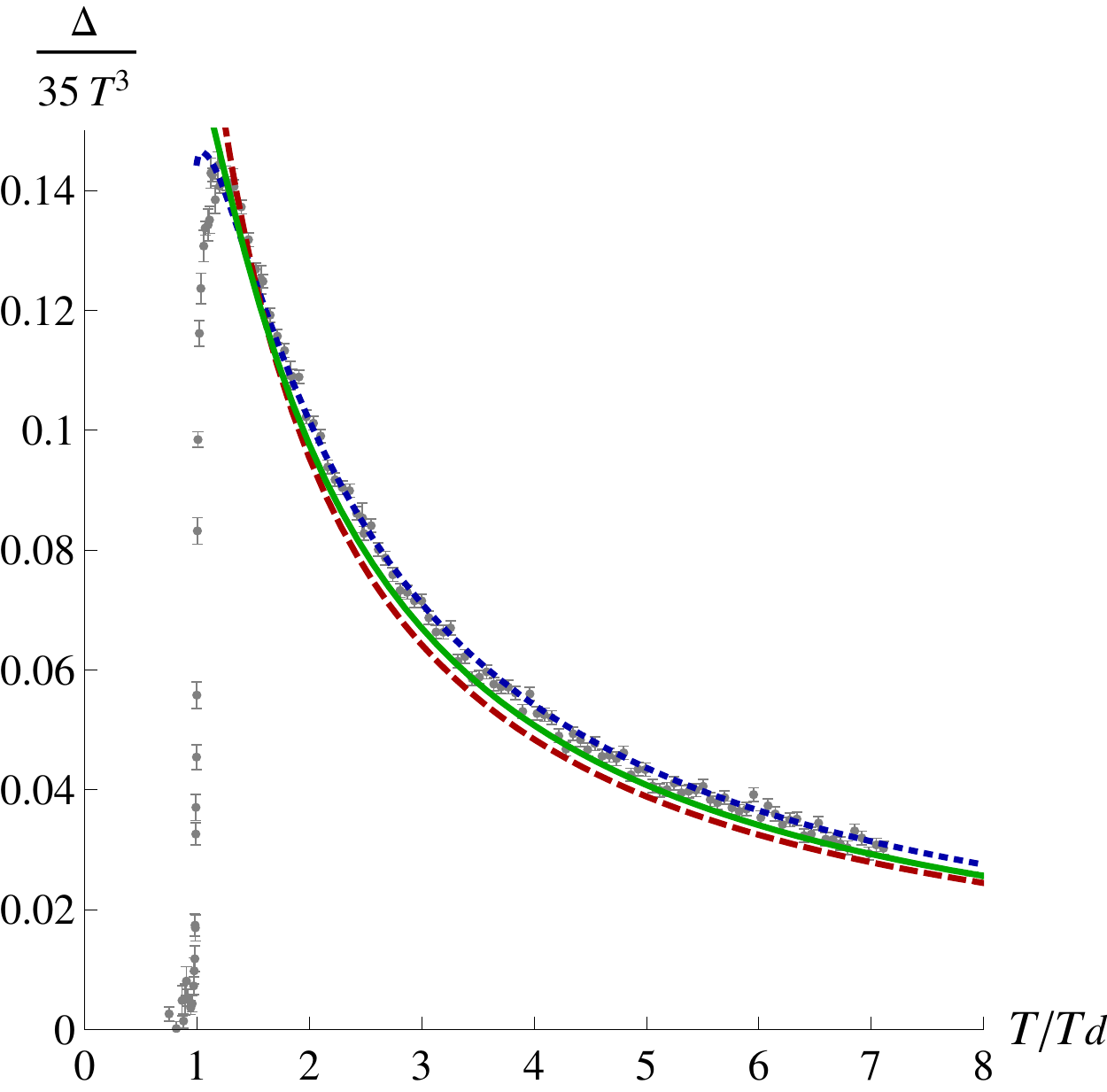}
\end{center}
\caption{Pressure (upper panels) and interaction measure (lower panels) scaled
by $N_{c}^{2}-1=35$ for $SU(N_{c}=6)$. Lattice data \cite{Caselle:2011mn} are
denoted by dots. The results of the matrix model are obtained by using the
Vandermonde determinant (left panel), and the linear term (right panel) within
the one-parameter model (dashed line), two-parameter model (solid line), and
four-parameter fit (dotted line). The horizontal lines represent the
perturbative limit of the pressure $c=\zeta\left(  3\right)  /2\pi$.}%
\label{ptr6}%
\end{figure}

\begin{figure}[t!]
\begin{center}
\includegraphics[width=0.49\textwidth]{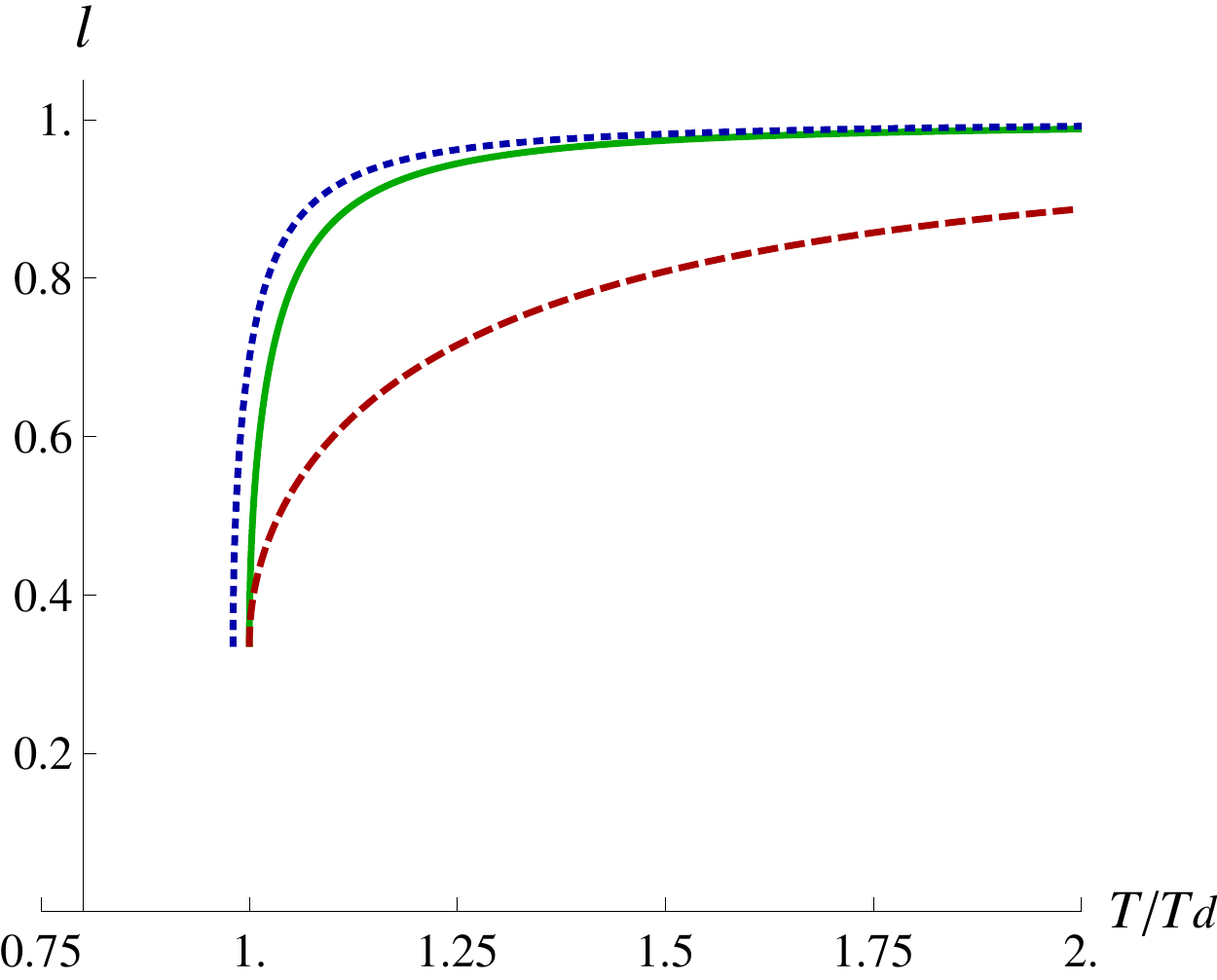}
\includegraphics[width=0.49\textwidth]{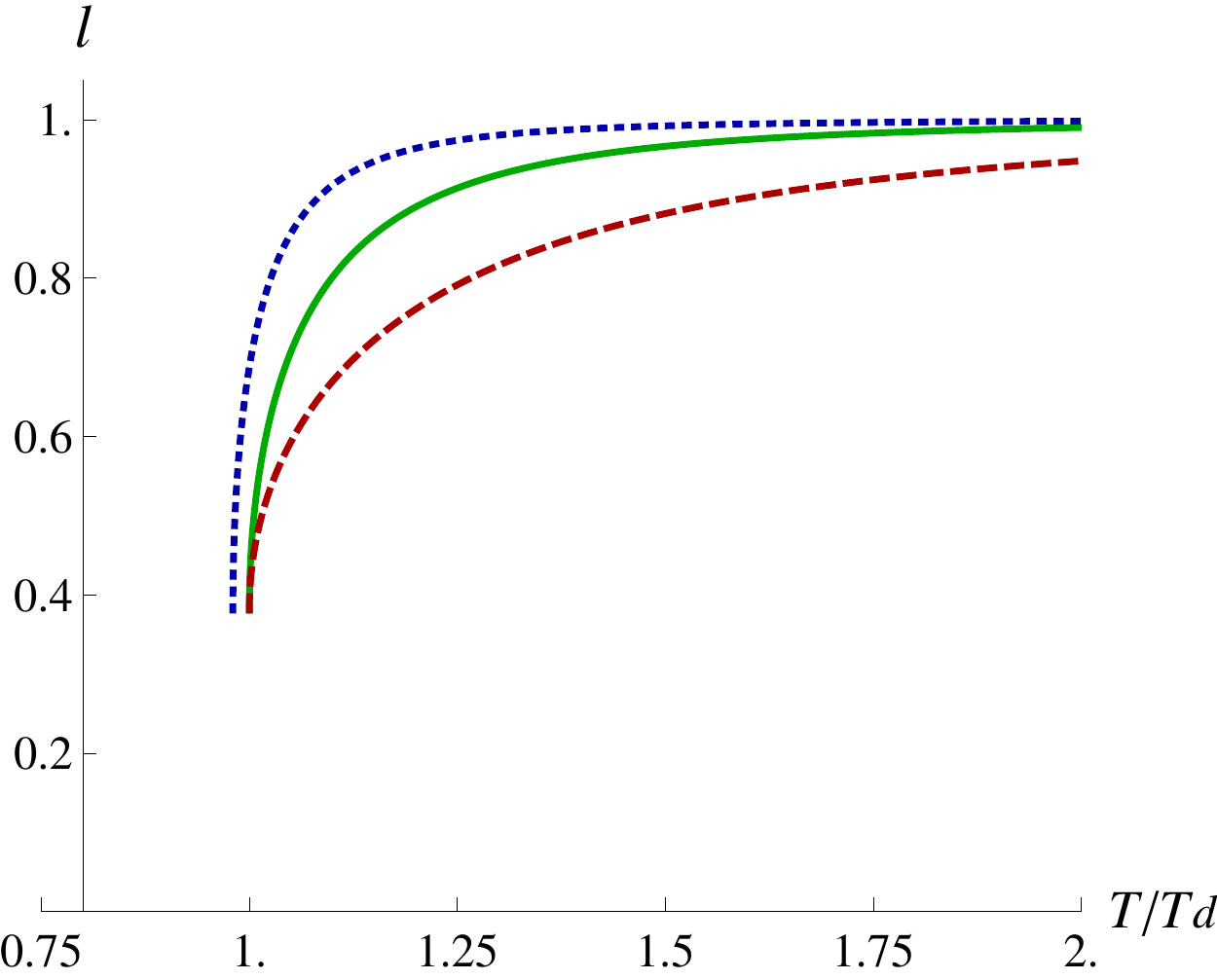}
\includegraphics[width=0.49\textwidth]{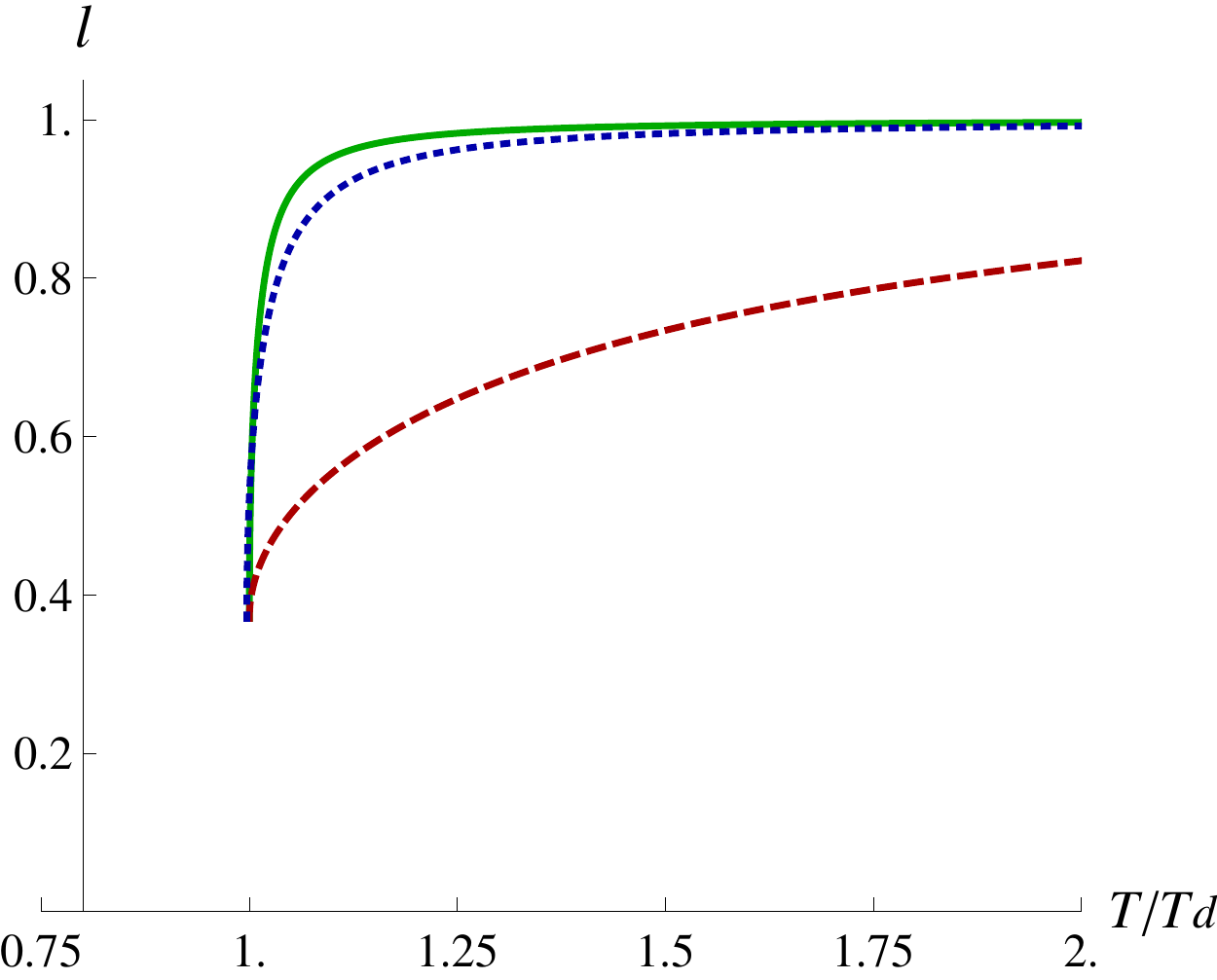}
\includegraphics[width=0.49\textwidth]{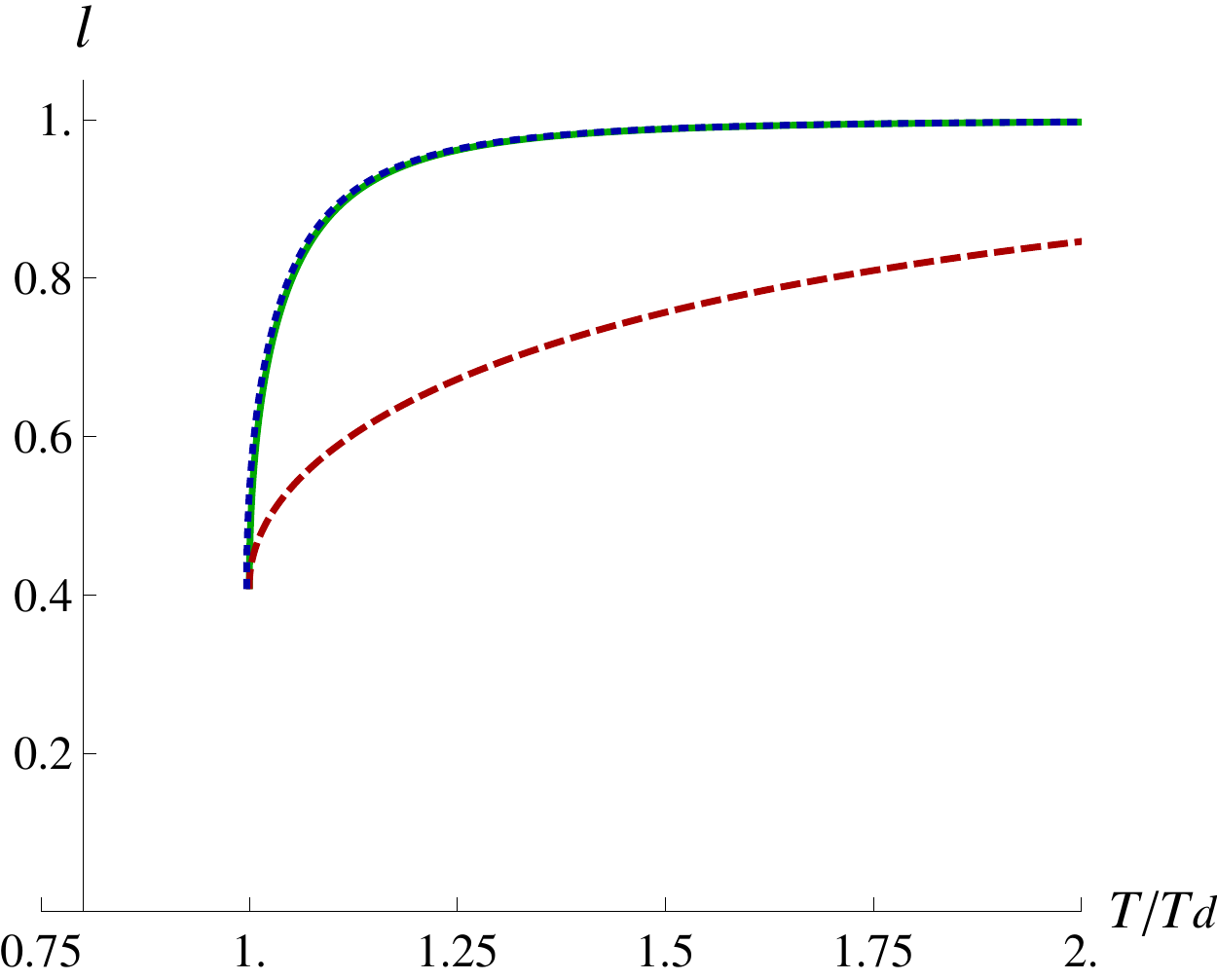}
\end{center}
\caption{The Polyakov loop for $SU(4)_{c}$ (upper panels) and $SU(5)_{c}$
(lower panels). The results are obtained for the Vandermonde determinant (left
panel), and the linear term (right panel) in the one-parameter model (dashed
line), two-parameter model (solid line), and four-parameter fit (dotted
line).}%
\label{pl45}%
\end{figure}

\begin{figure}[t!]
\begin{center}
\includegraphics[width=0.49\textwidth]{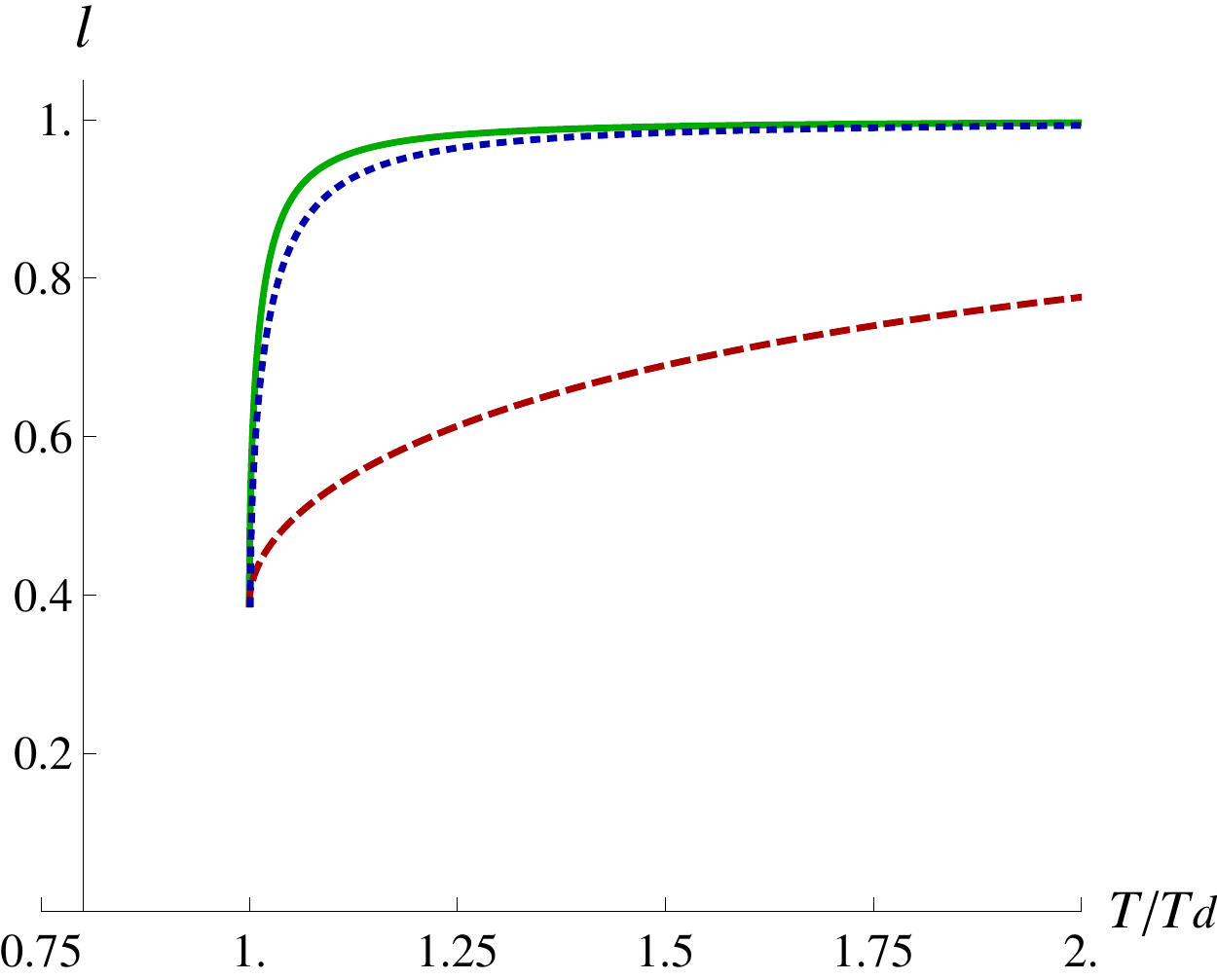}
\includegraphics[width=0.49\textwidth]{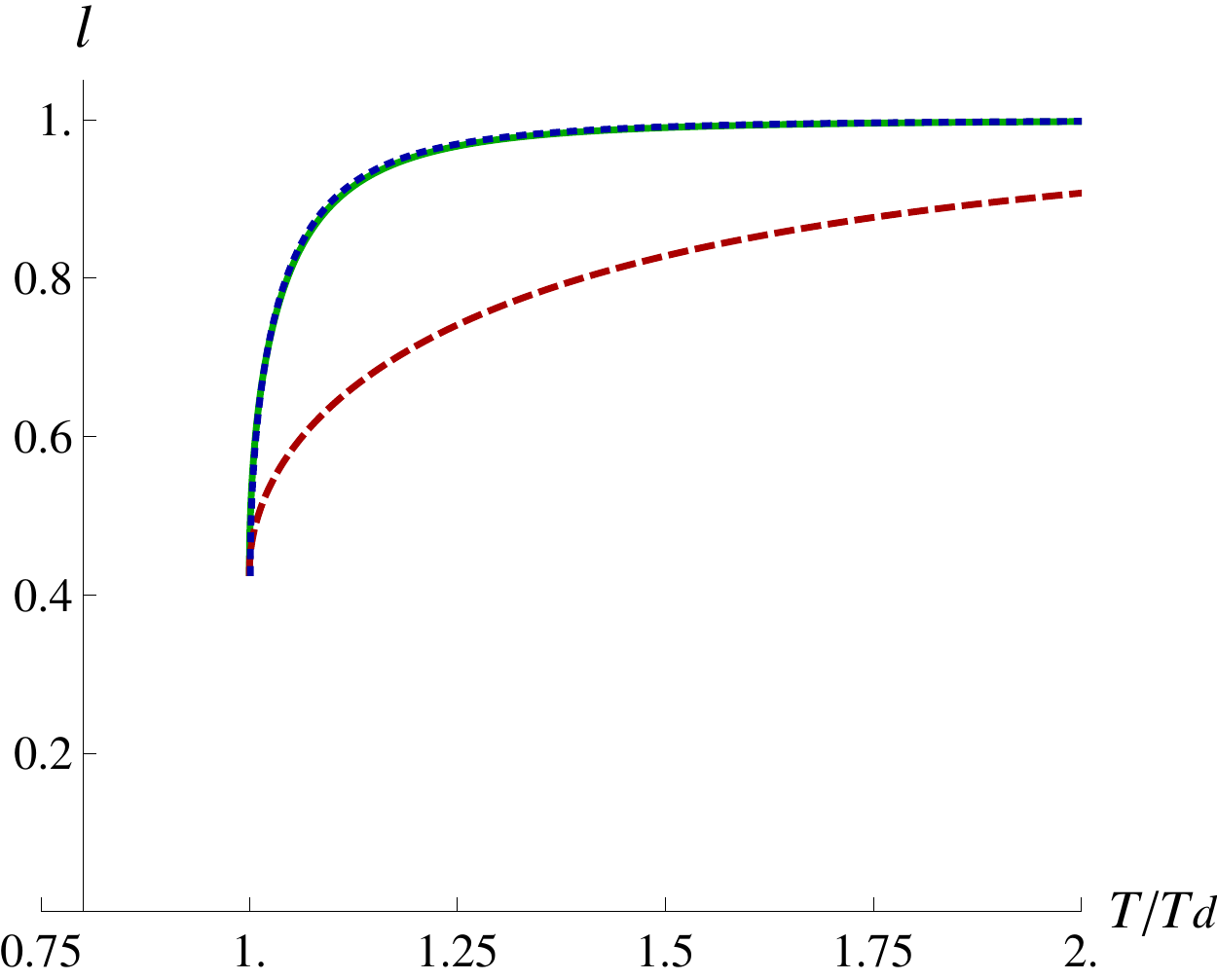}
\end{center}
\caption{The Polyakov loop for $SU(6)_{c}$ obtained for the Vandermonde
determinant (left panel), and the linear term (right panel) in the
one-parameter model (dashed line), two-parameter model (solid line), and
four-parameter fit (dotted line).}%
\label{pl6}%
\end{figure}

In this section we present the numerical results for the pressure and
interaction measure in Figs. \ref{ptr2}, \ref{ptr3}, \ref{ptr4}, \ref{ptr5},
and \ref{ptr6} for $N_{c}=2,3,4,5,6$ and compare to the lattice data of Ref.
\cite{Caselle:2011mn}. Moreover, we also numerically compute the Polyakov
loop, see Figs. \ref{pl23}, \ref{pl45}, and \ref{pl6}. For the renormalized
Polyakov loop in $d=2+1$ dimensions\ there are no lattice data available so
far. All results are obtained within the uniform eigenvalue Ansatz using the
one-parameter model, the two-parameter model, and the four-parameter fit. In
addition, we present the plots for two different nonperturbative terms, the
Vandermonde determinant (\ref{vdm22}) and the linear term (\ref{Vlin}).

\begin{table}[t!]
\vspace{0pt} \begin{adjustwidth}{-.9cm}{}
\begin{center}
\vspace{0pt}
\begin{tabular}
[c]{c|cccccc}\hline
 \hspace{.9cm} Non-pert. $V$ \hspace{.9cm} & \hspace{.7cm} $ C_{1}$\hspace{.7cm} &  \hspace{.7cm} $C_{2}$  \hspace{.7cm} & \hspace{.7cm}  $C_{3}$  \hspace{.7cm} &  \hspace{.6cm} $\delta C_{3}$  \hspace{.6cm} & \hspace{.1cm}  $T_{d}$
rescale  \hspace{.1cm}  & \hspace{.1cm}  $c$ rescale \hspace{.1cm}  \\\hline
\vspace{-2.5pt} SU(2)$_{c}$ Vdm 1 par & 0.000136 & 0.999951 & 0.999998 & 0 & 1 &
1 \\
\vspace{-2.5pt} SU(2)$_{c}$ Vdm 2 par & 0.000054 & 0.99998 & 0.999999 &
-0.0141610 & 1 & 1\\
\vspace{-2.5pt} SU(2)$_{c}$ Vdm 4 par & 0.216395 & 0.921952 & 0.997463 &
-0.111553 & 0.937285 & 1.03389 \\
\vspace{-2.5pt} SU(2)$_{c}$ lin 1 par & 0.000010 & 0.999996 & 1. & 0 & 1 & 1\\
\vspace{-2.5pt} SU(2)$_{c}$ lin 2 par & 0.000007 & 0.999997 & 1. & -0.0124482 &
1 & 1\\
\vspace{-2.5pt} SU(2)$_{c}$ lin 4 par & 0.958075 & 0.654448 & 0.988659 &
-0.0977655 & 0.933455 & 1.02394\\
\vspace{-2.5pt} SU(3)$_{c}$ Vdm 1 par & 0.0134234 & 0.992711 & 0.999587 & 0 &
1 & 1 \\
\vspace{-2.5pt} SU(3)$_{c}$ Vdm 2 par & 0.0103609 & 0.994374 & 0.999681 &
-0.0298485 & 1 & 1 \\
\vspace{-2.5pt} SU(3)$_{c}$ Vdm 4 par & 0.118785 & 0.935495 & 0.996459 &
-0.107797 & 0.968308 & 1.03236 \\
\vspace{-2.5pt} SU(3)$_{c}$ lin 1 par & 0.563232 & 0.75861 & 0.991713 & 0 & 1 &
1 \\
\vspace{-2.5pt} SU(3)$_{c}$ lin 2 par & 0.582769 & 0.750236 & 0.991426 & -
0.000000 & 1 & 1 \\
\vspace{-2.5pt} SU(3)$_{c}$ lin 4 par & 0.813344 & 0.651416 & 0.988387 &
-0.104560 & 0.967003 & 1.03043 \\
\vspace{-2.5pt} SU(4)$_{c}$ Vdm 1 par & 1.30689 & 0.159154 & 0.967712 & 0 & 1 &
1 \\
\vspace{-2.5pt} SU(4)$_{c}$ Vdm 2 par & 0.155114 & 0.900201 & 0.996168 &
-0.0678792 & 1 & 1 \\
\vspace{-2.5pt} SU(4)$_{c}$ Vdm 4 par & 0.118808 & 0.92356 & 0.997172 &
-0.159610 & 0.980361 & 1.03790 \\
\vspace{-2.5pt} SU(4)$_{c}$ lin 1 par & 3.40505 & -0.547142 & 0.965097 & 0 & 1 &
1 \\
\vspace{-2.5pt} SU(4)$_{c}$ lin 2 par & 1.26641 & 0.424584 & 0.987019 &
-0.0640259 & 1 & 1 \\
\vspace{-2.5pt} SU(4)$_{c}$ lin 4 par & 0.595675 & 0.729345 & 0.994108 &
-0.159541 & 0.980182 & 1.03626 \\
\vspace{-2.5pt} SU(5)$_{c}$ Vdm 1 par & 2.28362 & -0.612346 & 0.956678 & 0 & 1 &
1 \\
\vspace{-2.5pt} SU(5)$_{c}$ Vdm 2 par & 0.0504081 & 0.96441 & 0.999044 &
-0.0665565 & 1 & 1 \\
\vspace{-2.5pt} SU(5)$_{c}$ Vdm 4 par & 0.103883 & 0.926654 & 0.998076 &
-0.114687 & 0.996903 & 1.02406 \\
\vspace{-2.5pt} SU(5)$_{c}$ lin 1 par & 9.3767 & -3.37883 & 0.931681 & 0 & 1 &
1 \\
\vspace{-2.5pt} SU(5)$_{c}$ lin 2 par & 0.693992 & 0.675912 & 0.994944 &
-0.0642173 & 1 & 1 \\
\vspace{-2.5pt} SU(5)$_{c}$ lin 4 par & 0.689758 & 0.677889 & 0.995089 &
-0.114556 & 0.996670 & 1.02339 \\
\vspace{-2.5pt} SU(6)$_{c}$ Vdm 1 par & 3.26897 & -1.44726 & 0.951671 & 0 & 1 &
1 \\
\vspace{-2.5pt} SU(6)$_{c}$ Vdm 2 par & 0.0536514 & 0.959835 & 0.999207 &
-0.0722529 & 1 & 1 \\
\vspace{-2.5pt} SU(6)$_{c}$ Vdm 4 par & 0.0927683 & 0.93055 & 0.998666 &
-0.120177 & 1.00121 & 1.02807 \\
\vspace{-2.5pt} SU(6)$_{c}$ lin 1 par & 5.54761 & -1.63022 & 0.97006 & 0 & 1 &
1 \\
\vspace{-2.5pt} SU(6)$_{c}$ lin 2 par & 0.634597 & 0.699126 & 0.996575 &
-0.0711707 & 1 & 1 \\
\vspace{-2.5pt} SU(6)$_{c}$ lin 4 par & 0.598767 & 0.716114 & 0.996855 &
-0.120282 & 1.00109 & 1.02763 \\\hline
\end{tabular}
\end{center}
\par
\caption{Parameters determined by fitting the lattice pressure
for different $SU(N_{c})$ groups.}
\label{tablesun}
\end{adjustwidth}
\end{table}

The free parameters of the model are adjusted by fitting the lattice pressure
of Ref. \cite{Caselle:2011mn} for $T\geq1.0T_{d}$ for all $N_{c}$. Table
\ref{tablesun} lists the parameters for all cases studied in this work, where
we introduce the following notation: \textquotedblleft1 par\textquotedblright%
\ for the one-parameter model, \textquotedblleft2 par\textquotedblright\ for
the two-parameter model, and \textquotedblleft4 par\textquotedblright\ for the
four-parameter fit. Furthermore, \textquotedblleft lin\textquotedblright%
\ denotes the linear term and \textquotedblleft Vdm\textquotedblright\ the
Vandermonde determinant. The parameters $C_{2}$ and $C_{3}$ are not free, they
are functions of $C_{1}$. In \textquotedblleft1 par\textquotedblright\ we use
$C_{1}$ as the single free parameter to fit the lattice data.{\ In
\textquotedblleft2 par\textquotedblright\ we add a second free parameter,}%
\begin{equation}
{\delta C_{3}=C_{3}(T_{d})-C_{3}(T=\infty)}\text{ ,}%
\end{equation}
to include the effects of the bag constant $B$, see Sec. \ref{fixparam}. In
\textquotedblleft4 par\textquotedblright\ we further allow for small shifts in
$T_{d}$, and in the perturbative constant $c$ (\ref{Limit}), in order to
encompass other possible nonperturbative effects not included in our matrix model.

Overall, the change in $C_{1}$ and $C_{2}$ with the number of colors becomes
weaker with increasing number of parameters. Moreover, for the Vandermonde
determinant the variation of $C_{1}$ and $C_{2}$ with $N_{c}$ is milder than
for the linear nonperturbative term. In the one-parameter model $C_{1}$ and
$C_{2}$ change significantly with increasing $N_{c}$ and/or when switching
between the linear and the Vandermonde term. For the two-parameter model, we
find that as $N_{c}$ increases, $C_{1}$ and $C_{2}$ approach the same values
as in the four-parameter fit which gives quantitatively the best fits to the
lattice data. 

In the four-parameter fit $C_{1}$ tends to decrease with increasing number of
colors. Moreover, the value of $C_{1}$ in the four-parameter fit is
significantly smaller than for the one-parameter model. Typically, $C_{1}$
drops by an order of magnitude as one goes from one to four parameters.
Notably, for the Vandermonde determinant $C_{2}$ assumes approximately the
same value for all $N_{c}$ in the four-parameter fit, $C_{2}\simeq0.93$.
Finally, we notice that for all scenarios studied in this work, $C_{3}$ is
approximately constant, $C_{3}\simeq1,$ presumably because it is fixed by the
high-$T$ behavior.

\subsection{Pressure and interaction measure}

\subsubsection{One- and two-parameter model}

The one- and two-parameter matrix models reproduce the lattice data remarkably
well within the considered temperature region $T_{d}\leq T\leq8T_{d}$. An
important result is that with only one free parameter, we can already describe
the deviation from an ideal gluon gas observed on the lattice at
$1.2\,T_{d}\lesssim T\lesssim10$ $T_{d}$. Regarding the dependence on the
number of colors, we confirm the general trend observed on the lattice: when
scaled by the number of gluons, $N_{c}^{2}-1$, the pressure and the
interaction measure only marginally change with varying $N_{c}$. For all
$N_{c}$, we find that{\ including the effects of the bag constant $B$} allows
for a better agreement with the lattice data than the one-parameter model.

Despite the overall good correspondence, we notice that at high temperatures
the deviation between our results and the lattice pressure slightly increases.
In addition, we are not able to reproduce the correct shape for the peak of
the interaction measure residing near $T_{d}$. Therefore, we also present the
results using the four-parameter fit.

An interesting observation is that the two-parameter model can be seen as an
interpolation between the one-parameter model for $N_{c}=2,3$ and the
four-parameter fit for larger values of $N_{c}$.

\subsubsection{Four-parameter fit}

The four-parameter fit quantitatively improves the agreement with lattice data
for the pressure and interaction measure. The shift in $T_{d}$ is introduced
to account for possible glueballs effects not included in our model, while the
shift in $c$ is motivated by possible higher-order loop corrections to the
perturbative potential. Furthermore, finite-volume effects present on the
lattice close to $T_{d}$ and at high temperatures may also slightly shift the
values of $T_{d}$ and $c$.

\subsubsection{$Shift$ $in$ $T_{d}$}

\begin{figure}[t!]
\begin{center}
\includegraphics[width=0.5\textwidth]{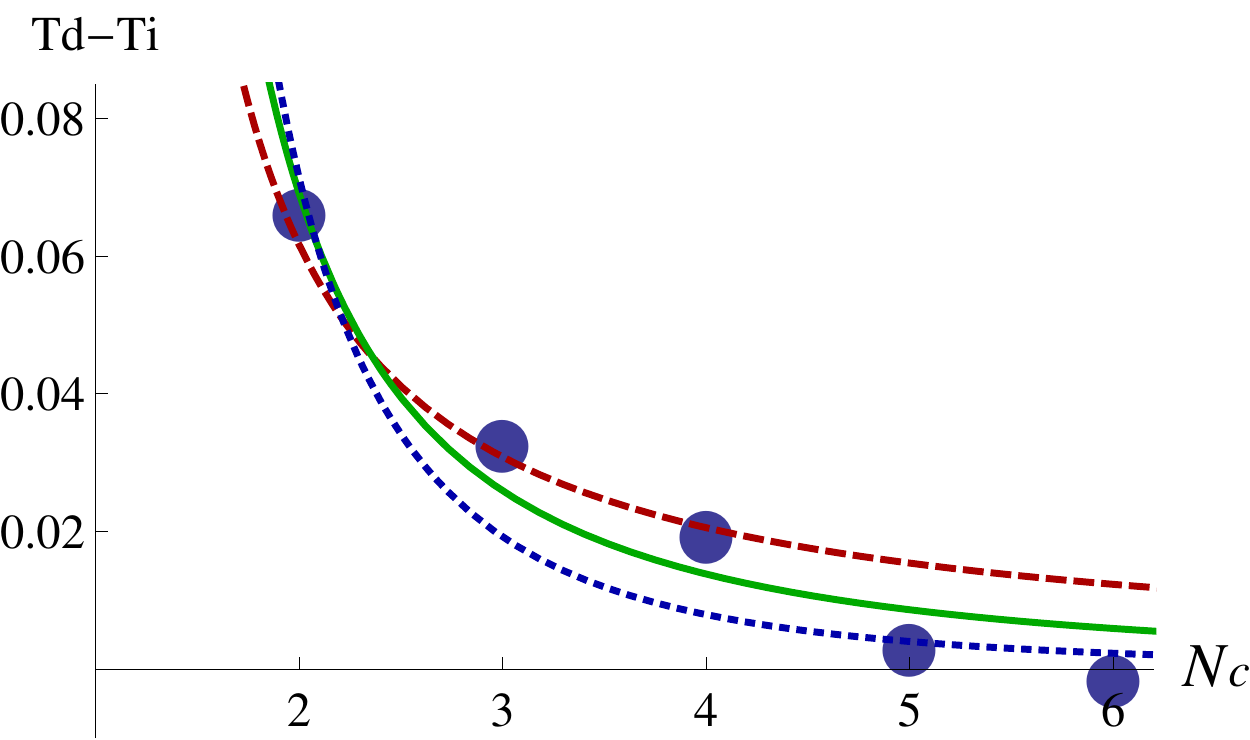}
\end{center}
\caption{Shift in the critical temperature, $T_{d}-T_{i}$, as a function of
$N_{c}$, where $T_{i}$ is the estimated temperature, at which the pressure
vanishes. We also show three different fits: $1/(N_{c}-1)$ (dashed line),
$1/(N_{c}^{2}-1)$ (solid line), and $1/(N_{c}^{3}-1)$ (dotted line). }%
\label{shift1}%
\end{figure}

In Fig. \ref{shift1} we plot the shift in the critical temperature,
$T_{d}-T_{i}$, as a function of $N_{c}$, together with three different fit
functions, $1/\left(  N_{c}-1\right)  ,$ $1/\left(  N_{c}^{2}-1\right)  ,$ and
$1/\left(  N_{c}^{3}-1\right)  $. The intercept temperature $T_{i}$
corresponds to the best estimate for the temperature where the pressure
vanishes. Notably, the values for $T_{d}-T_{i}$ are rather small and rapidly
decrease with $N_{c}$, approximately as $1/\left(  N_{c}^{2}-1\right)  ,$
indicating that any possible glueball contribution becomes less important for
larger $N_{c}$. This is in accordance with general expectations that glueballs
become suppressed by a factor $\sim1/\left(  N_{c}^{2}-1\right)  $ above
$T_{d}$.

\subsubsection{$Shift$ $in$ $c$}

Concerning the shift in the perturbative constant $c,$ it is interesting to
note that it remains approximately constant when varying the numbers of
colors, $c\simeq3\%$, supporting the expectation that including higher-order
loop perturbative corrections may account for the small deviation from lattice
data in the perturbative limit.

\subsection{Polyakov loop}

Using the parameters determined by fitting the lattice pressure, we also plot
the Polyakov loop.

\subsubsection{$N_{c}=2$}

\begin{figure}[t!]
\begin{center}
\includegraphics[width=0.49\textwidth]{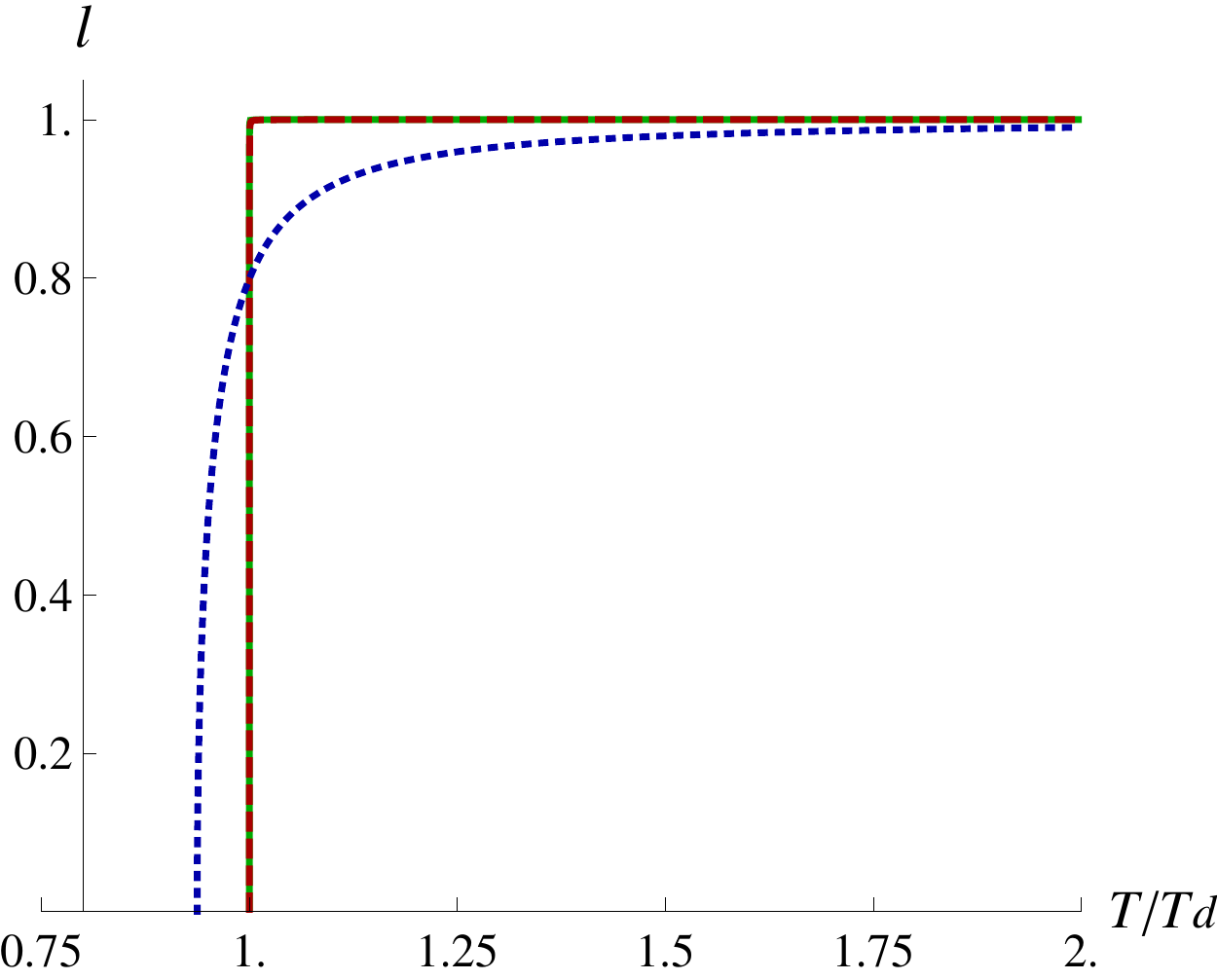}
\includegraphics[width=0.49\textwidth]{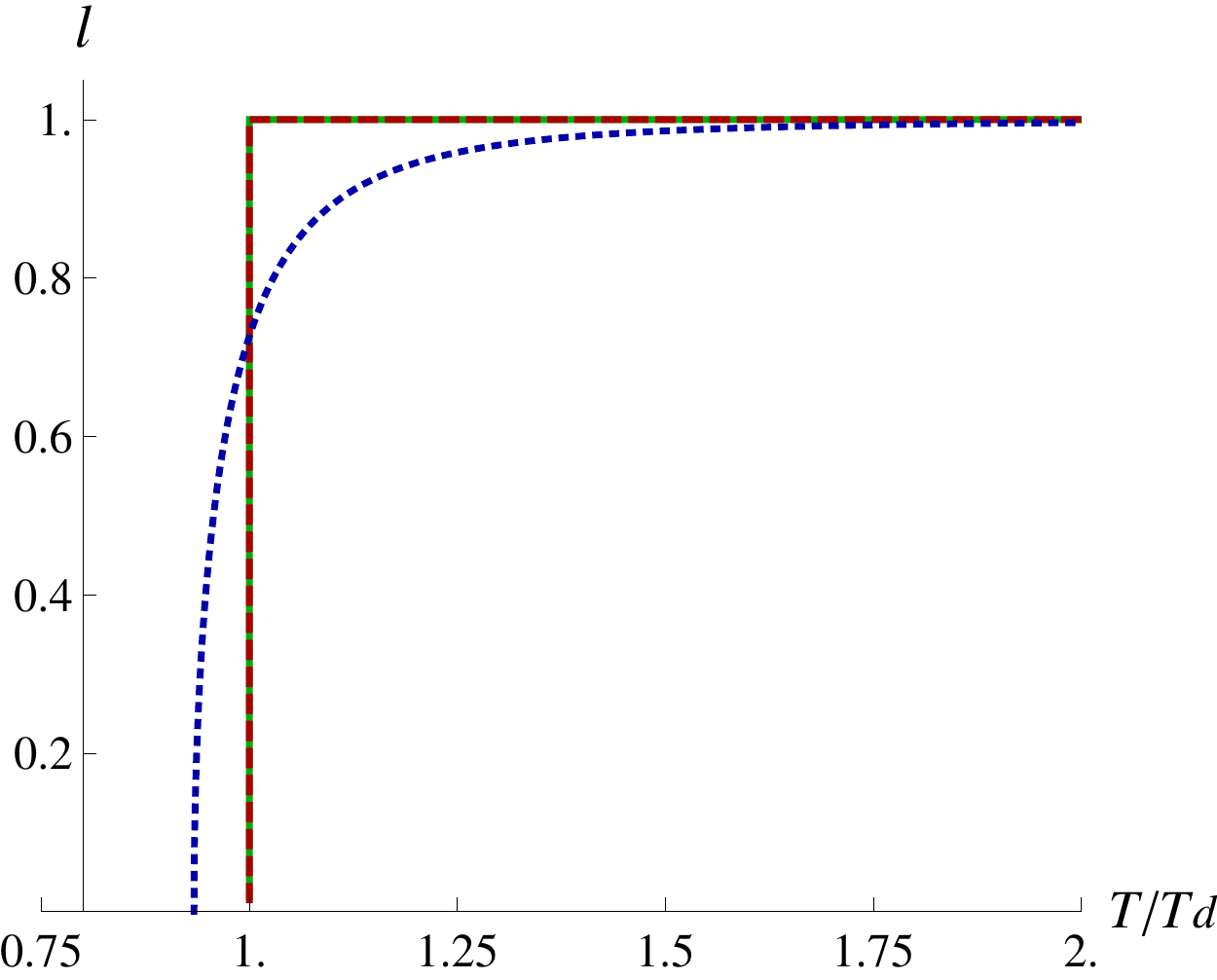}
\includegraphics[width=0.49\textwidth]{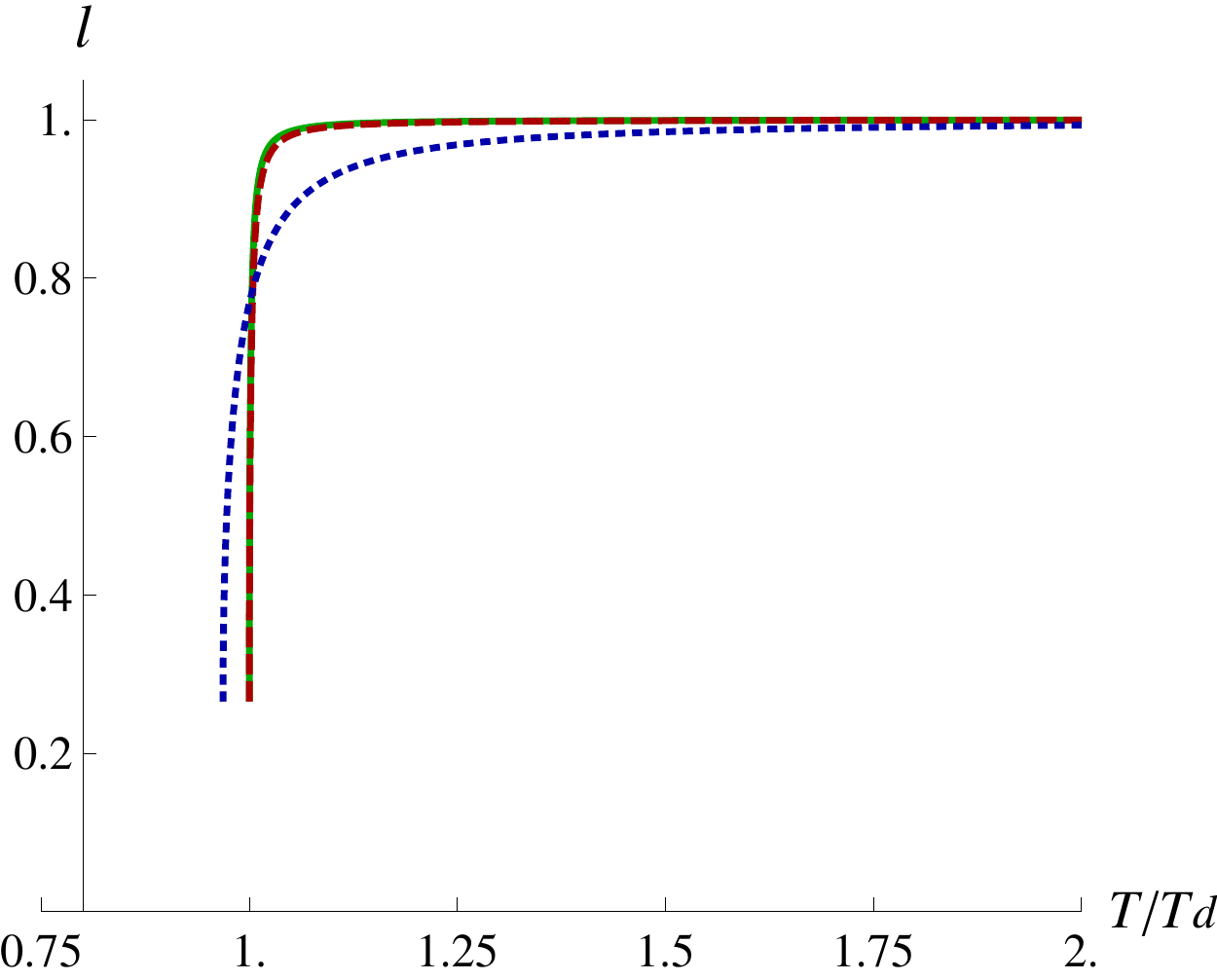}
\includegraphics[width=0.49\textwidth]{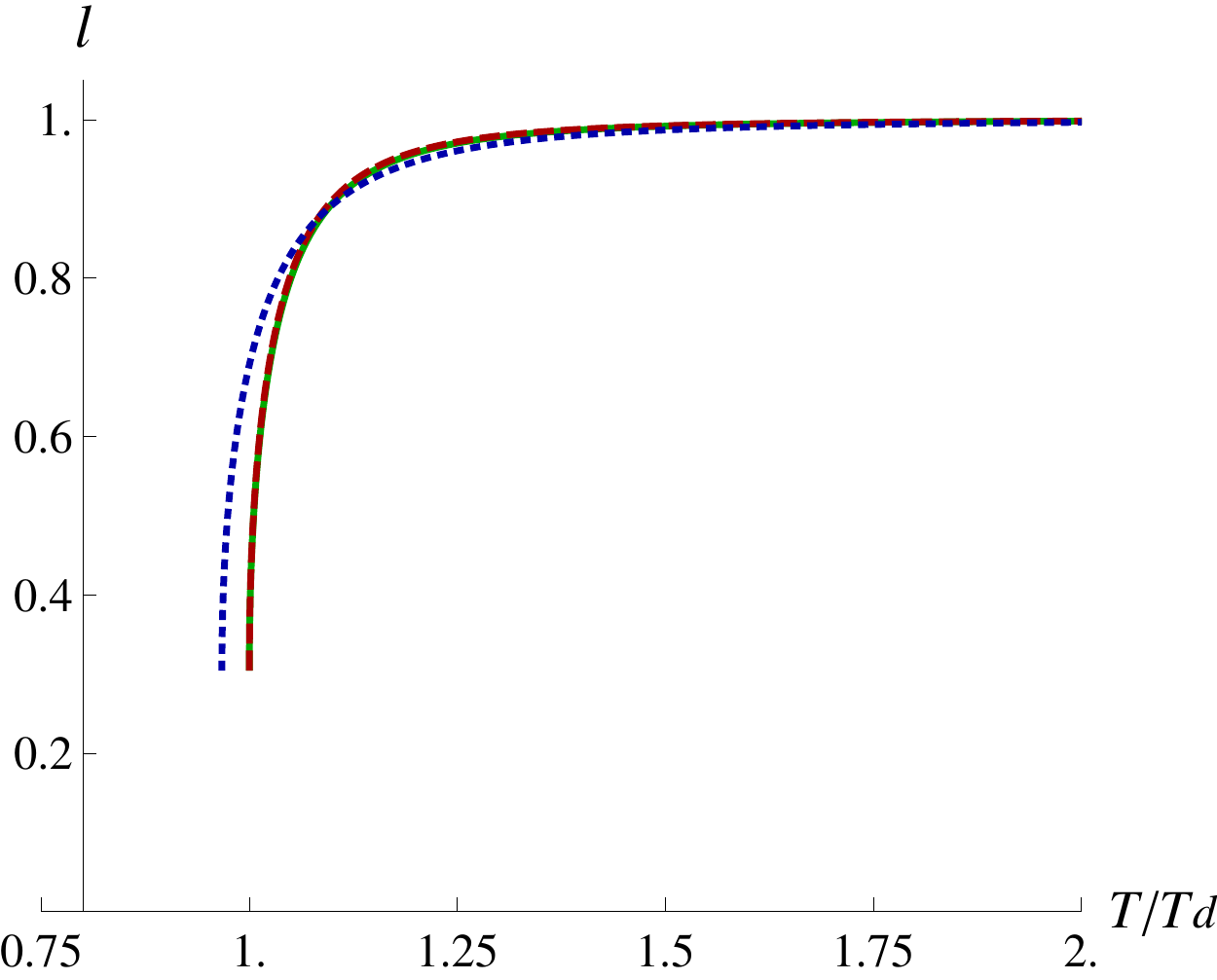}
\end{center}
\caption{The Polyakov loop for $SU(2)_{c}$ (upper panels) and $SU(3)_{c}$
(lower panels). The results are obtained for the Vandermonde determinant (left
panel), and the linear term (right panel) in the one-parameter model (dashed
line), two-parameter model (solid line), and four-parameter fit (dotted
line).}%
\label{pl23}%
\end{figure}

In the one- and two-parameter model, the best fits to the lattice pressure for
$N_{c}=2$ are obtained in the cases, where the system instantly merges from
the confining vacuum, $s_{c}=1$, to the perturbative vacuum, $s=0$, above
$T_{d}$. Consequently, for two colors the Polyakov loop rises abruptly from
$l_{c}=0$ to $l=1$, see Fig. \ref{pl23}. This is because in our model the
Polyakov loop differs from unity only for nonzero values of the condensate
$s_{min}(T)$, see Eqs. (\ref{pl1}) and (\ref{pl2}). Strictly speaking, in the
presence of a linear or a Vandermonde term the condensate never identically
vanishes. Still, the region where $s_{min}(T)$ is numerically large, is very
narrow. As explained in Sec. \ref{fixparam}, the reason for the shortcoming of
the one- and two-parameter model may be that for $N_{c}=2$ the glueball
contribution to the free energy is not negligible at the phase transition.
Therefore, we need to allow for a nonzero pressure at $T_{d}$, which is
equivalent to shifting the critical temperature, the way it is done in the
four-parameter fit.

\subsubsection{$N_{c}\geq3$}

For $N_{c}\geq3$ the phase transition is of first order. Thus, the expectation
value of the Polyakov loop jumps from zero to a nonzero value at the critical
temperature, and then approaches unity at asymptotically high temperatures.

In the one-parameter model, the region where the Polyakov loop notably differs
from unity increases with $N_{c}$. In the four-parameter fit, on the other
hand, the plots for the Polyakov loop only slightly change when varying
$N_{c}$, similar to other thermodynamical observables. Interestingly, the
two-parameter plots for the Polyakov loop essentially coincide with the
one-parameter model for $N_{c}=2,3$, while for $N_{c}=5,6$ they are very close
to the results of the four-parameter fit. Note that for $N_{c}\geq3$ any
possible glueball contribution near the phase transition is already
sufficiently suppressed by the factor $1/\left(  N_{c}^{2}-1\right)  $. As a
consequence, for $N_{c}\geq3,$ all models considered in this work provide
plausible plots for the Polyakov loop, while for $N_{c}=2$\ only the
four-parameter fit allows for reasonable results.

\subsubsection{Transition region}

From the Polyakov loop, we can directly determine the transition range where
the condensate $s_{min}(T)$ is nonvanishing. This is the region where the $s-
$dependent nonperturbative terms enter the thermodynamical quantities. In the
one-parameter model, the transition range is practically zero for $N_{c}=2$.
It becomes broader with increasing number of colors, and ends up approximately
at $4T_{d}$ for $N_{c}=6$. In the four-parameter fit, the transition region is
broadly independent of $N_{c}$ and extends up to $\sim1.25T_{d}$. Similar
results are obtained in the two-parameter model, except for the case $N_{c}%
=2$, where the transition range vanishes. Remarkably, this is very close to
the results in $d=3+1$ dimensions, where the Polyakov loop notably differs
from unity up to $\sim1.2T_{d}$.

\section{The potential for a general parametrization of the background field
\label{meta}}

In the previous sections \ref{ua} and \ref{results} we have presented a
detailed study of the region between the perturbative and confining vacuum,
$0\leq s\leq s_{c}$, where the Polyakov loop varies from zero to unity, using
the uniform eigenvalue Ansatz $\mathbf{q}\left(  s\right)  =s\mathbf{q}_{c}$
(\ref{ansatz}). In this section we extend the parametrization of the
background $\mathcal{A}_{0}$ field to a complete basis of diagonal generators
of the $SU(N_{c})$\ Lie algebra. Thus, the Polyakov loop may take any value
within the solid lines of Fig. \ref{flowers}.

\begin{figure}[t!]
\begin{center}
\includegraphics[width=0.45\textwidth]{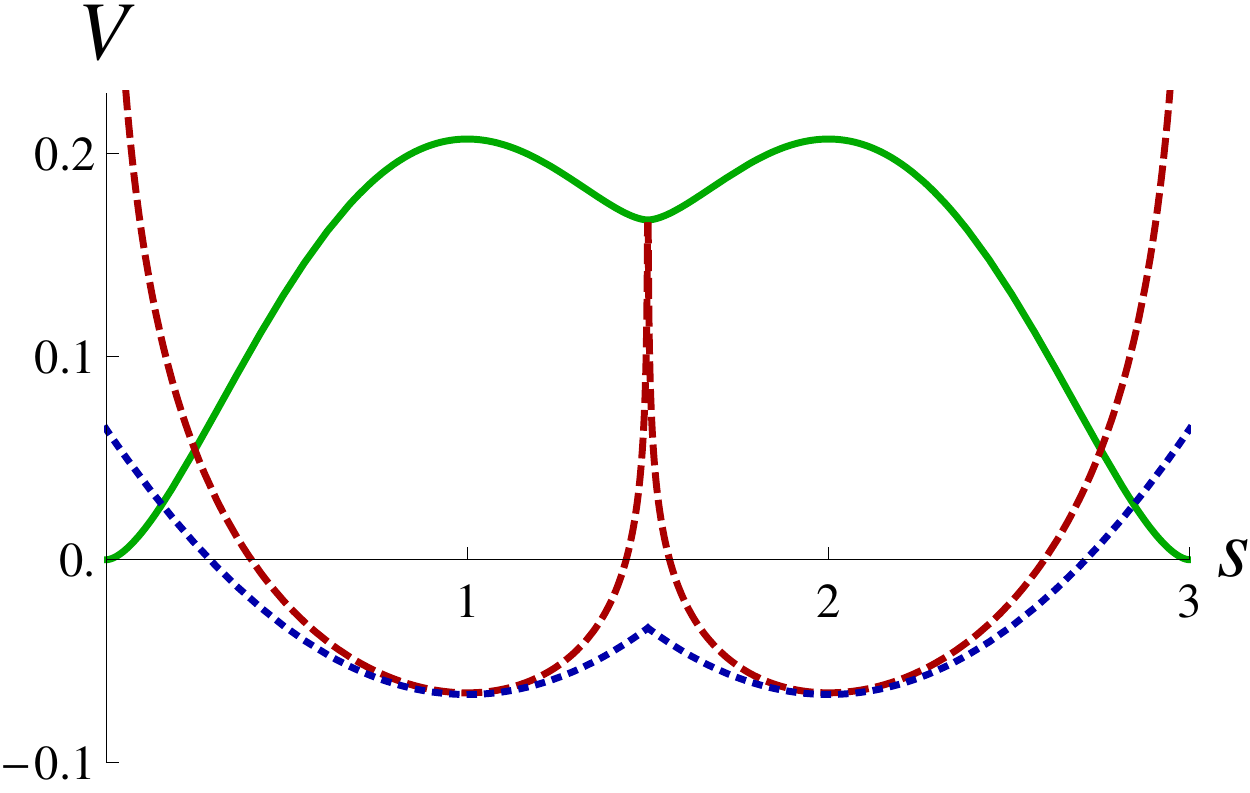}
\hspace{10pt}
\includegraphics[width=0.45\textwidth]{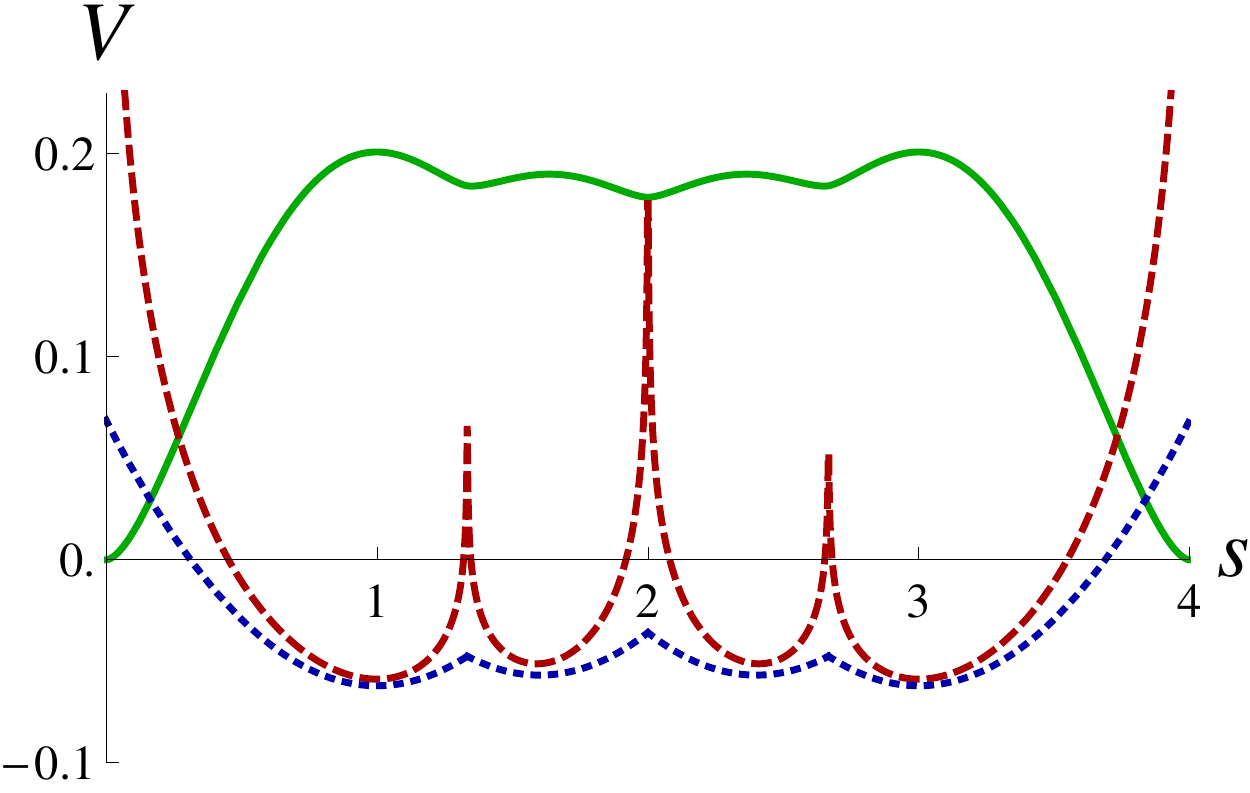}
\includegraphics[width=0.45\textwidth]{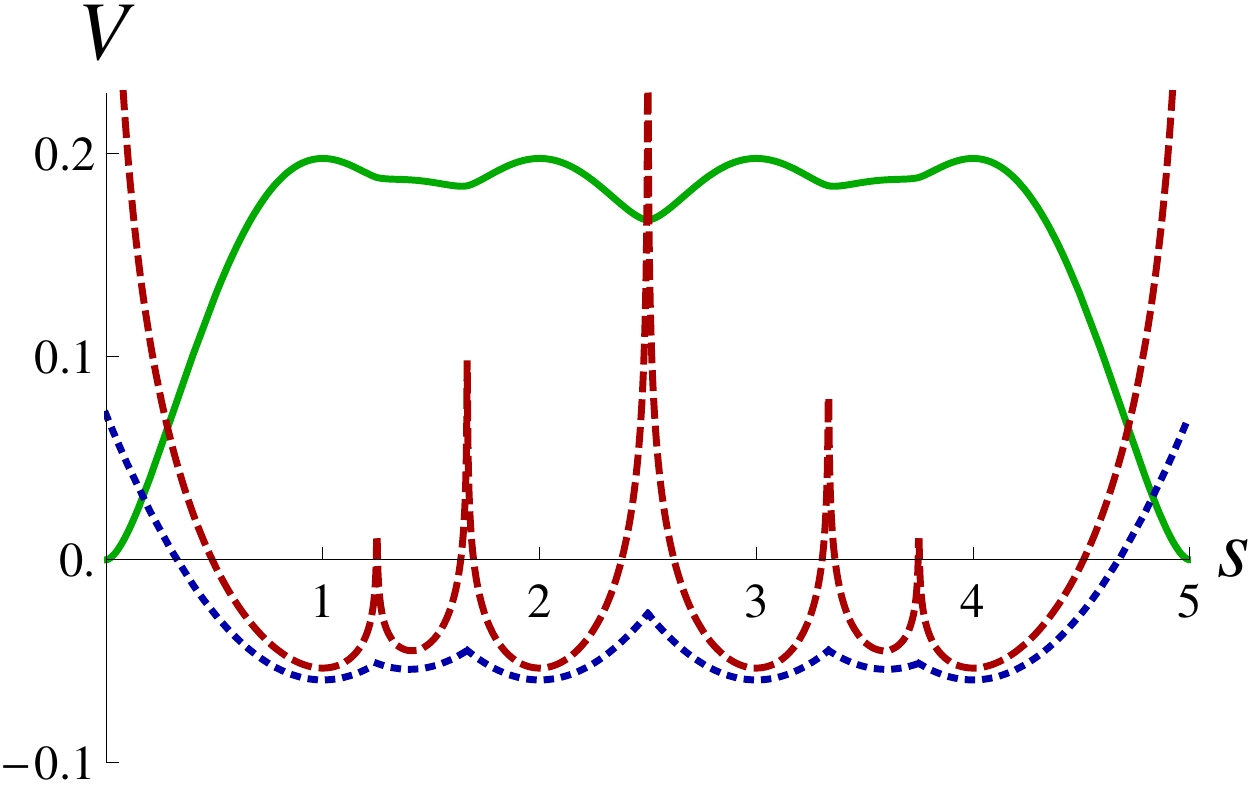}
\hspace{10pt}
\includegraphics[width=0.45\textwidth]{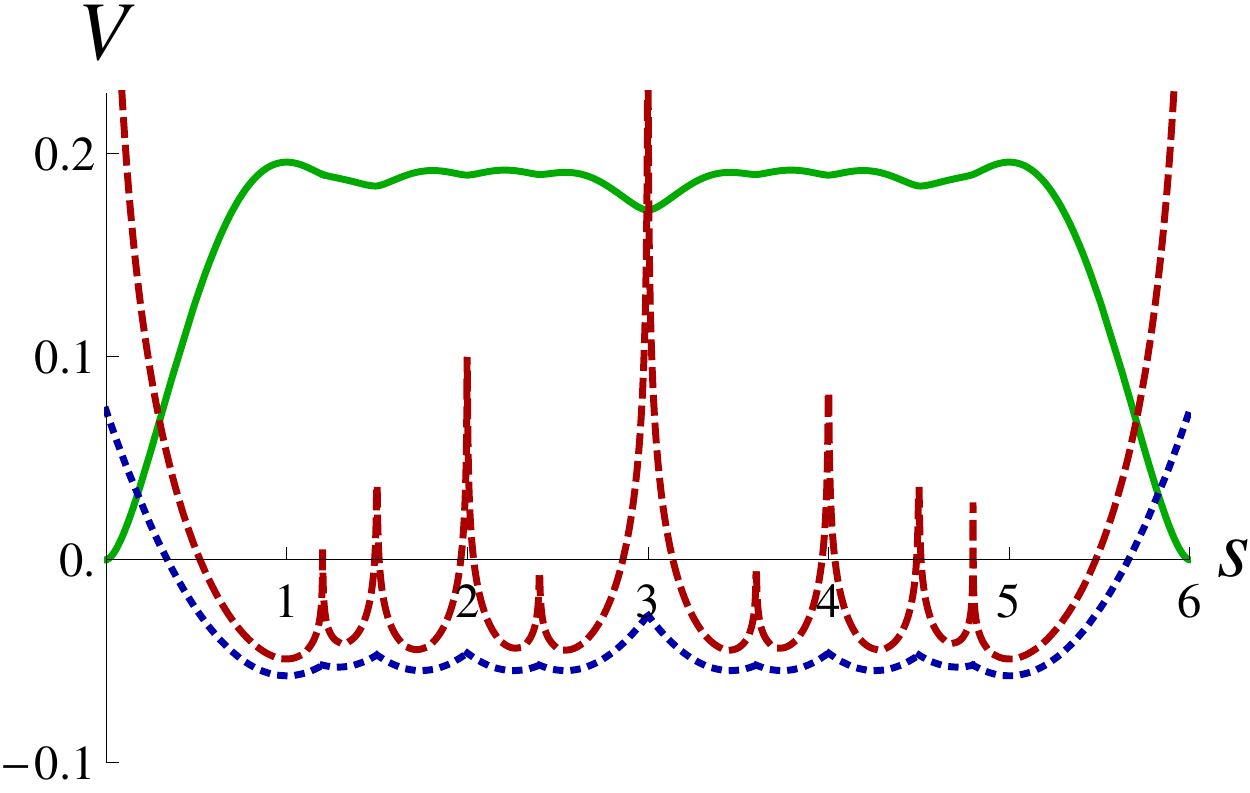}
\end{center}
\caption{ The perturbative potential $V_{pt}(s)$ (solid), the Vandermonde term
$V_{Vdm}(s)$ (dashed), and the linear term $V_{lin}(s)$ (dotted) as a function
of $s$\ in the uniform eigenvalue Ansatz, for $N_{c}=3$\ (left top), $N_{c}%
=4$\ (right top), $N_{c}=5$\ (left bottom), $N_{c}=6$\ (right bottom). }%
\label{Vall}%
\end{figure}

\begin{figure}[t!]
\begin{center}
\includegraphics[width=0.6\textwidth]{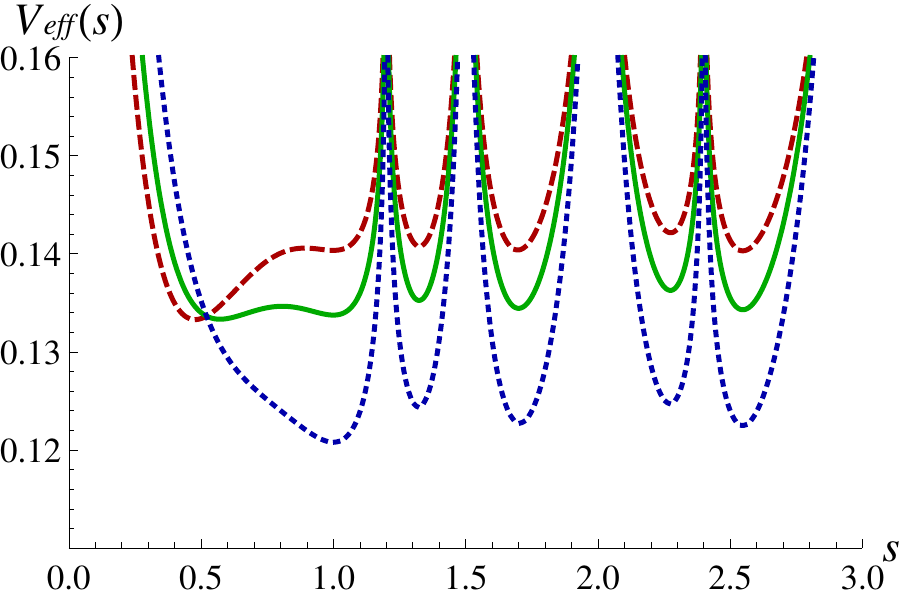}
\end{center}
\caption{The effective potential $V_{eff}(s)$ for $N_{c}=6$ using the
Vandermonde term $V_{Vdm}(s)$, in the semi-QGP (dashed), at the phase
transition (solid), and in the confined phase (dotted).}%
\label{veff6}%
\end{figure}

First we look at the potential along the uniform eigenvalue Ansatz beyond the
confining vacuum, $s>s_{c}$. Figure \ref{Vall} shows the perturbative
potential $V_{pt}(s)$\ (\ref{vpt_1}), the Vandermonde term $V_{Vdm}(s)$, and
the linear term $V_{lin}(s)$\ (\ref{Vlin}) for $N_{c}=3,4,5,6$ up to $s=N_{c}%
$. In addition, in Fig. \ref{veff6} we plot the effective potential in the
presence of the Vandermonde term for $N_{c}=6$. The functions $V_{pt}(s),$
$V_{Vdm}(s)$, $V_{lin}(s)$, and $V_{eff}(s)$ are all periodic in $s\rightarrow
s+N_{c},$ and symmetric with respect to $s=N_{c}/2$.

An interesting observation is that the perturbative potential exhibits local
minima at $s>s_{c}$. Notably, there appear also extrema in the nonperturbative
contributions and in the effective potential. For $1<s\leq N_{c}/2,$\ the
Vandermonde term exhibits $N_{c}-2$\ local minima and $N_{c}-2 $\ divergences
which partly coincide with the local minima in the perturbative potential. The
same arguments apply to the effective potential. For the linear term the
results are qualitatively similar, except that the divergences disappear.
However, it is instructive to focus first only on the perturbative potential
which, in contrast to the nonperturbative and effective potential, does not
depend on the details of the model.

As demonstrated in Fig. \ref{pl}, in the uniform eigenvalue Ansatz the
Polyakov loop $l(s)$ takes always values along the real axis within the solid
lines of Fig. \ref{flowers}. Therefore, the local minima in the perturbative
potential at $s>s_{c}$ represent possible physical solutions. It is important
to point out, however, that the extremal points present in the potential
beyond the confining vacuum\ do not affect the results of our study between
the perturbative and confining vacuum. Nevertheless, an interesting question
is if the local minima are metastable solutions, or just saddle-points. For
this purpose, we have to check if they are stable in the direction transverse
to the uniform eigenvalue Ansatz $\mathbf{q}\left(  s\right)  $ (\ref{ansatz}%
), i.e., if they correspond to local minima in the entire $\left(
N_{c}-1\right)  $-dimensional space spanned by the orthogonal diagonal
generators of $SU(N_{c})$.

This can be done by looking for local minima in the perturbative potential
$V_{pt}\left[  \mathbf{q}\left(  s,s_{i}\right)  \right]  $ in the presence of
a background field,%
\begin{equation}
\mathcal{A}_{0}=\;\frac{2\pi\,T}{g}\mathbf{\,\mathbf{q}}\left(  s,s_{i}%
\right)  \text{ ,} \label{a0n2}%
\end{equation}
where $\mathbf{\,\mathbf{q}}\left(  s,s_{i}\right)  $ is a diagonal matrix
parametrized as%
\begin{equation}
\mathbf{q}\left(  s,s_{i}\right)  =s\mathbf{q}_{c}+\sum_{i=1}^{N_{c}-2}%
s_{i}\mathbf{q}_{i}\text{ .}%
\end{equation}
The matrices $\mathbf{q}_{c}$ and $\mathbf{q}_{i}$ form a set of $N_{c}-1$
orthogonal diagonal generators of the $SU(N_{c})$\ Lie algebra, while $s$ and
$s_{i}$ are the associated parameters. As elements of the $SU(N_{c})$ Lie
algebra, the diagonal generators must be traceless and real. In order to
obtain an orthogonal basis, we also require that the inner product between two
generators must vanish. Moreover, it is convenient to normalize the generators
$\mathbf{q}_{i}$ in such a way that the perturbative potential exhibits the
same periodicity, $s_{i}\rightarrow$ $s_{i}+N_{c}$, and symmetry,
\begin{equation}
V_{pt}\left[  \mathbf{q}\left(  s,s_{i}\right)  \right]  =V_{pt}\left[
\mathbf{q}\left(  s,s_{i}+\frac{N_{c}}{2}\right)  \right]  \text{ \ , \ }%
\end{equation}
along all $\mathbf{q}_{i}$ as along $\mathbf{q}_{c}$. Therefore, it is
sufficient to restrict the search for local minima to the range $1<s,s_{i}\leq
N_{c}/2.$

\subsection{$N_{c}\ =3$}

First we consider $N_{c}=3$. The perturbative potential has a local minimum in
the $\mathbf{q}_{c}$ direction,
\begin{equation}
\mathbf{q}_{c}=\frac{1}{3}\operatorname{diag}(1,0,-1)\text{ ,}%
\end{equation}
at $s=1.5$, where $l=-1/3$. For the generator orthogonal to $\mathbf{q}_{c}$
it is convenient to choose%
\begin{equation}
\mathbf{q}_{1}=\frac{1}{3}\operatorname{diag}(1,-2,1)\text{ .}%
\end{equation}
In order to check if the local minimum at $s=1.5$ is stable in the direction
transverse to $\mathbf{q}_{c}$, we compute the perturbative potential in the
plane spanned by $\mathbf{q}_{c}$ and $\mathbf{q}_{1}.$

\begin{figure}[t!]
\begin{center}
\includegraphics[width=0.5\textwidth]{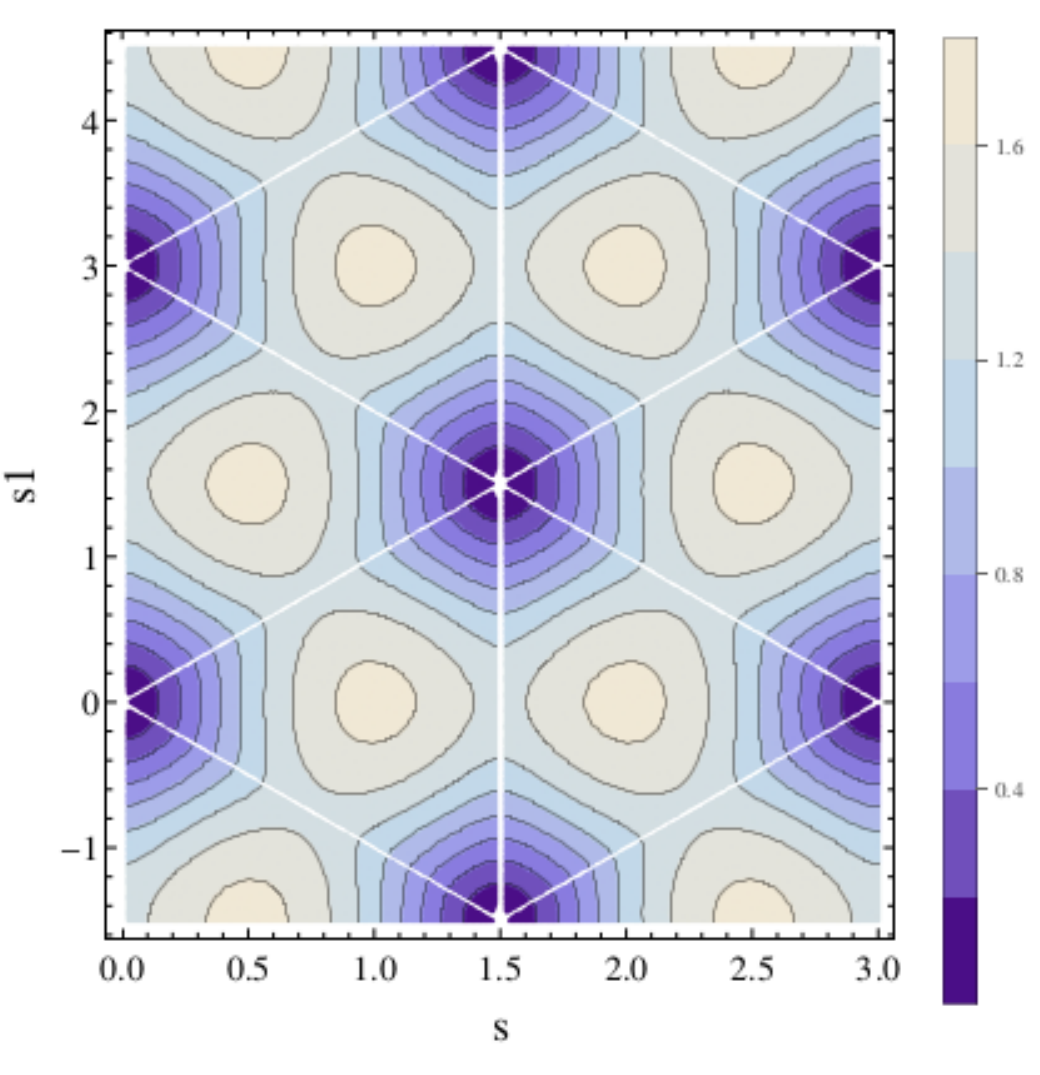}
\end{center}
\caption{ Density plot of the perturbative potential $V_{pt}\left[
\mathbf{q}\left(  s,s_{1}\right)  \right]  $ for $N_{c}=3.$ The $s$-axis
points in the direction of the uniform eigenvalue ansatz, while the $s_{1}%
$-axis points in the transverse direction. }%
\label{density3}%
\end{figure}

Figure \ref{density3} shows a density plot for $V_{pt}\left[  \mathbf{q}%
\left(  s,s_{1}\right)  \right]  $, where $\mathbf{q}\left(  s,s_{1}\right)  $
is a diagonal matrix for the background $\mathcal{A}_{0}$ field (\ref{a0n2})
parametrized as%
\begin{equation}
\mathbf{q}\left(  s,s_{1}\right)  =s\mathbf{q}_{c}+s_{1}\mathbf{q}_{1}\text{
.}%
\end{equation}
For $s_{1}=0$, there is one global minimum at $s=0$ corresponding to the
perturbative vacuum, where $l=1$. But there is no local minimum at $s_{1}=0$
and $s=1.5$. Taking the periodicity of the perturbative potential into
account, the other global minima in Fig. \ref{density3} are all degenerate
with the perturbative vacuum at $s=s_{1}=0$.

\subsection{$N_{c}=4$}

\subsubsection{Minima along the uniform eigenvalue Ansatz}

\begin{figure}[t!]
\begin{adjustwidth}{-.0cm}{}
\begin{center}
\begin{subfigure}[b]{0.45\textwidth}
\includegraphics[width=\textwidth]{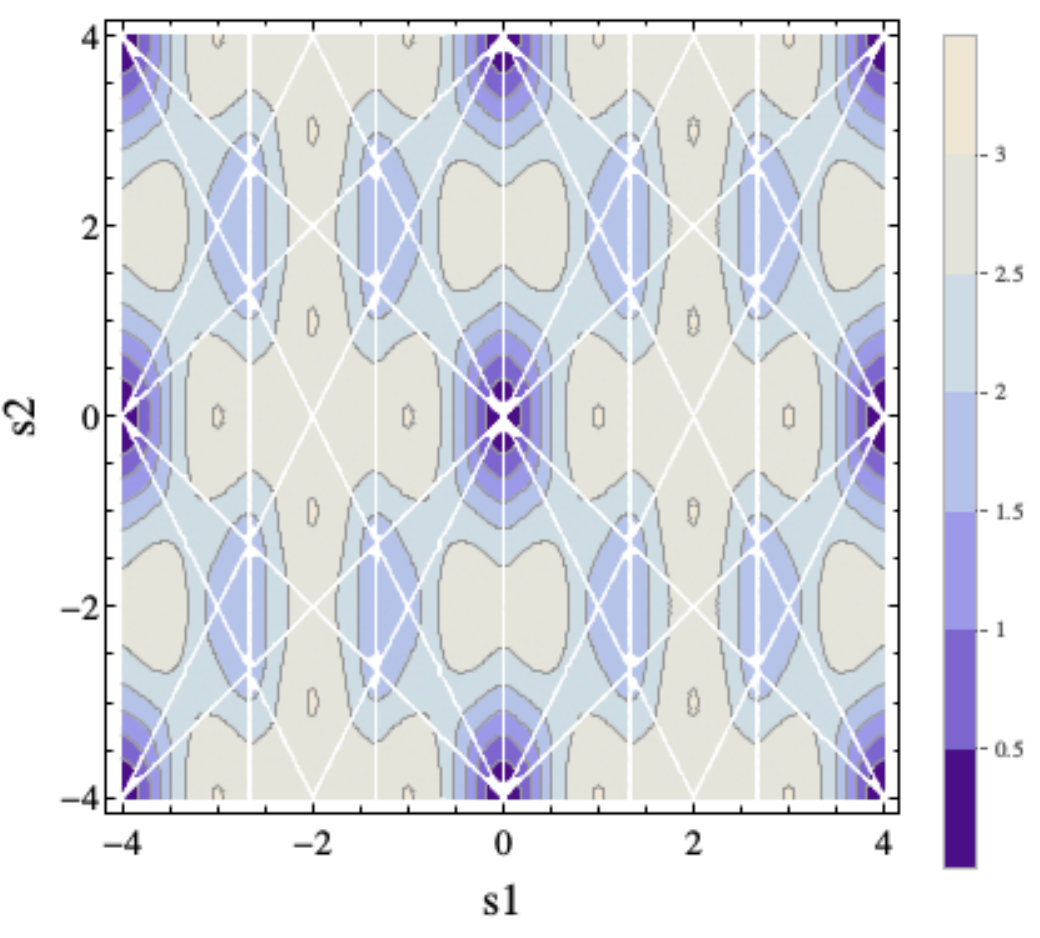}
\caption{
$V_{pt}\left[  \mathbf{q}\left( s=0,s_{1},s_{2}\right)  \right]  $
\label{density4a}
}
\end{subfigure}
\begin{subfigure}[b]{0.45\textwidth}
\includegraphics[width=\textwidth]{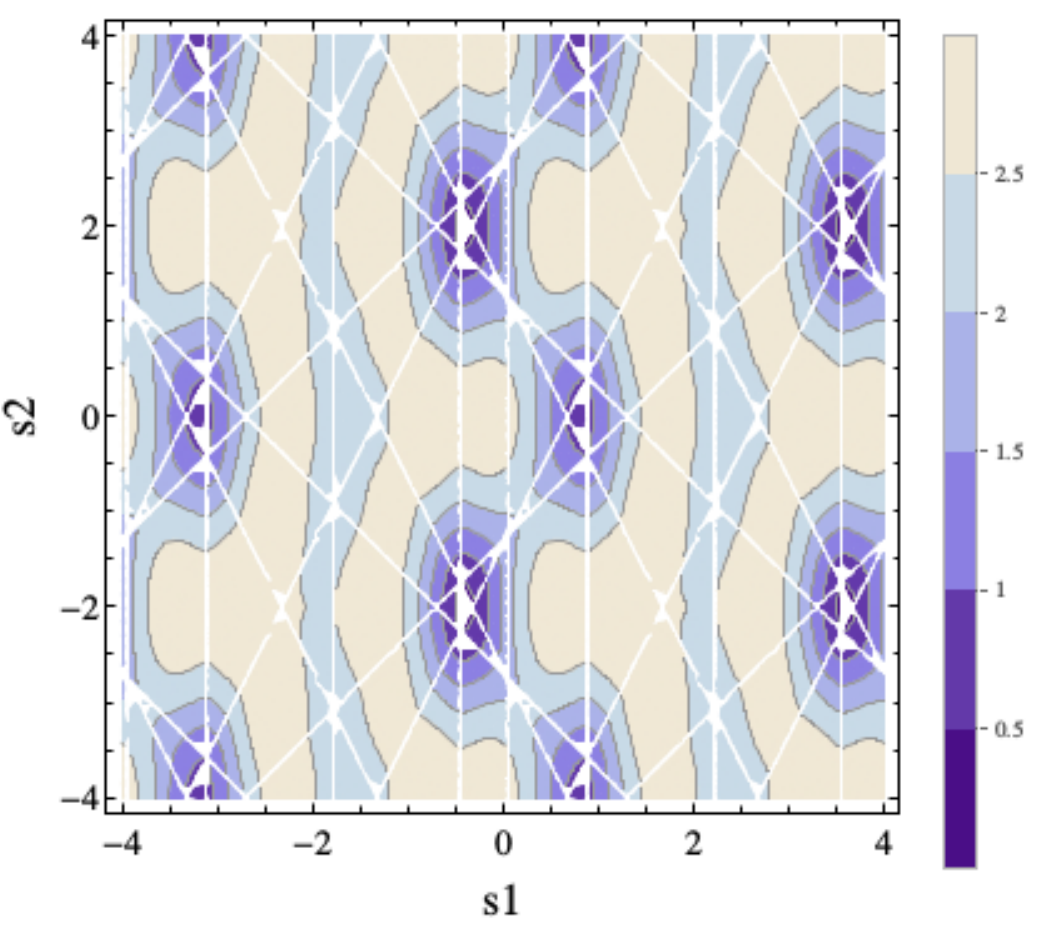}
\caption{
$V_{pt}\left[  \mathbf{q}\left( s=1.35,s_{1},s_{2}\right)  \right]  $
\label{density4b}
}
\end{subfigure}
\begin{subfigure}[b]{0.45\textwidth}
\includegraphics[width=\textwidth]{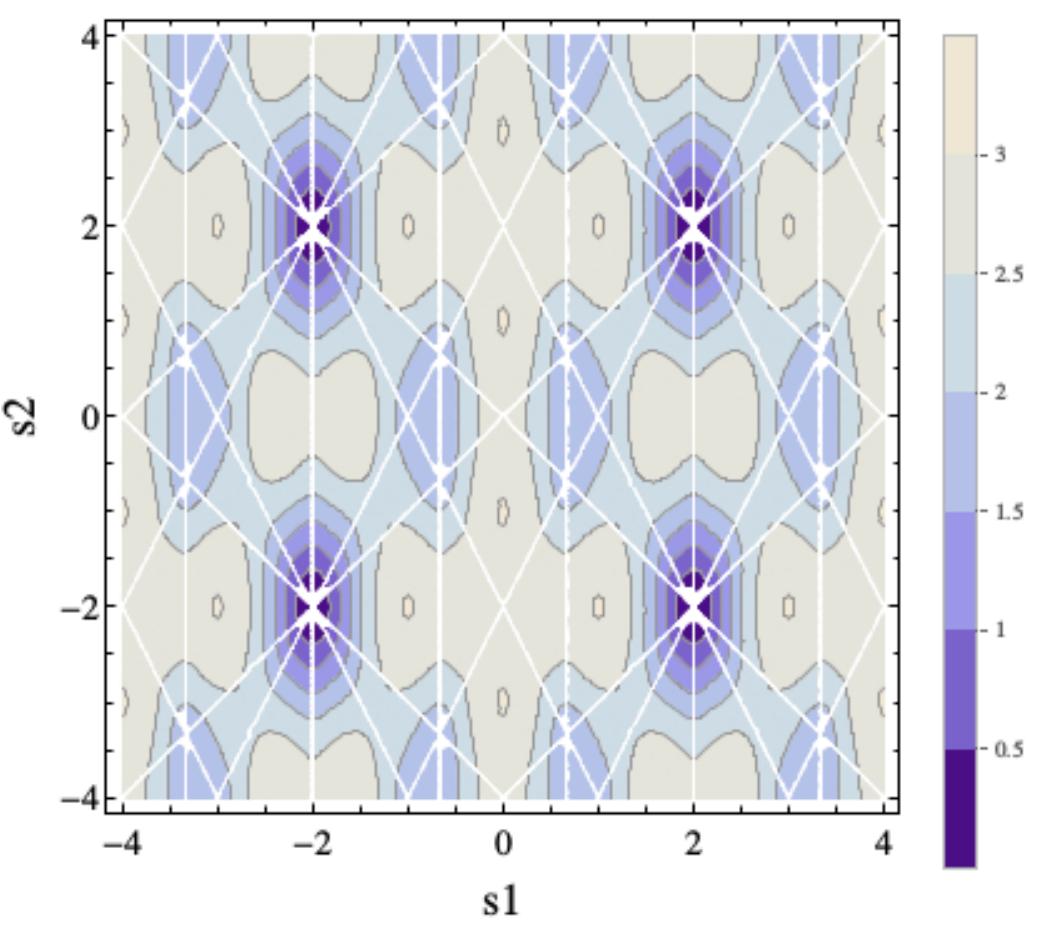}
\caption{
$V_{pt}\left[  \mathbf{q}\left(s=2,s_{1},s_{2}\right)  \right]  $
\label{density4c}
}
\end{subfigure}
\end{center}
\end{adjustwidth}
\caption{ Density plot of the perturbative potential for $N_{c}=4.$ The
potential is plotted for different slices in the $s_{1,}s_{2}$-plane
perpendicular to $\mathbf{q}_{c}$, at $s=0$ (left top), $s=1.35$ (right top)
and $s=2$ (bottom).}%
\end{figure}

For $N_{c}=4,$ we find two local minima in the region $1<s\leq2$\ along the
uniform eigenvalue Ansatz,
\begin{equation}
\mathbf{q}_{c}=\frac{1}{8}\operatorname{diag}(3,1,-1,-3)\text{ ,}%
\end{equation}
at $s=1.35$ and $s=2$. A symmetric diagonal basis for $SU(4)_{c}$ is obtained
by choosing
\begin{equation}
\mathbf{q}_{1}=\frac{1}{8}\operatorname{diag}(-1,3,-3,1)\text{ ,
\ \ \ }\mathbf{q}_{2}=\frac{1}{8}\operatorname{diag}(1,-1,-1,1)\text{ }%
\end{equation}
for the remaining two generators orthogonal to $\mathbf{q}_{c}$. In order to
see if the local minima along $\mathbf{q}_{c}$ are metastable or saddle
points, we parametrize the matrix $\mathbf{q}$ as
\begin{equation}
\mathbf{q}\left(  s,s_{1},s_{2}\right)  =s\mathbf{q}_{c}+s_{1}\mathbf{q}%
_{1}+s_{2}\mathbf{q}_{2}\text{ ,}%
\end{equation}
and plot the perturbative potential $V_{pt}\left[  \mathbf{q}\left(
s,s_{1,}s_{2}\right)  \right]  $ in the $s_{1}s_{2}$-plane perpendicular to
$\mathbf{q}_{c}$, at $s=0$ (Fig. \ref{density4a}), $s=1.35$ (Fig.
\ref{density4b}), and $s=2$ (Fig. \ref{density4c}). Considering the
periodicity of the perturbative potential, our analysis indicates that, except
for degenerate global minima which correspond to the perturbative vacuum at
$s=s_{1}=s_{2}=0$,\ there are only saddle points along the uniform eigenvalue Ansatz.

\subsubsection{Minima in the entire 3-dimensional volume}

Note that in order to take into account all possible metastable solutions in
the perturbative potential, we should ideally look for local minima in the
entire $\left(  N_{c}-1\right)  $-dimensional space. This can be done by
plotting slices of the perturbative potential in the $s,s_{1}$-plane along the
perpendicular direction for fixed values of $s_{2}$, see Figs. \ref{part1} and
\ref{part2}.

We start with Fig. \ref{slice0} which shows a density plot of the perturbative
potential in the $s,s_{1}$-plane, spanned by the generators $\mathbf{q}_{c}$
and $\mathbf{q}_{1},$ for $s_{2}=0$. The global and local minima reside at
intersections of the white lines which connect the regions where the
perturbative potential is minimized. From the intersections we can therefore
determine the exact positions of the minima.

Within the considered region $0\leq s,s_{1}\leq4,$ there are eight degenerate
global minima at $\left(  0,0\right)  $, $\left(  0,4\right)  $, $\left(
4,0\right)  $, $\left(  4,4\right)  $, $\left(  8/5,4/5\right)  $, $\left(
8/5,16/5\right)  $, $\left(  12/5,16/5\right)  $, and $\left(
4/5,12/5\right)  $. Each global minimum exhibits an octagonal structure and is
located in the center between eight neighboring maxima which are all
degenerate with the confining vacuum at $\left(  1,0\right)  $.

\begin{figure}[t!]
\begin{adjustwidth}{-.0cm}{}
\begin{center}
\begin{subfigure}[b]{0.45\textwidth}
\includegraphics[width=\textwidth]{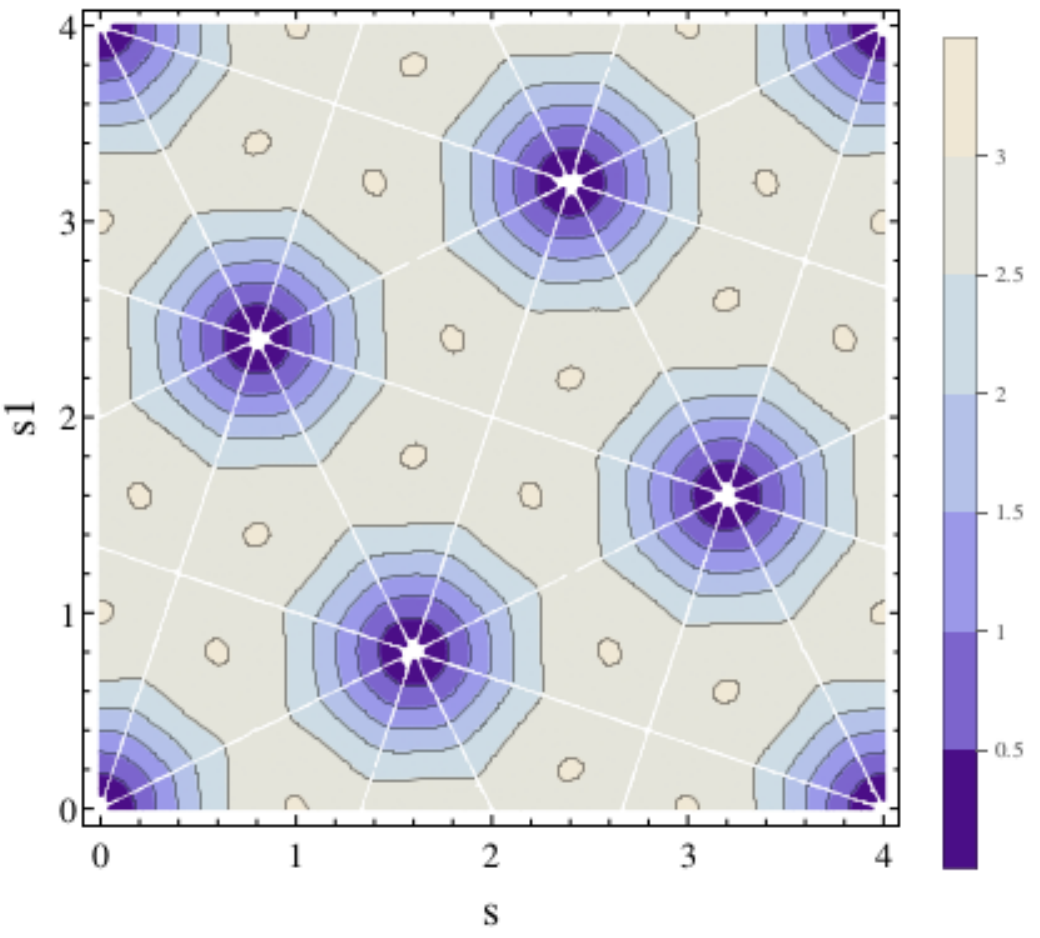}
\caption{
$V_{pt}\left[  \mathbf{q}\left(s,s_{1},s_{2}=0.0\right)  \right]  $
\label{slice0}
}
\end{subfigure}
\begin{subfigure}[b]{0.45\textwidth}
\includegraphics[width=\textwidth]{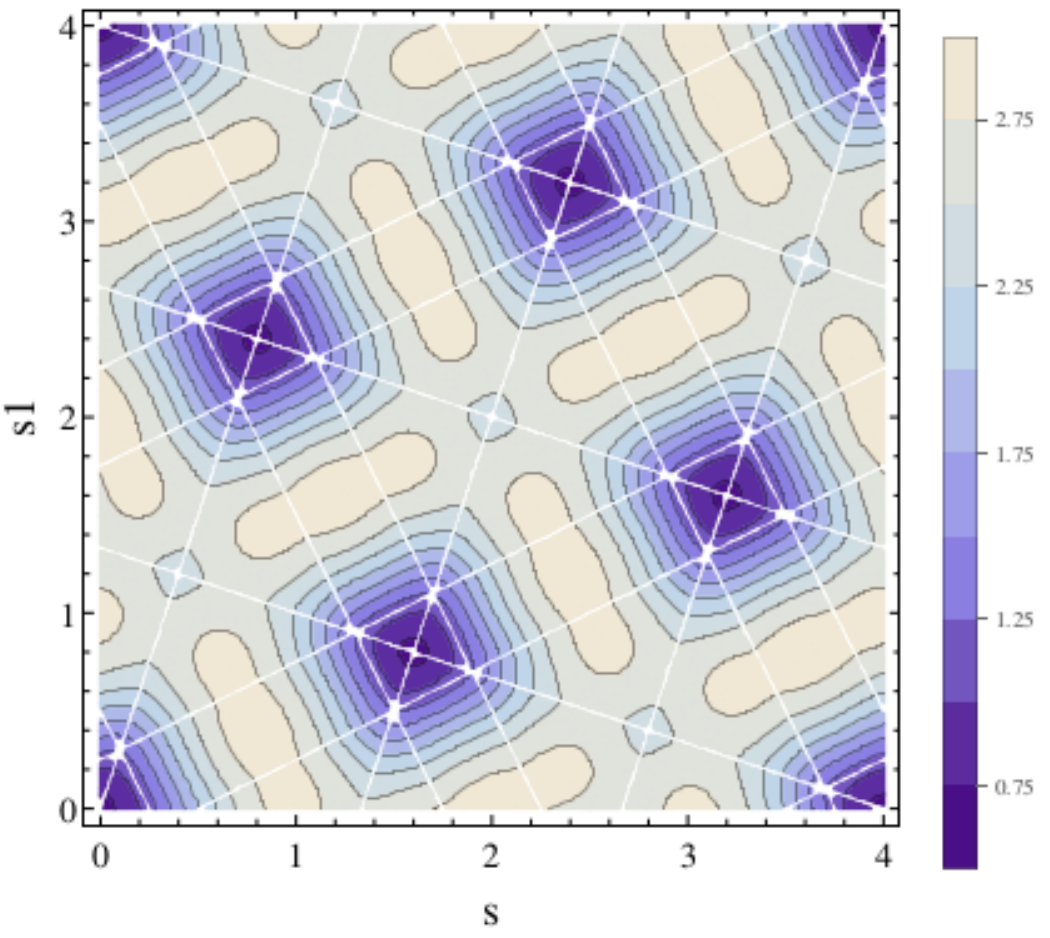}
\caption{
$V_{pt}\left[  \mathbf{q}\left(s,s_{1},s_{2}=0.5\right)  \right]  $
\label{slice5}
}
\end{subfigure}
\begin{subfigure}[b]{0.45\textwidth}
\includegraphics[width=\textwidth]{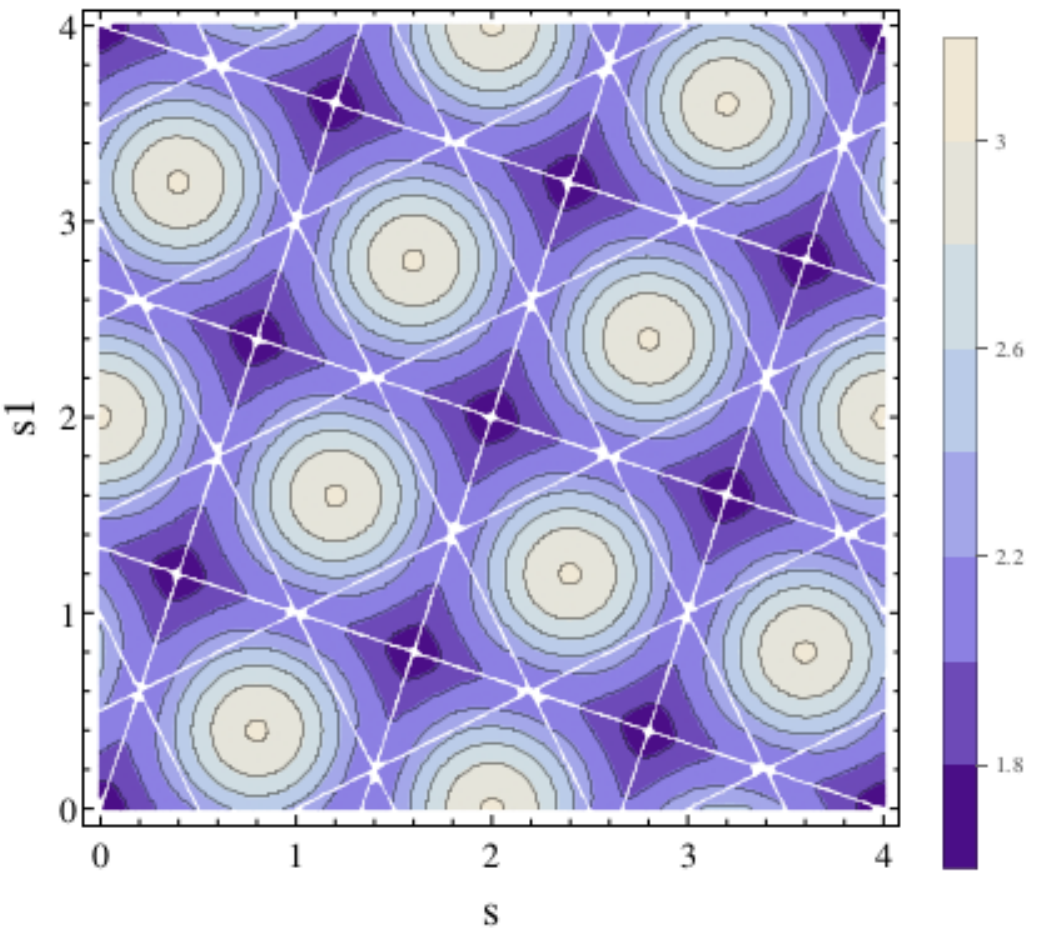}
\caption{
$V_{pt}\left[  \mathbf{q}\left(s,s_{1},s_{2}=1.0\right)  \right]  $
\label{slice10}
}
\end{subfigure}
\begin{subfigure}[b]{0.45\textwidth}
\includegraphics[width=\textwidth]{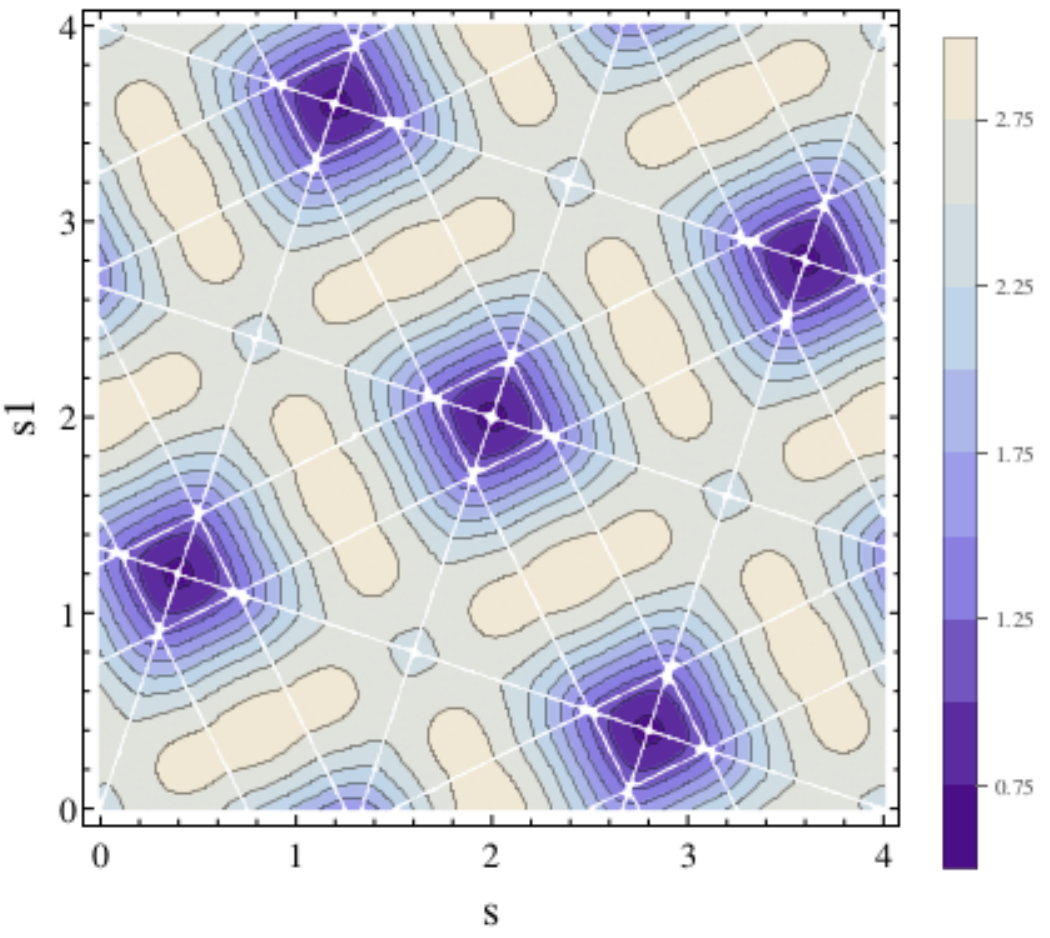}
\caption{
$V_{pt}\left[  \mathbf{q}\left(s,s_{1},s_{2}=1.5\right)  \right]  $
\label{slice15}
}
\end{subfigure}
\end{center}
\end{adjustwidth}
\caption{ Density plot of the perturbative potential for $N_{c}=4$. The
potential is plotted for different $s,s_{1}$-slices, varying the distance in
the perpendicular direction: $s_{2}=0$ (left top), $s_{2}=0.5$ (right top),
$s_{2}=1$ (left bottom), $s_{2}=1.5$ (right bottom). (To be continued in Fig.
\ref{part2}.)}%
\label{part1}%
\end{figure}

\begin{figure}[t!]
\begin{adjustwidth}{-.0cm}{}
\begin{center}
\begin{subfigure}[b]{0.45\textwidth}
\includegraphics[width=\textwidth]{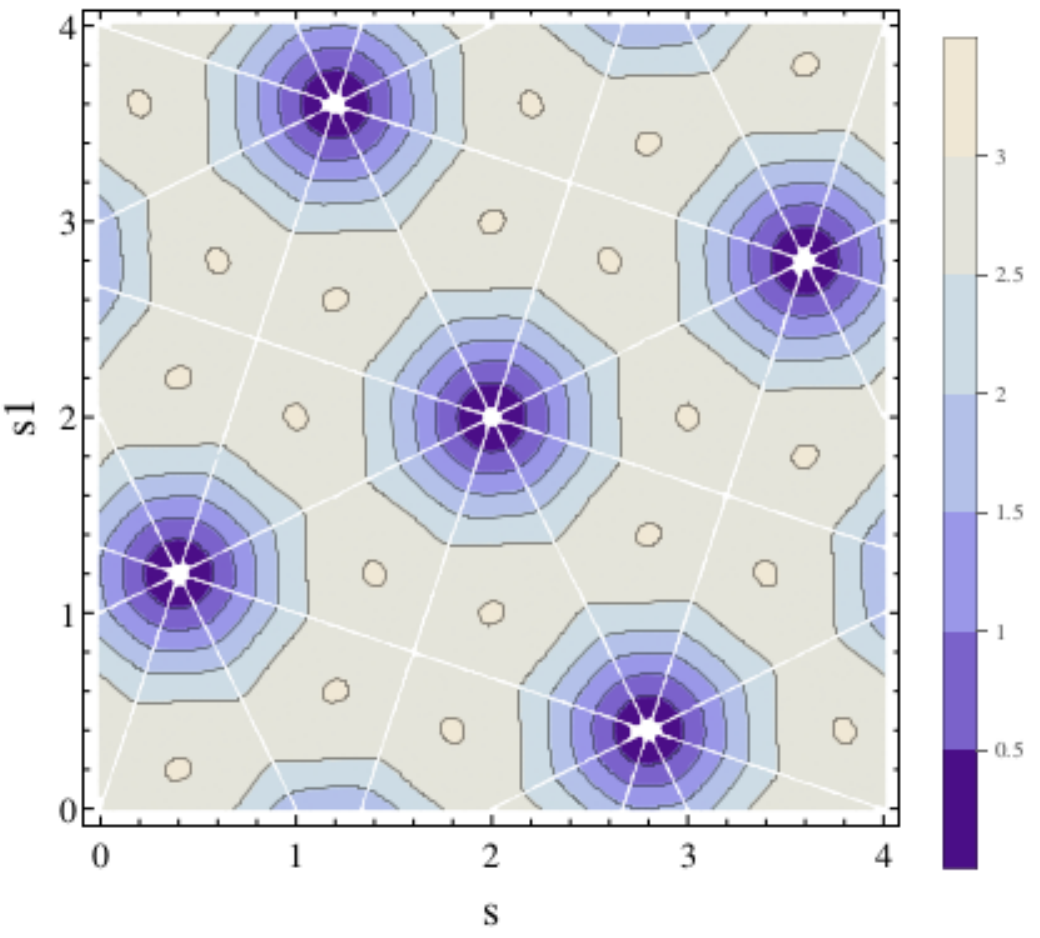}
\caption{
$V_{pt}\left[  \mathbf{q}\left(s,s_{1},s_{2}=2.0\right)  \right]  $
\label{slice20}
}
\end{subfigure}
\begin{subfigure}[b]{0.45\textwidth}
\includegraphics[width=\textwidth]{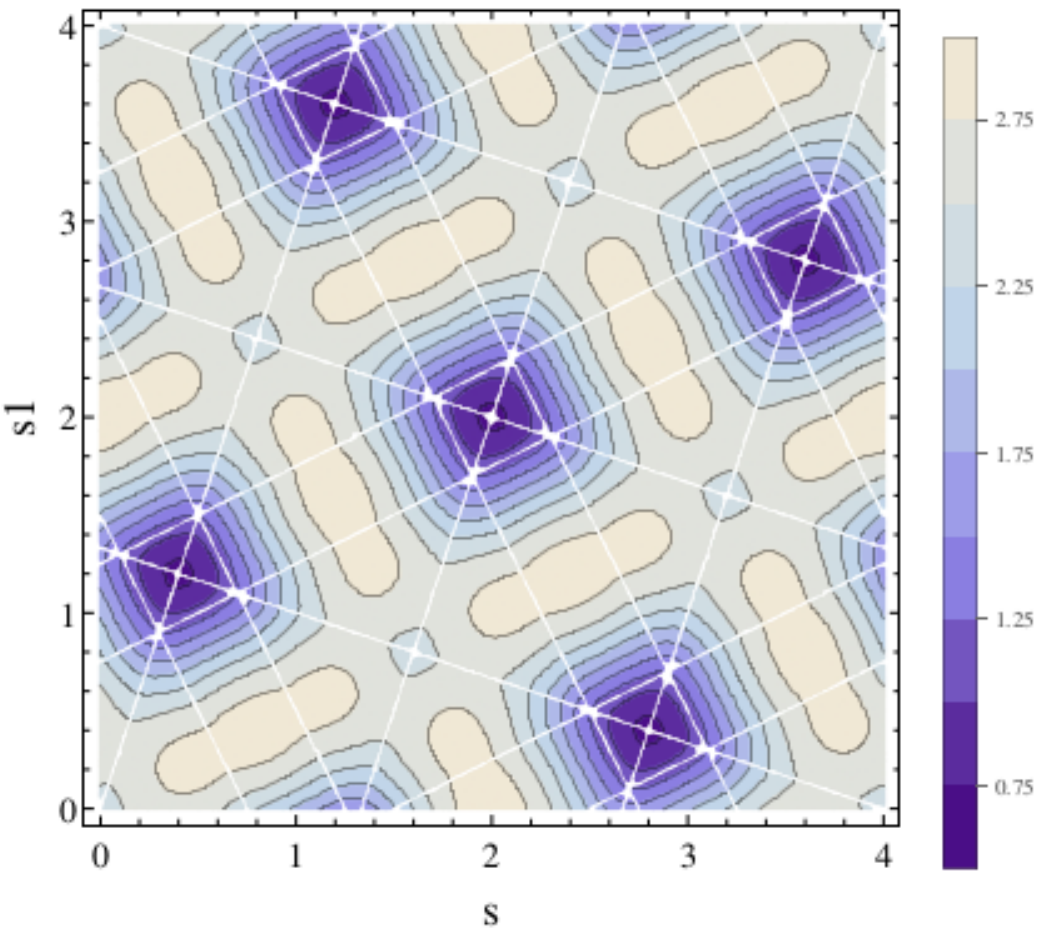}
\caption{
$V_{pt}\left[  \mathbf{q}\left(s,s_{1},s_{2}=2.5\right)  \right]  $
\label{slice25}
}
\end{subfigure}
\begin{subfigure}[b]{0.45\textwidth}
\includegraphics[width=\textwidth]{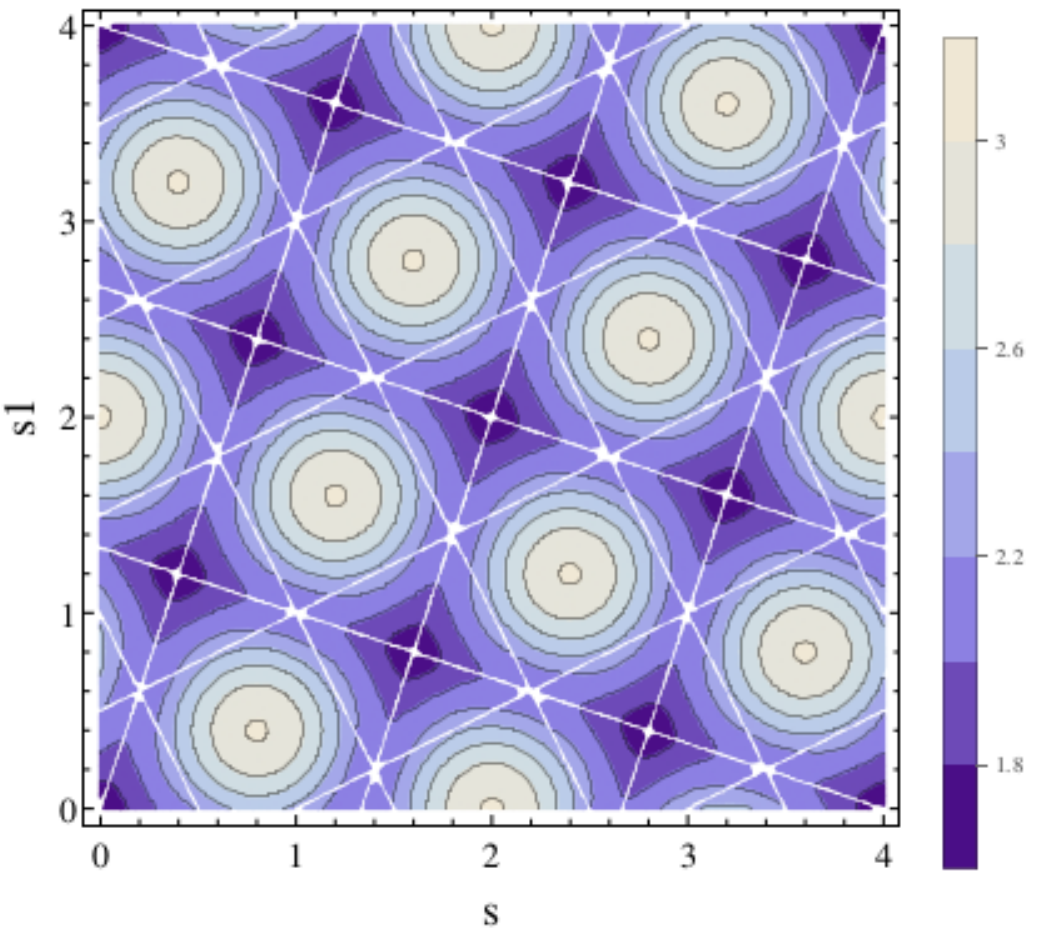}
\caption{
$V_{pt}\left[  \mathbf{q}\left(s,s_{1},s_{2}=3.0\right)  \right]  $
\label{slice30}
}
\end{subfigure}
\begin{subfigure}[b]{0.45\textwidth}
\includegraphics[width=\textwidth]{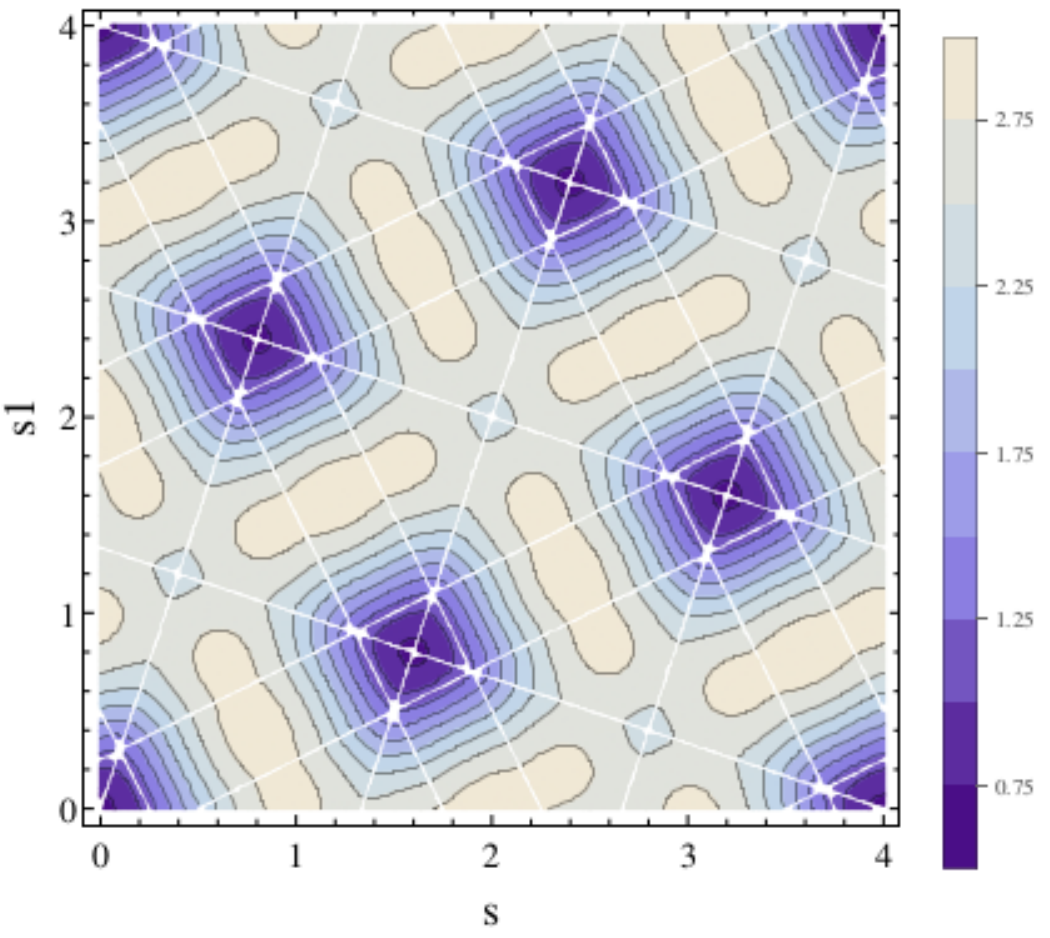}
\caption{
$V_{pt}\left[  \mathbf{q}\left(s,s_{1},s_{2}=3.5\right)  \right]  $
\label{slice35}
}
\end{subfigure}
\end{center}
\end{adjustwidth}
\caption{Continuation of Fig. \ref{part1} for $s_{2}=2$ (left top),
$s_{2}=2.5$ (right top), $s_{2}=3$ (left bottom), $s_{2}=3.5$ (right bottom).}%
\label{part2}%
\end{figure}

We also find five local minima at $\left(  2/5,6/5\right)  $, $\left(
14/5,2/5\right)  $, $\left(  10/5,10/5\right)  $, $\left(  6/5,18/5\right)  $,
and $\left(  18/5,14/5\right)  $. Their coordinates are offset from the global
minima by the vector $2/5\mathbf{q}_{c}+6/5\mathbf{q}_{1}$. Each local minimum
resides in the center between four neighboring global minima.

As we start moving along the $s_{2}$ direction, the global and local minima
begin to transform into one another: the global minima become shallower, while
the local minima get deeper, see Fig. \ref{slice5}. Consequently, at $s_{2}=1$
all minima become degenerate, see Fig. \ref{slice10}. Moving further to
$s_{2}=2$, Fig. \ref{slice20}, the roles of the minima interchange: the former
global minima are transformed into local minima and vice versa. Finally, at
$s_{2}=4$, we recover the same pattern as in the initial state at $s_{2}=0,$
see Fig. \ref{slice0}.

Overall, we find that the structure of the perturbative potential in the
entire three-dimensional space spanned by the diagonal generators
$\mathbf{q}_{c}$, $\mathbf{q}_{1}$, and $\mathbf{q}_{2}$ resembles a crystal,
and that the edges of the elementary cell are spanned by the three vectors
$\mathbf{v}_{1}=8/5\mathbf{q}_{c}+4/5\mathbf{q}_{1},$ $\mathbf{v}%
_{2}=-4/5\mathbf{q}_{c}+8/5\mathbf{q}_{1},$ and $\mathbf{v}_{3}=2/5\mathbf{q}%
_{c}+6/5\mathbf{q}_{1}+10/5\mathbf{q}_{2}.$ The periodicity of these
elementary cells is given by the norm of the vectors, $\left\vert
\mathbf{v}_{1}\right\vert =\left\vert \mathbf{v}_{2}\right\vert =4/\sqrt{5},$
$\left\vert \mathbf{v}_{3}\right\vert =2\sqrt{7/5}.$

Summarizing our study for $N_{c}=3$ and $N_{c}=4$ demonstrates that there are
only global minima in the perturbative potential which are all degenerate with
the perturbative vacuum of our uniform eigenvalue Ansatz at $s=s_{i}=0$, where
$l=1$. Moreover, all maxima in the perturbative potential are degenerate with
the confining vacuum at $s_{c}=1$ and $s_{i}=0$, where $l=0$. But we find no
indication of metastable solutions.

Thus, we verify that the perturbative potential exhibits no other stationary
points except for the ones at $s=0$ and $s_{c}=1$ along the uniform eigenvalue
Ansatz studied in this work. Consequently, we confirm that employing any other
parametrization of the background $\mathcal{A}_{0}$ field, which takes us from
the perturbative vacuum (minimum) to the confining vacuum (neighboring
maximum), will not change the physics.

An interesting outlook for future projects would be to extend the present
analysis to $N_{c}\geq5$. We can also look for stationary points by solving
the stationary conditions%
\begin{equation}
\frac{\partial V_{pt}\left[  \mathbf{q}\left(  s,s_{i}\right)  \right]
}{\partial s}=0\text{ , \ \ \ }\frac{\partial V_{pt}\left[  \mathbf{q}\left(
s,s_{i}\right)  \right]  }{\partial s_{i}}\ =0\text{ , \ \ \ }i=1,...,N_{c}%
-2\text{ .}%
\end{equation}
Furthermore, it would be certainly useful to look for local minima in the full
effective potential. In the present work, however, we do not further address
the question of possible metastable solutions.

\section{Conclusions \label{summary}}

We have used an effective matrix model to study the deconfinement phase
transition in pure $SU(N_{c})$ gauge theories in $d=2+1$ dimensions. The
effective potential was constructed as a sum of a perturbative and a
nonperturbative part. The perturbative potential was computed to one-loop
order in the presence of a constant background $\mathcal{A}_{0}$ field. In
order to model the transition to confinement, we then constructed appropriate
$\mathcal{A}_{0}$-dependent and $\mathcal{A}_{0}$-independent nonperturbative
terms, motivated by lattice results for the pressure and interaction measure.

The analytical calculations were performed for general $N_{c}$ and in the
large-$N_{c}$ limit. We also presented the numerical solution for the
potential using the uniform eigenvalue ansatz which guaranties that the
Polyakov loop is real. The free parameters of the model were adjusted by
fitting to the lattice pressure of Ref. \cite{Caselle:2011mn}. We have shown
the pressure and interaction measure for $N_{c}=2,3,4,5,6$ and compared to the
lattice data of Ref. {\cite{Caselle:2011mn}. Overall, the results exhibit a
mild sensitivity }with respect to the details of the $\mathcal{A}_{0}%
$-dependent nonperturbative terms.

Using one and two free parameters, we already obtain a good agreement with the
lattice data. Notably, with only one free parameter we reproduce the correct
temperature dependence\ for the deviation from an ideal gluon gas at
$1.2\,T_{d}\lesssim T\lesssim10$ $T_{d}$. As observed on the lattice, we also
find a small dependence on the number of colors for the pressure and the
interaction measure, except for the factor $N_{c}^{2}-1$.

In order to further improve the agreement with lattice close to the
deconfinement temperature $T_{d}$ and at high $T$, we introduced a
four-parameter fit. This fit is constructed from the two-parameter model by
allowing for a shift in $T_{d}$, and in the perturbative limit of the
pressure, $c$. The shift in $T_{d}$ is rather small and rapidly vanishes with
increasing number of colors. This supports the general expectation that any
possible glueball contribution becomes suppressed by a factor $\sim1/\left(
N_{c}^{2}-1\right)  $ at the phase transition. The shift in $c$ is
approximately constant for all $N_{c}$, $c\simeq3\%$.

The four-parameter fit allows for a very good agreement with the lattice
results at all temperatures. We stress however, that the four-parameter fit is
just an approximation to a more complete theory properly incorporating the
physics in the confined phase. The general trend we observe is that the
two-parameter model can be regarded as an interpolation between the
one-parameter model for $N_{c}=2,3$, and the four-parameter fit for larger
$N_{c}$.

Using the parameters determined by fitting the lattice pressure, we have also
plotted the Polyakov loop. We find that the transition region, where the
system exhibits a nonvanishing condensate for $\mathcal{A}_{0}$ is broadly
independent of the number of colors and extends up to $\sim1.25T_{d}$. This is
very close to the results obtained in $d=3+1$ \cite{Dumitru:2012fw}. So far,
the renormalized Polyakov loop has not yet been computed on the lattice in
$d=2+1$ dimensions. The corresponding lattice data could help to clarify the
role of nonperturbative effects in the deconfined phase.

It would be certainly useful, to extend the matrix model to a more general
effective theory including the physics in the confined phase. Another
interesting project would be to study the interface tension which gives the
tunneling probability between different vacua of the system. Finally, we could
include dynamical quarks, as was done in $d=3+1$ in Ref. \cite{Kashiwa:2012wa}.

\begin{acknowledgments}
The authors would like to thank Marco Panero
for kindly sharing the lattice data of Ref.
\cite{Caselle:2011mn}. We also thank Dirk H. Rischke, Nuno Cardoso and Marco
Panero for valuable discussions. The research of R.D.P. is supported
by the U.S. Department of Energy under contract \#DE-AC02-98CH10886.
E. S. thanks the hospitality of RIKEN/BNL and CFTP.
The research of P.B. is supported by the CFTP grant
PEST-OE/FIS/UI0777/2011, the FCT grant  CERN/FP/123612/2011, and the
CRUP/DAAD exchange A10/10.
\end{acknowledgments}

\bibliographystyle{apsrev4-1}
\bibliography{matrix_model.bib}

\begin{thebibliography}{61}%
\makeatletter
\providecommand \@ifxundefined [1]{%
 \@ifx{#1\undefined}
}%
\providecommand \@ifnum [1]{%
 \ifnum #1\expandafter \@firstoftwo
 \else \expandafter \@secondoftwo
 \fi
}%
\providecommand \@ifx [1]{%
 \ifx #1\expandafter \@firstoftwo
 \else \expandafter \@secondoftwo
 \fi
}%
\providecommand \natexlab [1]{#1}%
\providecommand \enquote  [1]{``#1''}%
\providecommand \bibnamefont  [1]{#1}%
\providecommand \bibfnamefont [1]{#1}%
\providecommand \citenamefont [1]{#1}%
\providecommand \href@noop [0]{\@secondoftwo}%
\providecommand \href [0]{\begingroup \@sanitize@url \@href}%
\providecommand \@href[1]{\@@startlink{#1}\@@href}%
\providecommand \@@href[1]{\endgroup#1\@@endlink}%
\providecommand \@sanitize@url [0]{\catcode `\\12\catcode `\$12\catcode
  `\&12\catcode `\#12\catcode `\^12\catcode `\_12\catcode `\%12\relax}%
\providecommand \@@startlink[1]{}%
\providecommand \@@endlink[0]{}%
\providecommand \url  [0]{\begingroup\@sanitize@url \@url }%
\providecommand \@url [1]{\endgroup\@href {#1}{\urlprefix }}%
\providecommand \urlprefix  [0]{URL }%
\providecommand \Eprint [0]{\href }%
\providecommand \doibase [0]{http://dx.doi.org/}%
\providecommand \selectlanguage [0]{\@gobble}%
\providecommand \bibinfo  [0]{\@secondoftwo}%
\providecommand \bibfield  [0]{\@secondoftwo}%
\providecommand \translation [1]{[#1]}%
\providecommand \BibitemOpen [0]{}%
\providecommand \bibitemStop [0]{}%
\providecommand \bibitemNoStop [0]{.\EOS\space}%
\providecommand \EOS [0]{\spacefactor3000\relax}%
\providecommand \BibitemShut  [1]{\csname bibitem#1\endcsname}%
\let\auto@bib@innerbib\@empty
\bibitem [{\citenamefont {Caselle}\ \emph
  {et~al.}(2011{\natexlab{a}})\citenamefont {Caselle}, \citenamefont
  {Castagnini}, \citenamefont {Feo}, \citenamefont {Gliozzi}, \citenamefont
  {Gursoy} \emph {et~al.}}]{Caselle:2011mn}%
  \BibitemOpen
  \bibfield  {author} {\bibinfo {author} {\bibfnamefont {M.}~\bibnamefont
  {Caselle}}, \bibinfo {author} {\bibfnamefont {L.}~\bibnamefont {Castagnini}},
  \bibinfo {author} {\bibfnamefont {A.}~\bibnamefont {Feo}}, \bibinfo {author}
  {\bibfnamefont {F.}~\bibnamefont {Gliozzi}}, \bibinfo {author} {\bibfnamefont
  {U.}~\bibnamefont {Gursoy}},  \emph {et~al.},\ }\href@noop {} {\  (\bibinfo
  {year} {2011}{\natexlab{a}})},\ \Eprint {http://arxiv.org/abs/1111.0580}
  {arXiv:1111.0580 [hep-th]} \BibitemShut {NoStop}%
\bibitem [{\citenamefont {Gopakumar}\ and\ \citenamefont
  {Gross}(1995)}]{Gopakumar:1994iq}%
  \BibitemOpen
  \bibfield  {author} {\bibinfo {author} {\bibfnamefont {R.}~\bibnamefont
  {Gopakumar}}\ and\ \bibinfo {author} {\bibfnamefont {D.~J.}\ \bibnamefont
  {Gross}},\ }\href {\doibase 10.1016/0550-3213(95)00340-X} {\bibfield
  {journal} {\bibinfo  {journal} {Nucl.Phys.}\ }\textbf {\bibinfo {volume}
  {B451}},\ \bibinfo {pages} {379} (\bibinfo {year} {1995})},\ \Eprint
  {http://arxiv.org/abs/hep-th/9411021} {arXiv:hep-th/9411021 [hep-th]}
  \BibitemShut {NoStop}%
\bibitem [{\citenamefont {Dumitru}\ \emph {et~al.}(2011)\citenamefont
  {Dumitru}, \citenamefont {Guo}, \citenamefont {Hidaka}, \citenamefont
  {Korthals~Altes},\ and\ \citenamefont {Pisarski}}]{Dumitru:2010mj}%
  \BibitemOpen
  \bibfield  {author} {\bibinfo {author} {\bibfnamefont {A.}~\bibnamefont
  {Dumitru}}, \bibinfo {author} {\bibfnamefont {Y.}~\bibnamefont {Guo}},
  \bibinfo {author} {\bibfnamefont {Y.}~\bibnamefont {Hidaka}}, \bibinfo
  {author} {\bibfnamefont {C.~P.}\ \bibnamefont {Korthals~Altes}}, \ and\
  \bibinfo {author} {\bibfnamefont {R.~D.}\ \bibnamefont {Pisarski}},\ }\href
  {\doibase 10.1103/PhysRevD.83.034022} {\bibfield  {journal} {\bibinfo
  {journal} {Phys.Rev.}\ }\textbf {\bibinfo {volume} {D83}},\ \bibinfo {pages}
  {034022} (\bibinfo {year} {2011})},\ \Eprint {http://arxiv.org/abs/1011.3820}
  {arXiv:1011.3820 [hep-ph]} \BibitemShut {NoStop}%
\bibitem [{\citenamefont {Dumitru}\ \emph {et~al.}(2012)\citenamefont
  {Dumitru}, \citenamefont {Guo}, \citenamefont {Hidaka}, \citenamefont
  {Altes},\ and\ \citenamefont {Pisarski}}]{Dumitru:2012fw}%
  \BibitemOpen
  \bibfield  {author} {\bibinfo {author} {\bibfnamefont {A.}~\bibnamefont
  {Dumitru}}, \bibinfo {author} {\bibfnamefont {Y.}~\bibnamefont {Guo}},
  \bibinfo {author} {\bibfnamefont {Y.}~\bibnamefont {Hidaka}}, \bibinfo
  {author} {\bibfnamefont {C.~P.~K.}\ \bibnamefont {Altes}}, \ and\ \bibinfo
  {author} {\bibfnamefont {R.~D.}\ \bibnamefont {Pisarski}},\ }\href {\doibase
  10.1103/PhysRevD.86.105017} {\bibfield  {journal} {\bibinfo  {journal}
  {Phys.Rev.}\ }\textbf {\bibinfo {volume} {D86}},\ \bibinfo {pages} {105017}
  (\bibinfo {year} {2012})},\ \Eprint {http://arxiv.org/abs/1205.0137}
  {arXiv:1205.0137 [hep-ph]} \BibitemShut {NoStop}%
\bibitem [{\citenamefont {Bicudo}\ \emph {et~al.}(2013)\citenamefont {Bicudo},
  \citenamefont {Pisarski},\ and\ \citenamefont {Seel}}]{Bicudo:2013yza}%
  \BibitemOpen
  \bibfield  {author} {\bibinfo {author} {\bibfnamefont {P.}~\bibnamefont
  {Bicudo}}, \bibinfo {author} {\bibfnamefont {R.~D.}\ \bibnamefont
  {Pisarski}}, \ and\ \bibinfo {author} {\bibfnamefont {E.}~\bibnamefont
  {Seel}},\ }\href@noop {} {\  (\bibinfo {year} {2013})},\ \Eprint
  {http://arxiv.org/abs/1306.2943} {arXiv:1306.2943 [hep-ph]} \BibitemShut
  {NoStop}%
\bibitem [{\citenamefont {Karsch}(2002)}]{Karsch:2001cy}%
  \BibitemOpen
  \bibfield  {author} {\bibinfo {author} {\bibfnamefont {F.}~\bibnamefont
  {Karsch}},\ }\href@noop {} {\bibfield  {journal} {\bibinfo  {journal}
  {Lect.Notes Phys.}\ }\textbf {\bibinfo {volume} {583}},\ \bibinfo {pages}
  {209} (\bibinfo {year} {2002})},\ \Eprint
  {http://arxiv.org/abs/hep-lat/0106019} {arXiv:hep-lat/0106019 [hep-lat]}
  \BibitemShut {NoStop}%
\bibitem [{\citenamefont {Teper}(1999)}]{Teper:1998te}%
  \BibitemOpen
  \bibfield  {author} {\bibinfo {author} {\bibfnamefont {M.~J.}\ \bibnamefont
  {Teper}},\ }\href {\doibase 10.1103/PhysRevD.59.014512} {\bibfield  {journal}
  {\bibinfo  {journal} {Phys.Rev.}\ }\textbf {\bibinfo {volume} {D59}},\
  \bibinfo {pages} {014512} (\bibinfo {year} {1999})},\ \Eprint
  {http://arxiv.org/abs/hep-lat/9804008} {arXiv:hep-lat/9804008 [hep-lat]}
  \BibitemShut {NoStop}%
\bibitem [{\citenamefont {Johnson}\ and\ \citenamefont
  {Teper}(2002)}]{Johnson:2000qz}%
  \BibitemOpen
  \bibfield  {author} {\bibinfo {author} {\bibfnamefont {R.~W.}\ \bibnamefont
  {Johnson}}\ and\ \bibinfo {author} {\bibfnamefont {M.~J.}\ \bibnamefont
  {Teper}},\ }\href {\doibase 10.1103/PhysRevD.66.036006} {\bibfield  {journal}
  {\bibinfo  {journal} {Phys.Rev.}\ }\textbf {\bibinfo {volume} {D66}},\
  \bibinfo {pages} {036006} (\bibinfo {year} {2002})},\ \Eprint
  {http://arxiv.org/abs/hep-ph/0012287} {arXiv:hep-ph/0012287 [hep-ph]}
  \BibitemShut {NoStop}%
\bibitem [{\citenamefont {Meyer}\ and\ \citenamefont
  {Teper}(2003)}]{Meyer:2003wx}%
  \BibitemOpen
  \bibfield  {author} {\bibinfo {author} {\bibfnamefont {H.~B.}\ \bibnamefont
  {Meyer}}\ and\ \bibinfo {author} {\bibfnamefont {M.~J.}\ \bibnamefont
  {Teper}},\ }\href {\doibase 10.1016/j.nuclphysb.2003.07.003} {\bibfield
  {journal} {\bibinfo  {journal} {Nucl.Phys.}\ }\textbf {\bibinfo {volume}
  {B668}},\ \bibinfo {pages} {111} (\bibinfo {year} {2003})},\ \Eprint
  {http://arxiv.org/abs/hep-lat/0306019} {arXiv:hep-lat/0306019 [hep-lat]}
  \BibitemShut {NoStop}%
\bibitem [{\citenamefont {Lucini}\ and\ \citenamefont
  {Teper}(2002)}]{Lucini:2002wg}%
  \BibitemOpen
  \bibfield  {author} {\bibinfo {author} {\bibfnamefont {B.}~\bibnamefont
  {Lucini}}\ and\ \bibinfo {author} {\bibfnamefont {M.}~\bibnamefont {Teper}},\
  }\href {\doibase 10.1103/PhysRevD.66.097502} {\bibfield  {journal} {\bibinfo
  {journal} {Phys.Rev.}\ }\textbf {\bibinfo {volume} {D66}},\ \bibinfo {pages}
  {097502} (\bibinfo {year} {2002})},\ \Eprint
  {http://arxiv.org/abs/hep-lat/0206027} {arXiv:hep-lat/0206027 [hep-lat]}
  \BibitemShut {NoStop}%
\bibitem [{\citenamefont {Bursa}\ and\ \citenamefont
  {Teper}(2006)}]{Bursa:2005tk}%
  \BibitemOpen
  \bibfield  {author} {\bibinfo {author} {\bibfnamefont {F.}~\bibnamefont
  {Bursa}}\ and\ \bibinfo {author} {\bibfnamefont {M.}~\bibnamefont {Teper}},\
  }\href {\doibase 10.1103/PhysRevD.74.125010} {\bibfield  {journal} {\bibinfo
  {journal} {Phys.Rev.}\ }\textbf {\bibinfo {volume} {D74}},\ \bibinfo {pages}
  {125010} (\bibinfo {year} {2006})},\ \Eprint
  {http://arxiv.org/abs/hep-th/0511081} {arXiv:hep-th/0511081 [hep-th]}
  \BibitemShut {NoStop}%
\bibitem [{\citenamefont {Bialas}\ \emph {et~al.}(2009)\citenamefont {Bialas},
  \citenamefont {Daniel}, \citenamefont {Morel},\ and\ \citenamefont
  {Petersson}}]{Bialas:2008rk}%
  \BibitemOpen
  \bibfield  {author} {\bibinfo {author} {\bibfnamefont {P.}~\bibnamefont
  {Bialas}}, \bibinfo {author} {\bibfnamefont {L.}~\bibnamefont {Daniel}},
  \bibinfo {author} {\bibfnamefont {A.}~\bibnamefont {Morel}}, \ and\ \bibinfo
  {author} {\bibfnamefont {B.}~\bibnamefont {Petersson}},\ }\href {\doibase
  10.1016/j.nuclphysb.2008.08.019} {\bibfield  {journal} {\bibinfo  {journal}
  {Nucl.Phys.}\ }\textbf {\bibinfo {volume} {B807}},\ \bibinfo {pages} {547}
  (\bibinfo {year} {2009})},\ \Eprint {http://arxiv.org/abs/0807.0855}
  {arXiv:0807.0855 [hep-lat]} \BibitemShut {NoStop}%
\bibitem [{\citenamefont {Liddle}\ and\ \citenamefont
  {Teper}(2008)}]{Liddle:2008kk}%
  \BibitemOpen
  \bibfield  {author} {\bibinfo {author} {\bibfnamefont {J.}~\bibnamefont
  {Liddle}}\ and\ \bibinfo {author} {\bibfnamefont {M.}~\bibnamefont {Teper}},\
  }\href@noop {} {\  (\bibinfo {year} {2008})},\ \Eprint
  {http://arxiv.org/abs/0803.2128} {arXiv:0803.2128 [hep-lat]} \BibitemShut
  {NoStop}%
\bibitem [{\citenamefont {Caselle}\ \emph
  {et~al.}(2011{\natexlab{b}})\citenamefont {Caselle}, \citenamefont
  {Castagnini}, \citenamefont {Feo}, \citenamefont {Gliozzi},\ and\
  \citenamefont {Panero}}]{Caselle:2011fy}%
  \BibitemOpen
  \bibfield  {author} {\bibinfo {author} {\bibfnamefont {M.}~\bibnamefont
  {Caselle}}, \bibinfo {author} {\bibfnamefont {L.}~\bibnamefont {Castagnini}},
  \bibinfo {author} {\bibfnamefont {A.}~\bibnamefont {Feo}}, \bibinfo {author}
  {\bibfnamefont {F.}~\bibnamefont {Gliozzi}}, \ and\ \bibinfo {author}
  {\bibfnamefont {M.}~\bibnamefont {Panero}},\ }\href {\doibase
  10.1007/JHEP06(2011)142} {\bibfield  {journal} {\bibinfo  {journal} {JHEP}\
  }\textbf {\bibinfo {volume} {1106}},\ \bibinfo {pages} {142} (\bibinfo {year}
  {2011}{\natexlab{b}})},\ \Eprint {http://arxiv.org/abs/1105.0359}
  {arXiv:1105.0359 [hep-lat]} \BibitemShut {NoStop}%
\bibitem [{\citenamefont {Bialas}\ \emph {et~al.}(2012)\citenamefont {Bialas},
  \citenamefont {Daniel}, \citenamefont {Morel},\ and\ \citenamefont
  {Petersson}}]{Bialas:2012qz}%
  \BibitemOpen
  \bibfield  {author} {\bibinfo {author} {\bibfnamefont {P.}~\bibnamefont
  {Bialas}}, \bibinfo {author} {\bibfnamefont {L.}~\bibnamefont {Daniel}},
  \bibinfo {author} {\bibfnamefont {A.}~\bibnamefont {Morel}}, \ and\ \bibinfo
  {author} {\bibfnamefont {B.}~\bibnamefont {Petersson}},\ }\href@noop {} {\
  (\bibinfo {year} {2012})},\ \Eprint {http://arxiv.org/abs/1211.3304}
  {arXiv:1211.3304 [hep-lat]} \BibitemShut {NoStop}%
\bibitem [{\citenamefont {Moshe}\ and\ \citenamefont
  {Zinn-Justin}(2003)}]{Moshe:2003xn}%
  \BibitemOpen
  \bibfield  {author} {\bibinfo {author} {\bibfnamefont {M.}~\bibnamefont
  {Moshe}}\ and\ \bibinfo {author} {\bibfnamefont {J.}~\bibnamefont
  {Zinn-Justin}},\ }\href {\doibase 10.1016/S0370-1573(03)00263-1} {\bibfield
  {journal} {\bibinfo  {journal} {Phys.Rept.}\ }\textbf {\bibinfo {volume}
  {385}},\ \bibinfo {pages} {69} (\bibinfo {year} {2003})},\ \Eprint
  {http://arxiv.org/abs/hep-th/0306133} {arXiv:hep-th/0306133 [hep-th]}
  \BibitemShut {NoStop}%
\bibitem [{\citenamefont {Boyd}\ \emph {et~al.}(1996)\citenamefont {Boyd},
  \citenamefont {Engels}, \citenamefont {Karsch}, \citenamefont {Laermann},
  \citenamefont {Legeland} \emph {et~al.}}]{Boyd:1996bx}%
  \BibitemOpen
  \bibfield  {author} {\bibinfo {author} {\bibfnamefont {G.}~\bibnamefont
  {Boyd}}, \bibinfo {author} {\bibfnamefont {J.}~\bibnamefont {Engels}},
  \bibinfo {author} {\bibfnamefont {F.}~\bibnamefont {Karsch}}, \bibinfo
  {author} {\bibfnamefont {E.}~\bibnamefont {Laermann}}, \bibinfo {author}
  {\bibfnamefont {C.}~\bibnamefont {Legeland}},  \emph {et~al.},\ }\href
  {\doibase 10.1016/0550-3213(96)00170-8} {\bibfield  {journal} {\bibinfo
  {journal} {Nucl.Phys.}\ }\textbf {\bibinfo {volume} {B469}},\ \bibinfo
  {pages} {419} (\bibinfo {year} {1996})},\ \Eprint
  {http://arxiv.org/abs/hep-lat/9602007} {arXiv:hep-lat/9602007 [hep-lat]}
  \BibitemShut {NoStop}%
\bibitem [{\citenamefont {Lucini}\ and\ \citenamefont
  {Teper}(2001)}]{Lucini:2001ej}%
  \BibitemOpen
  \bibfield  {author} {\bibinfo {author} {\bibfnamefont {B.}~\bibnamefont
  {Lucini}}\ and\ \bibinfo {author} {\bibfnamefont {M.}~\bibnamefont {Teper}},\
  }\href@noop {} {\bibfield  {journal} {\bibinfo  {journal} {JHEP}\ }\textbf
  {\bibinfo {volume} {0106}},\ \bibinfo {pages} {050} (\bibinfo {year}
  {2001})},\ \Eprint {http://arxiv.org/abs/hep-lat/0103027}
  {arXiv:hep-lat/0103027 [hep-lat]} \BibitemShut {NoStop}%
\bibitem [{\citenamefont {Teper}(2002)}]{Teper:2002kh}%
  \BibitemOpen
  \bibfield  {author} {\bibinfo {author} {\bibfnamefont {M.}~\bibnamefont
  {Teper}},\ }\href@noop {} {\ ,\ \bibinfo {pages} {108} (\bibinfo {year}
  {2002})},\ \Eprint {http://arxiv.org/abs/hep-ph/0203203}
  {arXiv:hep-ph/0203203 [hep-ph]} \BibitemShut {NoStop}%
\bibitem [{\citenamefont {Umeda}\ \emph {et~al.}(2009)\citenamefont {Umeda},
  \citenamefont {Ejiri}, \citenamefont {Aoki}, \citenamefont {Hatsuda},
  \citenamefont {Kanaya} \emph {et~al.}}]{Umeda:2008bd}%
  \BibitemOpen
  \bibfield  {author} {\bibinfo {author} {\bibfnamefont {T.}~\bibnamefont
  {Umeda}}, \bibinfo {author} {\bibfnamefont {S.}~\bibnamefont {Ejiri}},
  \bibinfo {author} {\bibfnamefont {S.}~\bibnamefont {Aoki}}, \bibinfo {author}
  {\bibfnamefont {T.}~\bibnamefont {Hatsuda}}, \bibinfo {author} {\bibfnamefont
  {K.}~\bibnamefont {Kanaya}},  \emph {et~al.},\ }\href {\doibase
  10.1103/PhysRevD.79.051501} {\bibfield  {journal} {\bibinfo  {journal}
  {Phys.Rev.}\ }\textbf {\bibinfo {volume} {D79}},\ \bibinfo {pages} {051501}
  (\bibinfo {year} {2009})},\ \Eprint {http://arxiv.org/abs/0809.2842}
  {arXiv:0809.2842 [hep-lat]} \BibitemShut {NoStop}%
\bibitem [{\citenamefont {Borsanyi}\ \emph {et~al.}(2011)\citenamefont
  {Borsanyi}, \citenamefont {Endrodi}, \citenamefont {Fodor}, \citenamefont
  {Katz},\ and\ \citenamefont {Szabo}}]{Borsanyi:2011zm}%
  \BibitemOpen
  \bibfield  {author} {\bibinfo {author} {\bibfnamefont {S.}~\bibnamefont
  {Borsanyi}}, \bibinfo {author} {\bibfnamefont {G.}~\bibnamefont {Endrodi}},
  \bibinfo {author} {\bibfnamefont {Z.}~\bibnamefont {Fodor}}, \bibinfo
  {author} {\bibfnamefont {S.}~\bibnamefont {Katz}}, \ and\ \bibinfo {author}
  {\bibfnamefont {K.}~\bibnamefont {Szabo}},\ }\href@noop {} {\  (\bibinfo
  {year} {2011})},\ \Eprint {http://arxiv.org/abs/1104.0013} {arXiv:1104.0013
  [hep-ph]} \BibitemShut {NoStop}%
\bibitem [{\citenamefont {Borsanyi}\ \emph {et~al.}(2012)\citenamefont
  {Borsanyi}, \citenamefont {Endrodi}, \citenamefont {Fodor}, \citenamefont
  {Katz}, \citenamefont {Krieg} \emph {et~al.}}]{Borsanyi:2012vn}%
  \BibitemOpen
  \bibfield  {author} {\bibinfo {author} {\bibfnamefont {S.}~\bibnamefont
  {Borsanyi}}, \bibinfo {author} {\bibfnamefont {G.}~\bibnamefont {Endrodi}},
  \bibinfo {author} {\bibfnamefont {Z.}~\bibnamefont {Fodor}}, \bibinfo
  {author} {\bibfnamefont {S.~D.}\ \bibnamefont {Katz}}, \bibinfo {author}
  {\bibfnamefont {S.}~\bibnamefont {Krieg}},  \emph {et~al.},\ }\href@noop {}
  {\  (\bibinfo {year} {2012})},\ \Eprint {http://arxiv.org/abs/1204.0995}
  {arXiv:1204.0995 [hep-lat]} \BibitemShut {NoStop}%
\bibitem [{\citenamefont {Lucini}\ and\ \citenamefont
  {Panero}(2012)}]{Lucini:2012gg}%
  \BibitemOpen
  \bibfield  {author} {\bibinfo {author} {\bibfnamefont {B.}~\bibnamefont
  {Lucini}}\ and\ \bibinfo {author} {\bibfnamefont {M.}~\bibnamefont
  {Panero}},\ }\href@noop {} {\  (\bibinfo {year} {2012})},\ \Eprint
  {http://arxiv.org/abs/1210.4997} {arXiv:1210.4997 [hep-th]} \BibitemShut
  {NoStop}%
\bibitem [{\citenamefont {Pisarski}(2000)}]{Pisarski:2000eq}%
  \BibitemOpen
  \bibfield  {author} {\bibinfo {author} {\bibfnamefont {R.~D.}\ \bibnamefont
  {Pisarski}},\ }\href {\doibase 10.1103/PhysRevD.62.111501} {\bibfield
  {journal} {\bibinfo  {journal} {Phys.Rev.}\ }\textbf {\bibinfo {volume}
  {D62}},\ \bibinfo {pages} {111501} (\bibinfo {year} {2000})},\ \Eprint
  {http://arxiv.org/abs/hep-ph/0006205} {arXiv:hep-ph/0006205 [hep-ph]}
  \BibitemShut {NoStop}%
\bibitem [{\citenamefont {Meisinger}\ \emph {et~al.}(2002)\citenamefont
  {Meisinger}, \citenamefont {Miller},\ and\ \citenamefont
  {Ogilvie}}]{Meisinger:2001cq}%
  \BibitemOpen
  \bibfield  {author} {\bibinfo {author} {\bibfnamefont {P.~N.}\ \bibnamefont
  {Meisinger}}, \bibinfo {author} {\bibfnamefont {T.~R.}\ \bibnamefont
  {Miller}}, \ and\ \bibinfo {author} {\bibfnamefont {M.~C.}\ \bibnamefont
  {Ogilvie}},\ }\href {\doibase 10.1103/PhysRevD.65.034009} {\bibfield
  {journal} {\bibinfo  {journal} {Phys.Rev.}\ }\textbf {\bibinfo {volume}
  {D65}},\ \bibinfo {pages} {034009} (\bibinfo {year} {2002})},\ \Eprint
  {http://arxiv.org/abs/hep-ph/0108009} {arXiv:hep-ph/0108009 [hep-ph]}
  \BibitemShut {NoStop}%
\bibitem [{\citenamefont {Meisinger}\ and\ \citenamefont
  {Ogilvie}(2002)}]{Meisinger:2001fi}%
  \BibitemOpen
  \bibfield  {author} {\bibinfo {author} {\bibfnamefont {P.~N.}\ \bibnamefont
  {Meisinger}}\ and\ \bibinfo {author} {\bibfnamefont {M.~C.}\ \bibnamefont
  {Ogilvie}},\ }\href {\doibase 10.1103/PhysRevD.65.056013} {\bibfield
  {journal} {\bibinfo  {journal} {Phys.Rev.}\ }\textbf {\bibinfo {volume}
  {D65}},\ \bibinfo {pages} {056013} (\bibinfo {year} {2002})},\ \Eprint
  {http://arxiv.org/abs/hep-ph/0108026} {arXiv:hep-ph/0108026 [hep-ph]}
  \BibitemShut {NoStop}%
\bibitem [{\citenamefont {Dumitru}\ \emph {et~al.}(2004)\citenamefont
  {Dumitru}, \citenamefont {Hatta}, \citenamefont {Lenaghan}, \citenamefont
  {Orginos},\ and\ \citenamefont {Pisarski}}]{Dumitru:2003hp}%
  \BibitemOpen
  \bibfield  {author} {\bibinfo {author} {\bibfnamefont {A.}~\bibnamefont
  {Dumitru}}, \bibinfo {author} {\bibfnamefont {Y.}~\bibnamefont {Hatta}},
  \bibinfo {author} {\bibfnamefont {J.}~\bibnamefont {Lenaghan}}, \bibinfo
  {author} {\bibfnamefont {K.}~\bibnamefont {Orginos}}, \ and\ \bibinfo
  {author} {\bibfnamefont {R.~D.}\ \bibnamefont {Pisarski}},\ }\href {\doibase
  10.1103/PhysRevD.70.034511} {\bibfield  {journal} {\bibinfo  {journal}
  {Phys.Rev.}\ }\textbf {\bibinfo {volume} {D70}},\ \bibinfo {pages} {034511}
  (\bibinfo {year} {2004})},\ \Eprint {http://arxiv.org/abs/hep-th/0311223}
  {arXiv:hep-th/0311223 [hep-th]} \BibitemShut {NoStop}%
\bibitem [{\citenamefont {Oswald}\ and\ \citenamefont
  {Pisarski}(2006)}]{Oswald:2005vr}%
  \BibitemOpen
  \bibfield  {author} {\bibinfo {author} {\bibfnamefont {M.}~\bibnamefont
  {Oswald}}\ and\ \bibinfo {author} {\bibfnamefont {R.~D.}\ \bibnamefont
  {Pisarski}},\ }\href {\doibase 10.1103/PhysRevD.74.045029} {\bibfield
  {journal} {\bibinfo  {journal} {Phys.Rev.}\ }\textbf {\bibinfo {volume}
  {D74}},\ \bibinfo {pages} {045029} (\bibinfo {year} {2006})},\ \Eprint
  {http://arxiv.org/abs/hep-ph/0512245} {arXiv:hep-ph/0512245 [hep-ph]}
  \BibitemShut {NoStop}%
\bibitem [{\citenamefont {Pisarski}(2006)}]{Pisarski:2006hz}%
  \BibitemOpen
  \bibfield  {author} {\bibinfo {author} {\bibfnamefont {R.~D.}\ \bibnamefont
  {Pisarski}},\ }\href {\doibase 10.1103/PhysRevD.74.121703} {\bibfield
  {journal} {\bibinfo  {journal} {Phys.Rev.}\ }\textbf {\bibinfo {volume}
  {D74}},\ \bibinfo {pages} {121703} (\bibinfo {year} {2006})},\ \Eprint
  {http://arxiv.org/abs/hep-ph/0608242} {arXiv:hep-ph/0608242 [hep-ph]}
  \BibitemShut {NoStop}%
\bibitem [{\citenamefont {Pisarski}(2007)}]{Pisarski:2006yk}%
  \BibitemOpen
  \bibfield  {author} {\bibinfo {author} {\bibfnamefont {R.~D.}\ \bibnamefont
  {Pisarski}},\ }\href {\doibase 10.1143/PTPS.168.276} {\bibfield  {journal}
  {\bibinfo  {journal} {Prog.Theor.Phys.Suppl.}\ }\textbf {\bibinfo {volume}
  {168}},\ \bibinfo {pages} {276} (\bibinfo {year} {2007})},\ \Eprint
  {http://arxiv.org/abs/hep-ph/0612191} {arXiv:hep-ph/0612191 [hep-ph]}
  \BibitemShut {NoStop}%
\bibitem [{\citenamefont {Hidaka}\ and\ \citenamefont
  {Pisarski}(2008)}]{Hidaka:2008dr}%
  \BibitemOpen
  \bibfield  {author} {\bibinfo {author} {\bibfnamefont {Y.}~\bibnamefont
  {Hidaka}}\ and\ \bibinfo {author} {\bibfnamefont {R.~D.}\ \bibnamefont
  {Pisarski}},\ }\href {\doibase 10.1103/PhysRevD.78.071501} {\bibfield
  {journal} {\bibinfo  {journal} {Phys.Rev.}\ }\textbf {\bibinfo {volume}
  {D78}},\ \bibinfo {pages} {071501} (\bibinfo {year} {2008})},\ \Eprint
  {http://arxiv.org/abs/0803.0453} {arXiv:0803.0453 [hep-ph]} \BibitemShut
  {NoStop}%
\bibitem [{\citenamefont {Hidaka}\ and\ \citenamefont
  {Pisarski}(2009{\natexlab{a}})}]{Hidaka:2009hs}%
  \BibitemOpen
  \bibfield  {author} {\bibinfo {author} {\bibfnamefont {Y.}~\bibnamefont
  {Hidaka}}\ and\ \bibinfo {author} {\bibfnamefont {R.~D.}\ \bibnamefont
  {Pisarski}},\ }\href {\doibase 10.1103/PhysRevD.80.036004} {\bibfield
  {journal} {\bibinfo  {journal} {Phys.Rev.}\ }\textbf {\bibinfo {volume}
  {D80}},\ \bibinfo {pages} {036004} (\bibinfo {year} {2009}{\natexlab{a}})},\
  \Eprint {http://arxiv.org/abs/0906.1751} {arXiv:0906.1751 [hep-ph]}
  \BibitemShut {NoStop}%
\bibitem [{\citenamefont {Hidaka}\ and\ \citenamefont
  {Pisarski}(2009{\natexlab{b}})}]{Hidaka:2009xh}%
  \BibitemOpen
  \bibfield  {author} {\bibinfo {author} {\bibfnamefont {Y.}~\bibnamefont
  {Hidaka}}\ and\ \bibinfo {author} {\bibfnamefont {R.~D.}\ \bibnamefont
  {Pisarski}},\ }\href {\doibase 10.1103/PhysRevD.80.074504} {\bibfield
  {journal} {\bibinfo  {journal} {Phys.Rev.}\ }\textbf {\bibinfo {volume}
  {D80}},\ \bibinfo {pages} {074504} (\bibinfo {year} {2009}{\natexlab{b}})},\
  \Eprint {http://arxiv.org/abs/0907.4609} {arXiv:0907.4609 [hep-ph]}
  \BibitemShut {NoStop}%
\bibitem [{\citenamefont {Hidaka}\ and\ \citenamefont
  {Pisarski}(2010)}]{Hidaka:2009ma}%
  \BibitemOpen
  \bibfield  {author} {\bibinfo {author} {\bibfnamefont {Y.}~\bibnamefont
  {Hidaka}}\ and\ \bibinfo {author} {\bibfnamefont {R.~D.}\ \bibnamefont
  {Pisarski}},\ }\href {\doibase 10.1103/PhysRevD.81.076002} {\bibfield
  {journal} {\bibinfo  {journal} {Phys.Rev.}\ }\textbf {\bibinfo {volume}
  {D81}},\ \bibinfo {pages} {076002} (\bibinfo {year} {2010})},\ \Eprint
  {http://arxiv.org/abs/0912.0940} {arXiv:0912.0940 [hep-ph]} \BibitemShut
  {NoStop}%
\bibitem [{\citenamefont {Pisarski}\ and\ \citenamefont
  {Skokov}(2012)}]{Pisarski:2012bj}%
  \BibitemOpen
  \bibfield  {author} {\bibinfo {author} {\bibfnamefont {R.~D.}\ \bibnamefont
  {Pisarski}}\ and\ \bibinfo {author} {\bibfnamefont {V.~V.}\ \bibnamefont
  {Skokov}},\ }\href {\doibase 10.1103/PhysRevD.86.081701} {\bibfield
  {journal} {\bibinfo  {journal} {Phys.Rev.}\ }\textbf {\bibinfo {volume}
  {D86}},\ \bibinfo {pages} {081701} (\bibinfo {year} {2012})},\ \Eprint
  {http://arxiv.org/abs/1206.1329} {arXiv:1206.1329 [hep-th]} \BibitemShut
  {NoStop}%
\bibitem [{\citenamefont {Kashiwa}\ \emph {et~al.}()\citenamefont {Kashiwa},
  \citenamefont {Pisarski},\ and\ \citenamefont {Skokov}}]{Kashiwa:2012wa}%
  \BibitemOpen
  \bibfield  {author} {\bibinfo {author} {\bibfnamefont {K.}~\bibnamefont
  {Kashiwa}}, \bibinfo {author} {\bibfnamefont {R.~D.}\ \bibnamefont
  {Pisarski}}, \ and\ \bibinfo {author} {\bibfnamefont {V.~V.}\ \bibnamefont
  {Skokov}},\ }\href@noop {} {\bibfield  {journal} {\bibinfo  {journal}
  {Phys.Rev.}\ }\textbf {\bibinfo {volume} {D85}},\ \bibinfo {pages}
  {114029}},\ \Eprint {http://arxiv.org/abs/1205.0545} {arXiv:1205.0545
  [hep-ph]} \BibitemShut {NoStop}%
\bibitem [{\citenamefont {Sasaki}\ and\ \citenamefont
  {Redlich}(2012)}]{Sasaki:2012bi}%
  \BibitemOpen
  \bibfield  {author} {\bibinfo {author} {\bibfnamefont {C.}~\bibnamefont
  {Sasaki}}\ and\ \bibinfo {author} {\bibfnamefont {K.}~\bibnamefont
  {Redlich}},\ }\href@noop {} {\  (\bibinfo {year} {2012})},\ \Eprint
  {http://arxiv.org/abs/1204.4330} {arXiv:1204.4330 [hep-ph]} \BibitemShut
  {NoStop}%
\bibitem [{\citenamefont {Ruggieri}\ \emph {et~al.}(2012)\citenamefont
  {Ruggieri}, \citenamefont {Alba}, \citenamefont {Castorina}, \citenamefont
  {Plumari}, \citenamefont {Ratti} \emph {et~al.}}]{Ruggieri:2012ny}%
  \BibitemOpen
  \bibfield  {author} {\bibinfo {author} {\bibfnamefont {M.}~\bibnamefont
  {Ruggieri}}, \bibinfo {author} {\bibfnamefont {P.}~\bibnamefont {Alba}},
  \bibinfo {author} {\bibfnamefont {P.}~\bibnamefont {Castorina}}, \bibinfo
  {author} {\bibfnamefont {S.}~\bibnamefont {Plumari}}, \bibinfo {author}
  {\bibfnamefont {C.}~\bibnamefont {Ratti}},  \emph {et~al.},\ }\href@noop {}
  {\  (\bibinfo {year} {2012})},\ \Eprint {http://arxiv.org/abs/1204.5995}
  {arXiv:1204.5995 [hep-ph]} \BibitemShut {NoStop}%
\bibitem [{\citenamefont {Diakonov}\ \emph {et~al.}(2012)\citenamefont
  {Diakonov}, \citenamefont {Gattringer},\ and\ \citenamefont
  {Schadler}}]{Diakonov:2012dx}%
  \BibitemOpen
  \bibfield  {author} {\bibinfo {author} {\bibfnamefont {D.}~\bibnamefont
  {Diakonov}}, \bibinfo {author} {\bibfnamefont {C.}~\bibnamefont
  {Gattringer}}, \ and\ \bibinfo {author} {\bibfnamefont {H.-P.}\ \bibnamefont
  {Schadler}},\ }\href@noop {} {\  (\bibinfo {year} {2012})},\ \Eprint
  {http://arxiv.org/abs/1205.4768} {arXiv:1205.4768 [hep-lat]} \BibitemShut
  {NoStop}%
\bibitem [{\citenamefont {Ogilvie}(2012)}]{Ogilvie:2012is}%
  \BibitemOpen
  \bibfield  {author} {\bibinfo {author} {\bibfnamefont {M.~C.}\ \bibnamefont
  {Ogilvie}},\ }\href {\doibase 10.1088/1751-8113/45/48/483001} {\bibfield
  {journal} {\bibinfo  {journal} {J.Phys.}\ }\textbf {\bibinfo {volume}
  {A45}},\ \bibinfo {pages} {483001} (\bibinfo {year} {2012})},\ \Eprint
  {http://arxiv.org/abs/1211.2843} {arXiv:1211.2843 [hep-th]} \BibitemShut
  {NoStop}%
\bibitem [{\citenamefont {Kashiwa}\ and\ \citenamefont
  {Pisarski}(2013)}]{Kashiwa:2013rm}%
  \BibitemOpen
  \bibfield  {author} {\bibinfo {author} {\bibfnamefont {K.}~\bibnamefont
  {Kashiwa}}\ and\ \bibinfo {author} {\bibfnamefont {R.~D.}\ \bibnamefont
  {Pisarski}},\ }\href@noop {} {\  (\bibinfo {year} {2013})},\ \Eprint
  {http://arxiv.org/abs/1301.5344} {arXiv:1301.5344 [hep-ph]} \BibitemShut
  {NoStop}%
\bibitem [{\citenamefont {Lin}\ \emph {et~al.}(2013)\citenamefont {Lin},
  \citenamefont {Pisarski},\ and\ \citenamefont {Skokov}}]{Lin:2013qu}%
  \BibitemOpen
  \bibfield  {author} {\bibinfo {author} {\bibfnamefont {S.}~\bibnamefont
  {Lin}}, \bibinfo {author} {\bibfnamefont {R.~D.}\ \bibnamefont {Pisarski}}, \
  and\ \bibinfo {author} {\bibfnamefont {V.~V.}\ \bibnamefont {Skokov}},\
  }\href@noop {} {\  (\bibinfo {year} {2013})},\ \Eprint
  {http://arxiv.org/abs/1301.7432} {arXiv:1301.7432 [hep-ph]} \BibitemShut
  {NoStop}%
\bibitem [{\citenamefont {Bialas}\ \emph {et~al.}(2000)\citenamefont {Bialas},
  \citenamefont {Morel}, \citenamefont {Petersson}, \citenamefont {Petrov},\
  and\ \citenamefont {Reisz}}]{Bialas:2000ev}%
  \BibitemOpen
  \bibfield  {author} {\bibinfo {author} {\bibfnamefont {P.}~\bibnamefont
  {Bialas}}, \bibinfo {author} {\bibfnamefont {A.}~\bibnamefont {Morel}},
  \bibinfo {author} {\bibfnamefont {B.}~\bibnamefont {Petersson}}, \bibinfo
  {author} {\bibfnamefont {K.}~\bibnamefont {Petrov}}, \ and\ \bibinfo {author}
  {\bibfnamefont {T.}~\bibnamefont {Reisz}},\ }\href {\doibase
  10.1016/S0550-3213(00)00263-7} {\bibfield  {journal} {\bibinfo  {journal}
  {Nucl.Phys.}\ }\textbf {\bibinfo {volume} {B581}},\ \bibinfo {pages} {477}
  (\bibinfo {year} {2000})},\ \Eprint {http://arxiv.org/abs/hep-lat/0003004}
  {arXiv:hep-lat/0003004 [hep-lat]} \BibitemShut {NoStop}%
\bibitem [{\citenamefont {Bialas}\ \emph {et~al.}(2005)\citenamefont {Bialas},
  \citenamefont {Morel},\ and\ \citenamefont {Petersson}}]{Bialas:2004gx}%
  \BibitemOpen
  \bibfield  {author} {\bibinfo {author} {\bibfnamefont {P.}~\bibnamefont
  {Bialas}}, \bibinfo {author} {\bibfnamefont {A.}~\bibnamefont {Morel}}, \
  and\ \bibinfo {author} {\bibfnamefont {B.}~\bibnamefont {Petersson}},\ }\href
  {\doibase 10.1016/j.nuclphysb.2004.10.045} {\bibfield  {journal} {\bibinfo
  {journal} {Nucl.Phys.}\ }\textbf {\bibinfo {volume} {B704}},\ \bibinfo
  {pages} {208} (\bibinfo {year} {2005})},\ \Eprint
  {http://arxiv.org/abs/hep-lat/0403027} {arXiv:hep-lat/0403027 [hep-lat]}
  \BibitemShut {NoStop}%
\bibitem [{\citenamefont {Bialas}\ \emph {et~al.}(2010)\citenamefont {Bialas},
  \citenamefont {Daniel}, \citenamefont {Morel},\ and\ \citenamefont
  {Petersson}}]{Bialas:2009pt}%
  \BibitemOpen
  \bibfield  {author} {\bibinfo {author} {\bibfnamefont {P.}~\bibnamefont
  {Bialas}}, \bibinfo {author} {\bibfnamefont {L.}~\bibnamefont {Daniel}},
  \bibinfo {author} {\bibfnamefont {A.}~\bibnamefont {Morel}}, \ and\ \bibinfo
  {author} {\bibfnamefont {B.}~\bibnamefont {Petersson}},\ }\href {\doibase
  10.1016/j.nuclphysb.2010.04.010} {\bibfield  {journal} {\bibinfo  {journal}
  {Nucl.Phys.}\ }\textbf {\bibinfo {volume} {B836}},\ \bibinfo {pages} {91}
  (\bibinfo {year} {2010})},\ \Eprint {http://arxiv.org/abs/0912.0206}
  {arXiv:0912.0206 [hep-lat]} \BibitemShut {NoStop}%
\bibitem [{\citenamefont {Bhattacharya}\ \emph {et~al.}(1992)\citenamefont
  {Bhattacharya}, \citenamefont {Gocksch}, \citenamefont {Korthals~Altes},\
  and\ \citenamefont {Pisarski}}]{Bhattacharya:1992qb}%
  \BibitemOpen
  \bibfield  {author} {\bibinfo {author} {\bibfnamefont {T.}~\bibnamefont
  {Bhattacharya}}, \bibinfo {author} {\bibfnamefont {A.}~\bibnamefont
  {Gocksch}}, \bibinfo {author} {\bibfnamefont {C.~P.}\ \bibnamefont
  {Korthals~Altes}}, \ and\ \bibinfo {author} {\bibfnamefont {R.~D.}\
  \bibnamefont {Pisarski}},\ }\href {\doibase 10.1016/0550-3213(92)90086-Q}
  {\bibfield  {journal} {\bibinfo  {journal} {Nucl.Phys.}\ }\textbf {\bibinfo
  {volume} {B383}},\ \bibinfo {pages} {497} (\bibinfo {year} {1992})},\ \Eprint
  {http://arxiv.org/abs/hep-ph/9205231} {arXiv:hep-ph/9205231 [hep-ph]}
  \BibitemShut {NoStop}%
\bibitem [{\citenamefont {Bhattacharya}\ \emph {et~al.}(1991)\citenamefont
  {Bhattacharya}, \citenamefont {Gocksch}, \citenamefont {Korthals~Altes},\
  and\ \citenamefont {Pisarski}}]{Bhattacharya:1990hk}%
  \BibitemOpen
  \bibfield  {author} {\bibinfo {author} {\bibfnamefont {T.}~\bibnamefont
  {Bhattacharya}}, \bibinfo {author} {\bibfnamefont {A.}~\bibnamefont
  {Gocksch}}, \bibinfo {author} {\bibfnamefont {C.~P.}\ \bibnamefont
  {Korthals~Altes}}, \ and\ \bibinfo {author} {\bibfnamefont {R.~D.}\
  \bibnamefont {Pisarski}},\ }\href {\doibase 10.1103/PhysRevLett.66.998}
  {\bibfield  {journal} {\bibinfo  {journal} {Phys.Rev.Lett.}\ }\textbf
  {\bibinfo {volume} {66}},\ \bibinfo {pages} {998} (\bibinfo {year}
  {1991})}\BibitemShut {NoStop}%
\bibitem [{\citenamefont {Korthals~Altes}\ \emph {et~al.}(1997)\citenamefont
  {Korthals~Altes}, \citenamefont {Michels}, \citenamefont {Stephanov},\ and\
  \citenamefont {Teper}}]{KorthalsAltes:1996xp}%
  \BibitemOpen
  \bibfield  {author} {\bibinfo {author} {\bibfnamefont {C.}~\bibnamefont
  {Korthals~Altes}}, \bibinfo {author} {\bibfnamefont {A.}~\bibnamefont
  {Michels}}, \bibinfo {author} {\bibfnamefont {M.~A.}\ \bibnamefont
  {Stephanov}}, \ and\ \bibinfo {author} {\bibfnamefont {M.}~\bibnamefont
  {Teper}},\ }\href {\doibase 10.1103/PhysRevD.55.1047} {\bibfield  {journal}
  {\bibinfo  {journal} {Phys.Rev.}\ }\textbf {\bibinfo {volume} {D55}},\
  \bibinfo {pages} {1047} (\bibinfo {year} {1997})},\ \Eprint
  {http://arxiv.org/abs/hep-lat/9606021} {arXiv:hep-lat/9606021 [hep-lat]}
  \BibitemShut {NoStop}%
\bibitem [{\citenamefont {Gross}\ \emph {et~al.}(1981)\citenamefont {Gross},
  \citenamefont {Pisarski},\ and\ \citenamefont {Yaffe}}]{Gross:1980br}%
  \BibitemOpen
  \bibfield  {author} {\bibinfo {author} {\bibfnamefont {D.~J.}\ \bibnamefont
  {Gross}}, \bibinfo {author} {\bibfnamefont {R.~D.}\ \bibnamefont {Pisarski}},
  \ and\ \bibinfo {author} {\bibfnamefont {L.~G.}\ \bibnamefont {Yaffe}},\
  }\href {\doibase 10.1103/RevModPhys.53.43} {\bibfield  {journal} {\bibinfo
  {journal} {Rev.Mod.Phys.}\ }\textbf {\bibinfo {volume} {53}},\ \bibinfo
  {pages} {43} (\bibinfo {year} {1981})}\BibitemShut {NoStop}%
\bibitem [{\citenamefont {D'Hoker}(1981)}]{D'Hoker:1980az}%
  \BibitemOpen
  \bibfield  {author} {\bibinfo {author} {\bibfnamefont {E.}~\bibnamefont
  {D'Hoker}},\ }\href {\doibase 10.1016/0550-3213(81)90425-9} {\bibfield
  {journal} {\bibinfo  {journal} {Nucl.Phys.}\ }\textbf {\bibinfo {volume}
  {B180}},\ \bibinfo {pages} {341} (\bibinfo {year} {1981})}\BibitemShut
  {NoStop}%
\bibitem [{\citenamefont {D'Hoker}(1982{\natexlab{a}})}]{D'Hoker:1981qp}%
  \BibitemOpen
  \bibfield  {author} {\bibinfo {author} {\bibfnamefont {E.}~\bibnamefont
  {D'Hoker}},\ }\href {\doibase 10.1016/0550-3213(82)90525-9} {\bibfield
  {journal} {\bibinfo  {journal} {Nucl.Phys.}\ }\textbf {\bibinfo {volume}
  {B200}},\ \bibinfo {pages} {517} (\bibinfo {year}
  {1982}{\natexlab{a}})}\BibitemShut {NoStop}%
\bibitem [{\citenamefont {D'Hoker}(1982{\natexlab{b}})}]{D'Hoker:1981us}%
  \BibitemOpen
  \bibfield  {author} {\bibinfo {author} {\bibfnamefont {E.}~\bibnamefont
  {D'Hoker}},\ }\href {\doibase 10.1016/0550-3213(82)90441-2} {\bibfield
  {journal} {\bibinfo  {journal} {Nucl.Phys.}\ }\textbf {\bibinfo {volume}
  {B201}},\ \bibinfo {pages} {401} (\bibinfo {year}
  {1982}{\natexlab{b}})}\BibitemShut {NoStop}%
\bibitem [{\citenamefont {Kajantie}\ \emph {et~al.}(1997)\citenamefont
  {Kajantie}, \citenamefont {Laine}, \citenamefont {Rummukainen},\ and\
  \citenamefont {Shaposhnikov}}]{Kajantie:1997tt}%
  \BibitemOpen
  \bibfield  {author} {\bibinfo {author} {\bibfnamefont {K.}~\bibnamefont
  {Kajantie}}, \bibinfo {author} {\bibfnamefont {M.}~\bibnamefont {Laine}},
  \bibinfo {author} {\bibfnamefont {K.}~\bibnamefont {Rummukainen}}, \ and\
  \bibinfo {author} {\bibfnamefont {M.~E.}\ \bibnamefont {Shaposhnikov}},\
  }\href {\doibase 10.1016/S0550-3213(97)00425-2} {\bibfield  {journal}
  {\bibinfo  {journal} {Nucl.Phys.}\ }\textbf {\bibinfo {volume} {B503}},\
  \bibinfo {pages} {357} (\bibinfo {year} {1997})},\ \Eprint
  {http://arxiv.org/abs/hep-ph/9704416} {arXiv:hep-ph/9704416 [hep-ph]}
  \BibitemShut {NoStop}%
\bibitem [{\citenamefont {Rischke}(2004)}]{Rischke:2003mt}%
  \BibitemOpen
  \bibfield  {author} {\bibinfo {author} {\bibfnamefont {D.~H.}\ \bibnamefont
  {Rischke}},\ }\href {\doibase 10.1016/j.ppnp.2003.09.002} {\bibfield
  {journal} {\bibinfo  {journal} {Prog.Part.Nucl.Phys.}\ }\textbf {\bibinfo
  {volume} {52}},\ \bibinfo {pages} {197} (\bibinfo {year} {2004})},\ \Eprint
  {http://arxiv.org/abs/nucl-th/0305030} {arXiv:nucl-th/0305030 [nucl-th]}
  \BibitemShut {NoStop}%
\bibitem [{\citenamefont {Belyaev}\ and\ \citenamefont
  {Eletsky}(1990)}]{Belyaev:1989bj}%
  \BibitemOpen
  \bibfield  {author} {\bibinfo {author} {\bibfnamefont {V.}~\bibnamefont
  {Belyaev}}\ and\ \bibinfo {author} {\bibfnamefont {V.}~\bibnamefont
  {Eletsky}},\ }\href {\doibase 10.1007/BF01549664} {\bibfield  {journal}
  {\bibinfo  {journal} {Z.Phys.}\ }\textbf {\bibinfo {volume} {C45}},\ \bibinfo
  {pages} {355} (\bibinfo {year} {1990})}\BibitemShut {NoStop}%
\bibitem [{\citenamefont {Enqvist}\ and\ \citenamefont
  {Kajantie}(1990)}]{Enqvist:1990ae}%
  \BibitemOpen
  \bibfield  {author} {\bibinfo {author} {\bibfnamefont {K.}~\bibnamefont
  {Enqvist}}\ and\ \bibinfo {author} {\bibfnamefont {K.}~\bibnamefont
  {Kajantie}},\ }\href {\doibase 10.1007/BF01552353} {\bibfield  {journal}
  {\bibinfo  {journal} {Z.Phys.}\ }\textbf {\bibinfo {volume} {C47}},\ \bibinfo
  {pages} {291} (\bibinfo {year} {1990})}\BibitemShut {NoStop}%
\bibitem [{\citenamefont {Unsal}\ and\ \citenamefont
  {Yaffe}(2008)}]{Unsal:2008ch}%
  \BibitemOpen
  \bibfield  {author} {\bibinfo {author} {\bibfnamefont {M.}~\bibnamefont
  {Unsal}}\ and\ \bibinfo {author} {\bibfnamefont {L.~G.}\ \bibnamefont
  {Yaffe}},\ }\href {\doibase 10.1103/PhysRevD.78.065035} {\bibfield  {journal}
  {\bibinfo  {journal} {Phys.Rev.}\ }\textbf {\bibinfo {volume} {D78}},\
  \bibinfo {pages} {065035} (\bibinfo {year} {2008})},\ \Eprint
  {http://arxiv.org/abs/0803.0344} {arXiv:0803.0344 [hep-th]} \BibitemShut
  {NoStop}%
\bibitem [{\citenamefont {von Smekal}\ \emph {et~al.}(2011)\citenamefont {von
  Smekal}, \citenamefont {Edwards},\ and\ \citenamefont
  {Strodthoff}}]{vonSmekal:2010du}%
  \BibitemOpen
  \bibfield  {author} {\bibinfo {author} {\bibfnamefont {L.}~\bibnamefont {von
  Smekal}}, \bibinfo {author} {\bibfnamefont {S.~R.}\ \bibnamefont {Edwards}},
  \ and\ \bibinfo {author} {\bibfnamefont {N.}~\bibnamefont {Strodthoff}},\
  }\href {\doibase 10.1063/1.3574980} {\bibfield  {journal} {\bibinfo
  {journal} {AIP Conf.Proc.}\ }\textbf {\bibinfo {volume} {1343}},\ \bibinfo
  {pages} {212} (\bibinfo {year} {2011})},\ \Eprint
  {http://arxiv.org/abs/1012.1712} {arXiv:1012.1712 [hep-ph]} \BibitemShut
  {NoStop}%
\bibitem [{\citenamefont {Strodthoff}\ \emph {et~al.}(2010)\citenamefont
  {Strodthoff}, \citenamefont {Edwards},\ and\ \citenamefont {von
  Smekal}}]{Strodthoff:2010dz}%
  \BibitemOpen
  \bibfield  {author} {\bibinfo {author} {\bibfnamefont {N.}~\bibnamefont
  {Strodthoff}}, \bibinfo {author} {\bibfnamefont {S.~R.}\ \bibnamefont
  {Edwards}}, \ and\ \bibinfo {author} {\bibfnamefont {L.}~\bibnamefont {von
  Smekal}},\ }\href@noop {} {\bibfield  {journal} {\bibinfo  {journal} {PoS}\
  }\textbf {\bibinfo {volume} {LATTICE2010}},\ \bibinfo {pages} {288} (\bibinfo
  {year} {2010})},\ \Eprint {http://arxiv.org/abs/1012.0723} {arXiv:1012.0723
  [hep-lat]} \BibitemShut {NoStop}%
\bibitem [{\citenamefont {Hagedorn}(1965)}]{Hagedorn:1965st}%
  \BibitemOpen
  \bibfield  {author} {\bibinfo {author} {\bibfnamefont {R.}~\bibnamefont
  {Hagedorn}},\ }\href@noop {} {\bibfield  {journal} {\bibinfo  {journal}
  {Nuovo Cim.Suppl.}\ }\textbf {\bibinfo {volume} {3}},\ \bibinfo {pages} {147}
  (\bibinfo {year} {1965})}\BibitemShut {NoStop}%
\bibitem [{\citenamefont {Cohen}\ and\ \citenamefont
  {Krejcirik}(2011)}]{Cohen:2011yx}%
  \BibitemOpen
  \bibfield  {author} {\bibinfo {author} {\bibfnamefont {T.~D.}\ \bibnamefont
  {Cohen}}\ and\ \bibinfo {author} {\bibfnamefont {V.}~\bibnamefont
  {Krejcirik}},\ }\href {\doibase 10.1007/JHEP08(2011)138} {\bibfield
  {journal} {\bibinfo  {journal} {JHEP}\ }\textbf {\bibinfo {volume} {1108}},\
  \bibinfo {pages} {138} (\bibinfo {year} {2011})},\ \Eprint
  {http://arxiv.org/abs/1104.4783} {arXiv:1104.4783 [hep-th]} \BibitemShut
  {NoStop}%
\end{thebibliography}%

\end{document}